\documentclass[12pt,fullpage,oneside,doublespace]{uc-thesis}

\usepackage{fullpage}
\usepackage{geometry}                
\usepackage{graphicx}
\usepackage{amssymb}
\usepackage{booktabs}
\usepackage{epstopdf}
\usepackage{setspace}
\usepackage{appendix}
\usepackage{listings}
\usepackage{titlesec}
\usepackage{rotating}
\usepackage{amsmath,amsthm}
\usepackage{amsmath} 

\usepackage{natbib}
\bibpunct{(}{)}{;}{a}{,}{,}
\titleformat{\subsection}{\normalsize\bfseries}{\thesubsection}{1em}{}
\titleformat{\section}{\large\bfseries}{\thesection}{1em}{}
\titleformat{\chapter}[display]{\huge\bfseries}{\chaptertitlename\ \thechapter}{0pt}{\Large}

\titlespacing*{\chapter}{0pt}{-21pt}{20pt}           

\begin{document}

\frontmatter

\setcounter{page}{1}

\title{FIRST DARK MATTER LIMITS FROM THE COUPP 4 kg BUBBLE CHAMBER \\ AT A DEEP UNDERGROUND SITE}
\author{Drew Anthony Fustin}
\month{March}
\year{2012}

\titlepage

\copyrightpage

\begin{dedication}
For my mother, the scientist

and

for Alexis --- my wife, my teammate\end{dedication}

\begin{abstract}

\noindent The COUPP 4 kg bubble chamber employs 4.0 kg of CF$_3$I as a WIMP scattering target for use as a dark matter direct detection search. This thesis reports the first experimental results from operating this bubble chamber at the deep underground site (6000 m.w.e.) of SNOLAB, near Sudbury, Ontario. Twenty dark matter candidate events were observed during an effective exposure of 553.0 kg-days, when operating the bubble chamber at three different bubble nucleation thresholds. These data are consistent with a neutron background internal to the detector. Characterization of this neutron background has led to the recommendation to replace two detector components to maximize dark matter signal sensitivity in a future run with this bubble chamber. A measurement of the gamma-ray flux has confirmed that this detector should not be sensitive to a gamma-induced background for more than three orders of magnitude below current sensitivity. The dark matter search data presented here set a new world-leading limit on the spin-dependent WIMP-proton scattering cross section and demonstrate significant sensitivity to spin-independent WIMP-nucleon scattering.

\end{abstract}

\begin{acknowledgments}

\noindent I'm not one to have an epigraph --- something like, ``To write a dissertation will be an awfully big adventure," (Sir James Barrie) least of which because that's not the actual quote (although, the replacement of the word ``die" with ``write a dissertation" \emph{does} seem rather fitting). But if I were to have one, it'd probably be something my advisor, Juan Collar, said to me after my defense: ``They say you have to be rather intelligent to get into a Ph.D. program, but quite the fool to stay in one." Me, I am doubly a fool because I escaped once, taking a year and a half leave of absence, before returning to complete my work, like Lot's wife turning back to Sodom and Gomorrah. Saltiness aside, finally finishing this work feels exceptionally gratifying and I am more than blessed for the opportunity to study at such a great institution. Although, graduate school has been a consistent struggle --- something that has always been more difficult than I had imagined, and something that has required quite the support structure for me to maintain my sanity. So, I would be remiss not to acknowledge everyone who's helped me through this. 

First and foremost, I must acknowledge the great women in my life --- my mother, Debra Minton, who got me into this whole science mess to begin with by providing the example of and encouragement towards the life of a scientist, and for always being the epitome of a fantastic parent; my grandmother, Nelda Balding, for being the perfect example of a loving caretaker and devoted spouse, while still maintaining a fiery spirit; and my wife, Alexis Wheeler (now Fustin, yay!), for being a constant helper, comforter, and security (beyond just being my sugar momma), my loving teammate who's not afraid to spur me on towards love and good deeds and help me become a better person. I love you all immensely. 

My brother, Matt Fustin, has always been there for me, from driving me to school to being my dungeon master (a phrase that helps paint myself into a rather nerdy corner). My step-father, Henry Minton, has loved my mother with such devotion that it's been quite the encouragement to me, despite his penchant for making me mow the lawn so frequently. And my Papaw, Neil Balding, whose striving to always learn more (be it in poring over maps or working the daily Jubble in the paper) and to solve any problem (the catwalk for the raccoons comes to mind) has got to be where I get it from. You were a good man in every sense of the word. 

My Ph.D. advisor, Juan Collar, is absolutely the best advisor I could ever have hoped for --- I appreciate everything you've done for me, from taking me in when I was a lowly options trader, to honoring your promise to get me out of here in 3 years (just missed the target, which was not your fault). I hope that one day you can enjoy running a nice blueberry farm. I'll miss all of your insightful Spanish wisdom and phrases. Thanks also to my fellow graduate students in the Collar group (Phil Barbeau, Matt Szydagis, Alan Robinson, and Nicole Fields) for making the office interesting and helping to solve a multitude of problems. And to my fellow 2004 Ph.D. program matriculants, particularly Tom Spears, Ali Brizius, Steph Wissel, Nathan Keim, Sophia Domokos, and Michael Mortonson: I never would have made it through this without the constant realization that grad school is hard for everyone, even people as smart as all of you. Finally, thanks to my professors at Drake University, particularly Athan Petridis, Larry Staunton, Bob Lutz, and Charles Nelson, for always encouraging me in my studies and teaching me so well. The physics program at Drake is fantastic, and I'd go to school there again in a heartbeat. 

The friendships I have developed in my years here in Chicago have been amazing. Phil Lynch, who knows everything I'm going through because he's already experienced it before (we are, after all, the same person, albeit temporally separated), has been an immensely helpful oracle, mentor, and friend. Thanks to the Bonner family (Ken and Anca) for taking me in, over and over, and being the best friends I've ever had. It's both a blessing and a curse to have people who are not afraid to tell it like it is when it pertains to your life. And also, thanks to all my friends from Drake who've kept in touch over the years, particularly Cody Johnson and Ryan Morgan, who've helped me with life issues, faith issues, and relationship issues --- most of which are hopefully a thing of the past now! 

Finally, thanks to the New England Patriots for giving me something to cheer about, even if they lost the Super Bowl days before my defense (I'm still not quite over this...), Steve Reich for writing Music for 18 Musicians (a 1.1-hour composition I listened to 100$\pm$10 times over the 700$\pm$50 hours of writing this dissertation, for an impressive 16$\pm$2\% listening ratio), and the city of Chicago for being my kind of town. 

\vspace{10 mm}

\center{Praise God from whom all blessing flow. \\ Thank You, thank You, thank You.}

\end{acknowledgments}

\tableofcontents

\newpage
\phantomsection \label{listoffig}
\addcontentsline{toc}{chapter}{List of Figures}
\listoffigures

\listoftables
\addcontentsline{toc}{chapter}{List of Tables}

\mainmatter

\singlespacing
\chapter{An Introduction to WIMP Dark Matter}
\label{ch:darkmatter}
\doublespacing

The principal goal of science is to understand the ingredients of the world around us. In the field of cosmology, this question primarily translates into determining the value and composition of the matter and energy densities of the universe. While a superficial observation reveals that the universe seems to be comprised almost entirely of stars, gas, and dust, it has become increasing apparent that visible components of the universe make up only a small fraction of its total composition. The missing component is therefore thought of as ``dark" --- consisting of invisible matter that does not interact electromagnetically. Precise theories of Big Bang nucleosynthesis go beyond this by requiring dark matter to be non-baryonic in nature \citep{schramm-98}, and our understanding of structure formation requires that it be non-relativistic as well \citep{blumenthal-84}. Dark matter is entirely unexplainable by the Standard Model of particle physics alone. 

The Chicagoland Observatory for Underground Particle Physics (COUPP) experiment is one of many searches seeking to directly detect dark matter in the form of Weakly Interacting Massive Particles (WIMPs), a candidate for dark matter predicted by many supersymmetric (SUSY) extensions to the Standard Model of particle interactions and strongly favored as the dark matter candidate. COUPP utilizes superheated liquids in bubble chambers as a target, a technique that allows for background rejection superior to many other WIMP direct detection experiments. This thesis summarizes the recent results of the first WIMP direct detection search performed by the COUPP 4 kg bubble chamber at a deep underground site. For this experiment, a superheated CF$_3$I bubble chamber was deployed at SNOLAB, a laboratory located 6000 meters water equivalent (m.w.e.) underground in the Vale Creighton Mine \#9 near Sudbury, Ontario.

In this chapter, a case is built for why dark matter is non-baryonic and non-relativistic, and specifically why WIMPs are favored. In Chapter \ref{ch:directdetection}, various methods for direct detection of WIMPs are described. Chapter \ref{ch:bubblechambers} highlights the theory of the bubble chamber and how it can be employed in a WIMP direct detection experiment, and Chapter \ref{ch:coupp} describes the COUPP 4 kg bubble chamber in particular, with dark matter search data sets defined in Chapter \ref{ch:datasets}. Chapter \ref{ch:efficiency} describes the neutron calibrations performed to determine the bubble nucleation efficiency on CF$_3$I. Chapter \ref{ch:backgrounds} contains a comprehensive study of the neutron backgrounds to the dark matter search data as well as how to discriminate or eliminate each neutron source, and Chapter \ref{ch:gammas} describes the gamma-ray background to the experiment, complete with a determination of the gamma rejection factor which limits this background to negligible importance. Finally, the WIMP-proton spin-dependent and WIMP-nucleon spin-independent dark matter limits established by this experiment are found in Chapter \ref{ch:limits}.

\section{The Mass Density of the Universe}
\label{sec:darkmatter_density}

Of primary concern to the field of cosmology is the value and characteristics of the energy density of the universe. Fritz Zwicky was the first to identify that a large part of the matter component was dark in nature when in 1933 he compared the mass of the Coma galaxy cluster implied by its luminosity with that required by the virial theorem to fit the velocity profile of objects on its edge \citep{zwicky-33}. Since this time, overwhelming evidence has been collected in support of the idea that dark matter is considerably more abundant than luminous matter, and that a third component, driving the accelerating expansion of the universe (known as ``dark energy"), is even more abundant.

An expanding universe with uniform energy density and curvature containing non-relativistic dark matter and a cosmological constant $\Lambda$ describing dark energy (known as the $\Lambda$CDM model) is described by the Friedmann equations

\begin{equation}
\label{eq:friedmann_1}
\left(\frac{H}{H_0}\right)^2 \equiv \frac{1}{H_0^2} \left(\frac{\dot{a}}{a}\right)^2 = \Omega_m \left(\frac{a_0}{a}\right)^3 + \Omega_k \left(\frac{a_0}{a}\right)^2 + \Omega_\Lambda
\end{equation}

\begin{equation}
\label{eq:friedmann_2}
\frac{1}{H_0^2} \frac{\ddot{a}}{a} = -\frac{1}{2} \Omega_m \left(\frac{a_0}{a}\right)^3 + \Omega_\Lambda,
\end{equation}

\noindent where $H$ and $a$ are the Hubble constant and scale factor in the metric for the universe at time $t$, $H_0 = 70.4^{+1.3}_{-1.4}$ km/s/Mpc \citep{jarosik-11} and $a_0$ are their present-day values, $\Omega_m$ and $\Omega_\Lambda$ are present-day energy densities of matter and cosmological constant, and $\Omega_k$ describes the spatial curvature of the universe. The $\Omega$s are defined such that 

\begin{equation}
\label{eq:Omega_m}
\Omega_m \equiv \Omega_b + \Omega_\chi \equiv \frac{8 \pi G}{3 H_0^2} \rho_0
\end{equation}

\begin{equation}
\label{eq:Omega_k}
\Omega_k \equiv -\frac{1}{H_0^2 a_0^2 R^2}
\end{equation}

\begin{equation}
\label{eq:Omega_Lambda}
\Omega_\Lambda \equiv \frac{\Lambda}{3 H_0^2}
\end{equation}

\noindent with curvature radius $R$, present-day matter density $\rho_0$, with matter density $\Omega_m$ composed of a baryonic matter component $\Omega_b$ and a dark matter component $\Omega_\chi$. The final term in Equation \ref{eq:Omega_m} is often written as $\rho_0 / \rho_c$, where $\rho_c \equiv 3 H_0^2 / 8 \pi G$ is defined as the critical density. By definition, these are confined to the relation $\Omega_m + \Omega_k + \Omega_\Lambda = 1$ (Equation \ref{eq:friedmann_1} with $t=t_0$). From the curvature $R$ in Equation \ref{eq:Omega_k}, a universe is flat when $\Omega_k = 0$, open when $\Omega_k > 0$, and closed when $\Omega_k < 0$. The total density $\Omega_\mathrm{tot} \equiv \Omega_m + \Omega_\Lambda$ of the universe today is then related to the curvature of the universe by the relation 

\begin{equation}
\label{eq:Omega_tot_vs_Omega_k}
\Omega_\mathrm{tot} = 1-\Omega_k.
\end{equation}

\subsection{Cosmic Microwave Background}
\label{sec:darkmatter_cmb}

\begin{figure} [t!]
\centering
\includegraphics[scale=0.5]{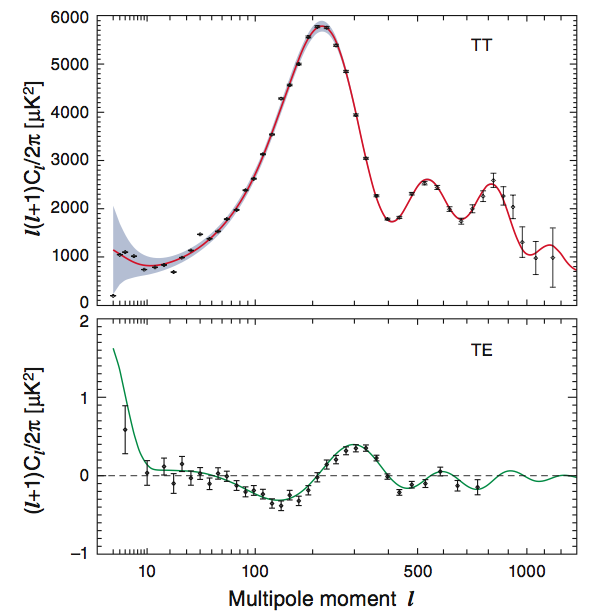}
\caption[CMB power spectrum]{Temperature (TT) and temperature-polarization (TE) CMB power spectra from the seven-year WMAP data. Best-fits to a minimal six-parameter flat $\Lambda$CDM model are shown. Figure from \cite{jarosik-11} reproduced by permission of the AAS.}
\label{fig:WMAP_power_spectrum}
\end{figure}

The most recent measurements of the anisotropies in the Cosmic Microwave Background (CMB) made by the Wilson Microwave Anisotropy Probe (WMAP) give a precise picture of the values of a minimal six-parameter flat $\Lambda$CDM model \citep{larson-11}. By measuring the temperature (TT) and temperature-polarization (TE) angular power spectra of the CMB (shown in Figure \ref{fig:WMAP_power_spectrum}),  and considering distance measurements from baryon acoustic oscillations in the distributions of galaxies made by the Sloan Digital Sky Survey \citep{percival-10} and recent Hubble constant measurements made by the Hubble Space Telescope \citep{reiss-09}, best-fits to a six-parameter $\Lambda$CDM model give baryon density of $\Omega_b = 0.0456\pm0.0016$, dark matter density of $\Omega_\chi = 0.227\pm0.014$, and dark energy density of $\Omega_\Lambda = 0.728^{+0.015}_{-0.016}$ \citep{jarosik-11}. The total density of the universe is then $\Omega_\mathrm{tot} = 1.0023^{+0.0056}_{-0.0054}$, suggesting that the universe is very nearly (if not exactly) flat. 

\subsection{Big Bang Nucleosynthesis}
\label{sec:darkmatter_bbn}

\begin{figure} [t!]
\centering
\includegraphics[scale=0.70]{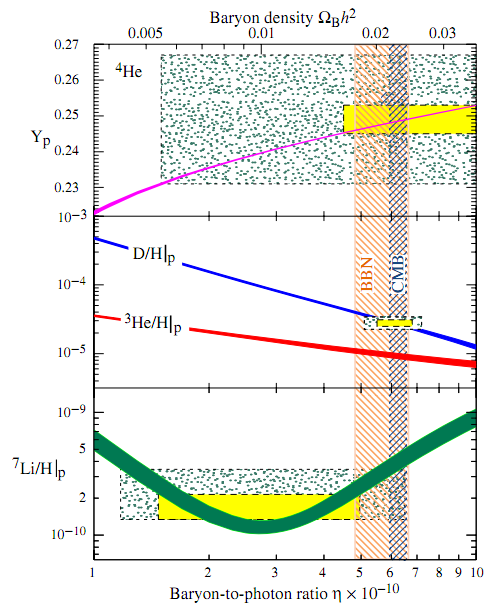}
\caption[Primordial abundances of light nuclei]{Primordial abundances of light nuclei as given by Big Bang nucleosynthesis (and the CMB power spectrum). Shaded regions depict different measurements taken, selecting out a preferred baryon density. Figure from \cite{amsler-08a} reproduced with permission.}
\label{fig:light_element_abundance}
\end{figure}

While the CMB power spectrum provides a value for the amount of baryonic matter density, a second, independent method confirms the results from the first. Big Bang nucleosynthesis (BBN) is a process used for calculating the primordial light-elemental abundance ($^1$H, $^2$H, $^3$He, $^4$He, and $^7$Li) at early times following a hot Big Bang. Early on ($t < 3$ min), the universe is still hot and in thermal equilibrium. But, as the universe expands and cools below nuclear binding energies ($t > 3$ min), the formation of free protons and neutrons into light nuclei occurs. This association continues as long as free neutrons are available to form nuclei --- the ratio of neutrons to protons being set by the temperature (and time) at which neutrinos decouple, halting the $p + e^- \leftrightarrow n + \nu_e$ reaction. Prior to stellar formation, there are no longer any processes that can alter the relative abundances of these light nuclei. By measuring the abundances of light elements in systems where little stellar processes have occurred (Figure \ref{fig:light_element_abundance}), one can accurately determine the baryonic density of the universe. The value of $\Omega_b$ found from BBN is consistent with that determined from measurement of the CMB power spectrum, confirming $\Omega_b = 0.0456$. 

\section{Evidence for Dark Matter}
\label{sec:darkmatter_evidence}

From these measurements, it is clear that a very small portion (4.6\%) of our universe is composed of ordinary matter. While the largest contribution (72.8\%) is from dark energy, it is not the purpose of this thesis to describe this component in any more detail. A sizable amount of the missing mass of the universe (22.7\%) is composed of dark matter, a form of matter for which no direct detection has ever been successfully observed. There has, however, been observed evidence for dark matter seen on large scales. Many of those signatures are described below.

\subsection{Galactic Rotation Curves}
\label{sec:darkmatter_rotation}

\begin{figure} [t!]
\centering
\includegraphics[scale=0.48]{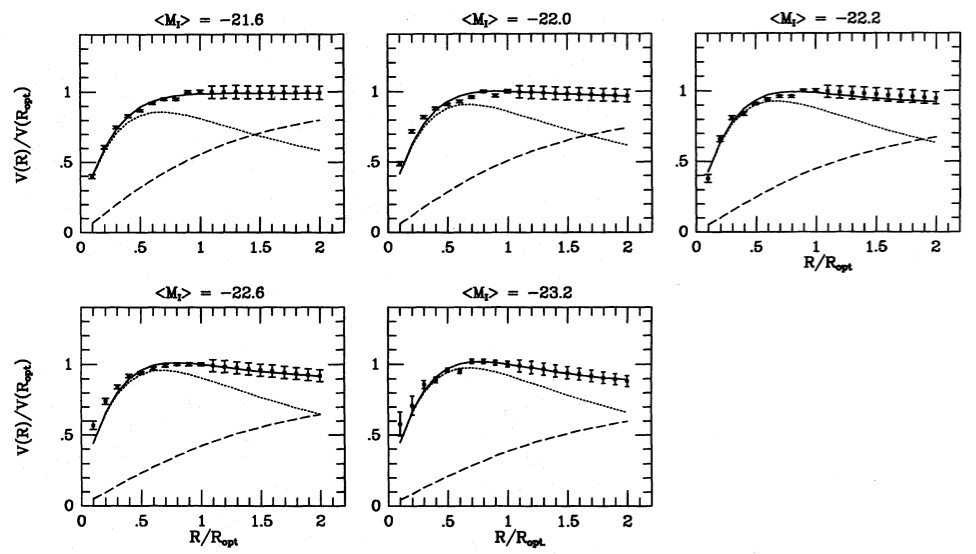}
\caption[Galactic rotation curves]{Examples of two-component (dotted: disk, dashed: halo) fits to galactic rotation curves. Each plot is a collection of measurements for different galactic luminosities, where bins contain cumulative information from several galaxies. Figure from \cite{persic-96} reproduced with permission.}
\label{fig:galactic_rotation_curves}
\end{figure}

Perhaps the most widely used evidence for dark matter on galactic scales is found in measuring the rotation curves of spiral galaxies, as shown in Figure \ref{fig:galactic_rotation_curves}.  The observable structure of a spiral galaxy is dominated by its luminous disk of stars, together with a center bulge and a cloud of atomic hydrogen. By measuring the velocities of stars and gas clouds using the Doppler shift of characteristic spectral lines, a galactic rotation curve can be plotted comparing the velocity of an object with its distance to the center of the galaxy. Assuming only the visible components of the galaxy (\emph{i.e.} the disk) contribute to the mass, this rotation curve would be expected to drop as the distance from the center increases (shown as the dotted lines in Figure \ref{fig:galactic_rotation_curves}). However, the defining characteristic of these curves is that they are nearly flat at distances beyond the optical radius of the galaxy. In order to explain this discrepancy, a spherical symmetric halo of invisible matter has to be present (shown as the dashed lines in Figure \ref{fig:galactic_rotation_curves}), a feature which clearly points to the existence of a galactic dark matter halo.

\subsection{Structure Formation}
\label{sec:darkmatter_structure}

\begin{figure} [t!]
\centering
\includegraphics[scale=0.64]{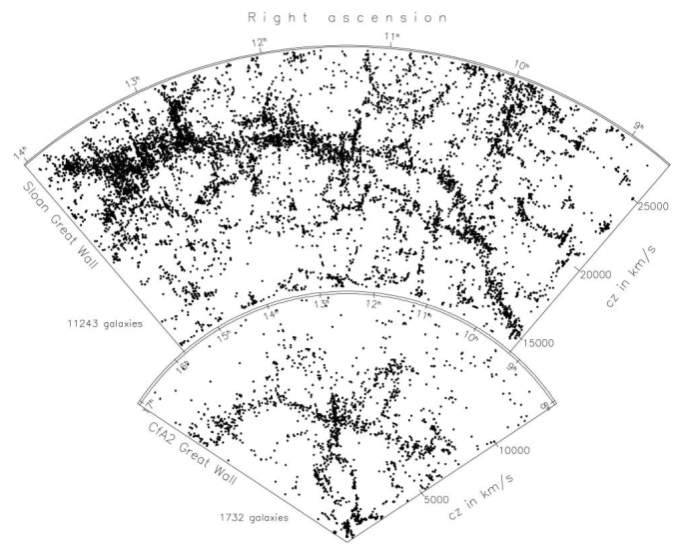}
\caption[Large scale structure in the Sloan Great Wall]{An example of large scale structure seen in the Sloan Great Wall, compared to another large scale structure, the CfA2 Great Wall. Figure from \cite{gott-05} reproduced by permission of the AAS.}
\label{fig:large_scale_structure}
\end{figure}

The universe is quite a structured place, from the formation of galaxies to the clustering of these galaxies to the formation of super clusters and even larger scale structures  (Figure \ref{fig:large_scale_structure}).  The processes required for each depend on the scale of the structure and point not only to the necessity of dark matter, but also to its being non-relativistic (\emph{i.e.} cold). Large scale structures require primordial density fluctuations at times earlier than would be possible with baryonic matter, since structure formation due to baryons cannot occur until after electron-proton combination occurs. Further, in a universe dominated by hot dark matter (presumably in the form relativistic neutrinos), density fluctuations are smoothed by the free streaming of relativistic particles, resulting in a delay to structure formation and size of galaxies not observed in the physical universe. Only in a universe dominated by cold dark matter can structure formation on all observed scales be possible \citep{blumenthal-84}. 

\subsection{Gravitational Lensing}
\label{sec:darkmatter_lensing}

\begin{figure} [t!]
\centering
\includegraphics[scale=0.6]{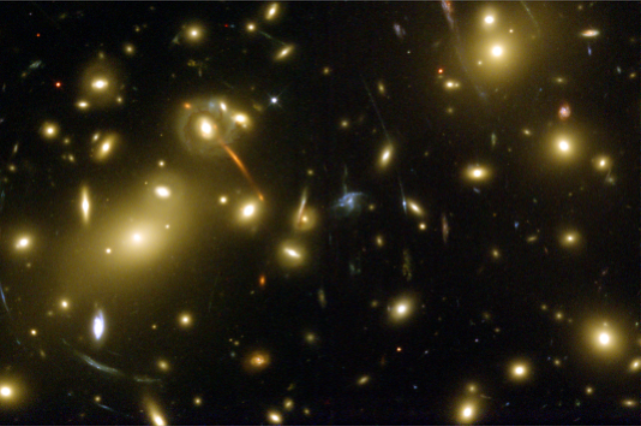}
\caption[Strong gravitational lensing in Abell 2218]{An example of strong gravitational lensing seen in galaxy cluster Abell 2218. The images of galaxies beyond the cluster are distorted into arcs by the gravitational field within the cluster. Public domain image courtesy of NASA.}
\label{fig:strong_lensing}
\end{figure}

Also on the scale of galaxy clusters, one can measure the matter density by looking at the gravitational lensing of objects beyond the cluster in question. As opposed to the straightforward gravitational lensing observable in the formation of arcs or multiple images via strong lensing (shown in Figure \ref{fig:strong_lensing}),  weak gravitational lensing is observable in the slight distortions of a galaxy's ellipticity and changes in its apparent magnitude \citep{bartelmann-01}. The effects of weak lensing are statistical in nature and cannot be seen in individual examples, but collecting large samples allows for a robust mapping of the matter distribution in galactic clusters that is more widely applicable than strong lensing observations (of which relatively few obvious examples exist). 

In comparison with weak gravitational lensing, which measures the total mass of a galaxy cluster, one can also estimate the total baryonic mass of a galaxy cluster by measuring the x-ray intensity from intercluster gas, which dominates the baryonic mass content of galaxy clusters. Assuming the matter content in galaxy clusters is a fair representation of the overall matter content of the universe, weak lensing and x-ray intensity measurements, combined with the baryonic density inferred from BBN, give a total matter density of the universe in very good agreement with the value determined from the CMB power spectrum \citep{allen-02}. Compared with the baryonic density, this too points to a large dark matter contribution to the total matter density of the universe. 

\subsection{The Bullet Cluster}
\label{sec:darkmatter_bullet}

\begin{figure} [t!]
\centering
\includegraphics[scale=0.55]{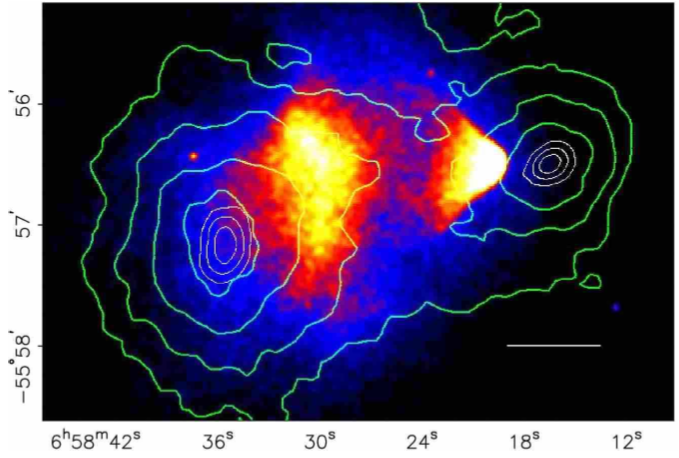}
\caption[The Bullet cluster]{The Bullet cluster, showing the matter distribution following a collision between two galaxy clusters. The green contours show the visible and invisible mass distributions taken from weak lensing reconstruction. The colored contour show the location of x-ray gas, experiencing a drag and subsequent shock wave. Figure from \cite{clowe-06} reproduced by permission of the AAS.}
\label{fig:bullet_cluster}
\end{figure}

One final evidence for dark matter is found in the Bullet cluster (1E 0657-56) (Figure \ref{fig:bullet_cluster}).  The Bullet cluster is presently observed shortly after the time of collision between two galaxy clusters. As before, comparisons between the total mass of the cluster (observable through weak lensing) and the baryonic mass of the cluster (seen in the x-ray emitting gas) paints a picture of a significant dark matter component to the galaxy cluster. However, in this instance, the location of these two mass distributions is key: the collision-less dark matter component was able to traverse the interaction unhindered while the baryonic component experienced drag passing through the intergalactic medium, resulting in a shock wave pattern (the colored contour of Figure \ref{fig:bullet_cluster}). Thus, the two centers of mass of the galaxy clusters in this collision (represented by the green contours of Figure \ref{fig:bullet_cluster}) have outpaced the visible, x-ray emitting matter component --- a clear indication that a majority of the mass is not only dark in nature, but also weakly interacting or collision-less. 

\section{WIMPs as Dark Matter}
\label{sec:darkmatter_wimps}

It is exceedingly apparent that there is a large component of the universe that is made up of a yet undiscovered form of matter, non-baryonic and non-relativistic in nature. Possible candidates include primordial black holes \citep{carr-10}, axions \citep{battesti-08}, and WIMPs. WIMPs are a strongly favored candidate not only because they satisfy all the necessary conditions for dark matter (\emph{i.e.} they interact weakly, can have an appropriately high density, can be stable, can be non-relativistic, and are by definition non-baryonic), but also because they naturally arise from many SUSY extensions to the Standard Model, representing the lightest SUSY particle (LSP) \citep{jungman-96}. While each candidate has merits, those of the WIMP dark matter model are discussed here. 

\begin{figure} [t!]
\centering
\includegraphics[scale=0.45]{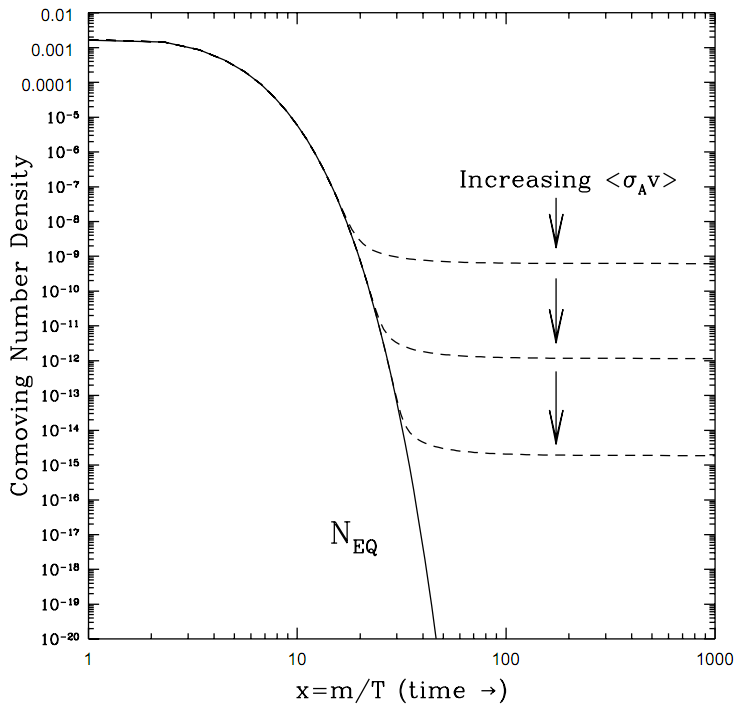}
\caption[Thermal relic WIMP density]{The thermal relic WIMP density after dropping out of thermal equilibrium. The solid line shows the density of matter in thermal equilibrium as the universe cools, while the dotted lines represent WIMP densities at various annihilation cross sections. Figure from \cite{kolb-89} reproduced with permission.}
\label{fig:wimp_density}
\end{figure}

Assuming WIMPs were in thermal equilibrium with other Standard Model particles in times just after the Big Bang, as the universe cools below the mass of the WIMP ($m_\chi$), and as the expansion rate of the universe $H$ exceeds the annihilation/creation rate, the WIMPs drop out of thermal equilibrium and are left as a thermal relic. The thermal relic density of WIMPs is then fixed, independent of $m_\chi$ and inversely proportional to $\left<\sigma_a v\right>$, 

\begin{equation}
\label{eq:Omega_chi}
\Omega_\chi = \frac{m_\chi n_\chi}{\rho_c H^2} \sim \frac{10^{-26} \mathrm{cm}^3 / \mathrm{s}}{\left<\sigma_a v\right>}
\end{equation}

\noindent where $\sigma_a$ is the average annihilation cross section, $v$ is the velocity (the chevrons indicate thermal averaging), $n_\chi$ is the number density of WIMPs (which is inversely proportional to $m_\chi$), and $\rho_c$ is as defined in Equation \ref{eq:Omega_m} (Figure \ref{fig:wimp_density}). See \cite{jungman-96} for a detailed explanation of this process. 

A WIMP as just described which arises as the stable particle associated with new physics at the electroweak scale (a hallmark of SUSY extensions to the Standard Model) has one surprising feature. Since the annihilation cross section is by definition set around the weak scale

\begin{equation}
\label{eq:sigma_a_v}
\left<\sigma_a v\right> \sim \alpha^2 (100\mathrm{GeV})^{-2} \sim 10^{-25} \mathrm{cm}^3 / \mathrm{s},
\end{equation}

\noindent then according to Equation \ref{eq:Omega_chi}, the thermal relic density of this WIMP naturally rests at $\Omega_\chi \sim 0.1$, on the same scale as that required by any cosmological argument. No assumptions have been made to arrive at this precise number, only that weak-scale interactions lead to our WIMP candidate. What this implies is that any stable particle arising from new physics at the electroweak scale naturally fills the role of dark matter. As any LSP from a SUSY extension to the Standard Model is a WIMP, they are therefore an extraordinarily natural candidate for the dark matter component missing from our understanding of the universe.

\singlespacing
\chapter{Direct Detection of WIMP Dark Matter}
\label{ch:directdetection}
\doublespacing

The case made in Chapter \ref{ch:darkmatter} for the existence of non-baryonic cold dark matter is significant. What remains is to observe the constituents of this dark matter so that its properties can be better understood. If it is formed by WIMPs, which are distributed within the halo of the galaxy, they will scatter elastically off target nuclei in a detector via the weak interaction with some frequency, depending on the strength of the WIMP-nucleon cross section. This cross section can be either spin-independent or spin-dependent (the latter of which having two fundamental cross sections to explore: WIMP-proton and WIMP-neutron), and so the specifics of the target nucleus become important. This chapter focuses on several techniques employed by different experiments that search for WIMPs by detecting nuclear recoils induced by their elastic scattering off target nuclei. The list below is by no means complete but rather provides a picture of the diversity of the field of WIMP direct detection experiments.

\section{WIMP Recoil Spectrum}
\label{sec:directdetection_recoil}

A review on the theory of elastic nuclear recoils of dark matter is provided by \cite{lewin-96}, and this section summarizes their derivation of the differential WIMP recoil spectrum $\mathrm{d}R / \mathrm{d}E_R$. Typically, experiments set upper limits on the rate of WIMP-nucleon scattering at particular recoil energies, and this translates to a limit on the total event rate (\emph{i.e.} dark matter signal) for each choice of dark matter mass $m_\chi$. This allows experiments to set limits on the largest elastic scattering cross sections allowed by the data, typically expressed as exclusion plots in cross section versus WIMP mass phase space.

\subsection{Differential Scattering Rate}
\label{sec:directdetection_scattering}

The event rate per unit target mass $m_T$ with scattering cross-section $\sigma_T$ is given by

\begin{equation}
\label{eq:dR_ini}
\mathrm{d}R = \frac{\sigma_T v \mathrm{d}n}{m_T},
\end{equation}

\noindent where $v$ is the WIMP velocity with respect to the target and $\mathrm{d}n$ is the differential WIMP particle density

\begin{equation}
\label{eq:dn}
\mathrm{d}n = \frac{1}{k} \frac{\rho_0}{m_\chi}f(\boldsymbol{v},\boldsymbol{v}_E)\mathrm{d}^3\boldsymbol{v},
\end{equation}

\noindent with $k$ a normalization such that $\int \mathrm{d}n = \rho_0 / m_\chi$, $\rho_0$ the local dark matter density, $\boldsymbol{v}$ the WIMP velocity with respect to the target, and $\boldsymbol{v}_E$ the velocity of the earth with respect to the rest frame of the galaxy. The function $f(\boldsymbol{v},\boldsymbol{v}_E)$ is assumed to follow a Maxwellian distribution around $\boldsymbol{v}_E$

\begin{equation}
\label{eq:f_v_vE}
f(\boldsymbol{v},\boldsymbol{v}_E) = e^{-(\boldsymbol{v}+\boldsymbol{v}_E)^2 / v_0^2},
\end{equation}

\noindent with $\left| \boldsymbol{v}+\boldsymbol{v}_E \right|$ truncated at the galactic escape velocity $v_\mathrm{esc}$ and $v_0$ the characteristic velocity of WIMPs in the galactic halo. For $v_\mathrm{esc} \rightarrow \infty$, the value of $k$ becomes $k_0 = \left(\pi v_0^2\right)^{3/2}$. When $v_\mathrm{esc}$ is finite, it is

\begin{equation}
\label{eq:k}
k = k_0 \left[ \mathrm{erf}\left(\frac{v_\mathrm{esc}}{v_0}\right) - \frac{2}{\sqrt{\pi}} \frac{v_\mathrm{esc}}{v_0} e^{-v_\mathrm{esc}^2 / v_0^2} \right].
\end{equation}

\noindent If the event rate per unit mass for $v_E = 0$ and $v_\mathrm{esc} \rightarrow \infty$ is defined as

\begin{equation}
\label{eq:R_0}
R_0 \equiv \frac{2}{\sqrt{\pi}} \frac{\rho_\chi}{m_\chi} \frac{v_0}{m_T} \sigma_T,
\end{equation}

\noindent Equation \ref{eq:dR_ini} can be rewritten (in light of Equation \ref{eq:dn}) as

\begin{equation}
\label{eq:dR_fin}
\mathrm{d}R = R_0 \frac{k_0}{k} \frac{1}{2 \pi v_0^4} v f(\boldsymbol{v},\boldsymbol{v}_E) \mathrm{d}^3 v.
\end{equation}

The recoil energy of a nucleus scattered at angle $\theta$ after interacting with a WIMP of energy $E = m_\chi v^2 / 2$ and kinematic factor $r = 4 m_\chi m_T / (m_\chi + m_T)^2$ is

\begin{equation}
\label{eq;E_R}
E_R = \frac{m_\chi m_T}{(m_\chi + m_T)^2} m_\chi v^2 (1-\cos{\theta}) = \frac{1}{2} E r (1 - \cos{\theta}).
\end{equation}

\noindent Isotropic scattering is assumed, and $\mathrm{d}R / \mathrm{d}E_R$ is built by integrating over $v$ limits (due the the $v$ functional dependence in Equation \ref{eq:dR_fin}), resulting in

\begin{equation}
\label{eq:dR_dER_ini}
\frac{\mathrm{d}R}{\mathrm{d}E_R} = \frac{R_0}{E_0 r} \frac{k_0}{k} \frac{1}{2 \pi v_0^2} \int_{v_\mathrm{min}}^{v_\mathrm{max}} \frac{f(\boldsymbol{v},\boldsymbol{v}_E)}{v} \mathrm{d}^3 v,
\end{equation}

\noindent with $E_0 = m_\chi v_0^2 / 2$. The physical case ($v_\mathrm{min} = v_E$, $v_\mathrm{max} = v_\mathrm{esc}$) is found to be

\begin{equation}
\label{eq:dR_dER_fin}
\frac{\mathrm{d}R}{\mathrm{d}E_R} = \frac{k_0}{k} \frac{R_0}{E_0 r} \left( \frac{\sqrt{\pi}}{4} \frac{v_0}{v_E} \left[ \mathrm{erf} \left( \sqrt{\frac{E_R}{E_0 r}} + \frac{v_E}{v_0} \right) - \mathrm{erf} \left( \sqrt{\frac{E_R}{E_0 r}} - \frac{v_E}{v_0} \right) \right] - e^{-v_\mathrm{esc}^2 / v_0^2} \right),
\end{equation}

\noindent with $v_0 = 230$ km/s, $v_\mathrm{esc} = 600$ km/s, and $v_E$ varying during the year as the Earth moves around the Sun:

\begin{equation}
\label{eq:v_E}
v_E \simeq \left[ 244 + 15 \sin(2 \pi y ) \right] \mathrm{km} / \mathrm{s},
\end{equation}

\noindent where $y$ is the elapsed time from March 2nd in years \citep{lewin-96}.

\subsection{Scattering Cross Sections}
\label{sec:directdetection_crosssections}

\begin{figure} [t!]
\centering
\includegraphics[scale=0.55]{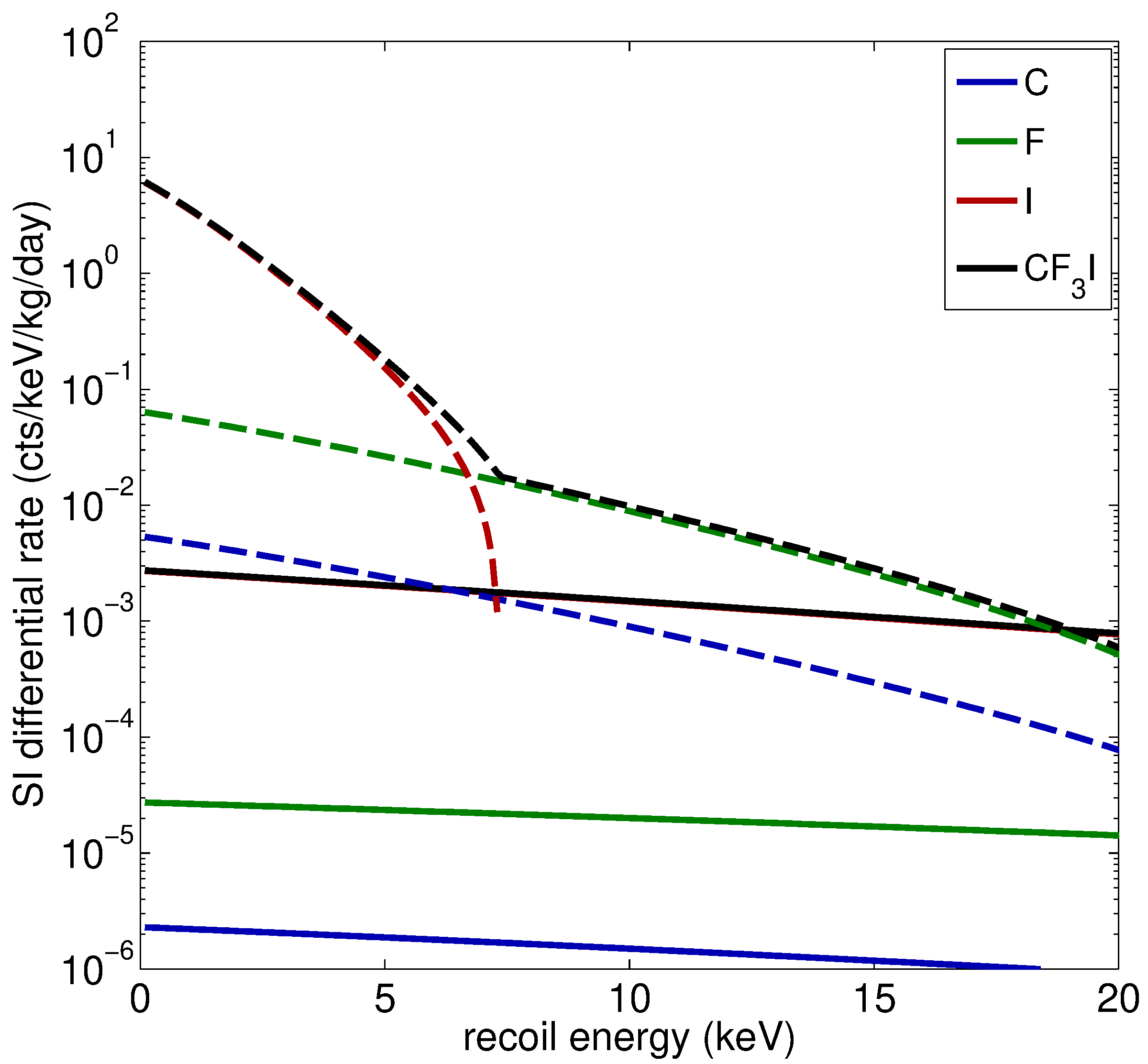}
\caption[Spin-independent CF$_3$I recoil spectrum from WIMP scattering]{The sensitivity of $^{12}$C, $^{19}$F, and $^{127}$I (and CF$_3$I) to spin-independent WIMP scattering in the case of $m_\chi = 100$ GeV and $\sigma_n = 10^{-43}$ cm$^2$ (solid lines) and $m_\chi =  10$ GeV and $\sigma_n = 2\times10^{-41}$ cm$^2$ (dashed lines). Iodine recoils dominate for high mass WIMPs, whereas fluorine can dominate the rate in certain spectral regions.}
\label{fig:SI_WIMP_recoil_CF3I}
\end{figure}

What remains is to determine the scattering cross-section per target $\sigma_T$ used in Equation \ref{eq:R_0}, considering nuclear structure corrections $\Phi$

\begin{equation}
\label{eq:sigma_T}
\sigma_T (q) = \sigma_T (0) \Phi (q),
\end{equation}

\noindent for momentum transfer $q = \sqrt{2 m_T E_R}$. This discussion follows \cite{cannoni-11a} and \cite{cannoni-11b}. Assuming the proton and neutron mass are approximately the same ($m_p \simeq m_n$) and that there is near equality in the proton and neutron spin-independent (SI) cross sections, the $q=0$ limit of the SI cross section is

\begin{equation}
\label{eq:sigma_0_SI}
\sigma_T^\mathrm{SI} (0) = \left( \frac{\mu_A}{\mu_n} \right)^2 A^2 \sigma_n^\mathrm{SI},
\end{equation}

\noindent where $\mu_A = A m_n m_\chi / (A m_n + m_\chi)$ and $\mu_n = m_n m_\chi / (m_n + m_\chi)$ are the reduced masses of the nucleus and nucleon, respectively, and $\sigma_n^\mathrm{SI}$ is the elementary WIMP-nucleon cross section. The SI nuclear structure corrections reduce to the square of the nuclear form factor $\Phi^\mathrm{SI} (q) = \left( F^\mathrm{SI} (q) \right)^2$, using the Helm formalism \citep{helm-56} from \cite{lewin-96}

\begin{equation}
\label{eq:F_SI}
F^\mathrm{SI} (q) = 3 \frac{j_1(q r_n)}{q r_n} e^{-(q s)^2 / 2},
\end{equation}

\noindent where $j_1$ is the first-order spherical Bessel function and $r_n^2 = c^2 + (7/3) \pi^2 a^2 - 5 s^2$, with $c \simeq (1.23 A^{1/3} - 0.60)$ fm, $a \simeq 0.52$ fm, and $s \simeq 0.9$ fm. Calculation of Equation \ref{eq:sigma_T} (given Equations \ref{eq:sigma_0_SI} and \ref{eq:F_SI}) is straightforward for all nuclei, which leads to the differential recoil spectrum in Equation \ref{eq:dR_dER_fin} shown in Figure \ref{fig:SI_WIMP_recoil_CF3I}. This form factor takes care of the previous isotropic scattering approximation leading to Equation \ref{eq:dR_dER_ini}.

The spin-dependent (SD) case is more complicated. The $q=0$ limit of the cross section is

\begin{equation}
\label{eq:sigma_0_SD}
\sigma_T^\mathrm{SD} (0) = \frac{1}{3} \left( \frac{\mu_A}{\mu_n} \right)^2 \left( \Upsilon_p^A \sqrt{\sigma_p^\mathrm{SD}} \pm \Upsilon_n^A \sqrt{\sigma_n^\mathrm{SD}} \right)^2,
\end{equation}

\noindent for $\sigma_{p,n}^\mathrm{SD}$ the elementary WIMP cross sections on the proton and neutron, respectively. The $\Upsilon$ terms are the spin nuclear matrix elements, defined as

\begin{equation}
\label{eq:Upsilon}
\Upsilon_{p,n}^A = 2 \sqrt{\frac{J+1}{J}} \left<\boldsymbol{S}_{p,n}\right>,
\end{equation}

\noindent with $J$ the total angular momentum of the nucleus in the ground state and $\left<\boldsymbol{S}_{p,n}\right>$ the expectation values of the spin of the proton and neutron groups.

\begin{figure} [t!]
\begin{minipage}[b]{0.5\linewidth}
\centering
\includegraphics[scale=0.45]{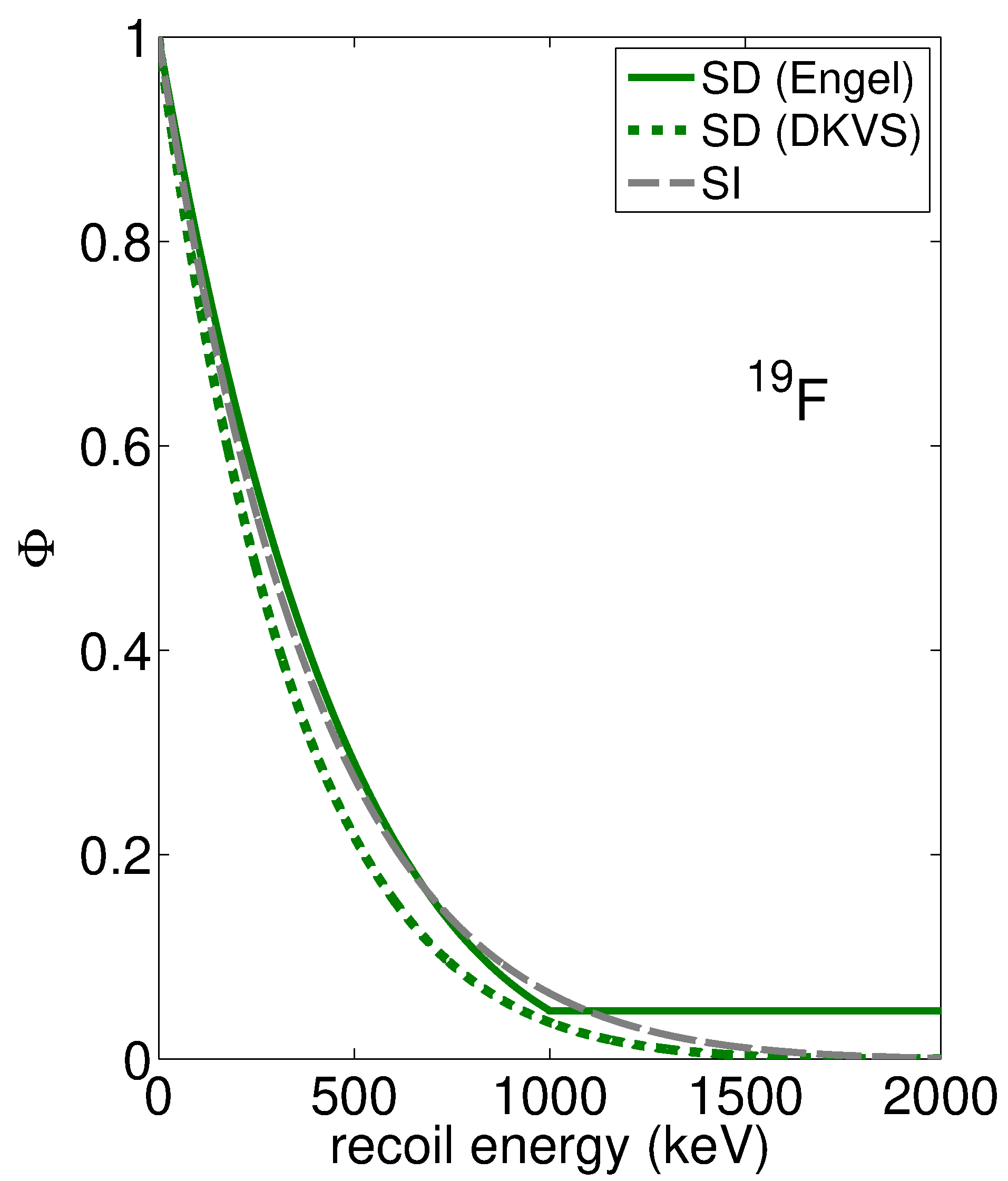}
\end{minipage}
\hspace{0.3cm}
\begin{minipage}[b]{0.5\linewidth}
\centering
\includegraphics[scale=0.45]{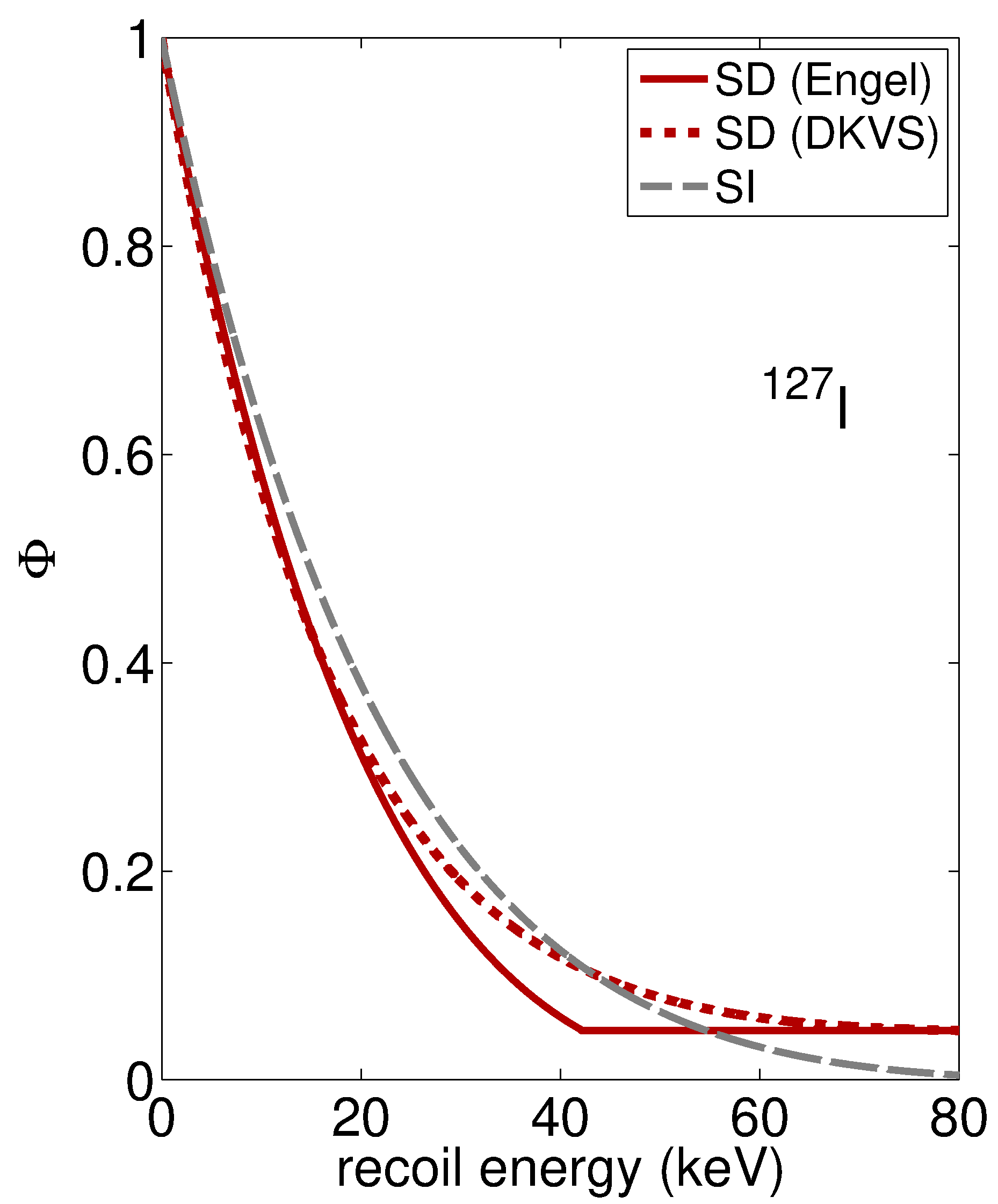}
\end{minipage}
\caption[Nuclear structure corrections from various formalisms]{Spin dependent nuclear structure corrections $\Phi^\mathrm{SD}$ for fluorine (left) and iodine (right). The Engel (solid lines) and DKVS (dotted lines) formalisms are shown (Equations \ref{eq:Phi_SD_LS} and \ref{eq:Phi_SD_DKVS}), along with the spin independent case $\Phi^\mathrm{SI} = \left( F^\mathrm{SI} \right)^2$ (from Equation \ref{eq:F_SI}), for comparison (dashed lines).}
\label{fig:SD_structure_functions}
\end{figure}

For the nuclear structure corrections, there is variation in formalisms, due to the complication of both the nuclear physics and particle physics being interconnected in whatever corrections $\Phi^\mathrm{SD}$ are chosen (structure corrections from different formalisms described here are shown in Figure \ref{fig:SD_structure_functions}).  One choice is the use of a SD analog to Equation \ref{eq:F_SI}

\begin{equation}
\label{eq:Phi_SD_LS}
\Phi^\mathrm{SD}_\mathrm{Engel} (q) = \left\{ \begin{array}{ll}
j_0^2(q r_n) & (q r_n < 2.55, q r_n > 4.5) \\
0.047 & (2.55 \leq q r_n \leq 4.5), \\
\end{array}
\right. \\
\end{equation}

\noindent with $j_0$ the zeroth-order spherical Bessel function and $r_n \simeq A^{1/3}$ fm \citep{engel-92}. The other is to represent the structure function as $\Phi^\mathrm{SD} (q) = \mathcal{F}(q) / \mathcal{F}(0)$ with spin structure functions $F_{ij}$ due to proton, neutron, and interference terms

\begin{equation}
\label{eq:script_F}
\mathcal{F} (q) = a_0^2 F_{00} (q) + 2 a_0 a_1 F_{01} (q) + a_1^2 F_{11} (q),
\end{equation}

\noindent where the spin structure functions $F_{ij}$ ($F_{ij}(0) = 1$ by construction) are computed using the shell model per nuclei and $a_0$ and $a_1$ are isoscalar and isovector WIMP-nucleon scattering amplitudes in the isospin basis \citep{divari-00}. With the observation that $F_{ij}$ are nearly identical for energy regions interesting to experiments \citep{cannoni-11b}, this simplifies to

\begin{equation}
\label{eq:Phi_SD_DKVS}
\Phi^\mathrm{SD}_\mathrm{DKVS} (q) = F_{11} (q).
\end{equation}

\begin{figure} [t!]
\centering
\includegraphics[scale=0.55]{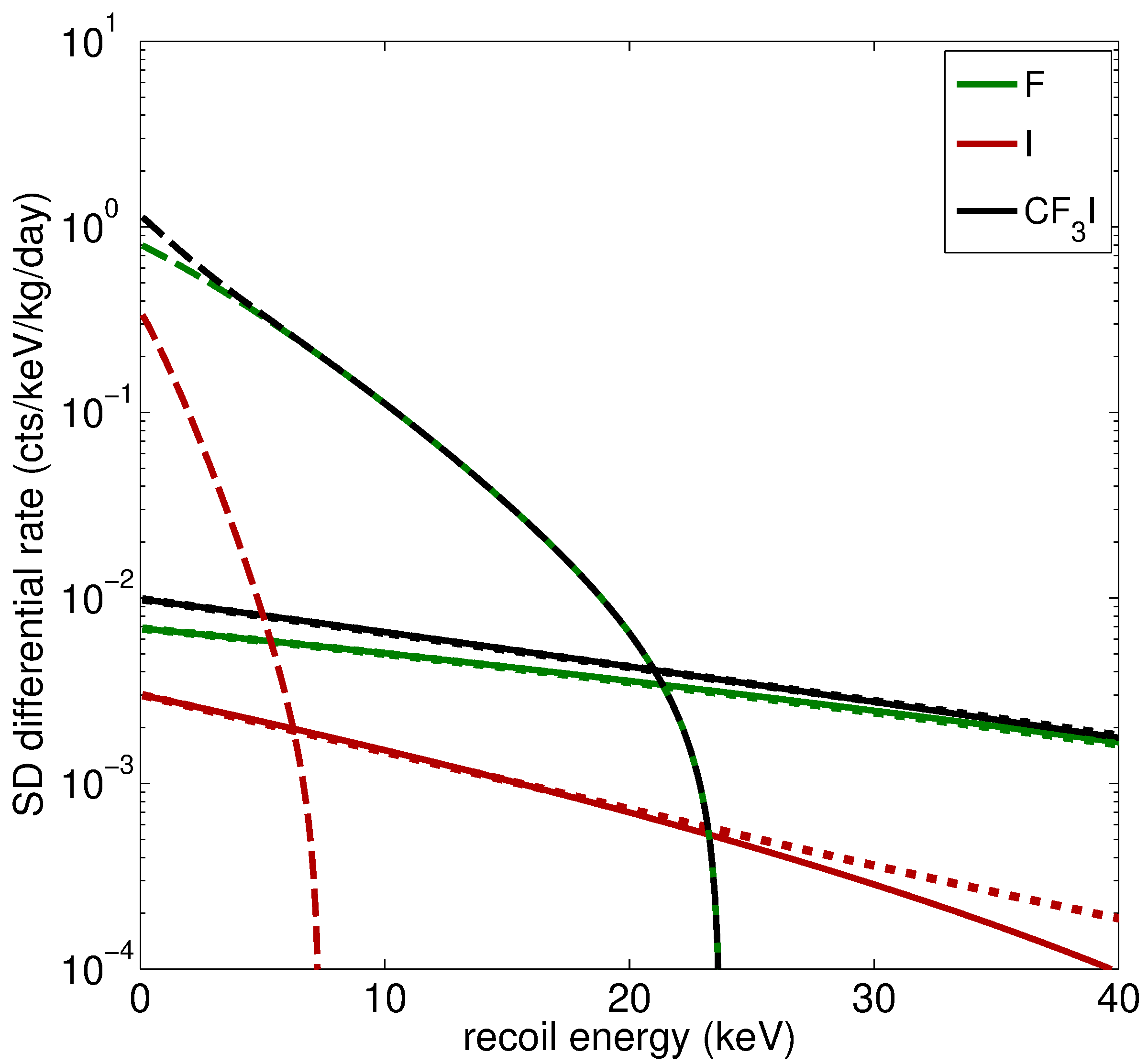}
\caption[Spin-dependent CF$_3$I recoil spectrum from WIMP scattering]{The sensitivity of $^{19}$F and $^{127}$I (and CF$_3$I) to spin-dependent WIMP-proton scattering in the case of $m_\chi = 100$ GeV and $\sigma_n = 10^{-38}$ cm$^2$ (solid lines) and $m_\chi =  10$ GeV and $\sigma_n = 10^{-37}$ cm$^2$ (dashed lines) in the Engel formalism (the DKVS formalism is shown as dotted lines, and only deviates for high-energy recoils on $^{127}$I). Fluorine recoils dominate in all cases.}
\label{fig:SD_WIMP_recoil_CF3I}
\end{figure}

For SD scattering, only nuclei with neutrons and/or protons that are uncoupled in spin (\emph{i.e.} odd number of neutrons and/or protons) contribute. For the COUPP experiment, these are $^{19}$F and $^{127}$I, with spin nuclear matrix elements of $\Upsilon_p^F = 1.646$ and $\Upsilon_n^F = -0.030$ for $^{19}$F \citep{divari-00} and $\Upsilon_p^I = 0.731$ and $\Upsilon_n^I = 0.177$ for $^{127}$I \citep{ressell-97}.  Note that for $^{19}$F, the neutron contribution to the SD rate can be neglected ($\Upsilon_n^F \ll \Upsilon_p^F$). For the DKVS formalism (Equation \ref{eq:Phi_SD_DKVS}), the spin structure function $F_{11}$ for $^{19}$F is found in \cite{divari-00} and for $^{127}$I is found in \cite{ressell-97}. The SD differential recoil spectrum from WIMP-proton scattering is shown in Figure \ref{fig:SD_WIMP_recoil_CF3I}.  It is notable that the choice of formalism makes very little difference to the recoil spectrum in regions of concern for the COUPP bubble chamber using CF$_3$I as a target, since the fluorine recoils dominate the spectrum and the structure function for $^{19}$F is very similar for $E_R < 1000$ keV (see Figure \ref{fig:SD_structure_functions}).

\section{Direct Detection Experimental Methods}
\label{sec:directdetection_experiments}

Direct detection experiments aim to observe the nuclear recoils from WIMP-nucleon elastic scattering in a detector target. This is a non-trivial undertaking due to the fact that these events are rare (the scattering cross sections are obviously quite small) and because any WIMP signal has to be distinguished from a significant radioactive background, either from the surrounding atmosphere or from detector components themselves. For most experiments, there is a dominant background from electromagnetically interacting particles (photons, electrons, and alphas) with a much lower contribution from neutrons, either cosmogenic (induced by cosmic muons) or from natural radioactivity. The goal of these experiments is to discriminate any WIMP signal from the expected background by using some method of background subtraction or by having a detector sensitive to only small parts of the total possible background (\emph{e.g.} having a detector sensitive to only nuclear recoils, as in COUPP). Any detector without backgrounds has a sensitivity directly proportional to $m_T t$ (for an exposure time $t$). If background subtraction is necessary, this sensitivity becomes proportional to $\sqrt{m_T t}$ \citep{gaitskell-96}. Several direct detection techniques are described herein.

\subsection{Scintillation Detectors}
\label{sec:directdetection_scintillation}

\begin{figure} [t!]
\centering
\includegraphics[scale=0.46]{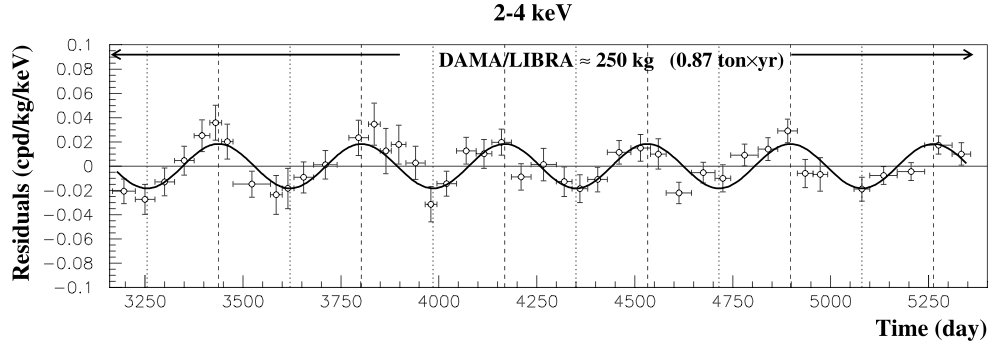}
\caption[Annual modulation observed by DAMA]{The annual modulation of a signal observed by DAMA/LIBRA, interpreted as a WIMP signal. Figure from \cite{bernabei-10} reproduced with permission.}
\label{fig:DAMA_annual_modulation}
\end{figure}

Scintillation detectors use photomultiplier tubes (PMTs) to observe the luminescence from excitation by ionizing radiation passing through a crystal. One example of an experiment using this technology as a WIMP direct detection method is DAMA, which uses 250 kg of highly radiopure thallium-doped NaI detectors. While DAMA can to some extent use pulse shape discrimination (PSD) to distinguish between electron recoil and nuclear recoil events \citep{bernabei-98}, there is little to no background rejection capability with this technique at low energies, like those expected from WIMP recoils. However, one key WIMP feature that can be searched for with DAMA is the annual modulation of a WIMP signal, resulting from the seasonal dependence of the Earth's velocity with respect to the galaxy, seen in Equation \ref{eq:v_E}. The most recent DAMA annual modulation results (Figure \ref{fig:DAMA_annual_modulation})  from 13 annual cycles with an exposure of 1.17 ton-years reports an $8.9\sigma$ evidence for an annual modulation compatible with dark matter interactions \citep{bernabei-10}. This result is, at first sight, incompatible with limits set by other dark matter experiments \citep{ahmed-09,aprile-11}, unless the WIMP mass is very modest ($\lesssim$ 10 GeV).

\subsection{Ionization Detectors}
\label{sec:directdetection_ionization}

\begin{figure} [t!]
\centering
\includegraphics[scale=0.35]{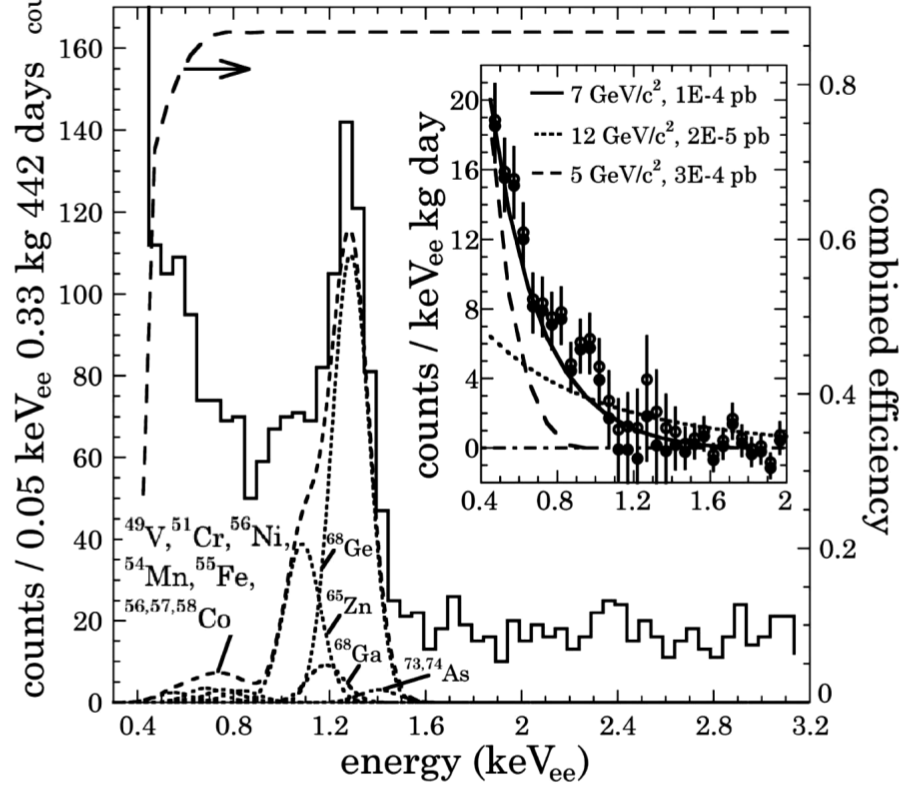}
\caption[Low recoil energy spectrum from CoGeNT]{Low-energy spectrum after surface event removal from CoGeNT, with a dashed line representing the threshold efficiency. Inset shows the spectra corrected by this efficiency without expected backgrounds, overlaid with examples of light WIMP signals. Figure from \cite{aalseth-11b} reproduced by permission of the APS, copyright 2011 by the American Physical Society.}
\label{fig:CoGeNT_signal}
\end{figure}

Ionization detectors employ semiconductors to detect nuclear recoils events as ionization, by means of measuring electron-hole pairs created between two collecting electrodes. One example of an experiment using semiconductors as a WIMP direct detection medium is CoGeNT, which operates $p$-type point contact high-purity Ge detectors, which offer a combination of high mass and very low electronic noise, ideal for sensitivity to low-energy nuclear recoils \citep{barbeau-07}. CoGeNT makes use of PSD for surface event rejection to reduce backgrounds \citep{aalseth-11a}. What remains is a spectrum of known energy peaks from cosmogenic activation of the Ge crystal as well as an exponentially rising component at the lowest ionization energy (Figure \ref{fig:CoGeNT_signal}).  This exponentially rising component also demonstrates an annual modulation with significance $\sim$2.8$\sigma$, compatible with a signal expected from a light mass WIMP, with best fit parameters of $m_\chi=7$ GeV and $\sigma_n=10^{-40}$ cm$^2$ \citep{aalseth-11b}. Again, this result is, at first sight, incompatible with other dark matter experiments \citep{ahmed-09,aprile-11}, although much discussion is available on the subject (see, \emph{e.g.}, \cite{hooper-11}).

\subsection{Cryogenic Detectors}
\label{sec:directdetection_cryogenic}

One could potentially look for a WIMP signal in a detector by looking for heat distribution in the form of phonons from nuclear recoils in a cryogenically-cooled target $<100$ mK. This method is promising because of its low energy thresholds, but without nuclear recoil discrimination, it is inadequate. However, by combining this detection method with the two previously described detection methods, successful experiments have been constructed. Two of interest are CRESST and CDMS.

\begin{figure} [t!]
\centering
\includegraphics[scale=0.45]{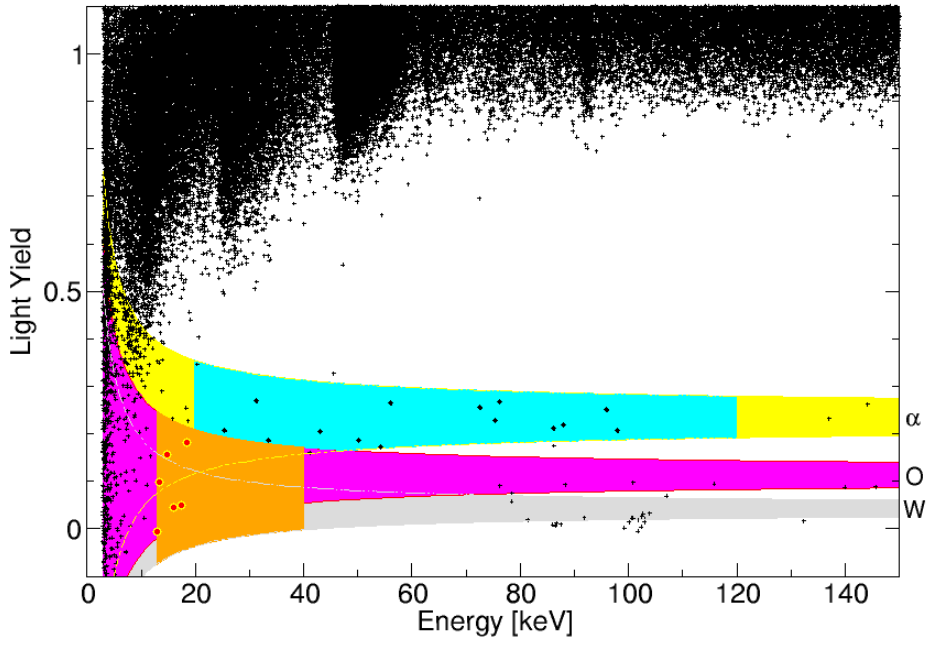}
\caption[Low recoil energy spectrum from CRESST]{Data from one module of CRESST, showing the bands from electron and gamma backgrounds (black) as well as those where alpha recoils (yellow), oxygen recoils (violet), and tungsten recoils (gray) are expected. The acceptance region for a WIMP signal is shown in orange, containing several candidate events. Figure from \cite{angloher-11} reproduced with permission.}
\label{fig:CRESST_signal}
\end{figure}

CRESST combines phonon measurement with scintillation techniques, employing eight CaWO$_4$ detector modules (300 g each) in the search. The phonon channel provides a measurement of the total recoil energy of an event. However, nuclear recoils are ``quenched" with respect to electron recoils\footnote{Nuclear recoils produce less ionization or scintillation than electron recoils. The quenching factor $QF$ is defined as the ratio of observed energy for nuclear recoils to that from electron recoils of the same energy. Units of keV$_\mathrm{ee}$ (electron-equivalent energy) are often used, as in $E(\mathrm{keV}_\mathrm{ee}) = QF \times E_R(\mathrm{keV})$.}, so the secondary channel from scintillation provides a way to discriminate between electron and nuclear recoils. The O and Ca components of the target make the experiment sensitive to light mass WIMPs, similar to those compatible with the CoGeNT result. In the most recent analysis of 730 kg-days of data, CRESST sees an excess of 67 low-energy nuclear recoil candidate events in the oxygen band (Figure \ref{fig:CRESST_signal}),  a signal with significance $>$ 4$\sigma$ \citep{angloher-11}. A discussion of the compatibility between DAMA, CoGeNT, and CRESST results is outlined in \cite{kelso-11}, with comments on the conflicting null results \citep{ahmed-09,aprile-11}.

One of these such null results comes from CDMS, which combines the techniques of phonon-mediated detectors with ionization detectors to provide analogous discrimination ability to that from CRESST. The most recent experiment employs 19 Ge (250 g each) and 11 Si (100 g each) solid-state detectors. Again, the phonon channel provides the total recoil energy of an event, while the secondary ionization channel allows for discrimination between electron and nuclear recoils. The most recent results from CDMS report no significant excess events above expected backgrounds in 397.8 kg-days of exposure \citep{ahmed-09}.

\subsection{Liquid Noble Detectors}
\label{sec:directdetection_noble}

Along with CDMS, the other conflicting null result is from XENON100, an experiment which combines the ionization and scintillation techniques described in Sections \ref{sec:directdetection_ionization} and \ref{sec:directdetection_scintillation} to provide discrimination. A recoiling particle in the detector's 161 kg of ultrapure liquid Xe deposits energy by direct electronic excitation of Xe$_2$ molecules resulting in scintillation and by ionization of Xe$_2$. A drift field is applied to extract electrons before they recombine. The ratio of charge to light signals can be used to distinguish electron and nuclear recoils. In 100.9 live days of data, XENON100 reports no dark matter signal \citep{aprile-11}.

\subsection{Superheated Liquid Detectors}
\label{sec:directdetection_superheated}

One final class of WIMP direct detection experiments described here is found in superheated liquid detectors, which employ superheated freons as a target, either in the form of droplets in a gel \citep{archambault-09} or as a bubble chamber (described herein). While all of the other experimental methods so far described require the statistical discrimination of electron recoil events from the nuclear recoil events that contribute to a WIMP signal, superheated liquid detectors have a dual threshold feature (described in detail in Chapter \ref{ch:bubblechambers}) which allows them to be insensitive to electron recoils. This is accomplished by tuning the temperature and pressure of the chamber (and thus the stopping power and energy thresholds) such that the energy deposition of electron recoils is too widely distributed to cross the stopping power threshold $\mathrm{d}E/\mathrm{d}x$, even if they are above the energy threshold. One early deficiency of such detectors was their inability to discriminate between bulk alpha decays and WIMP recoils, although both PICASSO \citep{aubin-08} and COUPP \citep{behnke-11} have demonstrated an excellent ability to discriminate the alpha background by their ultrasonic signature.

\singlespacing
\chapter{Theory of Bubble Chamber Operation}
\label{ch:bubblechambers}
\doublespacing

\begin{figure} [t!]
\centering
\includegraphics[scale=0.45]{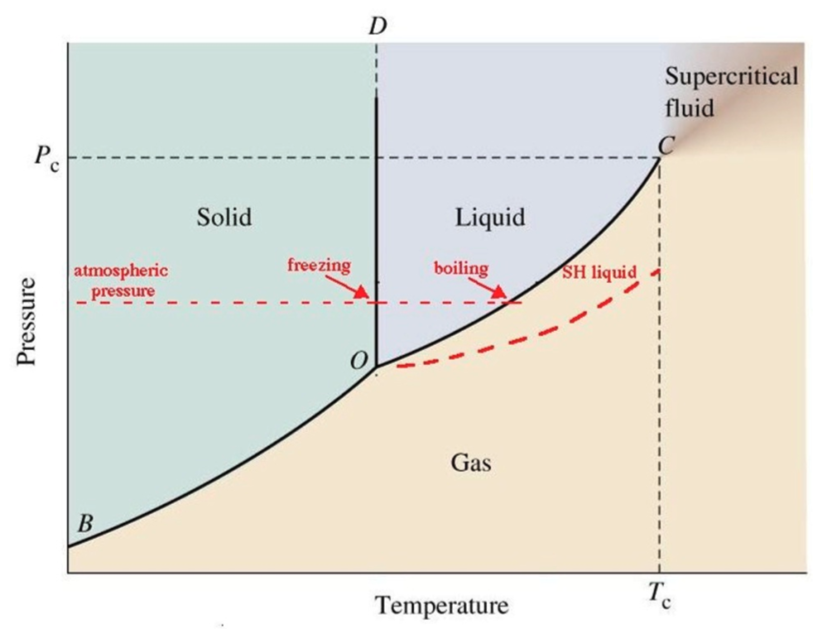}
\caption[Bubble chamber phase diagram]{A phase diagram depicting the thermodynamics of a superheated liquid. By decreasing the pressure or increasing the temperature, a superheated liquid can be formed within the region defined by the red dashed curve. Figure courtesy J.I. Collar.}
\label{fig:superheat_phase_diagram}
\end{figure}

Briefly described in Chapter \ref{ch:directdetection} for its potential as a superheated liquid detector for a WIMP direct detection experiment, the bubble chamber is described in more detail here. Originally developed as a particle detector for high energy physics experiments \citep{glaser-52}, the bubble chamber makes use of a liquid in a superheated state (\emph{i.e.} metastable equilibrium). Even when the temperature is above the boiling point or the pressure is below the vapor pressure, the superheated liquid will remain in a liquid state whose vaporization requires a nucleation site in the form of either a) impurities containing pockets of gas in the liquid or the liquid-container interface, or b) a particle interaction with energy and stopping power above thresholds defined by the thermodynamics of the liquid in its current state. At operating temperature, the difference between the vapor pressure and the pressure is defined as the ``degree of superheat." The higher the degree of superheat, the more unstable the superheated liquid. Figure \ref{fig:superheat_phase_diagram} shows an example phase diagram for a superheated liquid.

With newer techniques allowing for the development of high purity fluids and deactivation of container surfaces \citep{bolte-07}, bubble chambers can now be developed that do not have nucleation sites to lead to vaporization of the fluid. This allows for a bubble chamber to be superheated for nearly indefinite amounts of time, making them viable candidates for low-background, low-signal rate searches. The reason is that bubble chambers have two nucleation thresholds with regards to scattering: the energy deposition induced by a radiation has to be above a threshold energy and it has to be deposited in a region localized enough to nucleate a bubble. These energy and stopping power thresholds are tunable parameters of a bubble chamber and depend on the degree of superheat of the liquid. For liquids with high degrees of superheat, nearly any energy deposition induced by a radiation will nucleate a bubble, electron recoils included. But when the degree of superheat is modest, electron recoils no longer have enough stopping power to be above threshold, leaving only nuclear recoils (such as those produced via WIMP scattering) as capable of creating bubbles in the detector. This powerful built-in discrimination makes them ideal as dark matter detectors.

\section{Seitz ``Hot Spike" Model}
\label{sec:bubblechambers_seitz}

The original theory for how bubbles are created in a superheated liquid was first proposed by Seitz and is often referred to as the Seitz ``hot spike" model or just the Seitz model \citep{seitz-58}. At an operating temperature $T$, a superheated liquid is formed when its pressure is lowered below the vapor pressure of the liquid, resulting in a metastable equilibrium. This equilibrium is maintained until a potential barrier separating the liquid and its gas phase is overcome. This can occur if the liquid receives a spatially concentrated amount of energy in the form of a ``hot spike" resulting from the energy deposited in particle interactions with the liquid.

The foundation of the Seitz model is built around the concept of a critical bubble, a vapor bubble within and in unstable equilibrium with the surrounding superheated liquid, which is at a pressure $P_l$ below the vapor pressure $P_v$ at temperature $T$. Due to the unstable nature of the equilibrium, a proto-bubble with size less than some critical radius $r_c$ will collapse back onto itself as a result of surface tension before expanding to become a macroscopic bubble (thus the term ``proto-bubble"). A proto-bubble with size greater than $r_c$ will irreversibly expand into the fluid, evaporating it in the process, eventually vaporizing the entire superheated volume (a first-order phase transition).

\subsection{Critical Radius}
\label{sec:bubblechambers_rc}

The value of $r_c$ is tied to the equilibrium condition, which is defined such that the temperature and molar Gibbs chemical potential $\mu$ of the gas and liquid are equal, which in turn defines the pressure of the bubble gas $P_b$ relative to $P_l$ and $P_v$. To calculate $P_b$, note that the chemical potential $\mu$ is defined by the relation

\begin{equation}
\label{eq:dmu}
\mathrm{d}\mu = v\mathrm{d}P,
\end{equation}

\noindent for molar volume $v \propto 1/\rho$. Equilibrium requires that $\mu_b = \mu_l$, so

\begin{equation}
\label{eq:int_dmu}
\int_{P_l}^{P_v} \frac{\mathrm{d}P}{\rho_l} = \int_{P_b}^{P_v} \frac{\mathrm{d}P}{\rho_b},
\end{equation}

\noindent where $\rho_l$ and $\rho_b$ are the liquid and gas densities at temperature $T$ and pressure $P$. Assuming an incompressible fluid ($\rho_l = $ constant) and a constant isothermal compressibility for the gas ($\rho_b \propto P$),

\begin{equation}
\label{eq:int_dmu_solved}
\frac{P_v - P_l}{\rho_l} = \frac{P_v}{\rho_v} \ln \left(\frac{P_v}{P_b}\right),
\end{equation}

\noindent where $\rho_v$ is the saturated vapor density at temperature $T$. Solving for $P_b$,

\begin{equation}
\label{eq:P_b}
P_b = P_v e^{-\frac{P_v - P_l}{P_v} \frac{\rho_v}{\rho_l}} \approx P_v - (P_v - P_l)\frac{\rho_v}{\rho_l}.
\end{equation}

\begin{figure} [t!]
\centering
\includegraphics[scale=0.45]{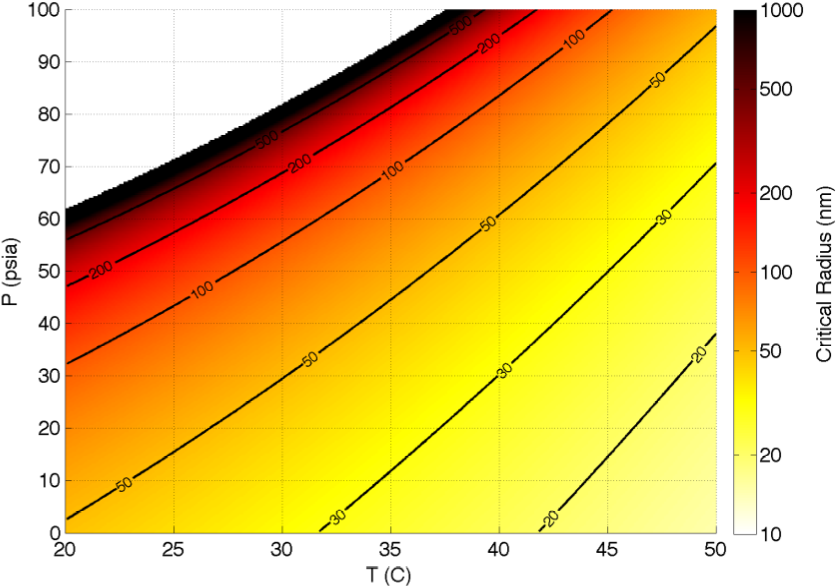}
\caption[Critical radius as a function of temperature and pressure for CF$_3$I]{Contours of the critical radius as a function of operating temperature and pressure of the superheated liquid for CF$_3$I. Figure from \cite{dahl-11a} reproduced with permission.}
\label{fig:rc_vs_PT}
\end{figure}

\noindent For criticality, the pressure differential between the gas bubble and the superheated liquid must be balanced by the pressure due to surface tension

\begin{equation}
\label{eq:P_gamma}
P_\gamma = \frac{2\gamma}{r},
\end{equation}

\noindent for $\gamma$ the surface tension and $r$ the radius of the bubble. In the critical case ($r = r_c$), setting Equation \ref{eq:P_gamma} equal to $P_b - P_l$ from Equation \ref{eq:P_b} yields for the critical radius

\begin{equation}
\label{eq:r_c}
r_c \approx \frac{2\gamma}{P_v - P_l} \frac{\rho_l}{\rho_l-\rho_v},
\end{equation}

\noindent which is shown in Figure \ref{fig:rc_vs_PT} as a function of operating temperature and pressure for a bubble chamber containing CF$_3$I.

Needed for this calculation is a value for the surface tension of a fluid at a temperature $T$. This has been well-determined for the case of CF$_3$I, the superheated target liquid chosen for the COUPP experiment. The surface tension $\gamma$ is a function of the temperature of the chamber $T$ and a critical temperature of the fluid $T_c$:

\begin{equation}
\label{eq:gamma}
\gamma = \gamma_0 \left(1-\frac{T}{T_c} \right)^n.
\end{equation}

\noindent For CF$_3$I, values for these parameters as given by \cite{duan-99} are $\gamma_0 = 0.057306$ N/m, $n=1.2933$, and $T_c=396.44$ K. With these values, the accuracy in $\gamma$ is about 1\% \citep{duan-99}. Other values required in any of these calculations (\emph{e.g.} the $P_v$ curve) are usually obtainable from the NIST REFPROP compilation \citep{lemmon-10}.

\subsection{Energy Threshold}
\label{sec:bubblechambers_Q}

The energy input required to form a critical bubble is often divided into thermodynamically reversible and irreversible work components. The reversible work components were first described in \cite{pless-56}, later summarized in \cite{peyrou-67}, as consisting of three parts: the works of vaporization, surface formation, and bubble expansion against an external liquid pressure. The irreversible work components account for the kinetic energy imparted by the expanding bubble against its wall, the energy lost through viscosity, the acoustic energy propagated as sound waves, and the thermal energy lost during expansion. These irreversible components contribute $<$ 1\% and are typically neglected \citep{harper-93}.

In reality, the energy for critical bubble formation is describable in two terms --- the work required for vaporization of the fluid ($W_b$) and the work required for formation of the surface ($W_s$). The vaporization of a liquid is accompanied by a transfer of heat, arising from a change in enthalpy during the transition from the liquid to the gas phase. The enthalpy $H$ (one of the thermodynamic potentials) is defined by the relation

\begin{equation}
\label{eq:dH}
\mathrm{d}H = T \mathrm{d} S + V \mathrm{d} P,
\end{equation}

\noindent where $S$ is the entropy. In terms of specific enthalpies, the work required for vaporization of the fluid is then

\begin{equation}
\label{eq:Q_b}
W_b \equiv \Delta H = \frac{4}{3} \pi r_c^3 \rho_b \left( h_b - h_l \right),
\end{equation}

\noindent with $\rho_b$ the gas density of the bubble ($\rho_b \approx \rho_v \left(P_b / P_v\right)$) and $h_b$ and $h_l$ the specific enthalpies of the gas and liquid, respectively, which are pressure-dependent. It should be noted that since the liquid is in a superheated state, it is not in pressure equilibrium, despite the fact that it is in (unstable) chemical equilibrium. The latent heat $h_{fg}$ (which is given in the literature as a property of different fluids) is defined as $\left( h_b - h_l \right)$ when both the liquid and the gas are at vapor pressure, which is not the case here. Despite the fact that this difference is quite small ($\sim$0.2\% at 34$^\circ$C and 30 psia) \citep{dahl-11a}, one should be cautious when using $h_{fg}$ instead of specific values of  $\left( h_b - h_l \right)$, which is commonly done \citep{peyrou-67,harper-93}.

From Equation \ref{eq:dH}, the change in enthalpy includes both the change in internal energy in the fluid as well as the work done by expanding the bubble against the liquid, by definition. \cite{peyrou-67} mistakenly treats the latent heat as only arising from the internal energy and adds a separate term for the work done by the bubble expanding against the liquid pressure. This term need not and should not be included separately, and only the heat from Equation \ref{eq:Q_b} should be used. Further, PICASSO actually subtracts the expansion piece from the latent heat, leaving only the change in internal energy \citep{archambault-11}. Although neither treatment is correct, the difference is negligible ($<$ 1\%), so this discrepancy should not matter in practice \citep{dahl-11a}.

The second work term required for the formation of a critical bubble arises from the formation of the surface. The surface tension $\gamma$ is defined as the proportionality constant between the free energy (d$F$) required to change the surface by an infinitesimal amount (d$A$), and so the free energy required to form a surface of area $4 \pi r_c^2$ is

\begin{equation}
\label{eq:F_s}
F_s = 4 \pi r_c^2 \gamma.
\end{equation}

\noindent But this is the free energy, which is not what is needed. The process of surface formation being adiabatic, one should use the internal energy $U \equiv F + TS$ with $S \equiv - \partial F / \partial T$ to find

\begin{equation}
\label{eq:Q_s}
W_s \equiv U_s = 4 \pi r_c^2 \left( \gamma - T \frac{\partial \gamma}{\partial T} \right).
\end{equation}

\noindent As stated, the first term is the free energy of the surface, the energy required to form the surface in the presence of a thermal reservoir. The second term describes the heat drawn from the thermal reservoir ($\partial \gamma / \partial T < 0$, so this term is also positive).

\begin{figure} [t!]
\centering
\includegraphics[scale=0.45]{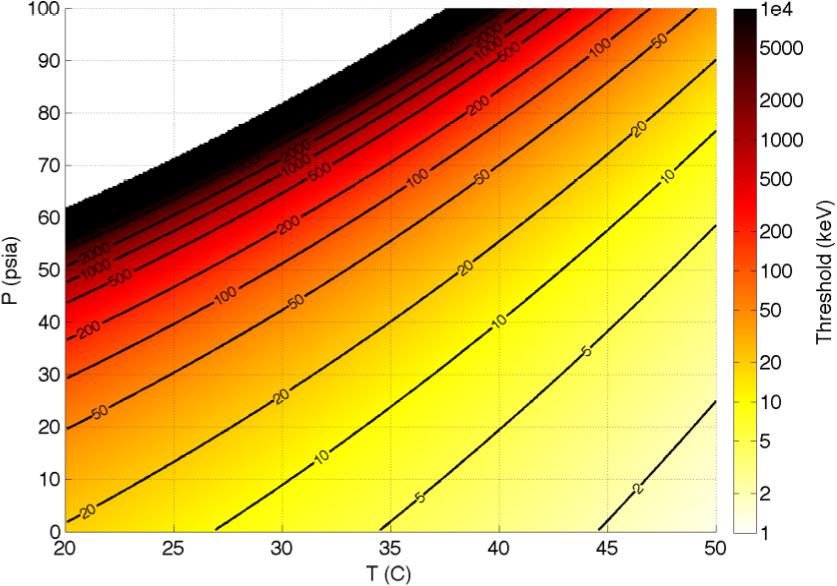}
\caption[Seitz energy threshold as a function of temperature and pressure for CF$_3$I]{Contours of the critical energy for bubble formation as a function of operating temperature and pressure of the superheated liquid for CF$_3$I. Figure from \cite{dahl-11a} reproduced with permission.}
\label{fig:Ec_vs_PT}
\end{figure}

Combining Equations \ref{eq:Q_b} and \ref{eq:Q_s}, the energy required to create a critical bubble is

\begin{equation}
\label{eq:Q}
E_c \equiv W_b + W_s = \frac{4}{3} \pi r_c^3 \rho_b \left( h_b - h_l \right) + 4 \pi r_c^2 \left( \gamma - T \frac{\partial \gamma}{\partial T} \right),
\end{equation}

\noindent which is shown in Figure \ref{fig:Ec_vs_PT} as a function of operating temperature and pressure for a bubble chamber containing CF$_3$I.

\subsection{Stopping Power Threshold}
\label{sec:bubblechambers_dEdx}

\begin{figure} [t!]
\centering
\includegraphics[scale=0.55]{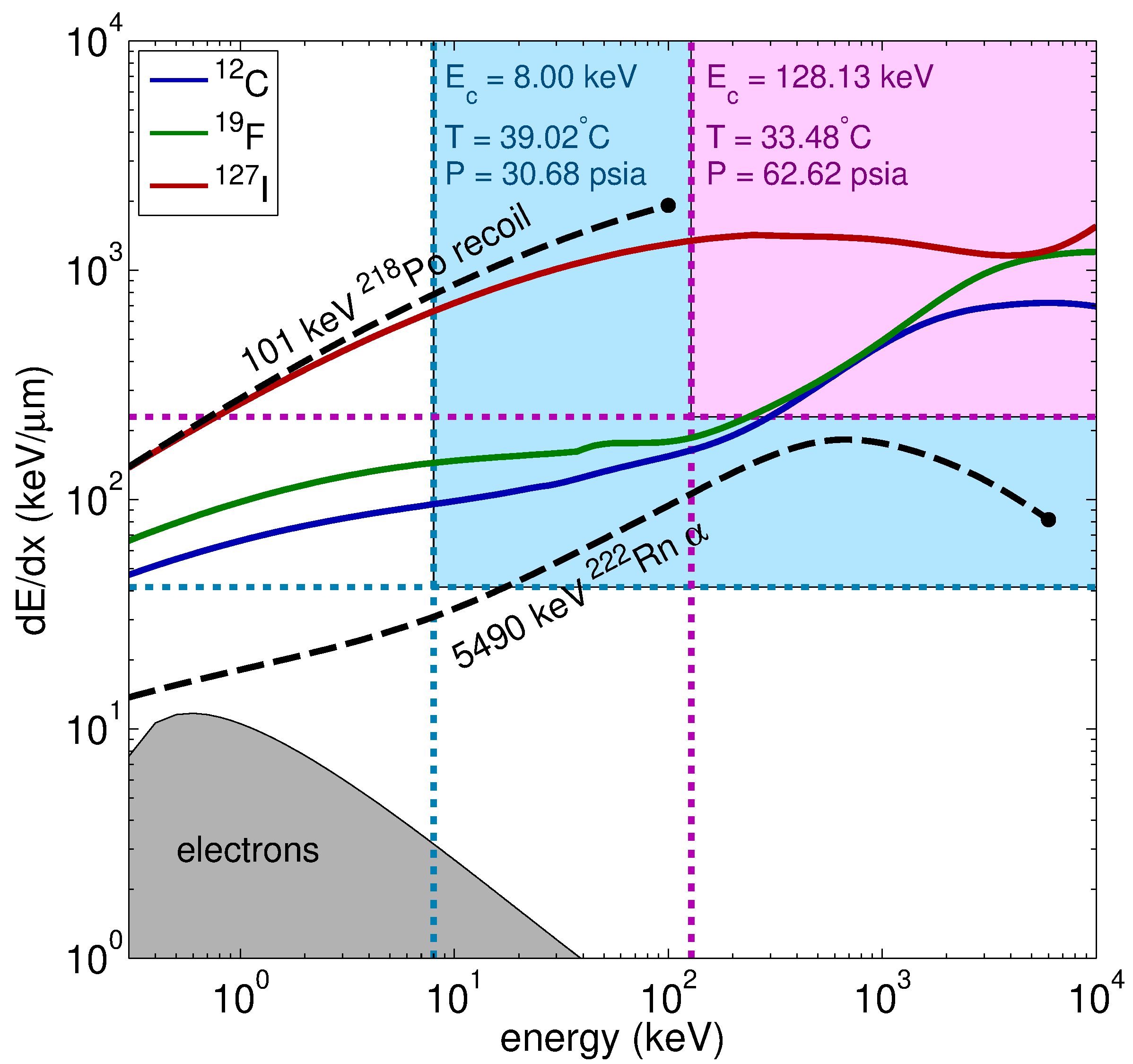}
\caption[Energy and stopping power thresholds for recoils in CF$_3$I]{Recoil curves for various nuclei in CF$_3$I. Thresholds are shown for the extrema of operation conditions (8.00 keV threshold at 39.02$^\circ$C, 30.68 psia and 128.13 keV at 33.48$^\circ$C and 62.62 psia), with a stopping power threshold given by Equation \ref{eq:dEdx} with $a=6.07$. Only recoils that cross both thresholds (shaded regions) can nucleate bubbles, and so the bubble chamber is mostly insensitive to electron recoils (ionization energy loss calculated with the Bethe formula). This idealized figure does not take into account extreme episodes of high $\mathrm{d}E/\mathrm{d}x$ from electrons, such as those mediated by Auger electron emission \citep{peyrou-67}.}
\label{fig:E_vs_dEdx}
\end{figure}

The energy threshold just described serves as one of two thresholds required by the Seitz model, the other being that an energy greater than $E_c$ must be deposited in a space much smaller than a critical range $L_c$, proportional to $r_c$, so as to ensure its local availability for proto-bubble formation. The simplest and most common formulation of this threshold is to add an instantaneous stopping power threshold,

\begin{equation}
\label{eq:dEdx}
\frac{\mathrm{d}E}{\mathrm{d}x} > \frac{E_c}{a r_c},
\end{equation}

\noindent where $a$ is a unitless scaling parameter with various interpretations and calculated or measured values \citep{bell-74,harper-93}. \cite{bell-74} obtains a theoretical value of $a = 6.07$ by using Rayleigh's criteria for instability of a vapor jet in a liquid, as a column of vapor breaks into individual bubbles while conserving vapor volume, although values of $a = 2$ \citep{apfel-79}, $a = 2\pi$ \citep{norman-63}, or even $a=12.96$ \citep{elnagdy-71} are used in the literature.  Based on the stopping power threshold given in Equation \ref{eq:dEdx} with a value of $a = 6.07$, the dual threshold nature of the bubble chamber is nicely illustrated in Figure \ref{fig:E_vs_dEdx}.  This figure relates the instantaneous stopping power of recoils with a given energy. As the recoiling particle slows down, it traces its respective curve shown in Figure \ref{fig:E_vs_dEdx}. Only those recoils with instantaneous stopping power curves within the shaded region are able to nucleate bubbles. Here, we can see the true power of the bubble chamber as a WIMP detector. By varying the temperature and pressure set-points of the bubble chamber, the energy and stopping power thresholds can be tuned, allowing for the elimination of electron recoils as a potential background.

The critical range $L_c$ is not a precisely known quantity and must therefore be experimentally constrained. In fact, it has been shown to vary not only with the properties of the liquid being superheated \citep{poesposoetjipto-70}, but also with the temperature \citep{das-04}. To account for the variation with fluid composition, it is common to replace the critical range used in Equation \ref{eq:dEdx} ($L_c = a r_c$) with one that varies with fluid density

\begin{equation}
\label{eq:L_c}
L_c = b \left(\frac{\rho_b}{\rho_l}\right)^{1/3} r_c,
\end{equation}

\noindent where $b$ is a unitless scaling parameter and  $r_c \left(\rho_b/\rho_l\right)^{1/3}$ is the radius of a fluid volume that when vaporized makes a critical bubble \citep{poesposoetjipto-70}. To account for the temperature-variation, $b$ should be allowed to scale with temperature.

\section{Extensions to the Seitz Model}
\label{sec:bubblechambers_extensions}

\begin{figure} [t!]
\centering
\includegraphics[scale=0.55]{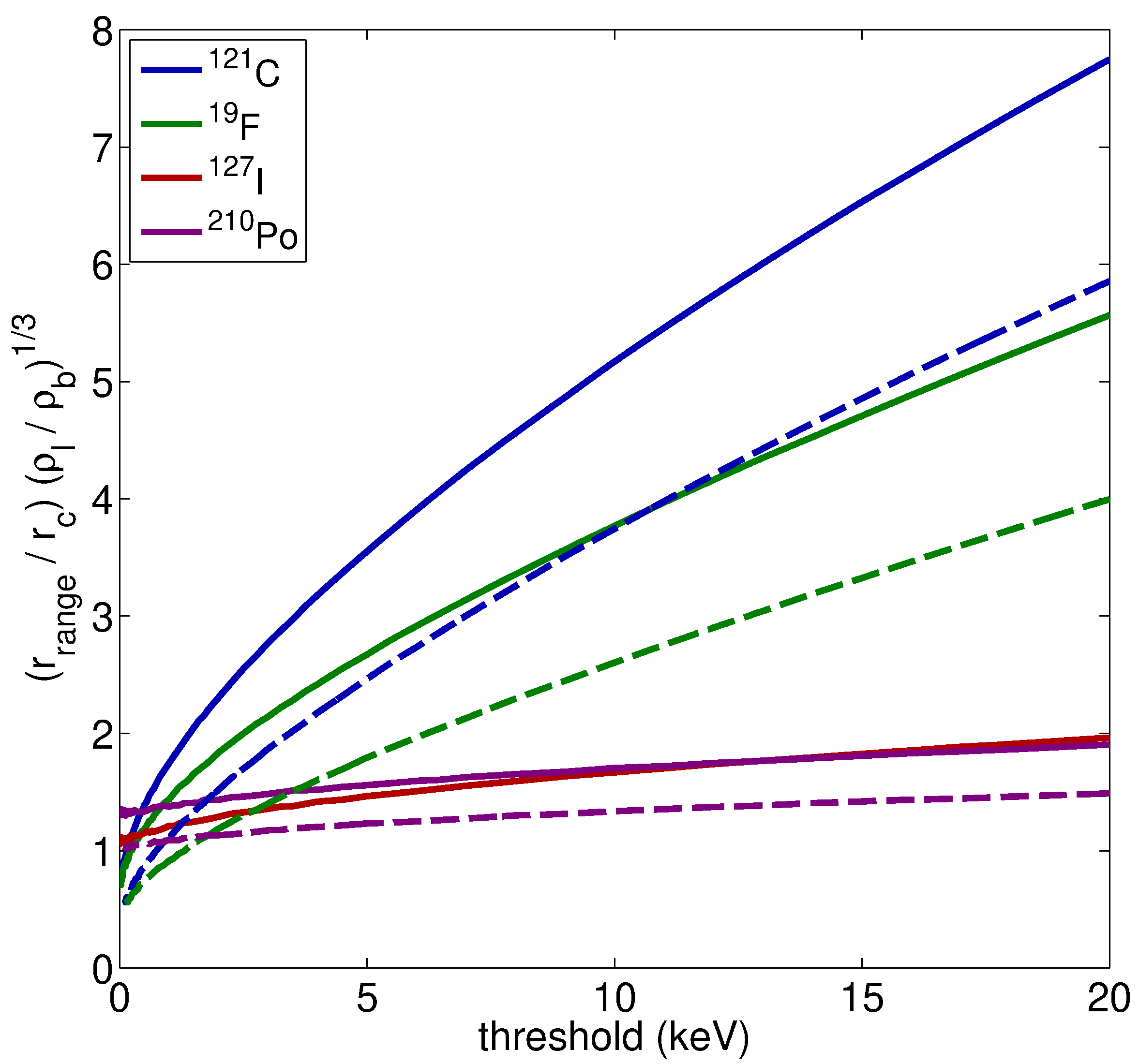}
\caption[Range of recoils in CF$_3$I and C$_4$F$_{10}$]{The mean projected range of recoils in CF$_3$I (solid) and C$_4$F$_{10}$ (dashed). The range of recoils in C$_4$F$_{10}$ is lower than in CF$_3$I for corresponding nuclei (the similar carbon and fluorine nuclei), indicating that CF$_3$I is closer to the region of stopping power threshold importance.}
\label{fig:range_in_CF3I_C4F10}
\end{figure}

To first order, the stopping power threshold is unimportant when the bubble chamber is operated at low energy thresholds (the stopping power threshold being inversely proportional to $E_c$). However, the amount to which the uncertainty in the stopping power matters is dependent not only upon $E_c$ but also upon how diffusely the recoil energy is deposited for different nuclear species in a superheated fluid. Using the TRIM program within the SRIM package, a Monte Carlo simulation packaged developed to calculate the interactions of ions with matter \citep{ziegler-96}, the mean projected range $r_\mathrm{range}$ of each recoil in CF$_3$I (and in PICASSO's target fluid C$_4$F$_{10}$, for reference) was calculated. As can be seen in Figure \ref{fig:range_in_CF3I_C4F10},  the lighter nuclei tend to disperse their energy over longer track lengths, a range which is seen to be longer for CF$_3$I than for C$_4$F$_{10}$. Further, since $r_\mathrm{range}$ is the mean value of the track length and there is some variation in this range, some individual recoils will have shorter track lengths while some will have longer. This dispersion further complicates the use of the simple Seitz model because it generates a non-zero probability of a recoil nominally being above $\mathrm{d}E/\mathrm{d}x$ threshold that still fails to nucleate a bubble due to the extension of its range.

\subsection{Bubble Nucleation Efficiency}
\label{sec:bubblechambers_efficiency}

\begin{figure} [t!]
\centering
\includegraphics[scale=0.55]{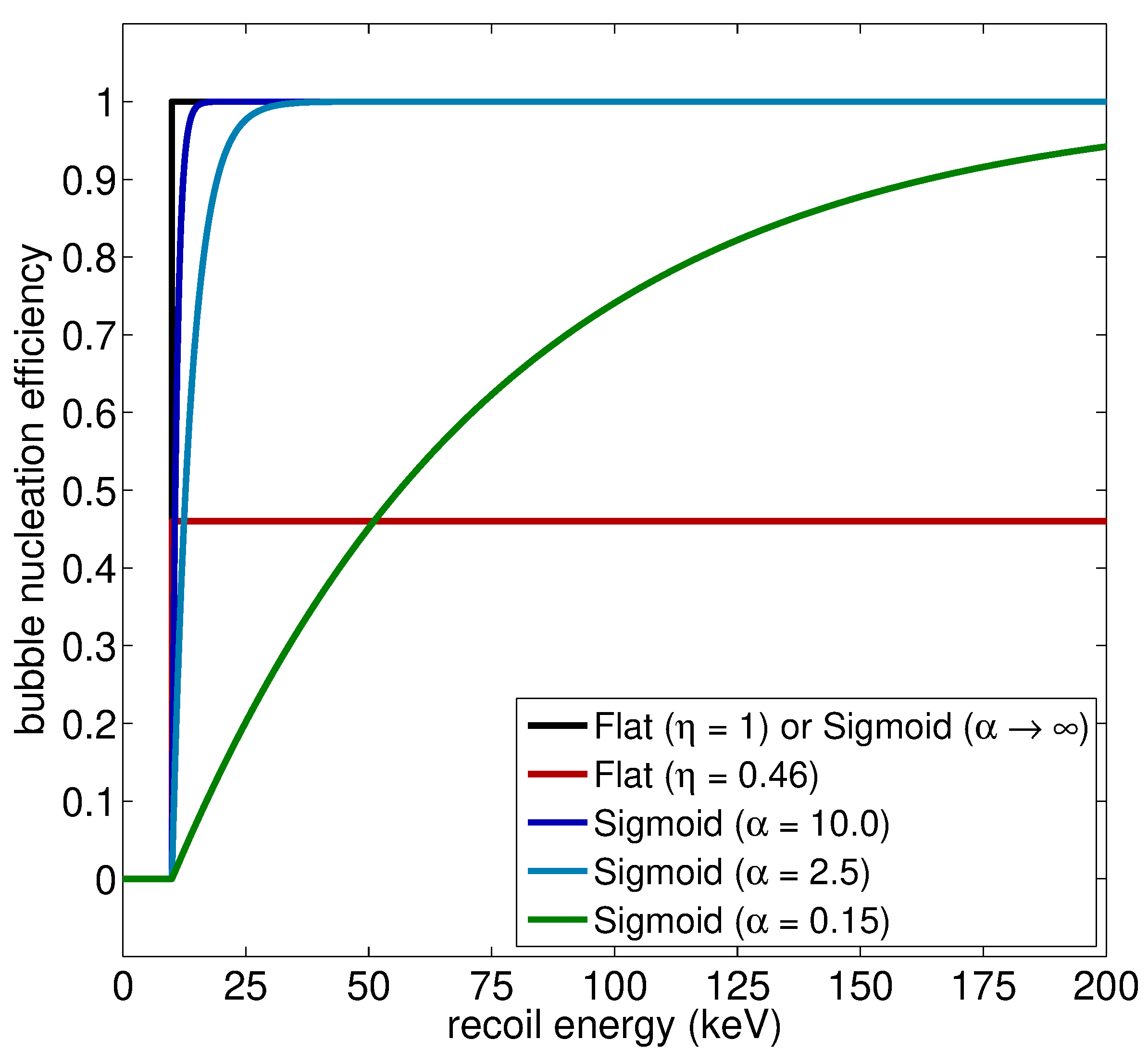}
\caption[Bubble nucleation probability functions for flat and PICASSO models]{The bubble nucleation probability functions for energy threshold $E_c = 10$ keV, given for a flat-efficiency model (Equation \ref{eq:P_flat}) with $\eta = 1$ or $\eta = 0.44$ and for the PICASSO sigmoid-effiency model (Equation \ref{eq:P_PICASSO}) with various choices for $\alpha$.}
\label{fig:nucleation_probability_function}
\end{figure}

The picture painted in Figure \ref{fig:E_vs_dEdx} is incomplete because it does not take into account the effect of these extended recoil track lengths. To incorporate the non-zero probability of nucleation failure, we introduce a bubble nucleation efficiency $\eta$ (Figure \ref{fig:nucleation_probability_function}).   One way of incorporating this second parameter (in addition to $a$ or $b$) is in what is called a flat-effiency model. So far, the two thresholds for bubble nucleation are taken to be step functions by the Seitz theory --- if a recoil is above both the energy threshold (Equation \ref{eq:Q}) and the stopping power threshold (Equation \ref{eq:dEdx}), then this recoil nucleates a bubble. In the flat-efficiency model, the parameter $\eta$ is introduced as a multiplicative factor in front of the bubble nucleation probability function for a recoil of energy $E_r$ and stopping power $\mathrm{d}E_r / \mathrm{d}x$

\begin{equation}
\label{eq:P_flat}
P\left(E_r,\frac{\mathrm{d}E_r}{\mathrm{d}x} ; E_c, L_c \right) = \eta \Theta\left(E_r - E_c\right) \Theta\left(\frac{\mathrm{d}E_r}{\mathrm{d}x} - \frac{E_c}{L_c}\right),
\end{equation}

\noindent where $\Theta$ is the Heaviside step function. The value of the nucleation efficiency $\eta$ is then best-fit using the calibration data found in Chapter \ref{ch:efficiency}.

A second model put forth by PICASSO \citep{archambault-11} replaces the step function turn-on for the stopping power threshold with a slower-rising function (sigmoid). A parameter $\alpha$ is introduced in lieu of $\eta$ that defines the width of the turn-on in the bubble nucleation probability function,

\begin{equation}
\label{eq:P_PICASSO}
P\left(E_r,\frac{\mathrm{d}E_r}{\mathrm{d}x} ; E_c, L_c \right) = \left( 1 - e^{-\alpha \left(E_r - E_c^\mathrm{eff}\right) / E_c^\mathrm{eff}} \right) \Theta \left(E_r - E_c\right) \Theta\left(\frac{\mathrm{d}E_r}{\mathrm{d}x} - \frac{E_c}{L_c}\right),
\end{equation}

\noindent where $E_c^\mathrm{eff}$ is the effective energy threshold given as the maximum of $E_c$ and the projection on the energy axis where the nuclear recoil crosses over the $\mathrm{d}E/\mathrm{d}x$ threshold. For the particular value of $E_c = 128.13$ keV shown as the pink shaded region in Figure \ref{fig:E_vs_dEdx}, $E_c^\mathrm{eff} = E_c$ for $^{127}$I recoils, while it is greater for both $^{12}$C and $^{19}$F recoils. As with $\eta$, the value for $\alpha$ is determined by fitting the calibration data in Chapter \ref{ch:efficiency}. So far, our data does not favor one model over the other. In this work, both will be considered.

\section{Acoustic Emission from Bubble Formation}
\label{sec:bubblechambers_acoustic}

Finally, one characteristic of the bubble chamber that is extremely important to the COUPP experiment is the acoustic energy emission from bubble formation, briefly mentioned in Section \ref{sec:bubblechambers_Q}. The PICASSO experiment has demonstrated the ability to discriminate alpha-induced bubbles from those generated by nuclear recoils by virtue of the alpha events being more intense in ultrasonic frequencies \citep{aubin-08}, a technique which has been confirmed by COUPP \citep{behnke-11}. The mechanism for the dissimilarity observed in the ultrasonic noise between neutron and alpha events is not understood with certainty, but a plausible explanation arises from the fact that individual nuclear recoils will generate one proto-bubble, while alpha decays will generate many (one from the recoiling daughter nucleus from the decay and several along the extended track-length of the alpha particle). These proto-bubbles are all localized within the size of one macroscopic bubble, and so visually there will be no recognizable difference between nuclear recoil events and alpha events. But, by analyzing the ultrasonic emission originating from the early stages of bubble formation, reliable identification of alpha events is possible (see Section \ref{sec:datasets_alpha}).

\singlespacing
\chapter{The COUPP 4 kg Bubble Chamber}
\label{ch:coupp}
\doublespacing

The potential in using a bubble chamber for a WIMP direct detection search is well-motivated. The COUPP experiment has a brief but successful history in operating bubble chambers as WIMP detectors. The first dark matter limits from COUPP were obtained using a 2 kg bubble chamber built at the University of Chicago and tested in the sub-basement laboratory (6 m.w.e. underground) in the Laboratory for Astrophysics and Space Research (LASR). For an engineering run, it was deployed to the MINOS near detector hall at the Fermi National Accelerator Center (Fermilab), at a depth of 225 m.w.e. \citep{michael-08}. This chamber used a natural quartz bell jar, which turned out to be prohibitively radioactive (for the purposes of COUPP), resulting in data with a large alpha background manifested as events on the interface between the CF$_3$I and the bell jar wall. Despite this, the COUPP 2 kg bubble chamber was able to set best-in-world spin-dependent WIMP-proton scattering cross section limits \citep{behnke-08}, motivating the development of future larger and improved chambers. 

\begin{figure} [t!]
\begin{minipage}[b]{0.54\linewidth}
\centering
\includegraphics[scale=0.35]{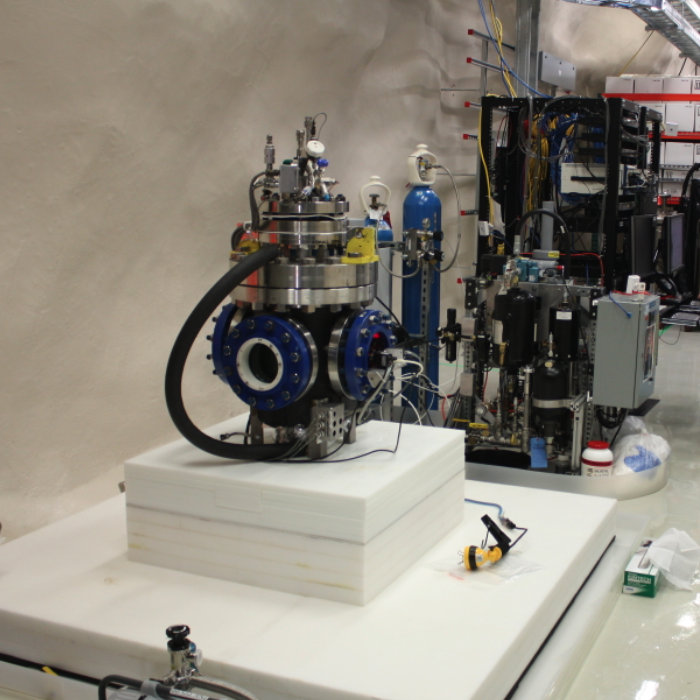}
\end{minipage}
\begin{minipage}[b]{0.46\linewidth}
\centering
\includegraphics[scale=0.35]{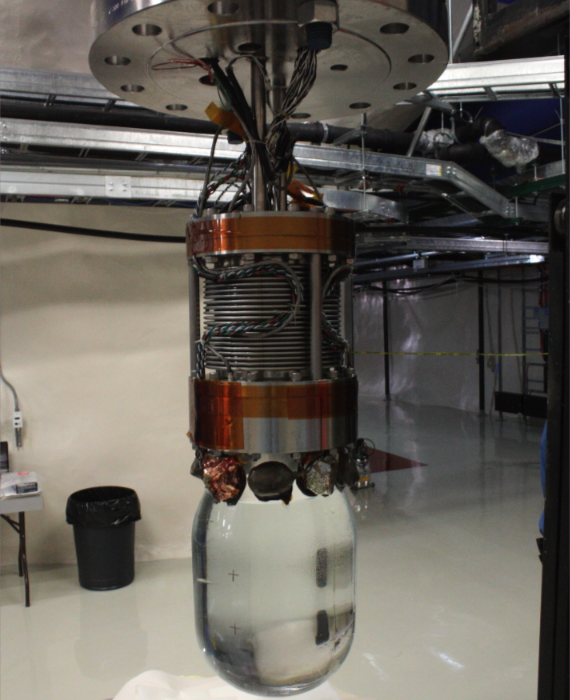}
\end{minipage}
\caption[The COUPP 4kg bubble chamber]{The COUPP 4kg bubble chamber (left) is shown on the platform for holding the water shield. The pressure control cart and computer racks are visible behind it to the right. The inner volume of the bubble chamber (right) is also shown, including the bell jar, bellows, piezoelectric acoustic sensors, and instrumentation all connected to the top flange of the pressure vessel (top flange on left).}
\label{fig:coupp_pictures}
\end{figure}

The COUPP 4 kg bubble chamber was the first such improvement, employing a larger synthetic fused silica bell jar inside the same pressure vessel used for the 2 kg chamber. Using synthetic silica for the bell jar served to reduce the alpha background wall rate from 0.7 events/cm$^2$/day in the natural quartz 2 kg chamber to immeasurably low levels in the 4 kg chamber \citep{behnke-11}. The 4 kg chamber also employed piezoelectric transducers to record the ultrasonic acoustic emission from bubble formation, which PICASSO has shown allows for event-by-event alpha-decay identification \citep{aubin-08}. An engineering run of the COUPP 4 kg bubble chamber at the MINOS near detector hall confirmed this acoustic alpha discrimination and demonstrated reliable remote operations of a bubble chamber, necessary for a deeper deployment. Using a 3.5 kg CF$_3$I target (resulting in 28.1 kg-days of data) in this engineering run, the COUPP 4 kg bubble chamber was again able to set best-in-world spin-dependent WIMP-proton scattering cross section limits \citep{behnke-11}. Figure \ref{fig:coupp_pictures}  shows the pressure vessel (left) and inner volume (right) of the COUPP 4 kg bubble chamber. 

Operating with a 225 m.w.e. overburden was sufficient to shield out the hadronic component of the cosmic ray flux and significantly reduce the cosmic ray muon flux (see Figure \ref{fig:muon_flux_vs_depth}). Residual muons passing near the 4 kg chamber (and thus posing a significant background) were tagged using a 1000 gallon Bicron 517L liquid scintillator muon veto equipped with 19 RCA-2425 PMTs, surrounding the bubble chamber on the sides and top \citep{behnke-11}. These muons are problematic in that they produce cosmogenic neutrons via the $(\mu,n)$ reaction. Although the backgrounds from cosmogenic neutrons can be tagged by the muon veto, the events they generate in the bubble chamber still limit the live time available for a dark matter search. The next step was therefore to eliminate the muon-induced background by deploying the 4 kg bubble chamber to the deep site at SNOLAB (6000 m.w.e.), in the Vale Creighton Mine \#9 near Sudbury, Ontario. The first results of this deployment with a 4.048 kg CF$_3$I target are contained herein. 

The future of COUPP is bright. A successful engineering run of a larger 60 kg bubble chamber in the MINOS near detector hall has just been completed. This bubble chamber likewise uses a synthetic fused silica bell jar and employs many of the same features as the 4 kg chamber, only on a larger scale. Deployment of the 60 kg chamber to SNOLAB is just underway at the time of this writing. Beyond this, a 500 kg bubble chamber is currently under development. Other endeavors of importance to the future of the COUPP experiment are the development of dedicated test chambers to specifically measure recoil thresholds. Two such measurements are currently being carried out. The COUPP Iodine Recoil Threshold Experiment (CIRTE) is designed to use the elastic scattering of pions to study iodine recoils directly. Another experiment using small bubble chambers at Argonne National Laboratory and the University of Chicago is being performed to measure the threshold efficiencies of CF$_3$I using a mono-energetic $^{88}$Y-Be neutron source. 

\section{Inner Volume Assembly}
\label{sec:coupp_inner}

The 4 kg bubble chamber makes use of modestly superheated CF$_3$I sensitive to WIMP-induced nuclear recoils. Though the ideal mechanism for bubble nucleation in the detector is initiated by nuclear recoils, nucleations can also occur on imperfections in the surfaces in contact with the superheated fluid, no matter how smooth the surface may be. These nucleations arise from gas entrapped in the cavities on the surface, which can catalyze the vaporization of the fluid, destabilizing the system. The deactivation of any nucleation sites is necessary to ensure long-term stability of the bubble chamber \citep{bolte-07}. A 150 mm diameter, 3 liter bell jar made of Suprasil synthetic fused silica manufactured by the Heraeus Corporation is used as a smooth-surfaced containment vessel for the superheated CF$_3$I, minimizing the available cavities that can act as nucleation sites. 

The pressure and temperature of the CF$_3$I volume are controlled by a propylene glycol bath, filling a stainless steel pressure vessel and surrounding the synthetic silica bell jar. Temperature of the glycol is controlled by a NESLAB RTE-740 heater/chiller unit, which in turn communicates temperature changes to the CF$_3$I via convection. Pressure of the glycol used as a hydraulic fluid is controlled by a custom-built hydraulic control cart (described below in Section \ref{sec:coupp_hydrauliccart}). Pressure changes to the glycol are transferred to the CF$_3$I volume through a flexible stainless steel bellows atop the synthetic silica bell jar. The bellows not only serves to ensure the contents of the bell jar are at the same pressure as the external hydraulic fluid, it also serves to seal off the inner volume CF$_3$I from the glycol system. However, the bellows is too rough to be in contact with the superheated CF$_3$I, and so approximately 2 liters of ultra-clean liquid (non-superheated) water resides atop the CF$_3$I volume to act as a buffer insulating the CF$_3$I from the metallic bellows.

\section{Pressure Vessel}
\label{sec:coupp_pressurevessel}

The roughly 20 liter glycol bath is contained within a large stainless steel pressure vessel made by Meyer Tool and Manufacturing, rated to 600 pounds per square inch gauge (psig) with pressure relief at 350 psig (operated only below 250 psig) and weighing less that 1500 pounds. The pressure vessel used for the initial 2 kg bubble chamber was recycled for use with the 4 kg bubble chamber, with some modifications to the top flange allowing for appropriate feedthroughs useful for operation and a stainless steel vertical extender to accommodate the larger inner volume accompanying the 4 kg chamber. During operation, the entire pressure vessel was thermally insulated by multiple layers of Reflectix double reflective insulation. 

One of the triggering mechanisms for events in the bubble chamber is video-based motion detection, and therefore 600 psig pressure-rated 6" diameter Canty Fuseview borosilicate glass viewports were installed at the vertical position of the CF$_3$I volume. Initially, orthogonal imaging of events was desired, and so four orthogonal viewports were built in to the pressure vessel. However, in order to properly expose the images with the 10 Luxeon Star high-powered red LEDs (necessary because shorter wavelength light degrades the chemical stability of CF$_3$I), the pressure vessel wall opposite of the cameras had to be equipped with 3M retroreflective backing, eliminating the ability to use all but one viewport. Still, two 100 frames per second Basler A602f firewire cameras were mountable within one viewport, with 20$^\circ$ stereo angle separation sufficient to provide good 3-dimensional position reconstruction of bubbles in the chamber. 

\section{Hydraulic Control Cart}
\label{sec:coupp_hydrauliccart}

The hydraulic control cart manages the pressure of the glycol in the pressure vessel which acts as a hydraulic fluid to regulate the operating pressure of the bubble chamber. The hydraulic cart is equipped with a hydraulic piston driven by a stepper motor which regulates the pressure when the chamber is expanded (the CF$_3$I is superheated). This maintains the level of superheat over the life of the expansion. A fast solenoid valve plus pneumatic-to-hydraulic converter manages the expansion and fast re-compression of the chamber. When an event produces a trigger (see Section \ref{sec:coupp_operations}), the fast pneumatic piston fires, re-compressing the chamber and thus collapsing the bubble back to its liquid state, before a significant amount of the CF$_3$I is vaporized. Operation of the hydraulic cart is in principle maintained by a LabVIEW program that is part of the data acquisition (DAQ) system, but protections are in place to park the bubble chamber in a safe condition (\emph{i.e.} compressed state) in the event of a power outage or communications failure. 

\section{Temperature, Pressure, and Acoustic Sensors}
\label{sec:coupp_sensors}

\begin{figure} [t!]
\centering
\includegraphics[scale=0.36]{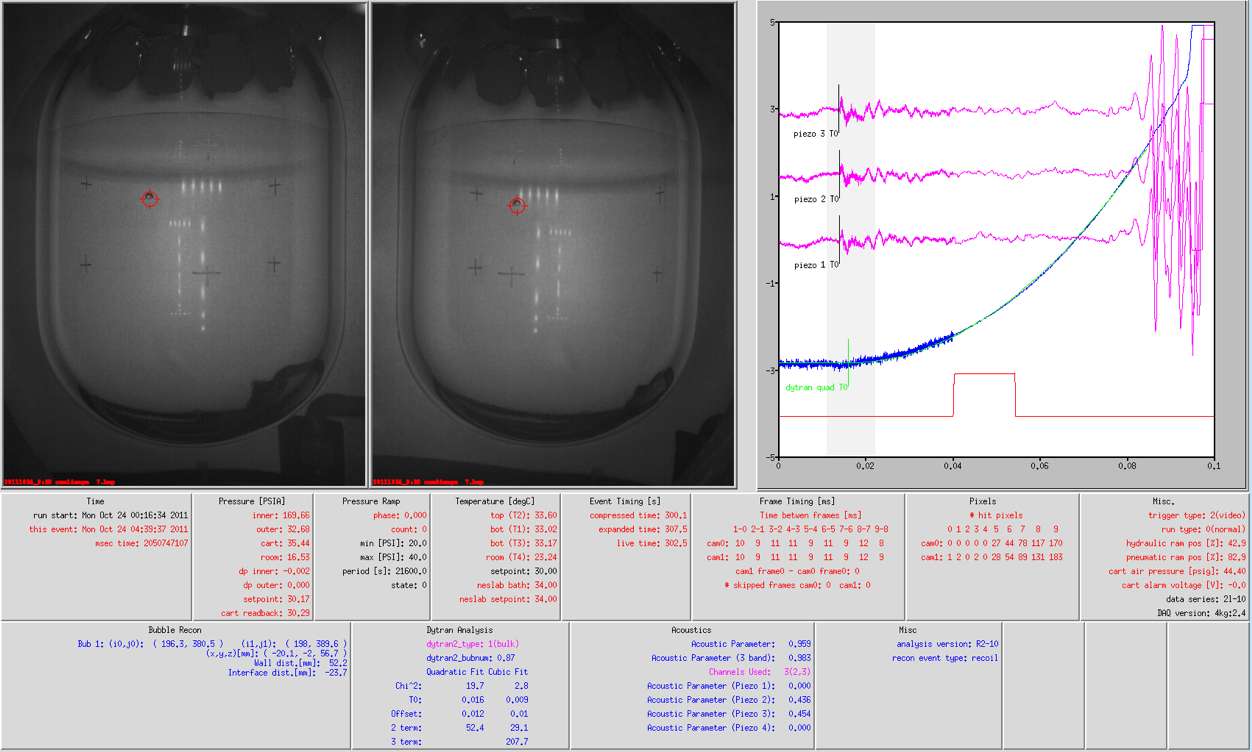}
\caption[Typical single-bubble event in the 4 kg chamber]{An example of a typical single-bubble event in the COUPP 4 kg bubble chamber. Shown on the left are the stereo images of the event, with a single bubble in the bulk and cross-hairs identifying the position reconstruction coordinates. The right shows the three piezo traces (magenta) and the Dytran trace (blue) with a quadratic fit (green). Information on the bottom lists the relevant temperature, pressure, acoustic, \emph{etc.} information for the event.}
\label{fig:coupp_event}
\end{figure}

In order to gather information necessary to perform a proper analysis of events in the 4 kg bubble chamber, many parameters need to be extracted for each event. Figure \ref{fig:coupp_event} shows a typical single-bubble event in the 4 kg chamber, along with the values of relevant variables useful for the analysis.  The sensors used to gather temperature, pressure, and acoustic information are described below.

The temperature of the system is controlled by a NESLAB RTE-740 heater/chiller unit, which is maintained by a computer in the DAQ system through a single serial communications cable. The heater/chiller controls the temperature using its own resistance temperature detectors (RTD) measurement circuit and an algorithmic proportional-integral-derivative (PID) control loop, with readings of its own bath temperature and from its external 3-wire RTD thermometer (T3), the latter of which is mounted on the bottom of the bellows. There are also two Omega PT100 class B 4-wire lead RTDs mounted on the bellows (T1 on the bottom and T2 on the bottom) which provide reference temperatures of the glycol system at different heights. A Sensortec 4-wire RTD is mounted external to the hydraulic cart to measure the room temperature (T4). These RTDs are fed into a National Instruments (NI) SCC-RTD01 module on an SC-2345 instrumentation chassis on the DAQ rack.

There are also five pressure sensors in the system. Two Setra GCT-225 ultra high-purity pressure transducers measure the pressure of the water/CF$_3$I volume (PT1) and the glycol pressure inside the pressure vessel (PT2).  Two Omegadyne PX309 pressure transducers measure the glycol pressure in the hydraulic cart (PT4) and the ambient pressure of the room (PT5). These transducers are fed into SCC-CI01 modules also on the SC-2345. An AC-coupled Dytran 2005V fast pressure transducer powered by a Dytran 4112B current source measures the pressure of the water/CF$_3$I volume (PT3) and is sampled at 10 kHz for 160 ms around the onset of bubble nucleation. It can therefore be used not only as a pressure gauge, but also as an event trigger and a data quality parameter. Analysis of the pressure rise of the Dytran trace (Figure \ref{fig:coupp_event}, right) shows that events in the bulk of the CF$_3$I volume follow a quadratic rise, while events on the CF$_3$I/bell jar wall or at the CF$_3$I/water interface are best-fit to a cubic function \citep{lippincott-11b}. This allows for event-by-event cross-checks of the 3-dimensional position reconstruction using camera images and universal event type identification. Further, the steepness of the pressure rise also scales with the number of bubbles in an event, allowing for the Dytran to provide information on bubble multiplicity for each event.

The rapid expansion of a bubble generates audible and ultrasonic emissions that provide valuable information on the nature of the bubble formation and allow for acoustic alpha event discrimination (see Section \ref{sec:bubblechambers_acoustic}). To record this acoustic information for each event, Ferroperm lead zirconate titanate (PZT) piezoelectric transducers (piezos) instrumented by collaborators at Indiana University South Bend were epoxied to the top of the bell jar. During the engineering run at Fermilab, four piezos were applied. These were left attached for the SNOLAB run, but four new piezos were applied and used in their place. Therefore, of the total eight piezos attached during this run, four were powered and three were used for the analysis (one of the new transducers failed shortly after commissioning). The piezos can be seen at the top of the jar in Figure \ref{fig:coupp_event}, above the CF$_3$I/water interface. 

\section{Data Acquisition Rack}
\label{sec:coupp_DAQ}

Signals from all of these sensors (and others) were fed into components of a DAQ rack. This rack contained the aforementioned NI SC-2345 chassis containing the modules reading the data from the temperature and pressure sensors. The acoustic sensors were powered by a distribution box, manufactured by the Fermilab Particle Physics Division Electrical Engineering Department. Their signal was managed by an NI PXI-6133 digitizer with a 2.5 MHz sampling rate. This digitizer was mounted on an NI PXI-1044 chassis, which also contained a processor for running the DAQ computer system (NI PXI-8106), a NI PXI-8252 firewire interface for the cameras, and a PXI-6221 to maintain the data input/output as a history buffer, specifically for pressure information from the hydraulic cart. The LEDs were powered by an LED driver distribution box manufactured at Fermilab, which managed the camera control signals and switched the LED driving voltages to power the array. A 100 Hz clock was provided to the cameras by a Hewlett-Packard 33120A arbitrary waveform generator.

\section{Water Shielding}
\label{sec:coupp_shielding}

\begin{figure} [t!]
\centering
\includegraphics[scale=0.45]{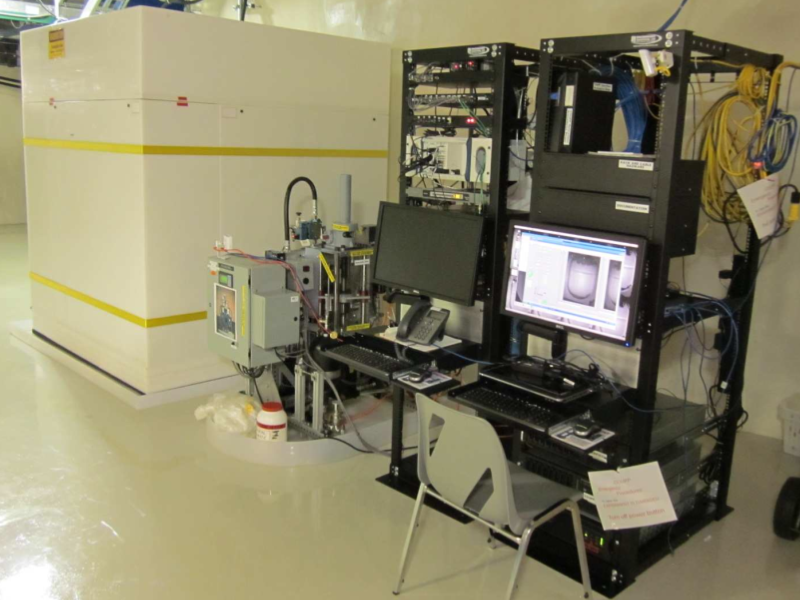}
\caption[COUPP 4 kg chamber with water shield installed]{The entire COUPP 4 kg bubble chamber setup, including the water shield surrounding the detector, the hydraulic cart, and the DAQ rack. Picture courtesy E. Vazquez-Jauregui.}
\label{fig:coupp_watershield}
\end{figure}

In order to shield the bubble chamber from the environmental neutron flux coming from radioactive decays in the cavern walls (see Section \ref{sec:backgrounds_rock}), a low-$Z$ moderating material of sufficient thickness was required. In SNOLAB, a large water shield was employed (Figure \ref{fig:coupp_watershield}).  Below the detector, a 20" thick high-density polyethylene (HDPE) pedestal was used to shield neutrons coming from the floor. This pedestal had a 6" thick base extending outward in each direction which held the walls of the shielding (the pedestal is visible in Figure \ref{fig:coupp_pictures}, left). The walls of the shielding were made of four water-filled HDPE containers, with 0.5" thick walls and a total thickness (HDPE plus water) of 20". The roof of the water shielding was made of a water-filled HDPE container, again with 0.5" thick walls and a total thickness of 20". The roof had an off-axis feedthrough built in to allow for the insertion of radioactive sources for calibrations. When not in use, this feedthrough was plugged with an appropriately-sized HDPE plug. 

\section{Bubble Chamber Operations}
\label{sec:coupp_operations}

As discussed in Chapter \ref{ch:bubblechambers}, by tuning the operating temperature and pressure of the bubble chamber, the threshold for bubble nucleation by incident radiation can be regulated. The superheat of the CF$_3$I is adjustable be either changing the temperature (relative to the boiling point) or the pressure (relative to the vapor pressure). However, it was seen during operation that at higher pressures, ultrasonic acoustic emission from bubble growth was reduced significantly. Therefore, during physics runs of the chamber, the recoil threshold was controlled by the temperature. 

When ready for a data-taking run, the hydraulic cart expands the bubble chamber to a specified pressure below the vapor pressure (at a constant temperature specified by the choice of threshold), thus superheating the CF$_3$I volume. Upon triggering on an event (be it a bubble nucleation or something systematic), the fast pneumatic piston re-compresses the system to around 200 psig, collapsing the bubble back to its liquid state before vaporization of large amounts of CF$_3$I can take place. It is left in a compressed state for 30 s (300 s every tenth event) to re-condense the CF$_3$I vapor. After re-expanding, valid live-time for the data begins 30 s after an expansion, allowing for equilibration of the CF$_3$I.

Events are triggered by multiple means. The primary trigger is from frame-by-frame differences in the image data (video trigger), which is able to detect a bubble shortly after becoming visible. A safe-guard trigger is in place in the event of multiple consecutive frames being skipped in the video buffer (frame-skip trigger), which could demonstrate that the video trigger is unreliable. The pressure rise measured by the Dytran can also be used as a trigger (pressure rise trigger). Finally, expansions that exceed a maximum 500 s time limit without bubbles are terminated (timeout trigger). 

\begin{figure} [t!]
\centering
\includegraphics[scale=0.58]{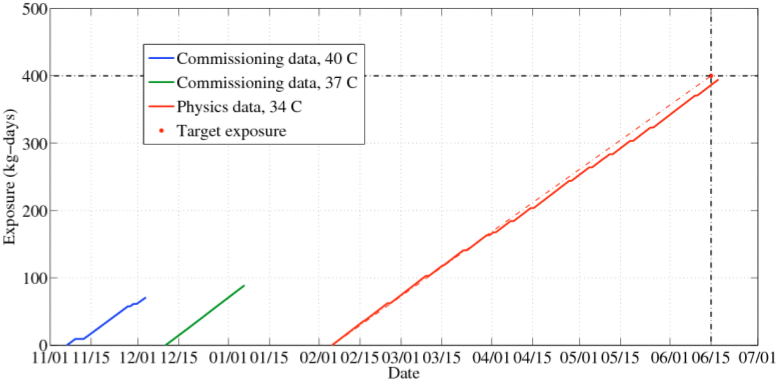}
\caption[Data run log and exposures for each run]{The dates over which the two commissioning runs and one physics run were compiled, along with the total exposure (in kg-days) of each. Figure courtesy C.E. Dahl.}
\label{fig:data_runs}
\end{figure}

Figure \ref{fig:data_runs} shows the time-frame of data-taking for the results reported here.  The water shield installation (the final step in the installation of the 4 kg chamber at SNOLAB) was completed on November 3, 2010. Two commissioning runs were performed at two different high degrees of superheat (lower thresholds), followed by a remedial trip to SNOLAB to install a de-gassing vessel to properly extract all gases dissolved in the glycol hydraulic fluid. A third run began in early February 2011 at a higher threshold, which included dedicated dark matter search runs as well as calibration runs with neutron and gamma sources in place. These three data sets provided dark matter search data at three different thresholds, resulting in total exposures of 17.5, 21.8, and 97.2 live days of data at 8, 11, and 16 keV thresholds for bubble nucleation by nuclear recoils, respectively. 

\singlespacing
\chapter{Dark Matter Search and Calibration Data Sets}
\label{ch:datasets}
\doublespacing

Each event in the dark matter search data for the COUPP 4 kg bubble chamber must undergo a series of data quality cuts to ensure that only proper events are accepted into the data set. This chapter highlights the cuts applied to single and multiple bubble events and the efficiency of each of these cuts in selecting only the proper WIMP-nucleon recoil-like events in the single-bubble data, as derived from neutron calibrations. These cuts make use of particular variables recorded for each event, each of which is introduced as needed below.

\section{Temperature and Pressure Variables}
\label{sec:datasets_variables}

The three different data sets described in Section \ref{sec:coupp_operations} (and the multiple configurations during calibration runs) require uniform temperature and pressure readings for each event to ensure that the threshold across the data set is consistent. In order to determine this threshold, the temperature and pressure variables used in this analysis must be introduced. As described in Section \ref{sec:coupp_sensors}, there are three temperature sensors mounted in the propylene glycol hydraulic fluid volume: T1 and T3 at the bottom of the bellows and T2 at the top. With the heating elements of the chamber being at the bottom of the glycol volume as opposed to the top, convection currents arise allowing for the even distribution of heat throughout the chamber. Therefore, it is assumed that the temperature within the pressure vessel is uniform. All three sensors have readings within a half a degree of each other, but it is unknown how well any is calibrated. Out of the three sensors, the reading for T3 (denoted $T_3$) has been chosen to represent the temperature of each event because it has the smallest standard deviation among the options (the smallest statistical uncertainty $\sigma_\mathrm{T3}$). The systematic uncertainty to the temperature measurement is estimated to be 0.5$^\circ$C, which is the average weighted over each dark matter search data set of the standard deviation of the median values from each of the three sensors ($\pm 0.25^\circ$C) plus an additional uncertainty assigned \emph{ad hoc} to take into account any temperature differential between the RTDs and the actual location of the bubble ($\pm 0.25^\circ$C). The temperature variable of a particular event is

\begin{equation}
\label{eq:T_event}
T_\mathrm{event} = T_3 \pm \sigma_\mathrm{T3} \:(\mathrm{stat.}) \pm 0.5^\circ\mathrm{C} \:(\mathrm{syst.})
\end{equation}

The pressure variable is more complicated. There are two pressure transducers capable of measuring the pressure of the inner volume: PT1 and PT2 as described in Section \ref{sec:coupp_sensors}. PT1 measures the pressure of the water/CF$_3$I volume and is located about 4 cm above the top flange. PT2 measures the pressure of the glycol volume and is located about 12 cm above the top flange. As described in Appendix \ref{ch:P1offset}, while these were initially calibrated and reading the same value, several offsets to the baseline of the PT1 reading occurred throughout running, all of which were determined to be inconsequential save for one, corresponding to an overextension of the bellows which could have resulted in a pressure differential arising between the glycol and the water/CF$_3$I volume. Because of these offset steps to PT1, it was determined that the best pressure measurement to use was that provided by PT2 (denoted $PT_2$), with the bellows overextension step factored in as in Equation \ref{eq:PT1_correction}. In addition to this, the pressure differential arising from the difference in liquid depth between the PT2 sensor and the CF$_3$I volume is appreciable --- 12 cm above the top flange to the center of the CF$_3$I volume (64 cm height differential) gives a total hydrostatic pressure differential ($\Delta P = \rho g h$) of 0.9 psia. With this measurement, a statistical uncertainty $\sigma_\mathrm{PT2}$ and \emph{ad hoc} systematic uncertainty of 0.5 psia is included, taking into account all the unknowns that are present with sensor calibrations and readings. The pressure variable of a particular event is

\begin{equation}
\label{eq:P_event}
P_\mathrm{event} = \left(PT_2 -2.84\:\mathrm{psia} + 0.9\:\mathrm{psia}\right) \pm \sigma_\mathrm{PT2} \:(\mathrm{stat.}) \pm 0.5\:\mathrm{psia} \:(\mathrm{syst.})
\end{equation}

\section{Data Set Identification Cuts}
\label{sec:datasets_identification}

There are three global types of cuts that should be applied to the data. The first are overall data set identification cuts, which are in place to ensure that the events considered as part of a data set are consistent across the set. The identification cuts are separated into two types: expansion time cuts and threshold cuts, described below. Only events that pass both of these data identification cuts are included in a data set and are considered when calculating the live-time of the set. Specifics for all dark matter search and calibration runs after both data identification cuts are shown in Table \ref{tab:data_sets}.

\begin{table}[p!]
\centering
\small{
\begin{tabular} {| c | c | c | c | c |}
\hline
\multicolumn{5}{| l |}{Dark matter search data} \\
\hline
Label & Temperature & Pressure & Threshold & Live-time \\
 & \small{($^\circ$C)} &  \small{(psia)} &  \small{(keV)} &  \small{(days)} \\
\hline
DM-34$^\circ$C & 33.48$\pm$0.04 & 30.64$\pm$0.17 & 15.83$^{+1.46}_{-1.31}$ & 97.24 \\
DM-37$^\circ$C & 36.24$\pm$0.01 & 30.67$\pm$0.15 & 11.21$^{+0.97}_{-0.88}$ & 21.84 \\
DM-40$^\circ$C & 39.02$\pm$0.01 & 30.68$\pm$0.25 & 8.00$^{+0.67}_{-0.61}$ & 17.53 \\
\hline
\hline
\multicolumn{5}{| l |}{Neutron calibration data} \\
\hline
Label & Temperature & Pressure & Threshold & Live-time \\
 & \small{($^\circ$C)} &  \small{(psia)} &  \small{(keV)} &  \small{(days)} \\
\hline
AmBe-36"-34$^\circ$C-30psia & 33.44$\pm$0.04 & 30.65$\pm$0.12 & 15.92$^{+1.46}_{-1.31}$ & 1.50 \\
AmBe-36"-34$^\circ$C-40psia & 33.46$\pm$0.01 & 41.23$\pm$0.27 & 26.67$^{+3.06}_{-2.68}$ & 0.49 \\
AmBe-36"-34$^\circ$C-50psia & 33.46$\pm$0.01 & 51.92$\pm$0.09 & 51.92$^{+7.48}_{-6.32}$ & 0.74 \\
AmBe-36"-34$^\circ$C-60psia & 33.48$\pm$0.01 & 62.62$\pm$0.12 & 128.13$^{+26.63}_{-20.98}$ & 0.82 \\
AmBe-36"-37$^\circ$C-30psia & 36.22$\pm$0.04 & 30.73$\pm$0.33 & 11.27$^{+1.02}_{-0.92}$ & 1.04 \\
AmBe-36"-37$^\circ$C-35psia & 36.20$\pm$0.01 & 36.13$\pm$0.17 & 14.10$^{+1.33}_{-1.19}$ & 0.42 \\
AmBe-36"-40$^\circ$C-30psia & 39.00$\pm$0.02 & 30.64$\pm$0.16 & 8.01$^{+0.66}_{-0.60}$ & 0.51 \\
AmBe-36"-41$^\circ$C-30psia & 39.90$\pm$0.01 & 30.68$\pm$0.16 & 7.21$^{+0.58}_{-0.53}$ & 0.04 \\
AmBe-34"-34$^\circ$C-30psia & 33.43$\pm$0.01 & 30.65$\pm$0.18 & 15.94$^{+1.47}_{-1.32}$ & 0.38 \\
AmBe-38"-34$^\circ$C-30psia & 33.44$\pm$0.01 & 30.56$\pm$0.15 & 15.86$^{+1.45}_{-1.31}$ & 0.38 \\
Cf252-42"-37$^\circ$C-29psia & 36.21$\pm$0.01 & 30.31$\pm$0.17 & 11.09$^{+0.96}_{-0.87}$ & 3.69 \\
Cf252-42"-37$^\circ$C-30psia & 36.21$\pm$0.02 & 30.98$\pm$0.24 & 11.39$^{+1.01}_{-0.91}$ & 9.38 \\
Cf252-54"-34$^\circ$C-30psia & 33.43$\pm$0.02 & 30.72$\pm$0.10 & 15.99$^{+1.46}_{-1.32}$ & 7.31 \\
Cf252-54"-34$^\circ$C-40psia & 33.47$\pm$0.01 & 41.33$\pm$0.09 & 26.78$^{+2.98}_{-2.62}$ & 2.67 \\
\hline
\hline
\multicolumn{5}{| l |}{Gamma calibration data} \\
\hline
Label & Temperature & Pressure & Threshold & Live-time \\
 & \small{($^\circ$C)} &  \small{(psia)} &  \small{(keV)} &  \small{(days)} \\
\hline
Mix-62"-37$^\circ$C-30psia & 36.22$\pm$0.01 & 30.75$\pm$0.10 & 11.27$^{+0.97}_{-0.88}$ & 1.85 \\
Mix-62"-40$^\circ$C-30psia & 38.99$\pm$0.04 & 30.57$\pm$0.09 & 8.00$^{+0.65}_{-0.59}$ & 1.01 \\
Mix-62"-41$^\circ$C-30psia & 39.91$\pm$0.01 & 30.76$\pm$0.12 & 7.22$^{+0.58}_{-0.53}$ & 1.75 \\
Co60-55"-34$^\circ$C-30psia & 33.52$\pm$0.02 & 30.67$\pm$0.11 & 15.77$^{+1.44}_{-1.30}$ & 3.83 \\
Co60-55"-40$^\circ$C-30psia & 39.04$\pm$0.02 & 30.61$\pm$0.12 & 7.96$^{+0.65}_{-0.59}$ & 0.07 \\
Co60-55"-40$^\circ$C-32psia & 39.06$\pm$0.01 & 33.24$\pm$0.15 & 8.73$^{+0.74}_{-0.67}$ & 0.16 \\
Co60-55"-40$^\circ$C-35psia & 39.06$\pm$0.01 & 35.90$\pm$0.11 & 9.65$^{+0.84}_{-0.76}$ & 0.54 \\
Co60-55"-40$^\circ$C-37psia & 39.06$\pm$0.01 & 38.53$\pm$0.12 & 10.68$^{+0.97}_{-0.87}$ & 0.22 \\
Ba133-55"-34$^\circ$C-30psia & 33.48$\pm$0.02 & 30.65$\pm$0.17 & 15.84$^{+1.46}_{-1.31}$ & 6.27 \\
Ba133-55"-37$^\circ$C-30psia & 36.25$\pm$0.02 & 30.65$\pm$0.09 & 11.19$^{+0.96}_{-0.87}$ & 2.96 \\
Ba133-55"-40$^\circ$C-30psia & 39.01$\pm$0.01 & 30.58$\pm$0.12 & 7.99$^{+0.65}_{-0.59}$ & 0.24 \\
\hline
\end{tabular} }
\caption[Data sets used for analysis]{The dark matter search and calibration data sets used for this analysis. The median temperature and pressure values for each set are listed only with statistical errors, while the threshold energies are calculated from these values (using Equation \ref{eq:Q}) with ranges from worst-case scenario of total errors from Equations \ref{eq:T_event} and \ref{eq:P_event} (in quadrature). The live-time for each set is also listed.}
\label{tab:data_sets}
\end{table}

\subsection{Expansion Time Cut}
\label{sec:datasets_expansion}

After each expansion of the bubble chamber, some window of time should be established to allow for the equilibration of the CF$_3$I. Events that occur too soon after the expansion of the chamber can result from any number of thermodynamic fluctuations or other such instabilities, leading to triggers that can not reliably be identified as bubbles occurring at a specific threshold. Thus, all events that occur less than 30 s after an expansion are cut out of the data. This expansion time cut will also affect the overall live-time of a data set, since the first 30 s of any expansion will be discounted, by definition. Therefore, when calculating the live-time of a set, 30 s is subtracted from the total expansion time for each event in the set.

\subsection{Threshold Cut}
\label{sec:datasets_threshold}

Data sets can include events recorded during which the temperature or pressure are still ramping towards their set-points or which have otherwise spurious $T_3$ or $PT_2$ readings. These events will contribute data which are not representative of the data set as a whole. Therefore, another cut should be established to ensure that all data sets only include events with operating threshold (calculated from $T_\mathrm{event}$ and $P_\mathrm{event}$) within some range from the median threshold from the set. The most conservative range that can be assigned to the threshold for a data set is to take the worst-case scenario threshold from the extrema of the temperature and pressure uncertainties in Equations \ref{eq:T_event} and \ref{eq:P_event} (in quadrature). All events that occur outside of this conservative threshold range are eliminated from the data set, prior to any further analysis.

\section{Fiducial Volume Cut}
\label{sec:datasets_fiducial}

\begin{figure} [t!]
\begin{minipage}[b]{1.0\linewidth}
\centering
\includegraphics[scale=0.45]{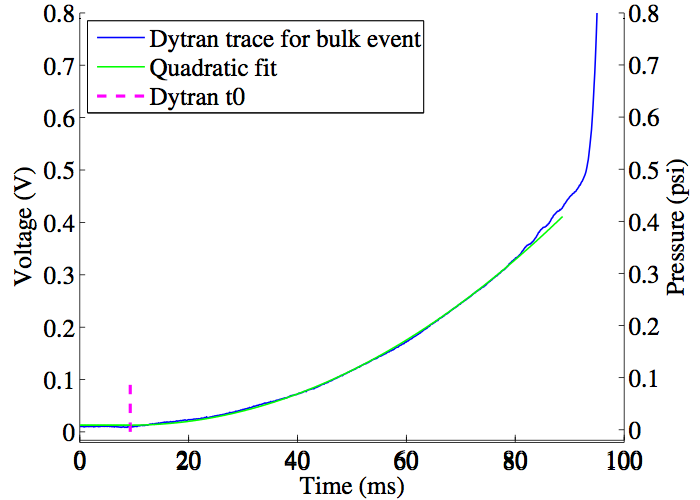}
\end{minipage} \\
\begin{minipage}[b]{0.5\linewidth}
\centering
\includegraphics[scale=0.41]{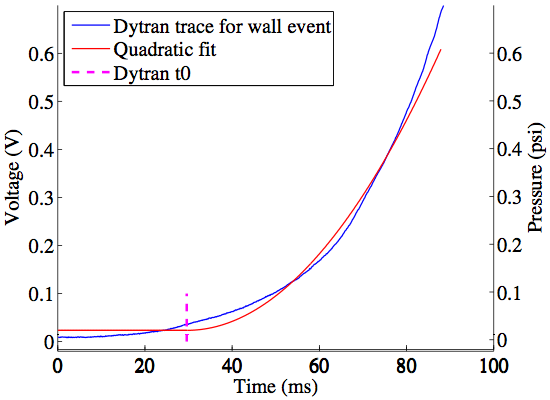}
\end{minipage}
\begin{minipage}[b]{0.5\linewidth}
\centering
\includegraphics[scale=0.41]{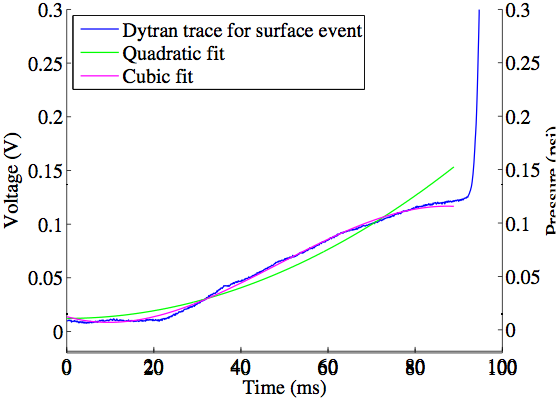}
\end{minipage}
\caption[Dytran traces from bulk, wall, and surface events]{Example characteristic Dytran traces for a bulk event (top), wall event (bottom left), and surface event (bottom right). The Dytran trace for bulk events is well-fit to a quadratic rise, while quadratic fit to either wall or surface events do not. A well-fitting cubic rise is shown in the case of the surface event. Figure from \cite{lippincott-11b} reproduced with permission.}
\label{fig:dytran_traces}
\end{figure}

The second type of cut placed on the data is the fiducial volume cut. For previous versions of the COUPP bubble chambers, a cut was put in place to ensure that all events considered take place within the bulk of the CF$_3$I volume, not on the water/CF$_3$I interface or on the wall of the bell jar. This fiducial volume cut was typically established by comparing the position of the bubble from the position reconstruction package of the data analysis library with some decided-upon spatial boundaries separating bulk events from surface or wall events. However, in the current run of the COUPP 4 kg bubble chamber at SNOLAB, it was observed that the trace of the pressure rise recorded by the Dytran 2005V fast pressure transducer (described in Section \ref{sec:coupp_sensors}) could accurately be used to identify bulk events \citep{lippincott-11b}. Specifically, bulk events follow a quadratic expansion while surface or wall events do not. In fact, surface events tend to follow a cubic expansion with a negative cubic coefficient (See Figure \ref{fig:dytran_traces}). 

\begin{figure} [t!]
\centering
\includegraphics[scale=0.55]{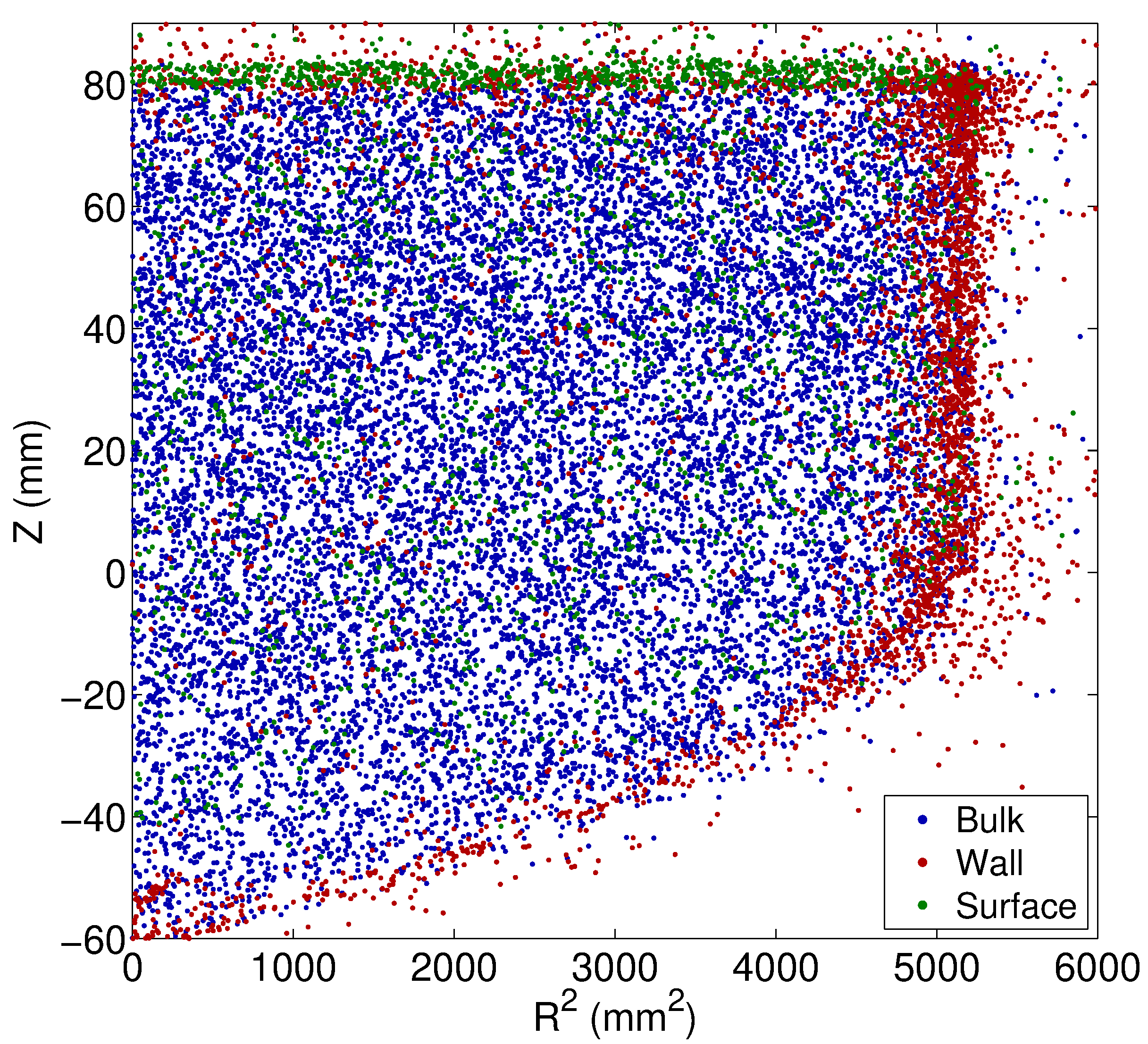}
\caption[Dytran event categorization]{Bubble locations in $Z$ vs. $R^2$ (from image position reconstruction package) for bulk, wall, and surface events as categorized by the Dytran analysis.}
\label{fig:dytran_categorization}
\end{figure}

In this analysis, the fiducial volume cut based on position reconstruction is replaced by a Dytran type cut \citep{lippincott-11b}, which is imposed by calculating the chi-square of a quadratic fit to the Dytran trace, taking into account the root mean square (rms) noise of the Dytran baseline before the pressure rise begins. If this chi-square is $<$ 100, the event is categorized as a bulk event. If the chi-square is $>$ 100 and the cubic coefficient to a best-fit cubic equation is between -2500 and -1060, the event is categorized as a surface event. Otherwise, the event is categorized as a wall event. Figure \ref{fig:dytran_categorization}  shows events classified as bulk, wall, or surface events by the Dytran analysis and their reconstructed location in $Z$ and $R^2$ from the position reconstruction package. It is apparent that using the Dytran trace type as a categorization of the bubble location is very efficient.

To determine the efficiency of this cut, the bulk, wall, and surface rates from the dark matter search data was calculated for each temperature and pressure available. These rates were subtracted out of neutron calibration data. Assuming that the bubble chamber was roughly bathed uniformly by the neutron source (an approximation supported by simulations), the efficiency is defined as the ratio of the rate of background-subtracted bulk events to the total background-subtracted rate. For data with a 30 s expansion time cut, the Dytran type cut efficiency was found to be 92.1$\pm$1.8\% \citep{lippincott-11b}.

\section{Pressure-Independent Acceptance Cuts}
\label{sec:datasets_acceptance_Pindep}

The third type of data cut is referred to here as acceptance cuts. These cuts are placed on the data after all the previously described cuts have been enforced. Their purpose is to ensure that all data employed have usable parameters required to perform a quality analysis. Acceptance cuts are further divided into pressure-independent and pressure-dependent cuts, since certain cuts are unusable for higher pressure ($>$ 30 psia) data.

In order to determine the efficiency of the acceptance cuts, the total number of single-bubble events must be known in a reliable way. Much of the data has undergone hand-scanning by individuals in the COUPP collaboration. The hand-scanner recorded the number of bubbles seen in each individual event, resulting in hand-scanned data which gives a trustworthy estimate of the total number of single-bubble events in a data set. The hand-scan data includes both dark matter search data (only from the DM-34$^\circ$C and DM-37$^\circ$C data) and calibration data. The efficiency of each cut is then determined by comparing the number of events passing the cut with the total number of single-bubble events determined by the hand-scan results, after data identification and fiducial volume cuts have been applied and independent of all other acceptance cuts. The cuts are then applied in series to give the total acceptance cut. The efficiencies of all acceptance cuts (pressure-independent and pressure-dependent) are shown in Table \ref{tab:acceptance_cuts}.

\begin{table}[t!]
\centering
\begin{tabular} {r | c | c |}
\cline{2-3}
 & Cut & Efficiency\\
\hline
\multicolumn{1}{| r |}{Pressure-independent:} & Video trigger & 100$^{+0.00}_{-0.02}$\% \\
\multicolumn{1}{| r |}{} & Frame skip & 98.38$\pm$0.17\% \\
\multicolumn{1}{| r |}{} & Getbub & 97.57$\pm$0.21\% \\
\multicolumn{1}{| r |}{} & Dytran baseline & 96.61$\pm$0.25\% \\
\multicolumn{1}{| r |}{} & Dytran bubble number & 96.98$\pm$0.23\% \\
\hline
\multicolumn{1}{| r |}{Pressure-dependent:} & Piezo noise & 99.38$\pm$0.11\% \\
\multicolumn{1}{| r |}{} & Piezo $t_0$ & 96.91$\pm$0.25\% \\
\hline
\hline
\multicolumn{2}{| r |}{All pressure-independent acceptance cuts} & 92.37$\pm$0.37\% \\
\multicolumn{2}{| r |}{All acceptance cuts combined} & 90.46$\pm$0.44\% \\
\hline
\end{tabular}
\caption[Acceptance cut efficiencies for single-bubble events]{The efficiency of each pressure-independent and pressure-dependent acceptance cut as well as the total efficiency of all cuts combined in series.}
\label{tab:acceptance_cuts}
\end{table}

\subsection{Video Trigger Cut}
\label{sec:datasets_video}

Each event is triggered in one of several ways. Most events that contain an actual bubble are video triggered, meaning that some threshold number of pixels are different in one camera image when compared to the previous image. Other triggers include pressure triggers (when the Dytran detects a pressure rise), timeout triggers (when the expansion has exceeded a maximum 500 s time limit), and frame skip triggers (when multiple consecutive frames are skipped by a camera). The video trigger cut ensures that each event in the final data set was triggered by the video. Of the total 5632 single-bubble events in the hand-scanned data, none failed to be video triggers. The efficiency of this cut is then 100$^{+0.00}_{-0.02}$\%.

\subsection{Frame Skip Cut}
\label{sec:datasets_frameskip}

Sometimes, the cameras fail to record each image in the video series to the data set, resulting in a frame skip. The frame skip cut ensures that each event did not experience any frame skipping, a more stringent requirement than ensuring the event is a video trigger because some video trigger events can still contain frame skips if the number of frames skipped is not above some threshold. This cut is important because events with frame skips tend to trigger very late after bubble formation, losing important acoustic and Dytran information. There were 91 hand-scanned single-bubble events that contained frame skips, so the efficiency of this cut is 98.38$\pm$0.17\%.

\subsection{Getbub Cut}
\label{sec:datasets_getbub}

The Getbub package of the COUPP data analysis library finds and reconstructs all bubbles in the images corresponding to an event. Because it is desirable to use all data to set dark matter limits (not just that which was hand-scanned), it is necessary for Getbub to accurately identify single-bubble events. The Getbub cut requires that Getbub returns a value of 1 for the number of bubbles in an event. There were 137 hand-scanned single-bubble events that Getbub failed to properly identify as single-bubble events. Therefore, the efficiency of this cut is 97.57$\pm$0.21\%.

\subsection{Dytran Baseline Cut}
\label{sec:datasets_dytranbaseline}

The Dytran trace provides significant amounts of information about each event, which can be used to properly identify the data. In order to use the Dytran trace in the analysis, the rms noise level of the trace has to be low ($<$ 0.02 V). The Dytran baseline cut ensures that the trace exists and that its noise is below this level. There were 191 hand-scanned single-bubble events that failed the Dytran baseline cut, so the efficiency of this cut is 96.61$\pm$0.25\%.

\subsection{Dytran Bubble Number Cut}
\label{sec:datasets_dytranbubnum}

As mentioned in Section \ref{sec:datasets_fiducial}, the Dytran trace for each event can be fit to either a quadratic or a cubic function, the parameters of which give significant information about the event  \citep{lippincott-11b}. The pressure rise of multiple bubble events is faster than for single-bubble events, and so the fit to the Dytran trace can provide an accurate measurement of the bubble multiplicity of an event. The Dytran bubble number cut only accepts events for which this parameter is between 0.5 and 1.5. There were 170 hand-scanned single-bubble events that failed the Dytran bubble number cut, so the efficiency of this cut is 96.98$\pm$0.23\%. It is interesting to note that there were 130 events which passed the Getbub cut that still failed the Dytran bubble number cut. The populations cut by these two multiplicity cuts are not greatly overlapping, and so each cut is quite important.

\section{Pressure-Dependent Acceptance Cuts}
\label{sec:datasets_acceptance_Pdep}

The piezoelectric transducers attached to the synthetic silica bell jar are necessary to provide acoustic information for each event (Section \ref{sec:bubblechambers_acoustic}). However, the acoustic emission becomes weaker at high pressures, because bubbles formed in a higher pressure environment tend to expand less violently, emitting significantly less ultrasonic noise in formation. Therefore, in the calibration runs at high pressure ($>$ 30 psia), the data provided by the piezos cannot be used for analysis. Without reliable acoustic information from the event, alpha discrimination is impossible, making a data set difficult to use. Therefore, all cuts relating to the piezos are only applied to data with the pressure set-point $\leq$ 30 psia, which includes all dark matter search data. The efficiency of each pressure-dependent acceptance cut and the total efficiency of all acceptance cuts combined is shown in Table \ref{tab:acceptance_cuts}.

\subsection{Piezo Noise Cut}
\label{sec:datasets_piezonoise}

There are three operable piezos functioning in the data set, and each of these provides an acoustic trace of the event. The piezo noise cut ensures that the rms noise level of the piezo traces is not too high ($<$ 0.01 V). Large noise levels in the piezos reduces the ability to determine the acoustic parameter necessary for alpha discrimination (Section \ref{sec:datasets_alpha}), and so high noise events should be cut. There were 30 hand-scanned single-bubble events that failed to pass the piezo noise cut but that passed all the data identification fiducial volume cuts and are in the $\leq$ 30 psia data (4823 events), so the efficiency of this cut is 99.38$\pm$0.11\%.

\subsection{Piezo $t_0$ Cut}
\label{sec:datasets_piezot0}

Another step in the data analysis identifies the initial time $t_0$ where each of the three piezo traces begins to deviate from the rms noise level (\emph{i.e.} when the bubble is first detected acoustically). If the finder does not identify a $t_0$ for each trace or the $t_0$ identified for each are not within 3 ms of each other, these events are cut out with the piezo $t_0$ cut. Again, this cut only applies to the low pressure data. Of the hand-scanned single-bubble events passing data identification and fiducial volume cuts in the $\leq$ 30 psia data, 149 failed to pass the piezo $t_0$ cut, resulting in an efficiency of 96.91$\pm$0.25\% for this cut.

\section{Alpha Discrimination}
\label{sec:datasets_alpha}

So far, cuts have been established to identify a data set constituted of bulk single-bubble events after a 30 s expansion time at a consistent threshold. While this data set is free from most systematic issues in generating false triggers, it is still not a representative set of recoil-like events needed for setting dark matter limits. A majority of the surviving single-bubble events are generated by alpha decays within the chamber originating from radon dissolved in the CF$_3$I (Section \ref{sec:datasets_alpharate}), resulting in a significant background that must be overcome. As demonstrated by PICASSO \citep{aubin-08} and confirmed by the engineering run of the COUPP 4 kg bubble chamber \citep{behnke-11}, the ultrasonic signature of bubble formation can be used to discriminate alpha decay events from recoil-like events, potentially due to the number of proto-bubbles initiated by each type of event (Section \ref{sec:bubblechambers_acoustic}). What would remain post-discrimination is a data set comprised solely of nuclear recoils generated by background neutrons, any residual sensitivity to environmental gammas, and WIMPs.

\subsection{Acoustic Parameter ($AP$)}
\label{sec:datasets_AP}

\begin{figure} [t!]
\centering
\includegraphics[scale=0.45]{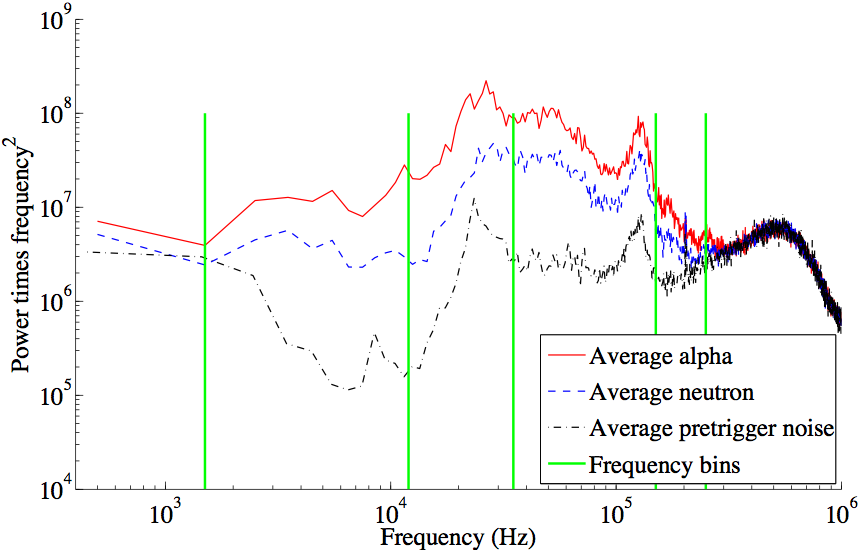}
\caption[Acoustic emission from alpha- and neutron-generated events]{The acoustic emission from alpha- and neutron-generated events, compared to background noise. Alphas tend to be louder than neutrons across the entire spectral range, up to 250 kHz. Figure from \cite{dahl-11b} reproduced with permission.}
\label{fig:acoustic_emission}
\end{figure}

The three piezoelectric transducer signals were digitized with a 2.5 MHz sampling rate and recorded for 40 ms for each event. These signals were filtered with a single-pole high-pass filter with cutoff at 500 Hz, and a low-pass anti-aliasing filter with cutoff at 600 kHz. A fast Fourier transform was performed for time between 1 ms before to 9 ms after the identified $t_0$ of the event. For frequencies below 250 kHz, the sound generated by bubble formation was significantly above background. Further, when comparing populations of neutron-generated events (from neutron calibration runs) to alpha-generated events (identified via time correlation in the alpha emission from the $^{222}$Rn-$^{218}$Po-$^{214}$Po chain, Figure \ref{fig:U238_decay_chain}), the acoustic emission was much higher for alphas than for neutrons (Figure \ref{fig:acoustic_emission}).  This is because while individual nuclear recoils will generate one proto-bubble, alpha decays will generate many along the extended track length of the alpha particle and one from the recoiling daughter nucleus.

Because the acoustic power varies slightly with the position of the bubble, and this dependence varies slightly with frequency, position correction terms were applied separately to four frequency bands (shown as the green lines in Figure \ref{fig:acoustic_emission}) --- 1.5-12 kHz, 12-35 kHz, 35-150 kHz, and 150-250 kHz.   Using these position corrections, a frequency-weighted acoustic power density integral denoted the acoustic parameter ($AP$) was defined as

\begin{equation}
\label{eq:AP}
AP = A(T) \sum_j G_j \sum_n C_n \left(\vec{x}\right) \sum_{f_\mathrm{min}^n}^{f_\mathrm{max}^n} f \times \mathrm{PSD}_f^j,
\end{equation}

\noindent where $A(T)$ is a temperature-dependent scale factor, $G_j$ is the gain of piezo $j$, $C_n \left(\vec{x}\right)$ is the correction factor for the bubble position dependence (position $\vec{x}$) in frequency bin $n$, $f$ is the frequency, $f_\mathrm{min}^n$ and $f_\mathrm{max}^n$ are the boundaries of frequency bin $n$, and $\mathrm{PSD}_f^j$ is the power spectral density for the bin with center frequency $f$ for piezo $j$ \citep{behnke-11}. The $AP$ calculated in Equation \ref{eq:AP} can be used to identify whether the bubble was generated by an alpha decay or whether it is recoil-like in nature for each event.

\subsection{Recoil-Like $AP$ Cut}
\label{sec:datasets_recoilAP}

\begin{figure} [t!]
\begin{minipage}[b]{0.48\linewidth}
\centering
\includegraphics[scale=0.4]{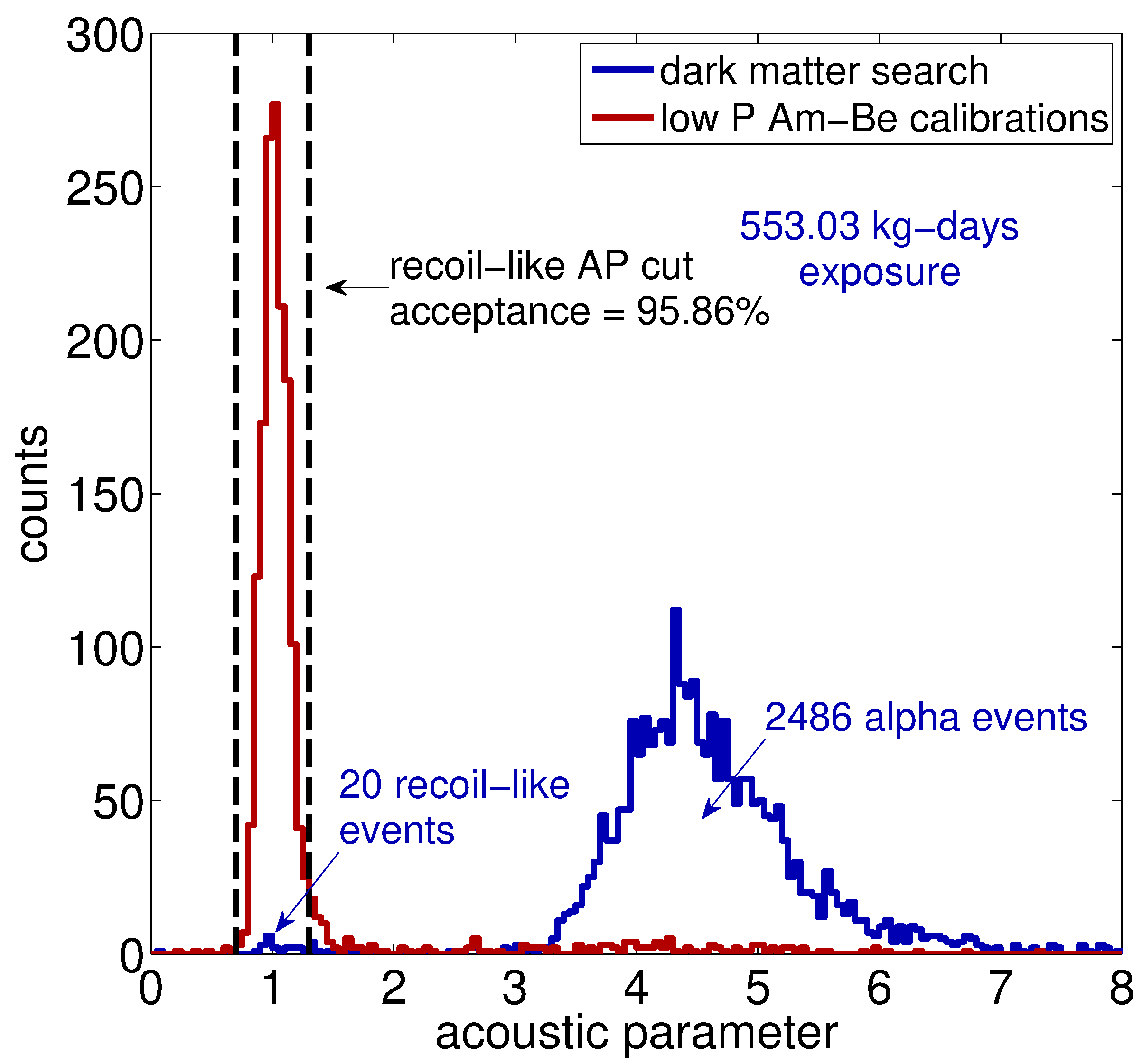}
\end{minipage}
\begin{minipage}[b]{0.52\linewidth}
\centering
\includegraphics[scale=0.4]{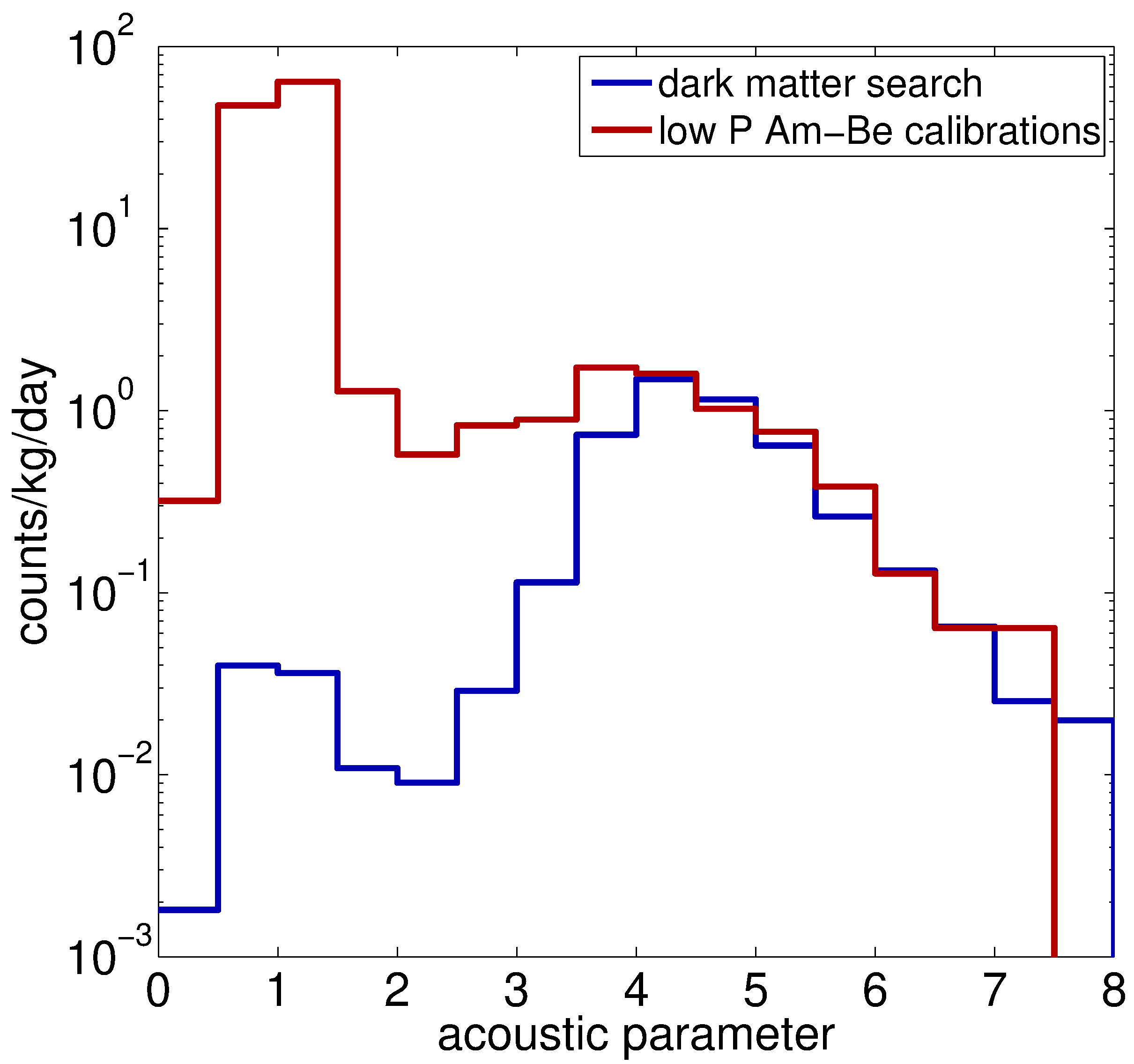}
\end{minipage}
\caption[$AP$ of dark matter search and neutron calibration data]{The distribution of acoustic parameter ($AP$) in the alpha-dominated dark matter search data compared to the low-pressure Am-Be neutron calibration data. The alpha population peaks above 4 in $AP$ and is clearly separated from the neutron population around $AP=1$. Raw counts are shown on the left and rates (in counts/kg/day) are shown on the right, with wider bins due to low sampling size at high $AP$ in the neutron calibration data. The rate of alpha decays is seen to be constant in both data sets.}
\label{fig:AP_alpha_neutron}
\end{figure}

When comparing the alpha-dominated dark matter search data with the neutron calibration data, the discriminatory power of using the $AP$ to identify alpha events is quite evident (Figure \ref{fig:AP_alpha_neutron}).  The dark matter search data has very few events with $AP<3$ and the neutron calibration data has very few events with $AP>2$. In the rate plot (Figure \ref{fig:AP_alpha_neutron}, right), it is seen that the alpha rate evident in the dark matter search data is also present in the neutron calibration data, and therefore the contribution arising from neutrons is contained within the region $AP<2$. An $AP$-based cut can be placed on the data that requires $AP$ to be between 0.7 and 1.3 for recoil-like events (Figure \ref{fig:AP_alpha_neutron}, left). If the number of events passing this cut in the low-pressure ($\leq$ 30 psia) Am-Be and $^{252}$Cf neutron calibration sets is compared to the number of events with $AP<2$ (or similarly, compared to the total number of events background-subtracted), 2197 of the total 2292 events survive the cut, resulting in a cut efficiency of 95.86$\pm$0.43\%.

\subsection{Alpha-Like $AP$ Cut}
\label{sec:datasets_alphaAP}

In Figure \ref{fig:AP_alpha_neutron} (right), a slow-rising tail for the alpha spectrum appears to begin at $AP\simeq2$. The alpha population is then identified by creating an alpha-like $AP$ cut requiring $AP>2$. Of the total 2500 events in the dark matter search data, 2470 of them pass this alpha-like $AP$ cut. Assuming that those that do not are indeed alpha-generated (which is probably not true as will be stated below, but this represents the conservative assumption), the efficiency of this cut is then 98.8$\pm$0.22\%.

\section{Alpha Event Rate}
\label{sec:datasets_alpharate}

Since bubble chambers are threshold detectors, they are expected to provide a constant response to monochromatic energy deposition once the corresponding threshold requirements have been surpassed. One example of such a monochromatic energy source is found in alpha decays, where for a particular isotope, the energies of the alpha and the recoiling daughter are fixed. Up to a cut-off threshold, the alpha event rate is therefore expected to be flat. All dark matter search and calibration data so far discussed are for thresholds below the lowest turn-off (corresponding to 101 keV $^{218}$Po recoil, see Figure \ref{fig:U238_decay_chain}), and so in each data set, the alpha rate should be a constant.

\begin{figure} [t!]
\centering
\includegraphics[scale=0.55]{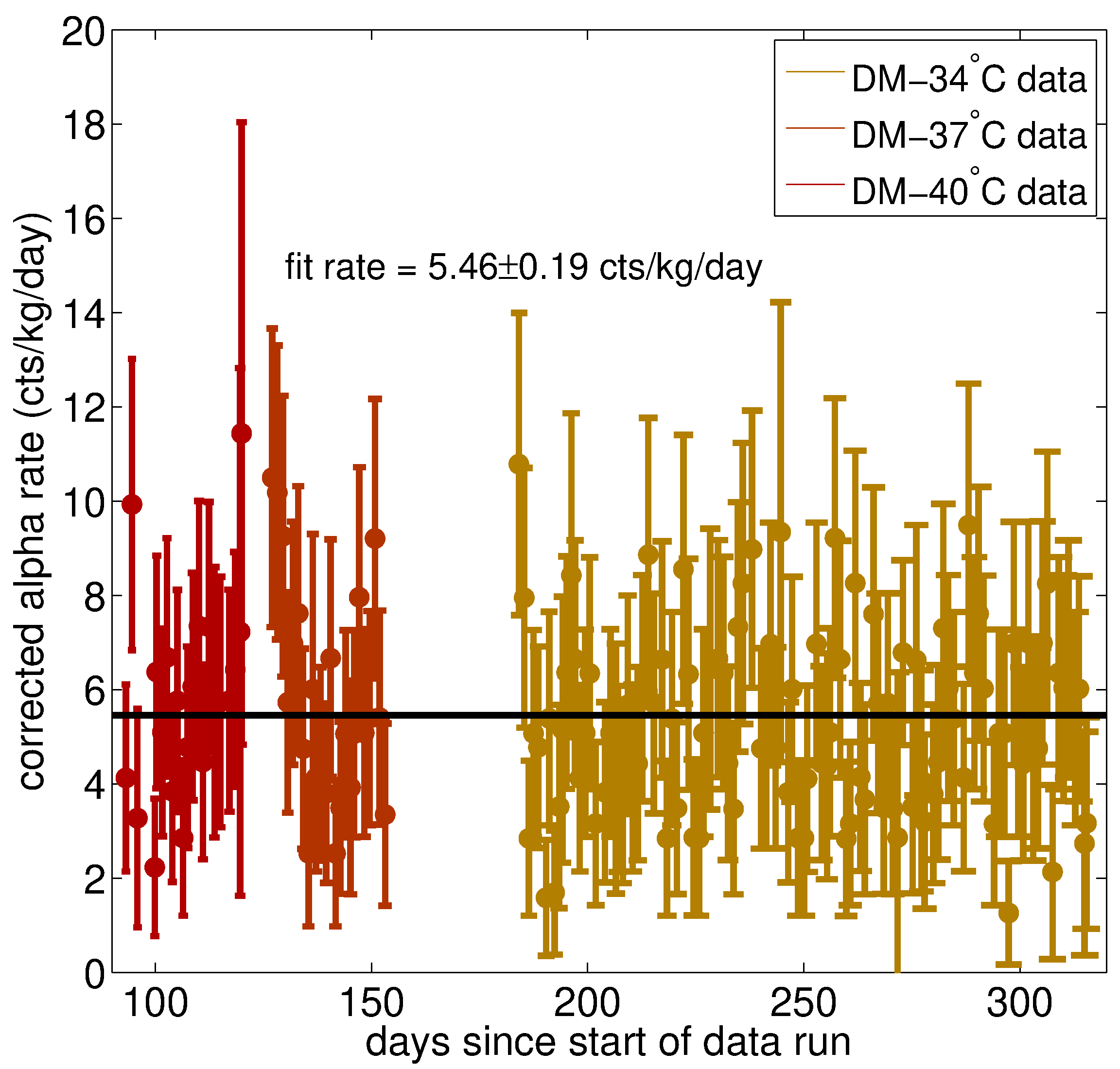}
\caption[Alpha event rates in dark matter search data sets]{The alpha event rates for each run in the dark matter search data, along with a flat fit for the rate. As expected, there is no evidence of a change of alpha event rate over time.}
\label{fig:alpha_rates}
\end{figure}

Figure \ref{fig:alpha_rates} shows the corrected alpha event rates in each run of the dark matter search data, with uncertainties arising not only from the statistics of the run, but also with an additional $\sqrt{3}$ factor due to the triplets of $^{222}$Rn alpha decays (Figure \ref{fig:U238_decay_chain}).  The rate plotted is corrected for the 98.8\% alpha-like $AP$ cut efficiency as well as the total 90.46\% acceptance cut efficiency and the 92.1\% Dytran type cut efficiency (for a total efficiency of 82.31\%). In order to determine the goodness-of-fit of the rate, the Pearson chi-squared test was used. The total $\chi^2$ is the sum of the $\chi_i^2$ for each run

\begin{equation}
\label{eq:chi2}
\chi^2 = \sum_i \frac{\left(O_i - E_i\right)^2}{\sigma_i^2},
\end{equation}

\noindent where $O_i$ is the rate for the run and $E_i$ is the fit rate. The $p$-value is found from the cumulative distribution function of the chi-squared distribution

\begin{equation}
\label{eq:p-value}
p\left(\chi^2,k\right) = 1 - \frac{\gamma\left(k/2,\chi^2/2\right)}{\Gamma\left(k/2\right)},
\end{equation}

\noindent where $k$ is the number of degrees of freedom (DOF), $\gamma$ is the lower incomplete Gamma function, and $\Gamma$ is the Gamma function. Taking each run as a DOF, the best-fit to a corrected flat rate of alpha events is 5.46$\pm$0.19 cts/kg/day with a $\chi^2$/DOF = 137.81/167 with a $p$-value of 95.19\%.

\section{Multi-Bubble Events}
\label{sec:datasets_multibubble}

One final selection criterion for a complete data set is the identification of multi-bubble events (for bubble multiplicity $N$). Due to the low nucleation efficiency for gammas (Section \ref{sec:gammas_calibrations}), the long interaction lengths of WIMPs, and the short track length (tens of microns) of alphas in CF$_3$I, the only events with multiple bubbles are those arising from neutron scattering. Therefore, a measurement of the number of multi-bubble events and a good understanding of the ratio of multi-bubble events to single-bubble neutron events can tell a lot about the total rate of neutrons in a chamber, and how many recoil-like events could be generated by neutrons.

Only some of the cuts described in this chapter apply to multi-bubble events as currently described. Both data identification cuts (expansion time and threshold cuts) should still apply without modification. The video trigger, frame skip, and Dytran baseline cuts also apply. Since no acoustic discrimination will be required, the information from piezo traces will be ignored for the analysis. The Dytran type cut does not prove to be a good cut for multi-bubble events (the efficiency for this cut is close to 50\%). However, a Dytran bubble number cut with boundaries $\pm$0.5 from the bubble number proves to be quite good at identifying bubble multiplicity. The Dytran bubble number cut is much better than the multiplicity returned by the Getbub image reconstruction package, which is quite poor as the multiplicity increases. When comparing data passing each of these cuts to the hand-scanned data, the efficiencies given in Table \ref{tab:acceptance_cuts_multi} are found. \begin{table}[t!]
\centering
\begin{tabular} {c | c | c | c |}
\cline{2-4}
 & \multicolumn{3}{| c |}{Efficiency} \\
\hline
\multicolumn{1}{| c |}{Cut} & ($N=2$) & ($N=3$) & ($N=4$) \\
\hline
\multicolumn{1}{| c |}{Video trigger} & 100$^{+0.00}_{-0.10}$\% &  100$^{+0.00}_{-0.34}$\% & 100$^{+0.00}_{-1.14}$\% \\
\multicolumn{1}{| c |}{Frame skip} & 98.50$^{+0.33}_{-0.38}$\% &  100$^{+0.00}_{-0.34}$\% & 100$^{+0.00}_{-1.14}$\%  \\
\multicolumn{1}{| c |}{Getbub} & 91.86$\pm$0.80\% & 82.68$\pm$2.13\% & 60.18$\pm$5.94\% \\
\multicolumn{1}{| c |}{Dytran baseline} & 97.23$\pm$0.47\% & 96.33$^{+0.97}_{-1.13}$\% & 98.23$^{+1.12}_{-1.99}$\% \\
\multicolumn{1}{| c |}{Dytran bubble number} & 93.91$\pm$0.69\% & 87.14$\pm$1.84\% & 84.07$^{+3.71}_{-4.27}$\% \\
\hline
\hline
\multicolumn{1}{| c |}{All cuts combined} & 84.27$\pm$1.12\% & 69.29$\pm$2.84\% & 53.10$\pm$6.44\% \\
\multicolumn{1}{| c |}{All cuts w/o Getbub} & 90.91$\pm$0.85\% & 83.46$\pm$2.08\% & 82.30$^{+3.69}_{-4.69}$\% \\
\hline
\end{tabular}
\caption[Acceptance cut efficiencies for multi-bubble events]{The efficiency of each pressure-independent and pressure-dependent acceptance cut for the multi-bubble data as well as the total efficiency of all cuts combined in series.}
\label{tab:acceptance_cuts_multi}
\end{table}

The problem with using these cuts in multi-bubble events is seen when comparing events identified with these cuts with those established in the hand-scanned calibration set. For each multiplicity, the data cuts identified $\sim$10\% more events than were seen in the hand-scanned data. In general, the best results are obtained by using the hand-scanning results, which of course becomes a cumbersome technique when dealing with large dark matter search data sets. This is a problem that will have to be addressed in future chambers.

\section{Data Set Summary}
\label{sec:datasets_summary}

The entirety of the first COUPP 4 kg bubble chamber data run at SNOLAB, including both the dark matter search and the neutron and gamma calibration data, is summarized in Tables \ref{tab:counts} and \ref{tab:rates}. \begin{table}[p!]
\centering
\small{
\begin{tabular} {| c | c | c | c c | c | c | c |}
\hline
\multicolumn{8}{| l |}{Dark matter search data} \\
\hline
Data Set & Threshold & Exposure & \multicolumn{5}{| c |}{Counts} \\
\cline{4-8}
 & \small{(kev)} & \small{(kg-days)} & \multicolumn{2}{| c |}{\small{$N=1$}} & \small{$N=2$} & \small{$N=3$} & \small{$N=4$} \\
& & & \small{alpha} & \small{recoil} & & & \\
\hline
DM-34$^\circ$C & 15.83$^{+1.46}_{-1.31}$ & 393.64 & 1734 & 8 & 1 & 0 & 0 \\
DM-37$^\circ$C & 11.21$^{+0.97}_{-0.87}$ & 88.41 & 424 & 6 & 0 & 2 & 0 \\
DM-40$^\circ$C & 8.00$^{+0.67}_{-0.61}$ & 70.98 & 328 & 6 & 0 & 0 & 0\\
\hline
\hline
\multicolumn{8}{| l |}{Neutron calibration data} \\
\hline
Data Set & Threshold & Exposure & \multicolumn{5}{| c |}{Counts} \\
\cline{4-8}
 & \small{(kev)} & \small{(kg-days)} & \multicolumn{2}{| c |}{\small{$N=1$}} & \small{$N=2$} & \small{$N=3$} & \small{$N=4$} \\
& & & \small{alpha} & \small{recoil} & & & \\
\hline
AmBe-36"-34$^\circ$C-30psia & 15.92$^{+1.46}_{-1.31}$ & 6.09 & 41 & 556 & 162 & 30 & 7 \\
AmBe-36"-34$^\circ$C-40psia & 26.67$^{+3.06}_{-2.68}$ & 2.00 & \multicolumn{2}{| c |}{165} & 29 & 8 & 1 \\
AmBe-36"-34$^\circ$C-50psia & 51.92$^{+7.48}_{-6.32}$ & 3.00 & \multicolumn{2}{| c |}{170} & 12 & 0 & 0 \\
AmBe-36"-34$^\circ$C-60psia & 128.13$^{+26.63}_{-20.98}$ & 3.33 & \multicolumn{2}{| c |}{70} & 2 & 1 & 0 \\
AmBe-36"-37$^\circ$C-30psia & 11.27$^{+1.02}_{-0.92}$ & 4.23 & 29 & 384 & 105 & 31 & 11 \\
AmBe-36"-37$^\circ$C-35psia & 14.10$^{+1.33}_{-1.19}$ & 1.69 & \multicolumn{2}{| c |}{150} & 29 & 11 & 2 \\
AmBe-36"-40$^\circ$C-30psia & 8.01$^{+0.66}_{-0.60}$ & 2.07 & 13 & 179 & 74 & 19 & 7 \\
AmBe-36"-41$^\circ$C-30psia & 7.21$^{+0.58}_{-0.53}$ & 0.17 & 1 & 20 & 8 & 2 & 2 \\
AmBe-34"-34$^\circ$C-30psia & 15.94$^{+1.47}_{-1.32}$ & 1.54 & 7 & 122 & 41 & 9 & 1 \\
AmBe-38"-34$^\circ$C-30psia & 15.86$^{+1.45}_{-1.31}$ & 1.55 & 16 & 195 & 42 & 8 & 1 \\
Cf252-42"-37$^\circ$C-29psia & 11.09$^{+0.96}_{-0.87}$ & 14.92 & 82 & 82 & 28 & 4 & 2 \\
Cf252-42"-37$^\circ$C-30psia & 11.39$^{+1.01}_{-0.91}$ & 37.96 & 161 & 210 & 70 & 12 & 3 \\
Cf252-54"-34$^\circ$C-30psia & 15.99$^{+1.46}_{-1.32}$ & 29.58 & 131 & 479 & 130 & 32 & 10 \\
Cf252-54"-34$^\circ$C-40psia & 26.78$^{+2.98}_{-2.62}$ & 10.81 & \multicolumn{2}{| c |}{194} & 26 & 8 & 1 \\
\hline
\hline
\multicolumn{8}{| l |}{Gamma calibration data} \\
\hline
Data Set & Threshold & Exposure & \multicolumn{5}{| c |}{Counts} \\
\cline{4-8}
 & \small{(kev)} & \small{(kg-days)} & \multicolumn{2}{| c |}{\small{$N=1$}} & \small{$N=2$} & \small{$N=3$} & \small{$N=4$} \\
& & & \small{alpha} & \small{recoil} & & & \\
\hline
Mix-62"-37$^\circ$C-30psia & 11.27$^{+0.97}_{-0.88}$ & 7.49 & 42 & 0 & 1 & 0 & 0 \\
Mix-62"-40$^\circ$C-30psia & 8.00$^{+0.65}_{-0.59}$ & 4.11 & 22 & 3 & 0 & 0 & 0 \\
Mix-62"-41$^\circ$C-30psia & 7.22$^{+0.58}_{-0.53}$ & 7.09 & 34 & 18 & 0 & 0 & 0 \\
Co60-55"-34$^\circ$C-30psia & 15.77$^{+1.44}_{-1.30}$ & 15.49 & 77 & 1 & 0 & 0 & 0 \\
Co60-55"-40$^\circ$C-30psia & 7.96$^{+0.65}_{-0.59}$ & 0.30 & 0 & 3 & 0 & 0 & 0 \\
Co60-55"-40$^\circ$C-32psia & 8.73$^{+0.74}_{-0.67}$ & 0.66 & \multicolumn{2}{| c |}{8} & 0 & 0 & 0 \\
Co60-55"-40$^\circ$C-35psia & 9.65$^{+0.84}_{-0.76}$ & 2.18 & \multicolumn{2}{| c |}{20} & 0 & 0 & 0 \\
Co60-55"-40$^\circ$C-37psia & 10.68$^{+0.97}_{-0.87}$ & 0.91 & \multicolumn{2}{| c |}{11} & 0 & 0 & 0 \\
Ba133-55"-34$^\circ$C-30psia & 15.84$^{+1.46}_{-1.31}$ & 25.40 & 23 & 0 & 0 & 0 & 0 \\
Ba133-55"-37$^\circ$C-30psia & 11.19$^{+0.96}_{-0.87}$ & 11.98 & 71 & 2 & 1 & 1 & 0 \\
Ba133-55"-40$^\circ$C-30psia & 7.99$^{+0.65}_{-0.59}$ & 0.97 & 2 & 22 & 0 & 0 & 0 \\
\hline
\end{tabular} }
\caption[Counts in dark matter search and calibration data sets]{The counts in each dark matter search and calibration data set for multiplicity $N$=1-4. For high-pressure ($>$ 30 psia) sets, no acoustic discrimination is possible, so there is no separation between alpha-like and recoil-like events in the $N$=1 case.}
\label{tab:counts}
\end{table} \begin{table}[p!]
\centering
\small{
\begin{tabular} {| c | c | c | c | c | c |}
\hline
\multicolumn{6}{| l |}{Dark matter search data} \\
\hline
Data Set & Threshold & \multicolumn{4}{| c |}{Rates (cts/kg/day)} \\
\cline{3-6}
 & \small{(kev)} & \small{$N=1$} & \small{$N=2$} & \small{$N=3$} & \small{$N=4$} \\
\hline
DM-34$^\circ$C & 15.83$^{+1.46}_{-1.31}$ & 0.025$^{+0.011}_{-0.009}$ & 0.003$^{+0.004}_{-0.002}$ & 0.000$^{+0.003}_{-0.000}$ & 0.000$^{+0.003}_{-0.000}$ \\
DM-37$^\circ$C & 11.21$^{+0.97}_{-0.87}$ & 0.085$^{+0.047}_{-0.031}$ & 0.000$^{+0.015}_{-0.000}$ & 0.023$^{+0.025}_{-0.014}$ & 0.000$^{+0.015}_{-0.000}$ \\
DM-40$^\circ$C & 8.00$^{+0.67}_{-0.61}$ & 0.106$^{+0.058}_{-0.039}$ & 0.000$^{+0.018}_{-0.000}$ & 0.000$^{+0.018}_{-0.000}$ & 0.000$^{+0.018}_{-0.000}$ \\
\hline
\hline
\multicolumn{6}{| l |}{Neutron calibration data} \\
\hline
Data Set & Threshold & \multicolumn{4}{| c |}{Rates (cts/kg/day)} \\
\cline{3-6}
 & \small{(kev)} & \small{$N=1$} & \small{$N=2$} & \small{$N=3$} & \small{$N=4$} \\
\hline
AmBe-36"-34$^\circ$C-30psia & 15.92$^{+1.46}_{-1.31}$ & 114.31$\pm$5.39 & 26.60$\pm$2.09 & 4.93$\pm$0.90 & 1.15$^{+0.54}_{-0.45}$ \\
AmBe-36"-34$^\circ$C-40psia & 26.67$^{+3.06}_{-2.68}$ & 91.52$\pm$7.80 & 14.50$\pm$2.69 & 4.00$^{+1.66}_{-1.35}$ & 0.50$^{+0.88}_{-0.32}$ \\
AmBe-36"-34$^\circ$C-50psia & 51.92$^{+7.48}_{-6.32}$ & 61.05$\pm$5.27 & 3.99$^{+1.43}_{-1.06}$ & 0.00$^{+0.43}_{-0.00}$ & 0.00$^{+0.43}_{-0.00}$ \\
AmBe-36"-34$^\circ$C-60psia & 128.13$^{+26.63}_{-20.98}$ & 19.24$\pm$3.00 & 0.60$^{+0.68}_{-0.38}$ & 0.30$^{+0.53}_{-0.19}$ & 0.00$^{+0.39}_{-0.00}$ \\
AmBe-36"-37$^\circ$C-30psia & 11.27$^{+1.02}_{-0.92}$ & 113.70$\pm$6.26 & 24.83$\pm$2.42 & 7.33$\pm$1.32 & 2.60$^{+0.90}_{-0.75}$ \\
AmBe-36"-37$^\circ$C-35psia & 14.10$^{+1.33}_{-1.19}$ & 98.96$\pm$8.78 & 17.17$\pm$3.19 & 6.51$^{+2.25}_{-1.89}$ & 1.18$^{+1.33}_{-0.75}$ \\
AmBe-36"-40$^\circ$C-30psia & 8.01$^{+0.66}_{-0.60}$ & 108.19$\pm$8.39 & 35.72$\pm$4.15 & 9.17$^{+2.33}_{-2.02}$ & 3.38$^{+1.59}_{-1.33}$ \\
AmBe-36"-41$^\circ$C-30psia & 7.21$^{+0.58}_{-0.53}$  & 149.05$^{+39.62}_{-31.23}$ & 47.61$^{+19.76}_{-16.07}$ & 11.90$^{+13.39}_{-7.50}$ & 11.90$^{+13.39}_{-7.50}$ \\
AmBe-34"-34$^\circ$C-30psia & 15.94$^{+1.47}_{-1.32}$ & 99.51$\pm$9.24 & 26.71$\pm$4.17 & 5.86$^{+2.47}_{-1.74}$ & 0.65$^{+1.14}_{-0.41}$ \\
AmBe-38"-34$^\circ$C-30psia & 15.86$^{+1.45}_{-1.31}$ & 157.08$\pm$11.71 & 27.02$\pm$4.17 & 5.15$^{+2.14}_{-1.73}$ & 0.64$^{+1.13}_{-0.41}$ \\
Cf252-42"-37$^\circ$C-29psia & 11.09$^{+0.96}_{-0.87}$ & 6.88$\pm$0.77 & 1.88$\pm$0.35 & 0.27$^{+0.18}_{-0.11}$ & 0.13$^{+0.15}_{-0.08}$ \\
Cf252-42"-37$^\circ$C-30psia & 11.39$^{+1.01}_{-0.91}$ & 6.93$\pm$0.50 & 1.84$\pm$0.22 & 0.32$^{+0.11}_{-0.08}$ & 0.08$^{+0.06}_{-0.05}$ \\
Cf252-54"-34$^\circ$C-30psia & 15.99$^{+1.46}_{-1.32}$ & 20.28$\pm$1.02 & 4.39$\pm$0.39 & 1.08$\pm$0.19 & 0.34$^{+0.13}_{-0.11}$ \\
Cf252-54"-34$^\circ$C-40psia & 26.78$^{+2.98}_{-2.62}$ & 15.63$\pm$1.58 & 2.40$\pm$0.47 & 0.74$^{+0.31}_{-0.25}$ & 0.09$^{+0.16}_{-0.06}$ \\
\hline
\hline
\multicolumn{6}{| l |}{Gamma calibration data} \\
\hline
Data Set & Threshold & \multicolumn{4}{| c |}{Rates (cts/kg/day)} \\
\cline{3-6}
 & \small{(kev)} & \small{$N=1$} & \small{$N=2$} & \small{$N=3$} & \small{$N=4$} \\
\hline
Mix-62"-37$^\circ$C-30psia & 11.27$^{+0.97}_{-0.88}$ & 0.00$^{+0.22}_{-0.00}$ & 0.13$^{+0.23}_{-0.08}$ & 0.00$^{+0.17}_{-0.00}$ & 0.00$^{+0.17}_{-0.00}$ \\
Mix-62"-40$^\circ$C-30psia & 8.00$^{+0.65}_{-0.59}$ &  0.91$^{+0.70}_{-0.58}$ &  0.00$^{+0.31}_{-0.00}$ & 0.00$^{+0.31}_{-0.00}$ & 0.00$^{+0.31}_{-0.00}$ \\
Mix-62"-41$^\circ$C-30psia & 7.22$^{+0.58}_{-0.53}$ & 3.18$^{+0.85}_{-0.74}$ & 0.00$^{+0.18}_{-0.00}$ & 0.00$^{+0.18}_{-0.00}$ & 0.00$^{+0.18}_{-0.00}$ \\
Co60-55"-34$^\circ$C-30psia & 15.77$^{+1.44}_{-1.30}$ & 0.08$^{+0.14}_{-0.05}$ & 0.00$^{+0.08}_{-0.00}$ & 0.00$^{+0.08}_{-0.00}$ & 0.00$^{+0.08}_{-0.00}$ \\
Co60-55"-40$^\circ$C-30psia & 7.96$^{+0.65}_{-0.59}$ & 12.56$^{+9.63}_{-7.96}$ & 0.00$^{+4.31}_{-0.00}$ & 0.00$^{+4.31}_{-0.00}$ & 0.00$^{+4.31}_{-0.00}$ \\
Co60-55"-40$^\circ$C-32psia & 8.73$^{+0.74}_{-0.67}$ & 8.80$^{+5.93}_{-4.83}$ & 0.00$^{+1.96}_{-0.00}$ & 0.00$^{+1.96}_{-0.00}$ & 0.00$^{+1.96}_{-0.00}$ \\
Co60-55"-40$^\circ$C-35psia & 9.65$^{+0.84}_{-0.76}$ & 5.34$^{+2.88}_{-2.27}$ & 0.00$^{+0.59}_{-0.00}$ & 0.00$^{+0.59}_{-0.00}$ & 0.00$^{+0.59}_{-0.00}$ \\
Co60-55"-40$^\circ$C-37psia & 10.68$^{+0.97}_{-0.87}$ & 8.82$^{+4.97}_{-4.16}$ & 0.00$^{+1.42}_{-0.00}$ & 0.00$^{+1.42}_{-0.00}$ & 0.00$^{+1.42}_{-0.00}$ \\
Ba133-55"-34$^\circ$C-30psia & 15.84$^{+1.46}_{-1.31}$ & 0.00$^{+0.06}_{-0.00}$ & 0.00$^{+0.05}_{-0.00}$ & 0.00$^{+0.05}_{-0.00}$ & 0.00$^{+0.05}_{-0.00}$ \\
Ba133-55"-37$^\circ$C-30psia & 11.19$^{+0.96}_{-0.87}$ & 0.21$^{+0.24}_{-0.13}$ & 0.08$^{+0.15}_{-0.05}$ & 0.08$^{+0.15}_{-0.05}$ & 0.00$^{+0.11}_{-0.00}$ \\
Ba133-55"-40$^\circ$C-30psia & 7.99$^{+0.65}_{-0.59}$ & 28.27$\pm$6.06 & 0.00$^{+1.32}_{-0.00}$ & 0.00$^{+1.32}_{-0.00}$ & 0.00$^{+1.32}_{-0.00}$ \\
\hline
\end{tabular} }
\caption[Event rates in dark matter search and calibration data sets]{The recoil-like and multi-bubble event rates in each calibration data set for multiplicity $N=1-4$, corrected for the efficiencies of the cuts used. For high-pressure ($>$ 30 psia) sets where acoustic alpha discrimination is impossible, a flat rate of 5.46 cts/kg/day was subtracted from the $N=1$ rate to account for the alpha rate calculated in Section \ref{sec:datasets_alpharate}.}
\label{tab:rates}
\end{table} Corrected rates for recoil-like and multi-bubble events in each data set are calculated as the ratio of counts to effective exposure, where the latter is given by the product of the total exposure (4.048 kg CF$_3$I target mass $\times$ live-time, shown in Table \ref{tab:counts} for each set) and the efficiencies of the cuts on each set. For $N$=1 low-pressure ($\leq$ 30 psia) data, the total acceptance cut (90.46$\pm$0.44\%), Dytran type cut (92.1$\pm$1.8\%), and recoil-like $AP$ cut (95.86$\pm$0.43\%) were used, yielding an effective exposure that is 79.86$\pm$1.90\% of the total exposure. For $N=1$ high-pressure ($>$ 30 psia) data, only the pressure-independent cut (92.37$\pm$0.37\%) and the Dytran type cut were used, yielding an effective exposure that is 85.07$\pm$1.84\% of the total exposure. The flat alpha rate of 5.46$\pm$0.19 cts/kg/day found in Section \ref{sec:datasets_alpharate} was subtracted from the rate calculated for the $N=1$ high-pressure data to account for the alpha-generated events in this data set, which cannot be acoustically discriminated. Finally, for all multi-bubble ($N>1$) events, the effective exposure is simply the total exposure, since these events were identified via a hand-scan with (nominally) 100\% efficiency.

The dark matter search data were accumulated over three different bubble nucleation thresholds --- 8.00 keV, 11.21 keV, and 15.83 keV. The 8.00 keV data were collected from November 6 to December 3, 2010 for a total exposure of 70.98 kg-days and yielded 6 recoil-like events ($N=1$ events passing the recoil-like $AP$ cut). The 11.21 keV data were collected from December 12, 2010 to January 5, 2011 for an exposure of 88.41 kg-days and yielded 6 recoil-like events and 2 three-bubble events. The 15.83 keV data were collected from February 5 to June 16, 2011 for an exposure of 393.64 kg-days and yielded 8 recoil-like events and 1 two-bubble event. Therefore, in the total 553.03 kg-days exposure in the dark matter search data, 20 candidate nuclear recoil events were observed, with 3 multi-bubble events. Since only neutrons can nucleate multiple bubbles in an event, the 3 multi-bubble events in the dark matter search data are the unambiguous signature of a neutron background. The origins of this residual background are explored thoroughly in Chapter \ref{ch:backgrounds}.

\begin{figure} [t!]
\centering
\includegraphics[scale=0.55]{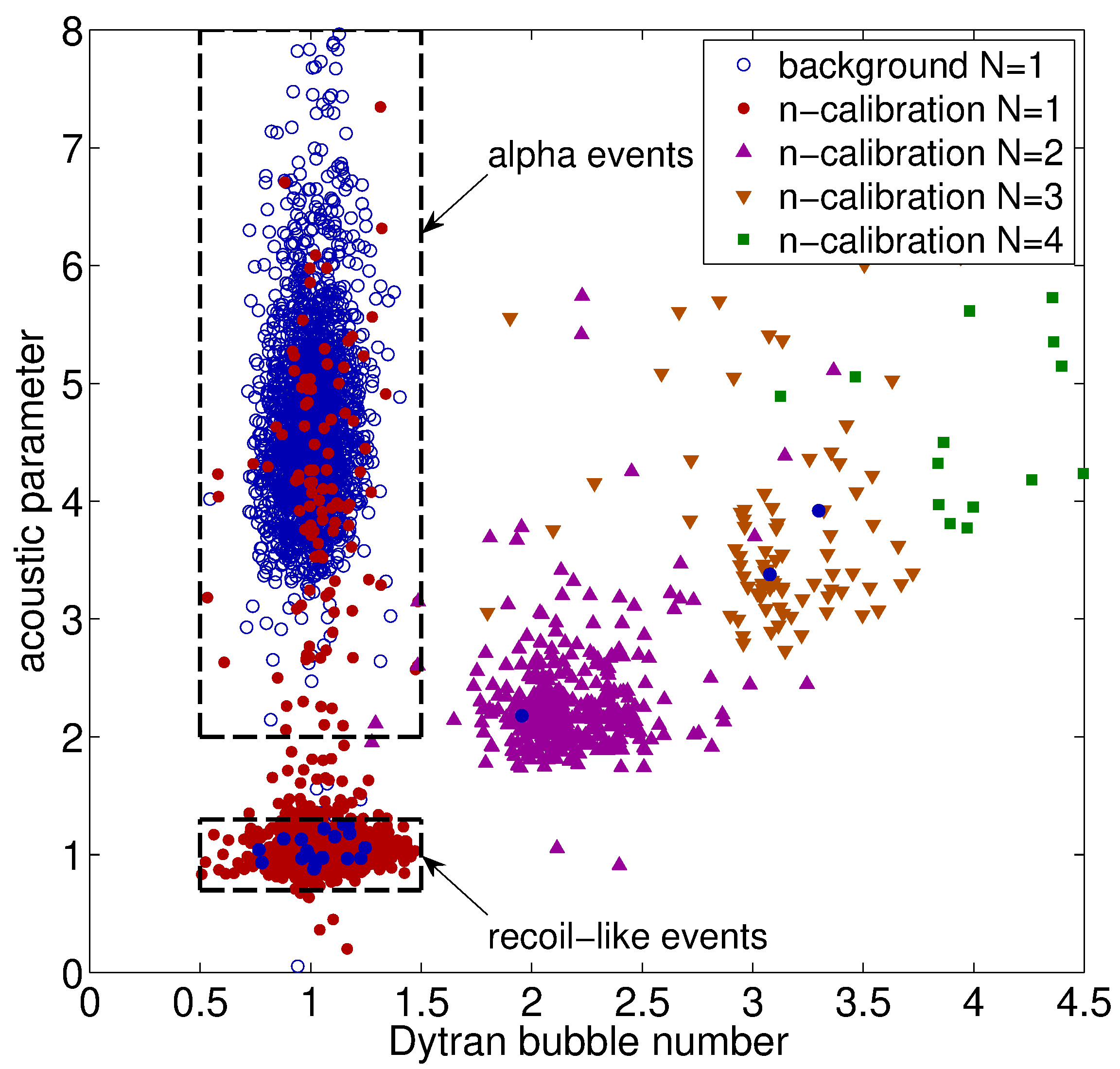}
\caption[The dark matter search and Am-Be calibration data]{The dark matter search data (blue) with the cut regions in Dytran bubble number and $AP$ identifying recoil-like and alpha events. The low-pressure Am-Be calibration data is also shown to illustrate the distribution of neutron-generated events. The 20 recoil-like and the 3 multi-bubble events in the dark matter search data are shown as filled blue circles.}
\label{fig:bubble_number_cuts}
\end{figure}

Figure \ref{fig:bubble_number_cuts}  shows the dark matter search data (blue, with filled blue circles representing the non-alpha events just described), with the Dytran bubble number and $AP$ cuts employed to identify the recoil-like and alpha population. The low-pressure Am-Be calibration data is also shown with clusters representing nuclear recoils for bubble multiplicities $N=1$ through $N=4$. As can be seen, the 20 recoil-like events are indistinguishable from neutron-generated events. These events can be categorized in one of three ways: background arising from failure of alpha discrimination, neutron background, or possible WIMP signal. This will be discussed in detail in the remainder of this work.

\singlespacing
\chapter{Bubble Nucleation Efficiency}
\label{ch:efficiency}
\doublespacing

The physics introduced in Chapter \ref{ch:bubblechambers} is both complicated and somewhat incomplete. According to the classical Seitz ``hot spike" model, a thermal deposition of energy above a critical energy

\begin{equation}
\label{eq:Q_repeat}
E > E_c = \frac{4}{3} \pi r_c^3 \rho_b \left( h_b - h_l \right) + 4 \pi r_c^2 \left( \gamma - T \frac{\partial \gamma}{\partial T} \right)
\end{equation}

\noindent is enough to nucleate a bubble in a superheated liquid, given this energy is deposited in a small enough region

\begin{equation}
\label{eq:dEdx_repeat}
\frac{\mathrm{d}E}{\mathrm{d}x} > \left.\frac{\mathrm{d}E}{\mathrm{d}x}\right|_c = \frac{E_c}{L_c},
\end{equation}

\noindent with $L_c = a r_c$ or $L_c = b \left( \rho_b / \rho_l \right)^{1/3} r_c$, and with the variables in Equations \ref{eq:Q_repeat} and \ref{eq:dEdx_repeat} as defined in Chapter \ref{ch:bubblechambers}. In terms of recoils from elastic scatters in a target superheated volume, the former requirement places a constraint on the energy of the recoil while the latter enforces a constraint on the stopping power of the recoiling particle. For properly chosen operating temperature and pressure, electron recoils from the abundant gamma and electron background do not nucleate bubbles \citep{behnke-08}, a concept explored in more detail in Chapter \ref{ch:gammas}. On the other hand, nuclear recoils resulting from the neutron and alpha background (as well as from WIMP-nucleon scattering) will nucleate bubbles for recoils above an effective threshold $E_c^\mathrm{eff}$, the maximum of $E_c$ and the projection on the energy axis where the instantaneous stopping power from a nuclear recoil crosses the $\mathrm{d}E/\mathrm{d}x$ threshold (Figure \ref{fig:E_vs_dEdx}).

Things become more complicated when considering the stopping power threshold further. Most commonly, this threshold is defined as the ratio of the critical energy $E_c$ to a critical distance $L_c$. The value of $b$ in Equation \ref{eq:L_c} is known only roughly, and changes with the temperature \citep{das-04, harper-91}. Further, the mean projected range $r_\mathrm{range}$ of each recoil track is just that --- an average. Each individual track has variation in range (width to the distribution), giving rise to the non-zero probability that some recoils will deposit their energy over too great a range to nucleate bubbles, despite having an energy above $E_c$ and a nominal stopping power above $\left.\mathrm{d}E/\mathrm{d}x\right|_c$.

\begin{figure} [t!]
\centering
\includegraphics[scale=0.35]{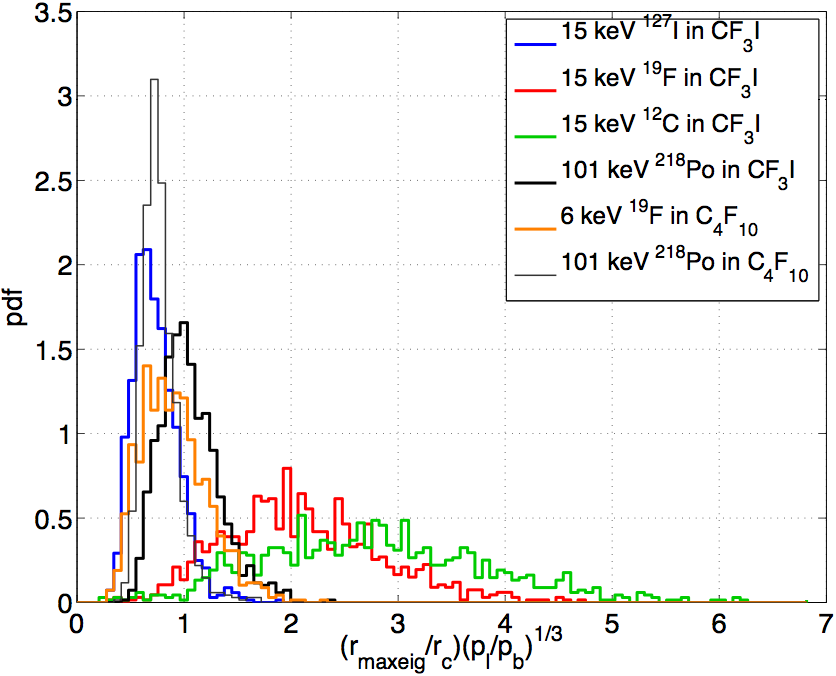}
\caption[Track size of $^{218}$Po and $^{19}$F recoils in CF$_3$I and C$_4$F$_{10}$]{The distribution of track sizes divided by the Seitz range for various recoils in COUPP (CF$_3$I) and PICASSO (C$_4$F$_{10}$). Figure from \cite{dahl-11a} reproduced with permission.}
\label{fig:tracks_CF3I_C4F10}
\end{figure}

This effect is statistical in nature and is not considered in the simple Seitz model. To study it more, simulations of individual recoils for different species of nuclei in CF$_3$I and C$_4$F$_{10}$ were performed using TRIM/SRIM \citep{ziegler-96}, giving track length measurements $r_\mathrm{track}$ that can be compared to the Seitz range $r_c \left(\rho_b/\rho_l\right)^{1/3}$. First, populations that are known to closely follow the Seitz model were considered (Figure \ref{fig:tracks_CF3I_C4F10}).  As seen in Section \ref{sec:efficiency_alphas}, the efficiency of 101 keV $^{218}$Po recoils (resulting from $^{222}$Rn alpha decays in the CF$_3$I) is $>$ 74\%, and so these recoils very closely follow the Seitz model. PICASSO similarly reports sharp turn-on in sensitivity to $^{218}$Po recoils in C$_4$F$_{10}$. On the other hand, $^{19}$F recoils in COUPP are expected to have lower nucleation efficiencies (Section \ref{sec:efficiency_neutrons}), compared to the still sharp turn-on of 6 keV $^{19}$F recoils in PICASSO \citep{archambault-11}. Indeed, as seen in Figure \ref{fig:tracks_CF3I_C4F10}, the track length of $^{19}$F recoils in CF$_3$I is much larger than either $^{218}$Po recoils or $^{19}$F recoils in C$_4$F$_{10}$, with a much larger variance, reinforcing the need for understanding beyond the Seitz model for $^{19}$F recoils in CF$_3$I.

It is apparent that the $^{12}$C and $^{19}$F recoils have larger track lengths with wider distributions than the $^{127}$I recoils for similar recoil energies in CF$_3$I. Therefore, for the remainder of this work, it will be assumed that $^{127}$I has a 100\% bubble nucleation efficiency, while that for the $^{12}$C and $^{19}$F (for simplicity, assumed to be equal) will be determined from a fit to the neutron calibrations. Both the flat-efficiency model

\begin{equation}
\label{eq:P_flat_repeat}
P\left(E_r,\frac{\mathrm{d}E_r}{\mathrm{d}x} ; E_c, L_c \right) = \eta \Theta\left(E_r - E_c\right) \Theta\left(\frac{\mathrm{d}E_r}{\mathrm{d}x} - \frac{E_c}{L_c}\right),
\end{equation}

\noindent and the PICASSO sigmoid-efficiency model \citep{archambault-11}

\begin{equation}
\label{eq:P_PICASSO_repeat}
P\left(E_r,\frac{\mathrm{d}E_r}{\mathrm{d}x} ; E_c, L_c \right) = \left( 1 - e^{-\alpha \left(E_r - E_c^\mathrm{eff}\right) / E_c^\mathrm{eff}} \right) \Theta \left(E_r - E_c\right) \Theta\left(\frac{\mathrm{d}E_r}{\mathrm{d}x} - \frac{E_c}{L_c}\right)
\end{equation}

\noindent are considered, and both $\eta_\mathrm{C,F}$ and $\alpha_\mathrm{C,F}$ (with assumed $\eta_\mathrm{I}=1$ and $\alpha_\mathrm{I}\rightarrow\infty$) are determined from a fit to the neutron calibration data.

\section{Alpha Turn-Off Calibration}
\label{sec:efficiency_alphas}

The 101 keV nuclear recoil of $^{218}$Po resulting from the alpha decay of $^{222}$Rn is among the population of nuclear recoils with short mean projected ranges in Figure \ref{fig:tracks_CF3I_C4F10}, so it too should be nearly 100\% efficient in nucleating bubbles. As explained in Section \ref{sec:datasets_alpharate}, monochromatic energy depositions such as those arising from this recoil induce a constant response in bubble chambers, up to the threshold energy where these recoils are no longer able to nuclear bubbles. In the decay chain from $^{222}$Rn (see Figure \ref{fig:U238_decay_chain}), there are 3 such alpha decays and associated recoils that are time-correlated --- including the 112 keV $^{218}$Pb recoil and the 146 keV $^{210}$Pb recoil. As these 3 alpha decays are in equilibrium, the event rate from each should be equivalent.

By analyzing the alpha event rate around where each of these recoils fall below threshold and subsequently turn off, two pieces of information can be obtained --- the relative bubble nucleation efficiency of these recoils can be determined and the energy threshold of the bubble chamber can be calibrated. The former is possible because the 3 recoils turn off at different energies, so the individual rate from each can be determined by scanning over high thresholds. The latter is possible by relating the threshold at turn-off with the energy of the recoiling nucleus. Though the alpha can contribute some amount of energy to that needed for bubble formation, a majority of the energy in the event comes from the recoil. A detailed understanding of how much energy an alpha can contribute to an event is not known. However, calculations performed by \cite{dahl-11a} similar to those that produced the theoretical $a=6.07$ value for the scaling factor in Equation \ref{eq:dEdx} \citep{bell-74} show that the amount of energy made available by the alpha to the recoil bubble is $\sim$20 keV. This value depends on the instantaneous stopping power of the alpha, and the critical energy and critical radius of the superheated fluid.

\begin{figure} [t!]
\centering
\includegraphics[scale=0.45]{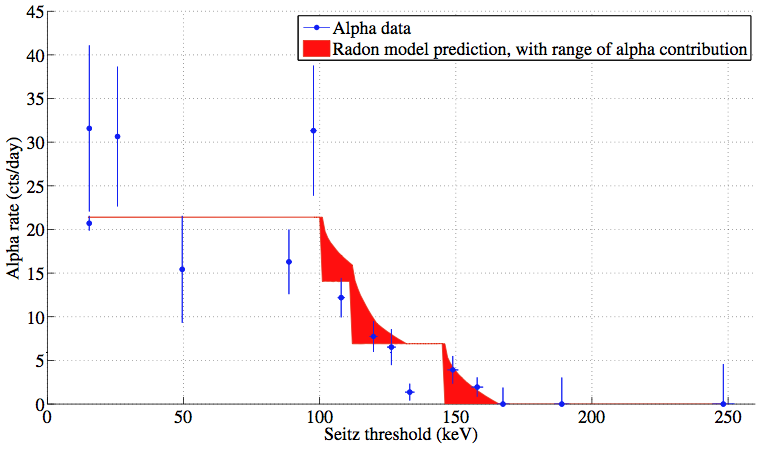}
\includegraphics[scale=0.45]{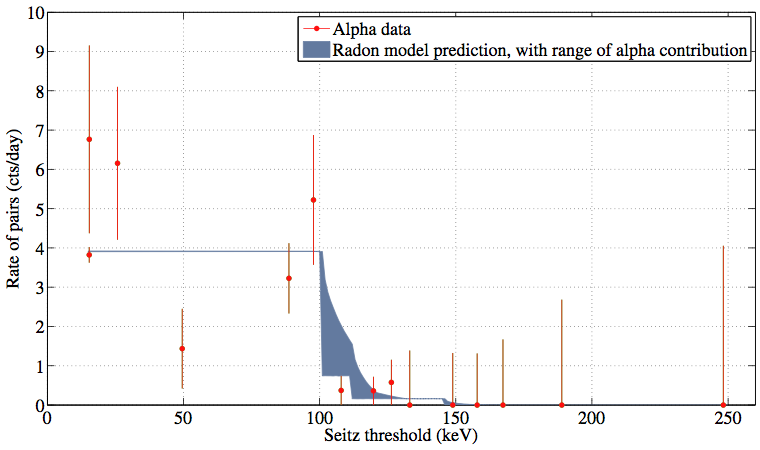}
\caption[High threshold scan to search for alpha event cutoff]{Event rates from the pressure scan data along with predictions based on the $^{222}$Rn event simulation, for single-event (top) and time-correlated (bottom) predictions. Figure from \cite{lippincott-11b} reproduced with permission.}
\label{fig:alpha_cutoff}
\end{figure}

Calibration data were taken from June 28 to July 28, 2011 to study the alpha turn-off region by pressure-scanning at high thresholds. A simulation based on the nominal Seitz model was developed by C.E. Dahl was used to predict the number of events generated in the chamber from the $^{222}$Rn decay chain, taking as a prior the number of alpha events observed in the dark matter search data (Section \ref{sec:datasets_alpharate}). These predictions were compared with the single-event and time-correlated data from the pressure-scan, shown in Figure \ref{fig:alpha_cutoff}.  The lower limit of the shaded region represents a model where the alpha track contributes no energy to the total recoil energy available to nucleated a bubble. The upper limit randomly assigns the total energy used in bubble formation from a quadratic extrapolation between the recoil energy of the daughter and this recoil energy plus 20 keV from the alpha contribution.

From comparing these simulations to data, it is seen that the nominal Seitz model accurately predicts the location of turn-off for each of the recoils in the $^{222}$Rn chain as well as in the time-correlated pairs. The sharp cutoff seen at the 101 keV threshold in the time-correlated pressure scan data shows that the Seitz model is good to within 10\% at these energies. Since the alpha event rate was found to be constant in time and threshold (Section \ref{sec:datasets_alpharate}), it is known that the nucleation efficiency is close to 100\% at low thresholds. As the threshold approaches the recoil energy of the daughters, this efficiency remains close to 100\%, with a conservative nucleation efficiency limit $>$ 75\% \citep{lippincott-11b}.

\section{Neutron Calibration Sources}
\label{sec:efficiency_sources}

\begin{figure} [t!]
\begin{minipage}[b]{0.5\linewidth}
\centering
\includegraphics[scale=0.35]{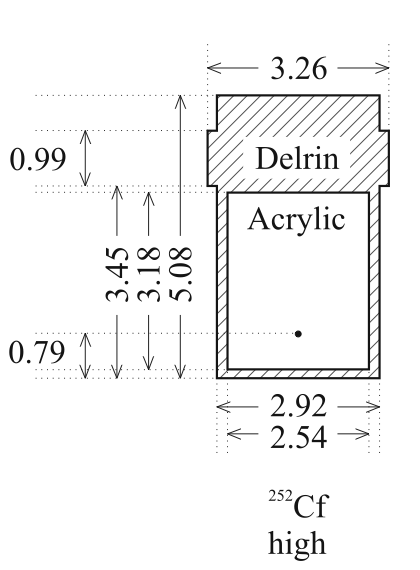}
\end{minipage}
\begin{minipage}[b]{0.5\linewidth}
\centering
\includegraphics[scale=0.35]{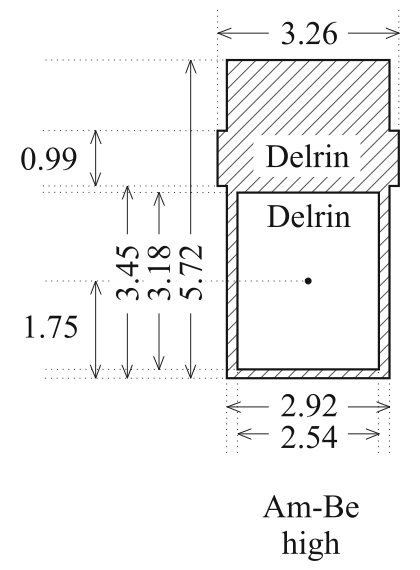}
\end{minipage}
\caption[$^{252}$Cf and Am-Be neutron source diagrams]{Diagrams of the $^{252}$Cf and Am-Be neutron sources used for calibrations. The location of the source within the container is designated by a dot. All measurements in cm. Figure from \cite{loach-08} reproduced with permission.}
\label{fig:Cf_AmBe_sources}
\end{figure}

Two neutron sources were used for calibration runs at SNOLAB (Figure \ref{fig:Cf_AmBe_sources}).  The first was an Am-Be ($\alpha$,n) neutron source with a yield of 68.7$\pm$0.74 n/s (measured towards the end of the SNO NCD phase, around the middle of 2006). The second was a $^{252}$Cf spontaneous fission neutron source with a yield of 16.55$\pm$0.11 n/s (measured on June 12, 2001) \citep{loach-08}. When comparing rates produced by simulations with measured rates, one must be careful to use the appropriate source intensities corrected for the date on which the data were collected. This is particularly true for the $^{252}$Cf source which has a half-life of only 2.645 years, resulting in a roughly 2\% loss in activity per month. The longer half-life of $^{241}$Am in the Am-Be source (432.2 years) makes this less important, and a current rate of 68.2$\pm$0.73 n/s is used.

When last measured, the $^{252}$Cf source was found to have 1.92$\pm$0.02\% of the neutrons coming from $^{250}$Cf, an isotope with a longer half-life (13.08 years) than $^{252}$Cf \citep{loach-08}. Propagating the $^{250}$Cf concentration to the time of each calibration run with the COUPP 4 kg bubble chamber (runs were performed in January, May, and June of 2011), and taking into account the loss of activity due to the half-life of each isotope, the appropriate activity of the $^{252}$Cf source was determined (Table \ref{tab:cf_activity}). For simplicity in the simulations, the source was assumed to have the spectrum of $^{252}$Cf, the neutron spectra from both $^{250}$Cf and $^{252}$Cf being approximately equivalent (both follow a Watts fission spectrum). The neutron multiplicity of $^{250}$Cf and $^{252}$Cf is 3.511$\pm$0.037 and 3.768$\pm$0.005, respectively \citep{loach-08}, and this was taken into account in the simulations given the relative abundance of each isotope at the time of calibration. See Appendix \ref{ch:mcnp} for a description of the Monte Carlo method used to produce expected count rates for each calibration source.

\begin{table}[t!]
\centering
\begin{tabular} {| c | c | c | c |}
\hline
Date & $^{252}$Cf activity & $^{250}$Cf activity & Total activity\\
 & \small{(n/s)} & \small{(n/s)} & \small{(n/s)} \\
\hline
\hline
June 2001 & 16.24$\pm$0.11 & 0.318$\pm$0.004 & 16.55$\pm$0.11 \\
\hline
January 2011 & 1.31$\pm$0.01 & 0.191$\pm$0.002 & 1.51$\pm$0.01 \\
May 2011 & 1.21$\pm$0.01 & 0.188$\pm$0.002 & 1.39$\pm$0.01 \\
June 2011 & 1.17$\pm$0.01 & 0.187$\pm$0.002 & 1.36$\pm$0.01 \\
\hline
\end{tabular}
\caption[Activity of $^{252}$Cf neutron source]{The activity of the $^{252}$Cf source at the time of calibration runs. The concentrations of both $^{252}$Cf and $^{250}$Cf isotopes were propagated forward in time to find the total activity of the source. For simplicity in simulations, the source was taken to be entirely $^{252}$Cf with an activity given by the total activity shown.}
\label{tab:cf_activity}
\end{table}

\begin{figure} [t!]
\centering
\includegraphics[scale=0.55]{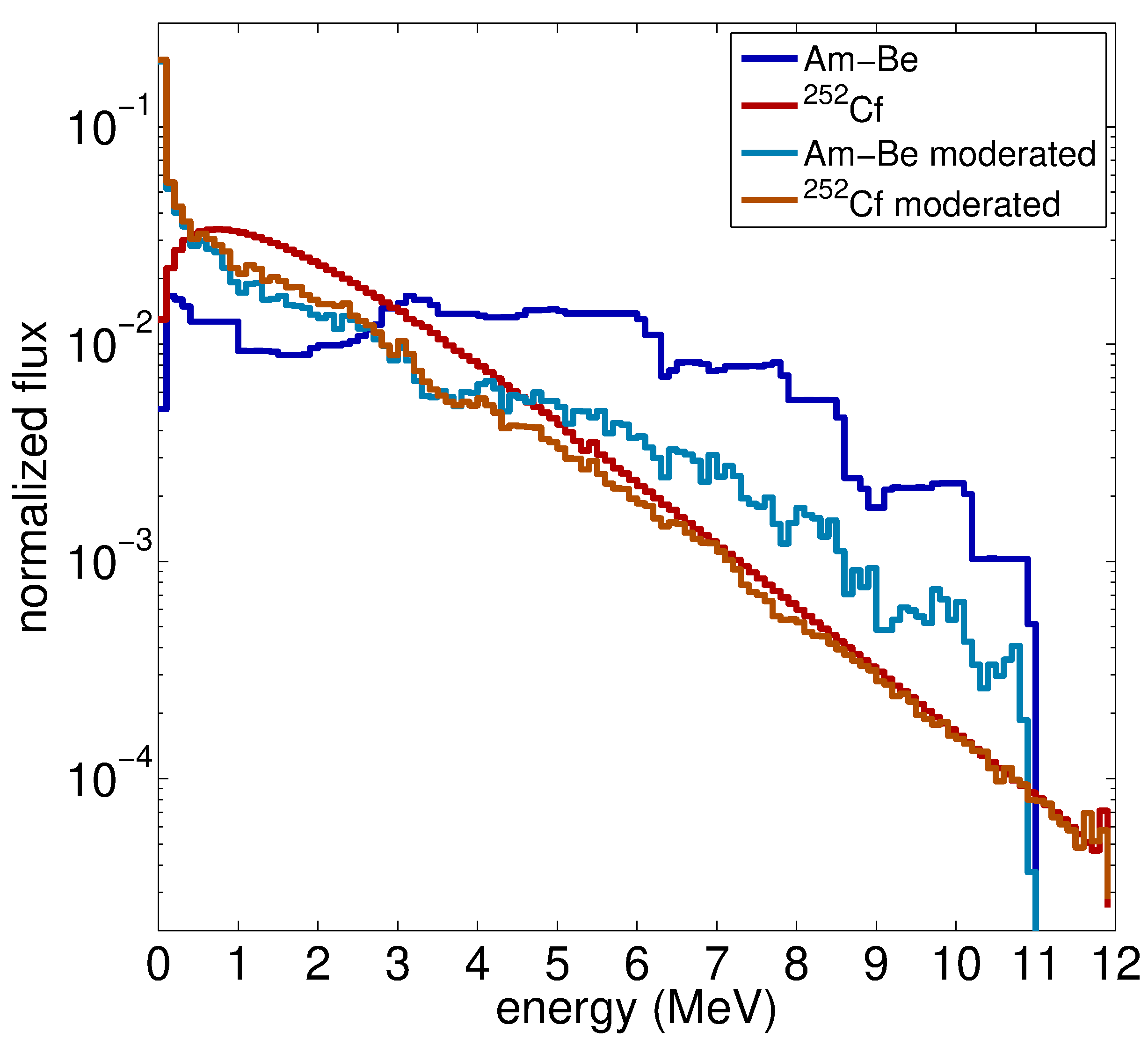}
\caption[Am-Be and $^{252}$Cf initial and moderated neutron spectra]{The neutron spectra from both Am-Be and $^{252}$Cf sources used in calibration simulations. Both spectra are built in to MCNP-PoliMi , with correct multiplicity where appropriate. Also shown is the spectrum of each as it reaches the CF$_3$I volume after being moderated by the material between source and target. While the Am-Be initially has a much harder spectrum than $^{252}$Cf, after moderation they are fairly similar, peaking sharply below 1 MeV.}
\label{fig:source_spectra}
\end{figure}

Because of safety standards at SNOLAB, the sources were required to be double encapsulated --- the Am-Be source in two layers of Delrin, also known as polyoxymethylene (CH$_2$O)$_n$, and the $^{252}$Cf source in a layer of acrylic surrounded by a layer of Delrin (Figure \ref{fig:Cf_AmBe_sources}). While the source containers are expected to moderate the neutrons slightly (which somewhat softens spectrum), simulations done with both the encapsulation present and absent show that this effect is only of order $\sim$2\%. The initial normalized neutron emission spectrum of each source used in the calibration simulations that is built in to MCNP-PoliMi \citep{padovani-02} is shown in Figure \ref{fig:source_spectra}. 

Despite the Am-Be source having a harder neutron spectrum than the $^{252}$Cf source, both sources produced roughly the same spectrum of neutrons incident on the CF$_3$I target. This is because the amount of moderator between the Am-Be source and the target volume was greater than that between the $^{252}$Cf source and target --- resulting in a more moderated spectrum. The neutron calibration sources were lowered through a hole in the top of the water shielding and placed adjacent to the pressure vessel of the bubble chamber. Since the Am-Be source has a higher neutron intensity than the $^{252}$Cf source, it was placed further away from the chamber than the $^{252}$Cf. Am-Be calibrations were mostly performed with the source lowered 36" below the top of the water shield, although smaller runs were carried out with the source at 34" and 38" below, as well. $^{252}$Cf calibrations were performed with the source at either 54" or 46" below the top of the water shield.

\begin{figure} [t!]
\centering
\includegraphics[scale=0.55]{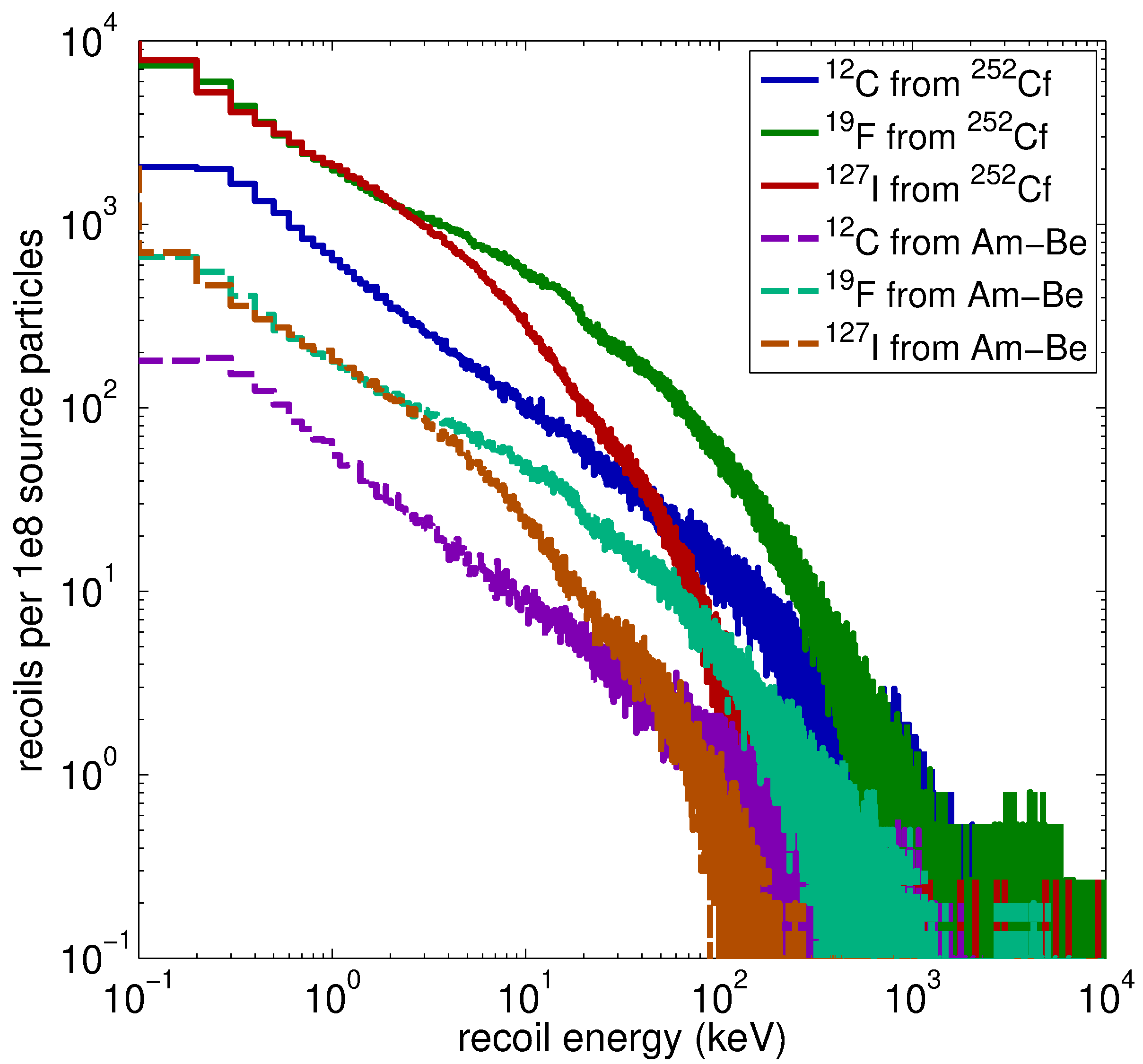}
\caption[$^{12}$C, $^{19}$F, and $^{127}$I recoil spectra from calibration neutron scatters]{The recoil energy spectra per recoiling species in CF$_3$I resulting from elastic scatters from the moderated calibration source neutrons, both in the calibrations with $^{252}$Cf (54" below the top of the water shield) and Am-Be (36" below).}
\label{fig:recoil_spectra}
\end{figure}

Both target-incident spectra are sharply peaked below 1 MeV (Figure \ref{fig:source_spectra}), resulting in the recoil energy spectra per recoiling species shown in Figure \ref{fig:recoil_spectra}.  These recoil spectra, produced by either the $^{252}$Cf source (at 54") or the Am-Be source (at 36"), are roughly the same except for the higher incidence of scatters from the $^{252}$Cf source being closer, an effect diminished by the fact that the $^{252}$Cf is roughly 50 times weaker than the Am-Be source.

\section{Neutron Calibration Results}
\label{sec:efficiency_neutrons}

\begin{figure} [t!]
\centering
\includegraphics[scale=0.55]{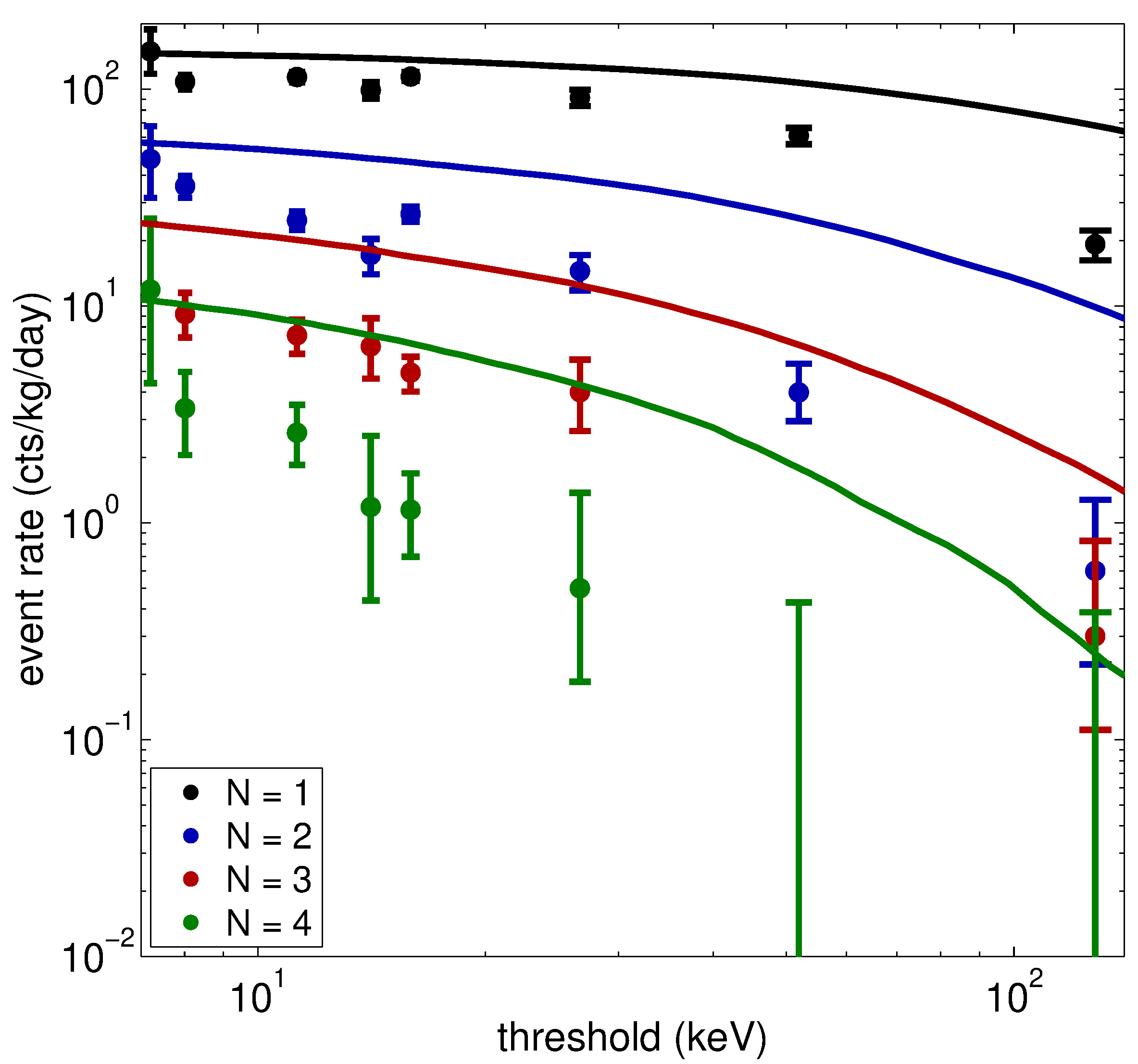}
\caption[Measured rates versus uncorrected MCNP predictions for Am-Be calibrations]{The observed rates for each calibration run with Am-Be at 36" (data points) compared to the MCNP predictions (solid lines), for multiplicity $N=1$ through $N=4$, with 100\% bubble nucleation efficiency and $b\rightarrow\infty$ (\emph{i.e.} no stopping power threshold).}
\label{fig:ambe_rates}
\end{figure}

Calibration runs were performed with the Am-Be source at three vertical positions --- 36", 34", and 38" below the top of the water shield (for reference, the center of the CF$_3$I volume is at 54.3"). The activity of this source was too high to be lowered further without risk of over-saturating the chamber the bubble nucleations. Most runs were with the source at 36", where data were taken at a variety of thresholds. Comparisons were made at each threshold with predicted rates from an MCNP-PoliMi simulation, a software package for simulating transport of gamma, neutron, and electron radiation in materials \citep{padovani-02}. Assuming an infinite $L_c$ scaling factor (\emph{i.e.} no stopping power threshold) and with a 100\% nucleation efficiency, the predicted rates versus the measured rates for all 36" Am-Be calibration runs are shown in Figure \ref{fig:ambe_rates}.  From this figure, the need for both a stopping power threshold and a bubble nucleation efficiency are quite evident. For realistic values of the scaling factor ($2 \leq a \leq 12$) from Equation \ref{eq:dEdx}, the stopping power threshold is consequential only for higher thresholds --- from Figure \ref{fig:E_vs_dEdx}, only for high thresholds does the instantaneous stopping power cross the stopping power threshold at energies higher than the critical energy. In other words, for low thresholds, the effective critical energy $E_c^\mathrm{eff}$ is equal to the critical energy $E_c$. Therefore, in addition to a finite scaling factor, a bubble nucleation efficiency lower than 100\% is required for predictions to match measurements at low thresholds ($\sim$10 keV).

Because the $^{252}$Cf source has a much lower neutron yield than the Am-Be source, calibrations with the $^{252}$Cf source were performed at vertical positions of 42" and 54" below the top of the water shield (the latter being approximately in line with the center of the CF$_3$I target volume). These calibrations provided a nice cross-check to ensure there were not source-specific errors affecting the calibrations and that the extra moderation of the Am-Be neutrons from the longer path-length did not cause the simulations to deviate too much from observations. The raw number of counts observed for each run at different thresholds for bubble multiplicity $N=1$ through $N=4$, as well as the predicted number of counts (taking into account live-time and cut efficiencies), are shown in Table \ref{tab:neutron_counts}. \begin{table}[t!]
\centering
\small{
\begin{tabular} {| c || c | c | c | c || c | c | c | c |}
\hline
\multicolumn{9}{| l |}{Neutron calibration data} \\
\hline
Data Set & \multicolumn{4}{| c ||}{Observed Counts} & \multicolumn{4}{| c |}{Predicted Counts} \\
\cline{2-9}
 & \small{$N=1$} & \small{$N=2$} & \small{$N=3$} & \small{$N=4$} & \small{$N=1$} & \small{$N=2$} & \small{$N=3$} & \small{$N=4$} \\
\hline
AmBe-36"-34$^\circ$C-30psia & 556 & 162 & 30 & 7 & 664.70 & 280.61 & 103.22 & 40.73 \\
AmBe-36"-34$^\circ$C-40psia & 165 & 29 & 8 & 1 & 225.79 & 76.62 & 24.84 & 8.64 \\
AmBe-36"-34$^\circ$C-50psia & 170 & 12 & 0 & 0 & 290.07 & 76.20 & 19.93 & 5.34 \\
AmBe-36"-34$^\circ$C-60psia & 70 & 2 & 1 & 0 & 210.72 & 32.88 & 5.52 & 0.85 \\
AmBe-36"-37$^\circ$C-30psia & 384 & 105 & 31 & 11 & 478.47 & 216.64 & 85.30 & 35.83 \\
AmBe-36"-37$^\circ$C-35psia & 150 & 29 & 11 & 2 & 208.68 & 80.82 & 30.73 & 12.36 \\
AmBe-36"-40$^\circ$C-30psia & 179 & 74 & 19 & 7 & 240.11 & 114.77 & 47.67 & 20.97 \\
AmBe-36"-41$^\circ$C-30psia & 20 & 8 & 2 & 2 & 19.62 & 9.48 & 4.02 & 1.78 \\
AmBe-34"-34$^\circ$C-30psia & 122 & 41 & 9 & 1 & 142.96 & 59.64 & 22.96 & 8.60 \\
AmBe-38"-34$^\circ$C-30psia & 195 & 42 & 8 & 1 & 208.67 & 88.68 & 32.92 & 13.23 \\
Cf252-42"-37$^\circ$C-29psia & 82 & 28 & 4 & 2 & 93.95 & 43.74 & 17.86 & 7.82 \\
Cf252-42"-37$^\circ$C-30psia & 210 & 70 & 12 & 3 & 238.52 & 110.54 & 44.95 & 19.50 \\
Cf252-54"-34$^\circ$C-30psia & 479 & 130 & 32 & 10 & 688.66 & 301.31 & 114.54 & 45.28 \\
Cf252-54"-34$^\circ$C-40psia & 194 & 26 & 8 & 1 & 306.09 & 90.31 & 30.67 & 10.49 \\
\hline
\end{tabular} }
\caption[Observed and predicted counts for uncorrected neutron calibrations]{The counts in each neutron calibration data set for multiplicity $N=1$ through $N=4$, along with those predicted by the MCNP-PoliMi simulation, with $\eta = 1$ and $a\rightarrow\infty$. The simulated number of counts were scaled by the exposure of the corresponding data set, considering the cut efficiencies and the residual alpha rate in the $N=1$ data for high-pressure sets.}
\label{tab:neutron_counts}
\end{table} The predicted number of counts are too high to match a simple Seitz model without a bubble nucleation efficiency, particularly in the case $N>1$. However, for nucleation efficiency $\eta$, the expected number of counts roughly scales as $\eta^N$ for multiplicity $N$, allowing for the simulated predictions to match the measured number of counts for a proper choice of efficiency.

\section{Efficiency and Scaling Parameter Fits}
\label{sec:efficiency_fits}

For CF$_3$I at the operating temperatures and pressures of the calibration sets, little variation in the scaling factors $a$ and $b$ for Equation \ref{eq:dEdx_repeat} is observed (with $b \simeq 3 a$). For this reason, only the simple scaling factor $a$ in the critical range $L_c = a r_c$ will be fit. Given a particular efficiency model, the parameters $\eta$ and $\alpha$ (Equations \ref{eq:P_flat_repeat} and \ref{eq:P_PICASSO_repeat}) are best-fit by comparing the efficiency-corrected predictions from MCNP-PoliMi corresponding to the exposure of each data set in Table \ref{tab:neutron_counts}. For high-pressure calibrations ($>$ 30 psia), the expected number of alphas in the calibration set (Section \ref{sec:datasets_alpharate}) is added to the $N=1$ prediction, since the measured number of counts for these sets contains these events.

Since potentially small numbers of raw counts with non-Gaussian error bars are being considered in particular calibration runs (Table \ref{tab:counts}), rather than using a standard $\chi^2$ based on a Gaussian distribution of counts (as in Section \ref{sec:datasets_alpharate}), a modified $\chi^2$ arising from the Poisson distribution of counts is used. For generic fit parameters $\theta$ (here, $a$ and $\eta$ or $\alpha$), in lieu of using the Gaussian $\chi^2$ (Equation \ref{eq:chi2}), it is appropriate to use the Poisson $\chi^2$

\begin{equation}
\label{eq:chi2_poisson}
\chi^2_P = \sum_i \left\{ \begin{array}{cl}
2 \left[ \nu_i - n_i + n_i \ln \left(\frac{n_i}{\nu_i}\right) \right] & n_i > 0 \\
2 \nu_i & n_i = 0 \\
\end{array} \right.
\end{equation}

\noindent with counts $n_i$ and expectations $\nu_i$. The process of determining the best-fit parameters $\hat{\theta}$ is then the same as usual --- find the value of $\theta$ which minimizes $\chi^2_P$. The 1$\sigma$ uncertainty of a parameter $\theta$ with best-fit value $\hat{\theta}$ and $\chi^2_P(\hat{\theta}) = \chi^2_\mathrm{min}$ is the value $\theta$ for which $\chi^2_P(\theta) = \chi^2_\mathrm{min} + 1$ \citep{amsler-08b}. However, the total error budget for the fits will be dominated by the fact that the MCNP predictions come with an inherent 10\% uncertainty (Section \ref{sec:mcnp_uncertainty}). To determine the 1$\sigma$ error bar on each fit parameter, the MCNP-predicted number of counts $\nu_i$ was scaled by 10\% in either direction, and the best fit was determined at these limits. This uncertainty was then added in quadrature with the uncertainty derived from the $\chi^2$ best-fit to determine the total uncertainty.

\begin{figure} [t!]
\centering
\includegraphics[scale=0.55]{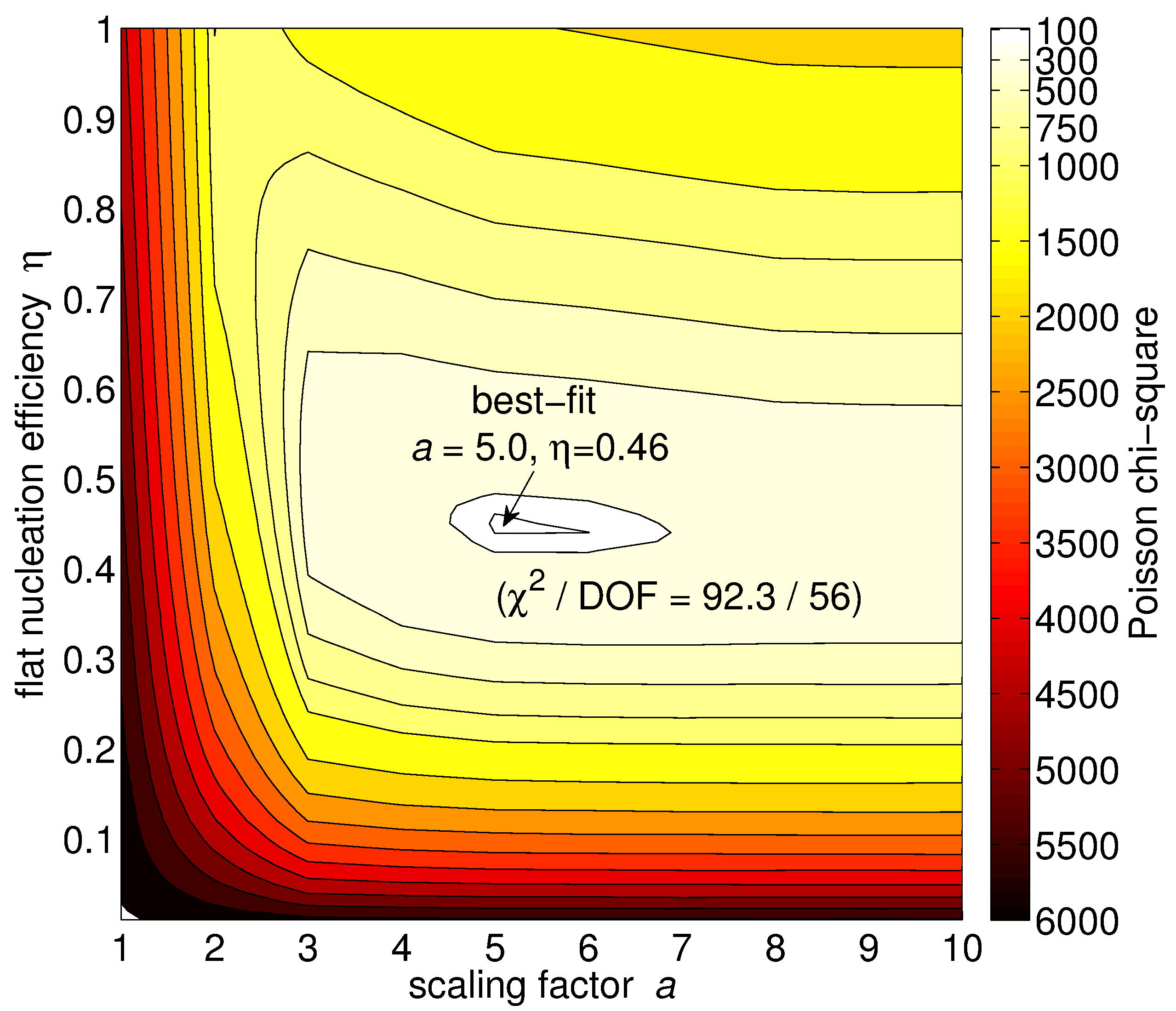}
\caption[Best-fit contours for scaling factor $a$ and flat parameter $\eta$]{Contour plot of the $\chi^2_P$ values (with 54 DOF) calculated for various scaling factors $a$ and nucleation efficiencies $\eta$ in the flat-efficiency model (Equation \ref{eq:P_flat_repeat}). The $p$-values calculated in Equation \ref{eq:p-value} does not represent goodness-of-fit directly because the $\chi^2_P$ does not take into account the $\sim$10\% uncertainty in the MCNP prediction (see text).}
\label{fig:contour_a_eta}
\end{figure}

\begin{figure} [t!]
\centering
\includegraphics[scale=0.55]{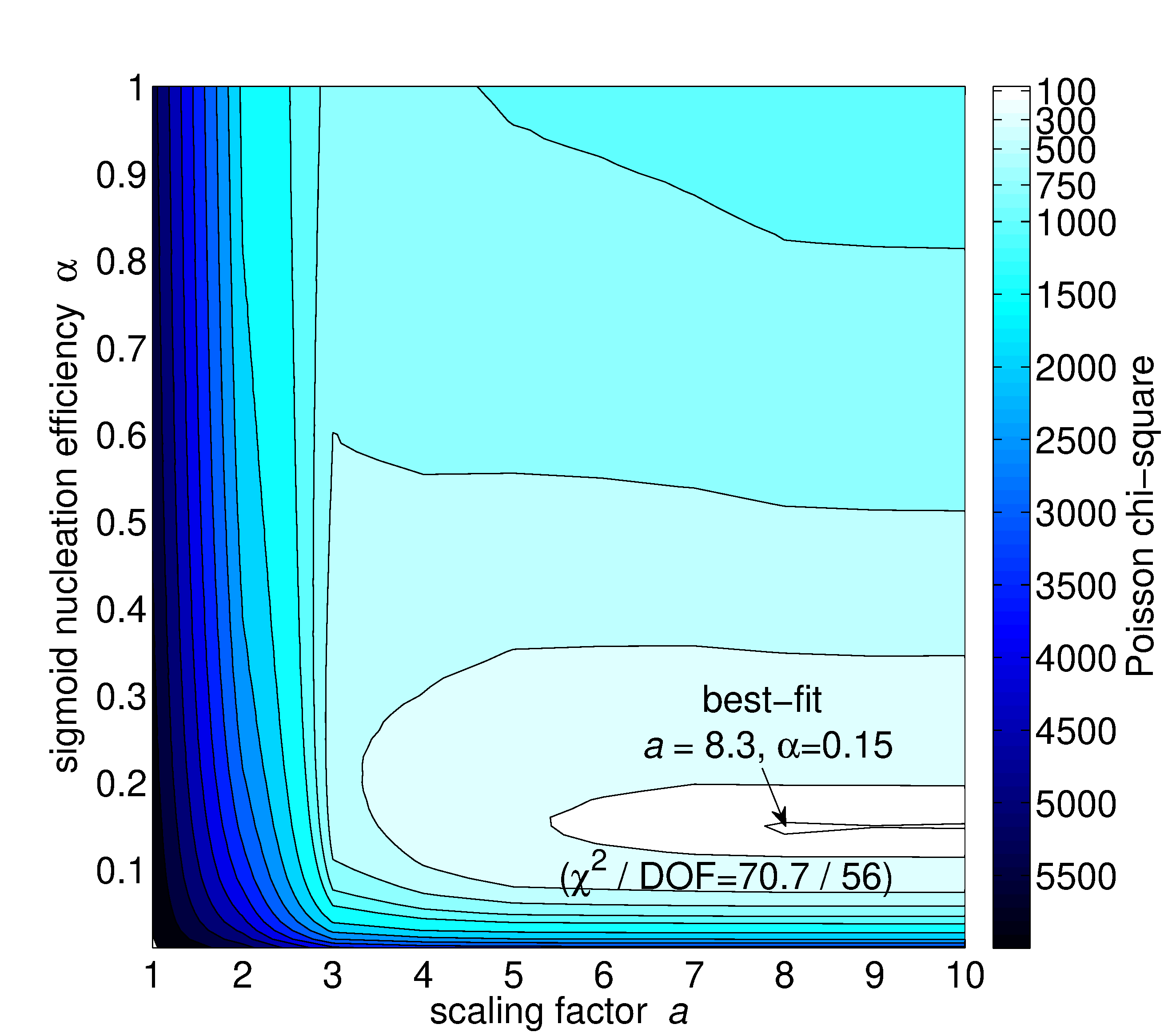}
\caption[Best-fit contours for scaling factor $a$ and sigmoid parameter $\alpha$]{Contour plot of the $\chi^2_P$ values (with 54 DOF) calculated for various scaling factors $a$ and sigmoid parameters $\alpha$ in PICASSO's sigmoid-efficiency model (Equation \ref{eq:P_PICASSO_repeat}). The $p$-values calculated in Equation \ref{eq:p-value} does not represent goodness-of-fit directly because the $\chi^2_P$ does not take into account the $\sim$10\% uncertainty in the MCNP prediction (see text).}
\label{fig:contour_a_alpha}
\end{figure}

In the flat-efficiency model (Equation \ref{eq:P_flat_repeat}), the values of the predicted number of counts $\nu_i$ were generated for each calibration data set $i$, with count number $n_i$.  Minimizing the $\chi^2_P$ resulted in best-fit parameters $a = 5.0^{+0.3}_{-0.1}$ and $\eta=0.46^{+0.04}_{-0.06}$ (Figure \ref{fig:contour_a_eta}).  The value for $a$ is in agreement with expectations mentioned in Section \ref{sec:efficiency_neutrons} and with those used in \cite{behnke-08}, and $\eta$ matches the 50\% efficiency found in \cite{behnke-11}. The same process was repeated using PICASSO's sigmoid model (Equation \ref{eq:P_PICASSO_repeat}), yielding best-fit parameters  $a = 8.3^{+0.2}_{-0.3}$ and $\alpha=0.15^{+0.05}_{-0.03}$ (Figure \ref{fig:contour_a_alpha}).  In both cases, the low scaling factors ($a \leq 3$) were excluded by the fits, but little information was available for fitting higher values, because there was not as much calibration data in the high threshold regime which would help define the shape of the fall-off dictated by $a$. Also, while earlier results from COUPP have suggested a temperature-dependence to the nucleation efficiency \citep{behnke-08}, no increase to the best-fit efficiency was observed when only considering the high-temperature data.

\begin{figure} [t!]
\centering
\includegraphics[scale=0.55]{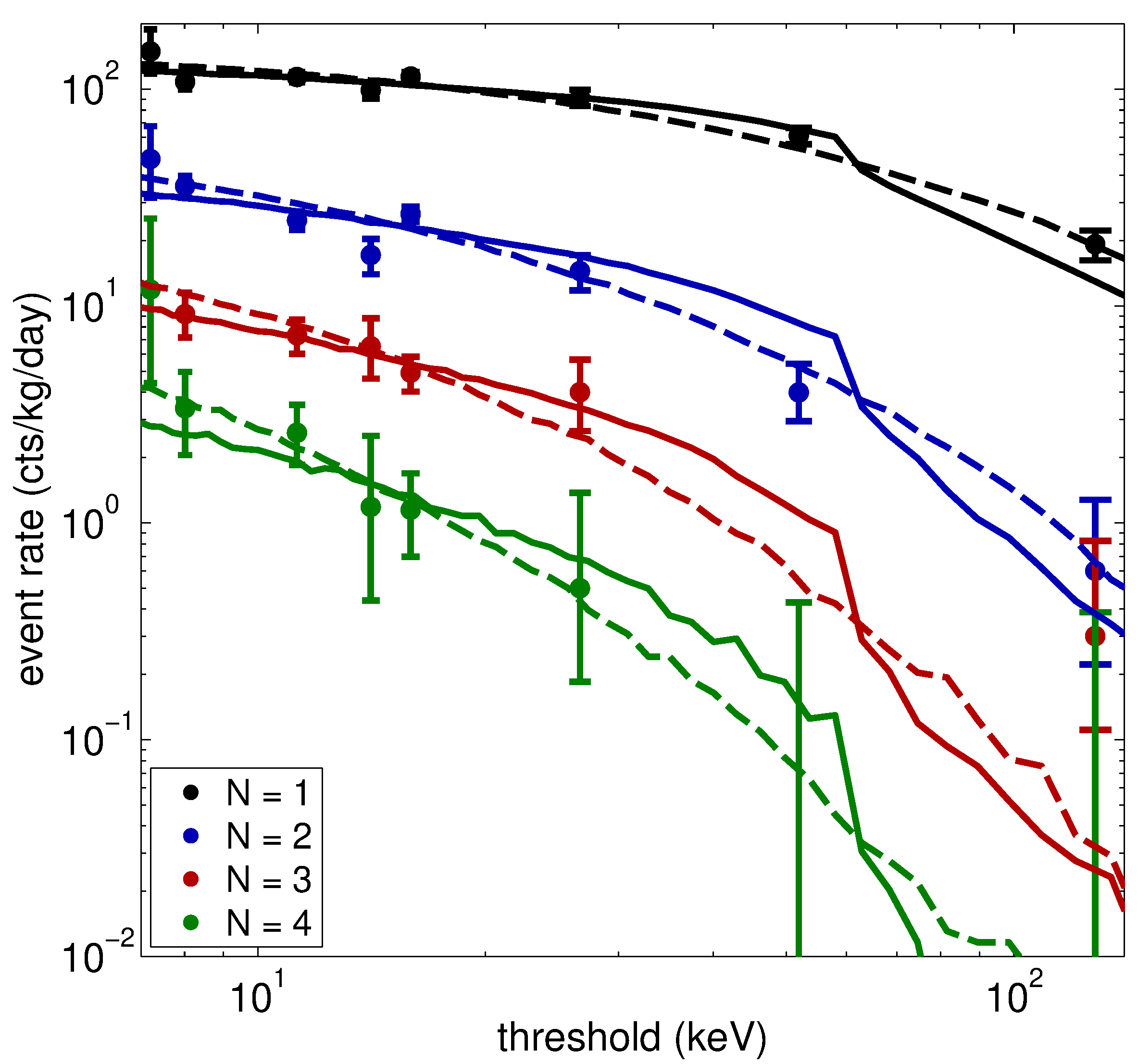}
\caption[Measured rates versus corrected MCNP predictions for Am-Be calibrations]{The observed rates for each calibration run with Am-Be at 36" (data points) compared to the MCNP predictions, for multiplicity $N=1$ through $N=4$, with $a=5.0$ and $\eta=0.46$ in the flat-efficiency model (solid lines) and $a=8.3$ and $\alpha=0.15$ in the sigmoid-efficiency model (dashed lines).}
\label{fig:ambe_rates_corrected}
\end{figure}

Applying the best-fit parameters found in each model, the MCNP-predicted rates provide a satisfactory fit to the calibration data, a large improvement with respect to the uncorrected comparison (Figure \ref{fig:ambe_rates}). Figure \ref{fig:ambe_rates_corrected}  shows these improved predictions for the flat-efficiency model (solid lines) and the sigmoid-efficiency model (dashed lines), compared to the Am-Be calibrations runs with the source located 36" below the top of the water shield. It is therefore evident that the best-fit parameters found here provide a good match to the available calibration information.

\singlespacing
\chapter{Neutron Background Characterization}
\label{ch:backgrounds}
\doublespacing

Because of the dual threshold nature of the bubble chamber, the COUPP experiment is particularly well-suited to reduce many of the backgrounds that arise in other rare-event searches. As can be seen in Figure \ref{fig:E_vs_dEdx}, by tuning the temperature and pressure of the bubble chamber, recoiling electrons from interactions with betas, gammas, x-rays, or muons can be ensured to be below nucleation threshold. Even if the recoiling electron is above energy threshold, parameters can be chosen such that it still will not exceed the stopping power threshold required to nucleate a bubble, except in the rarest of cases (see Section \ref{sec:gammas_calibrations}). Another potential background arises from nuclear recoils initiated by alpha decays in the CF$_3$I, but these can be identified via acoustic discrimination (Section \ref{sec:datasets_alpha}) and eliminated as a background. Therefore, nuclear recoil events resulting from neutron elastic scattering (and, in principle, the occasional photonuclear interaction from rare high energy gammas with $E_\gamma \gtrsim 10$ MeV, see Section \ref{sec:gammas_highenergy}) remains as the principal background for the COUPP experiment. These events may be similar in nature to the nuclear recoils resulting from WIMP scattering that would comprise a dark matter signal and must be reduced to maximize the detector's sensitivity to dark matter.

\begin{figure} [t!]
\centering
\includegraphics[scale=0.15]{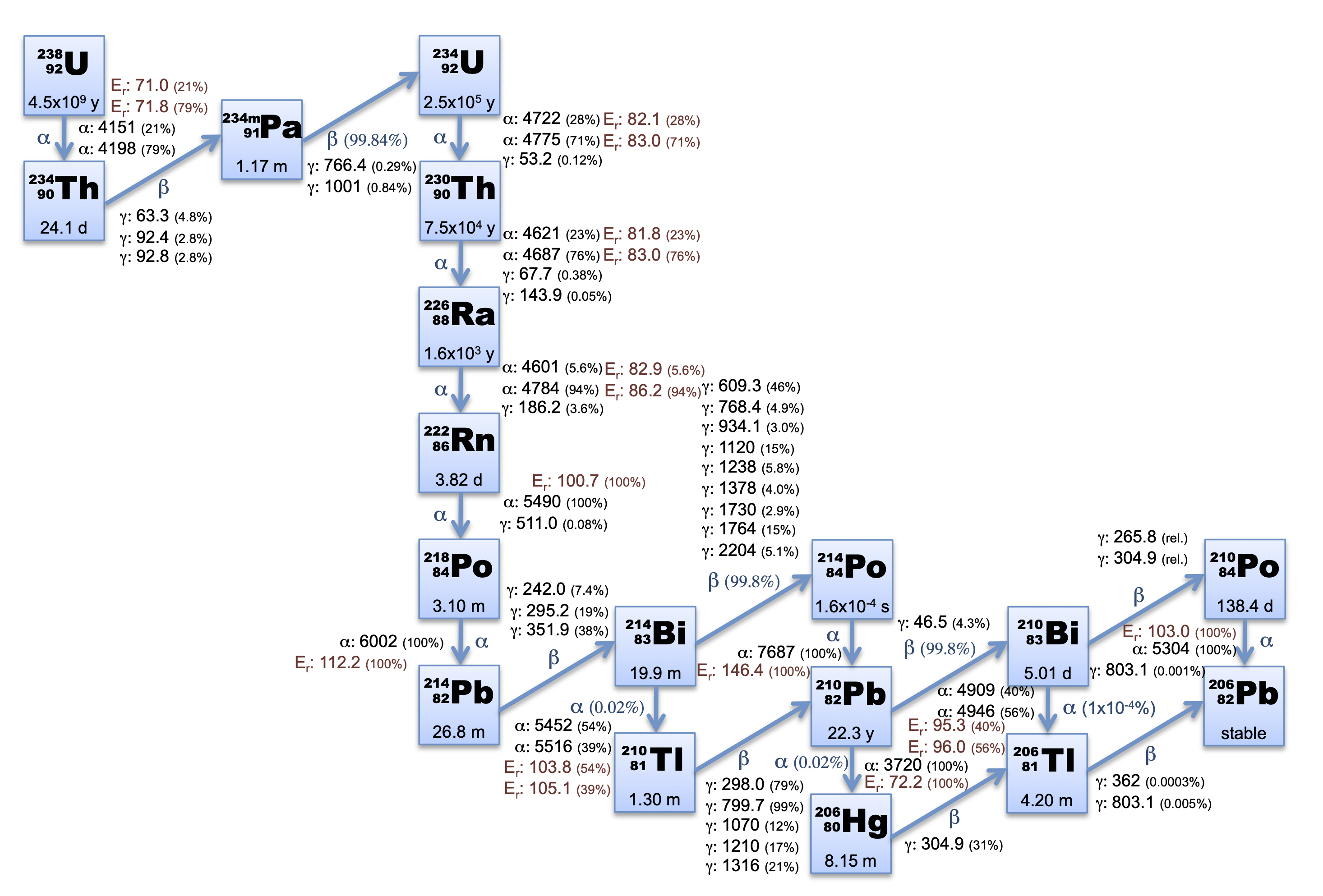}
\caption[$^{238}$U decay chain]{$^{238}$U decay chain, showing the half-life of each radioisotope in the chain, its decay mode (and the branching ratios for different modes, when appropriate), and the energies (and intensities) of all alphas and the most common gammas (in keV). For each alpha decay, the recoil energy $E_r$ (keV) of the daughter nucleus is also shown. Data from \cite{chu-99}.}
\label{fig:U238_decay_chain}
\end{figure}

\begin{figure} [t!]
\centering
\includegraphics[scale=0.13]{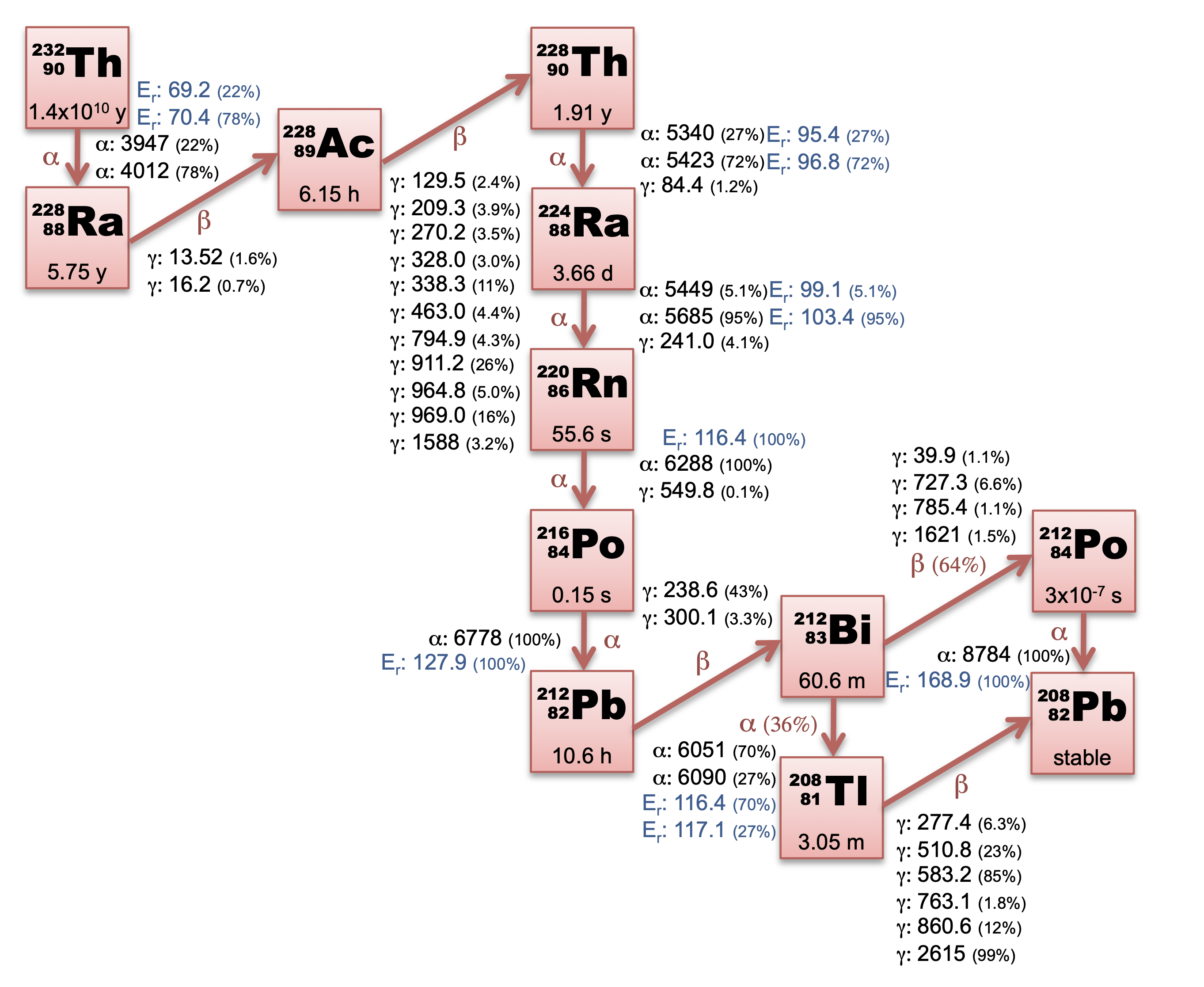}
\caption[$^{232}$Th decay chain]{$^{232}$Th decay chain, showing the half-life of each radioisotope in the chain, its decay mode (and the branching ratios for different modes, when appropriate), and the energies (and intensities) of all alphas and the most common gammas (in keV). For each alpha decay, the recoil energy $E_r$ (keV) of the daughter nucleus is also shown. Data from \cite{chu-99}.}
\label{fig:Th232_decay_chain}
\end{figure}

The nuclear recoil background is generated from multiple sources and can be handled or understood in different ways. Cosmogenic neutrons generated by the ($\mu$,n) reaction in the detector components and the surrounding materials can be minimized by running the experiment in a deep site, allowing for a significant overburden to shield the detector from most cosmic rays. Neutrons arising from spontaneous fission and ($\alpha$,n) reactions in the surrounding rock can be moderated by low-$Z$ shielding materials and mostly eliminated. Further, ($\alpha$,n) neutrons from concentrations of radon gas dissolved in the detector liquids can be minimized by proper fluid handling. Finally, radioactive contamination of the detector components themselves from natural abundances of $^{238}$U and $^{232}$Th and their respective daughters (the decay chains for both are shown in Figures \ref{fig:U238_decay_chain} and \ref{fig:Th232_decay_chain})  can potentially be eliminated by screening for radiopurity and proper handling. The study of each of the neutron-induced backgrounds as well as the ways of minimizing or eliminating them from our dark matter search data is the subject of this chapter.

In order to determine the expected event rate generated by each of these neutron backgrounds, the following procedure was conducted. For a given background, the energy spectrum and source strength were determined via means outlined in each section below. The energy spectrum was used to define a source in an MCNP-PoliMi simulation of the bubble chamber (Section \ref{sec:mcnp_mcnp}), where the information from each individual interaction (typically nuclear recoils) within the target CF$_3$I volume was recorded. The number of histories simulated was converted into a time scale using the measured or calculated source strength. The expected event rate is then simply the number of MCNP-predicted recoils above the effective threshold divided by the time scale. In order to compare with data, the bubble nucleation efficiencies and data cut efficiencies found in Chapters \ref{ch:datasets} and \ref{ch:efficiency} were taken into account when generating expected event rates.

\section{Cosmogenic Neutrons}
\label{sec:backgrounds_cosmogenic}

Cosmic rays consist of both primary particles generated from high-energy extraterrestrial sources and secondary particles generated by the interaction of these primary particles with the top layer of the atmosphere. They represent a significant hurdle for any rare-event search by quickly background-limiting any experiment without a significant shielding of these particles. While most cosmic ray particles are diminished rather rapidly by increasing the overburden of the underground experiment, muons generated as secondary cosmic rays have a deep penetration depth through the Earth's surface due to the their low rate of energy loss via bremsstrahlung and direct ionization because of their relatively high mass. These muons can also scatter off nuclei within detector components and generate a neutron background via the ($\mu$,n) reaction. Identifying these cosmogenic neutron events by tagging them using a muon veto can eliminate them from the rare-event search data, but the high rate of such events and finite veto efficiency when at a shallow site will limit the sensitivity of the detector. Therefore, performing experiments in deep-underground sites, such as at SNOLAB, is essential in eliminating the background associated with cosmogenic neutrons and maximizing the sensitivity of a detector.

The COUPP 4 kg bubble chamber was deployed in two steps: first at a shallow site to demonstrate stable remote operations and then at a deep site for a dedicated low-background dark matter search. The former deployment was to the MINOS near detector hall at Fermilab, located three meters off axis of the NuMI neutrino beam, in a cavern at a depth of 225 meters water equivalent (m.w.e.) \citep{michael-08}. This overburden was sufficient to eliminate the hadronic cosmic ray flux and significantly shielded the detector from the cosmic muon flux. Those muons still reaching this depth were tagged using a muon veto system consisting of 1000 gallon Bicron 517L liquid scintillator counter equipped with 19 RCA-2425 photomultiplier tubes, surrounding the detector on the top and sides \citep{behnke-11}. Events arising from neutrons generated by these muons were eliminated from the data set through the muon veto tagging.

\begin{figure} [t!]
\centering
\includegraphics[scale=0.55]{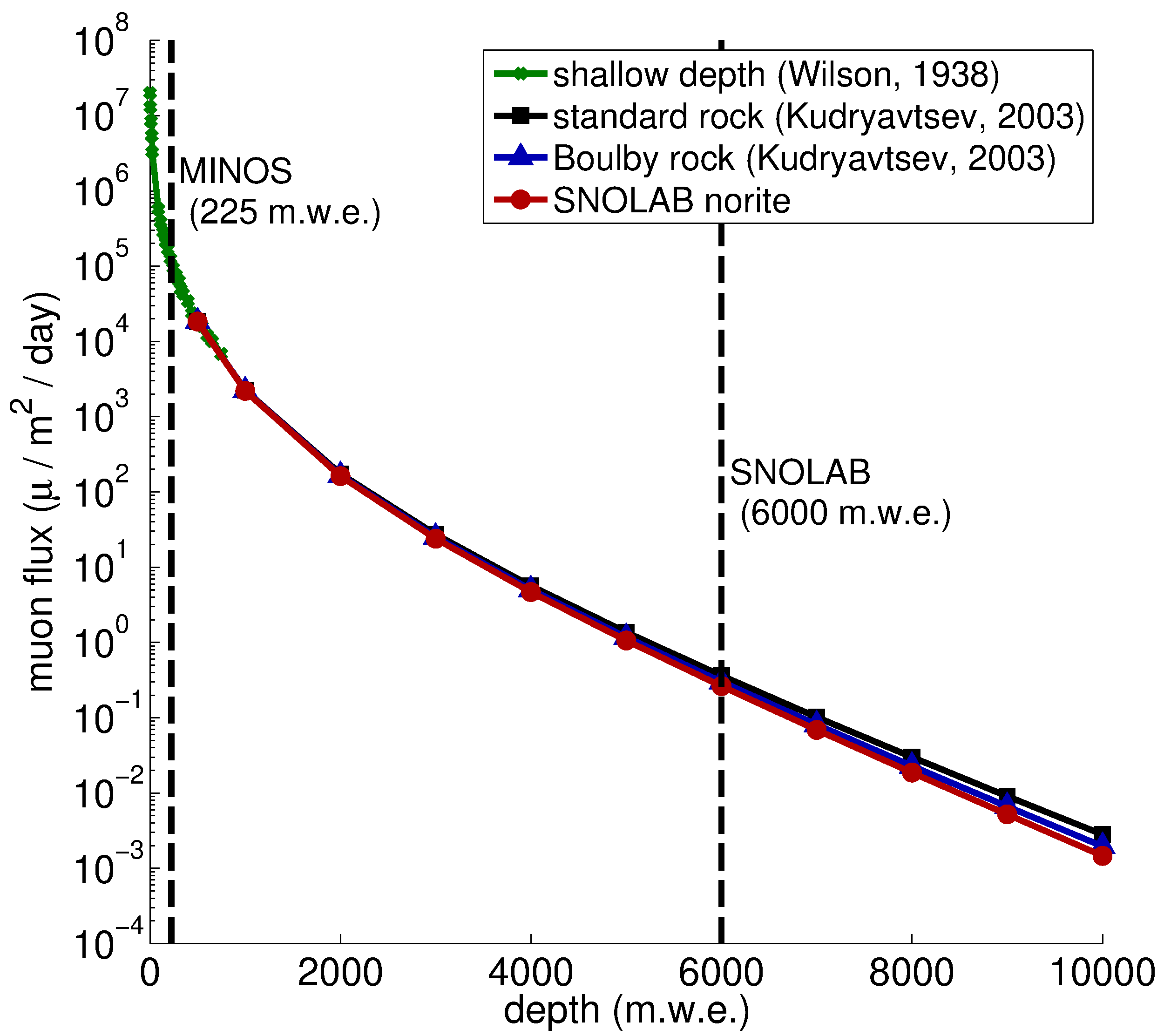}
\caption[Muon flux vs. depth]{Cosmic muon flux as a function of depth. Shallow depth data was taken from \cite{wilson-38}. Values for standard rock and from Boulby rock were taken from \cite{kudryavtsev-03}. The values for SNOLAB norite were calculated given the composition of norite found in \cite{snolab-handbook}.}
\label{fig:muon_flux_vs_depth}
\end{figure}

In order to minimize the cosmic muon flux, it was necessary to move the bubble chamber to a deep site. This second stage of deployment was to the SNOLAB underground laboratory located at the 6800 Level of the Vale Creighton Mine \#9 near Sudbury, Ontario. This location has a 6000 m.w.e. overburden, resulting in a cosmic muon flux over five orders of magnitude lower than that at the MINOS near detector hall (Figure \ref{fig:muon_flux_vs_depth}). Due to this low flux, no muon veto was installed around the bubble chamber at the deep site.

To estimate the total recoil-like background rate from cosmogenic neutrons generated in each detector component, the total rate of neutrons produced via the ($\mu$,n) reaction for that component and the energy spectrum of these neutrons must be calculated. These quantities are dependent on the total flux of muons at depth $h$, the energy characteristics of these muons, and the composition of target materials in which the ($\mu$,n) neutrons are generated. Once known, the neutron rates and energy spectra for each detector component can be passed to an MCNP-PoliMi simulation, where the detector components themselves act as sources of background neutrons.

\subsection{($\mu$,n) Production Rate}
\label{sec:backgrounds_mu_n_rate}

Figure \ref{fig:muon_flux_vs_depth} shows the muon flux as a function of depth, given specific local rock composition. The data for the muon flux at shallow depths was collected by a four-fold Geiger-M\"{u}ller tube telescope \citep{wilson-38}, which is in good agreement with the other shallow depth data reported in \cite{bugaev-98}. The values for the muon flux in standard rock ($A=22$, $Z=11$) and from Boulby rock ($A=23.6$, $Z=11.7$) were calculated in \cite{kudryavtsev-03}, which also states that the muon energy loss is proportional to $\left<Z^2/A\right>$. SNOLAB is located in a norite rock formation ($A=24.77$, $Z=12.15$) \citep{mei-06} and so proper scaling by $Z^2/A$ gives the muon flux values plotted in Figure \ref{fig:muon_flux_vs_depth}. From these trends, it follows that the muon flux at the MINOS near detector hall (225 m.w.e.) is $1.2\times10^5$ $\mu$/m$^2$/day, while at SNOLAB (6000 m.w.e.), it is 0.26 $\mu$/m$^2$/day, in good agreement with the value stated in the \cite{snolab-handbook}.

The differential muon energy spectrum at the surface follows a power-type law

\begin{equation}
\label{eq:mu_energy_surface}
N_\mu(E,0) = BE^{-(\gamma+1)}
\end{equation}

\noindent with $\gamma=2.77$ \citep{groom-01}. Given muon energy loss that is linear in energy, the energy spectrum at depth $h$ can be found to be

\begin{equation}
\label{eq:mu_energy_depth}
N_\mu(E,h) = Be^{-\gamma bh}\left[E+\frac{a}{b}\left(1-e^{-bh}\right)\right]^{-(\gamma+1)}
\end{equation}

\noindent with an average energy value of

\begin{equation}
\label{eq:mu_energy_average}
\left<E_\mu\right> = \frac{a}{b(\gamma-1)}\left(1-e^{-bh}\right)
\end{equation}

\noindent for parameters $a$ and $b$ that are dependent on the composition of the rock through which the muons are being attenuated --- approximately proportional to $\left<Z/A\right>$ and $\left<Z(Z+1)/A\right>$, respectively \citep{castagnoli-95}. For standard rock, $a = 0.277$ GeV/m.w.e. and $b = 4.0\times10^{-4}$ m.w.e.$^{-1}$ \citep{groom-01, battistoni-98}. Scaling by $A$ and $Z$ for norite, values of $a = 0.272$ GeV/m.w.e. and $b = 4.3\times10^{-4}$ m.w.e.$^{-1}$ will be used here. With these parameter choices and $h = 6000$ m.w.e., the average muon energy at SNOLAB is found to be $\left<E_\mu\right>_\mathrm{SNOLAB} = 330$ GeV.

This is important because the ($\mu$,n) neutron production rate $R_n$ in units of (n/$\mu$)/(g/cm$^2$) scales with both the atomic mass $A$ of the target material (or mean atomic mass in the case of a compound) and the average energy of the muon $\left<E_\mu\right>$ in GeV:

\begin{equation}
\label{eq:mu_n_rate}
R_n\left(\left<E_\mu\right>,A\right) = k \left<E_\mu\right>^\alpha A^\beta,
\end{equation}

\noindent with various values of $\alpha$, $\beta$, and $k$ in the literature --- $\alpha = 0.85$, $\beta = 0.81$, and $k = 3.9\times10^{-7}$ (n/$\mu$)/(g/cm$^2$) \citep{mei-06} will be used here. At SNOLAB, where $\left<E_\mu\right>_\mathrm{SNOLAB} = 330$ GeV and the muon flux is 0.26 $\mu$/m$^2$/day, the ($\mu$,n) neutron production rate is then

\begin{equation}
\label{eq:mu_n_rate_SNOLAB}
R_n\left(A\right) = 1.6\times10^{-14} A^{0.81} \rho \: \frac{\mathrm{n}}{\mathrm{s} \cdot \mathrm{cm}^3},
\end{equation}

\noindent for muons incident on a target with density $\rho$ (in g/cm$^3$) and atomic mass $A$. The ($\mu$,n) neutron production rate in each part of the detector simulated is shown in Table \ref{tab:mu_n_rates}.

\begin{table}[t!]
\centering
\begin{tabular} {| l | c | c | c |}
\hline
Material & $A$ & $\rho$ & $R_n$ \\
 & \small{(mean)} & \small{(g/cm$^3$)} & \small{(n/s/cm$^3$)} \\
\hline
\hline
304L Stainless Steel & 55.48 & 8.00 & $3.31\times10^{-12}$ \\
Water & 6.01 & 1.00 & $6.83\times10^{-14}$ \\
Propylene Glycol & 5.85 & 1.04 & $6.94\times10^{-14}$ \\
CF$_3$I & 39.18 & 1.99 & $6.21\times10^{-13}$ \\
\hline
\end{tabular}
\caption[Rate of ($\mu$,n) neutron production for various materials]{The rate of ($\mu$,n) neutron production for various materials at SNOLAB, calculated from Equation \ref{eq:mu_n_rate_SNOLAB}.}
\label{tab:mu_n_rates}
\end{table}

\subsection{($\mu$,n) Energy Spectrum}
\label{sec:backgroudns_mu_n_energy}

The differential energy spectrum of ($\mu$,n) neutrons is a function of both the energy of the muon as well as parameters which vary with depth. Here, a parameterization defined in \cite{mei-06} is used:

\begin{equation}
\label{eq:mu_n_energy}
\frac{\mathrm{d}N}{\mathrm{d}E_n} = A_\mu \left[ \frac{e^{-a_0 E_n}}{E_n} + B_\mu\left(\left<E_\mu\right>\right) e^{-a_1 E_n} \right] + a_2 E_n^{-a_3},
\end{equation}

\noindent with $A_\mu$ a normalization, $a_0 = 7.774$, $a_1 = 2.134$, $a_2 = -2.939\times10^{-16}$, $a_3 = 2.859$, and $\left<E_\mu\right>$ defined as above. The function $B\left(\left<E_\mu\right>\right)$, a function of muon energy in GeV, is given by

\begin{equation}
\label{eq:mu_n_B}
B\left(\left<E_\mu\right>\right) = 0.324 - 0.641 e^{-0.014 \left<E_\mu\right>}.
\end{equation}

\noindent Again, there are various parameterizations used in the literature. For consistency with the rest of the results, the one from \cite{mei-06} is used, which is consistent with \cite{formaggio-04}. No variation with target atomic mass $A$ is parameterized for in this spectrum --- this follows the results in \cite{singer-74} that very little variation is seen between target materials.

\subsection{($\mu$,n)-Generated Event Rates}
\label{sec:backgrounds_mu_n_events}

\begin{figure} [t!]
\centering
\includegraphics[scale=0.55]{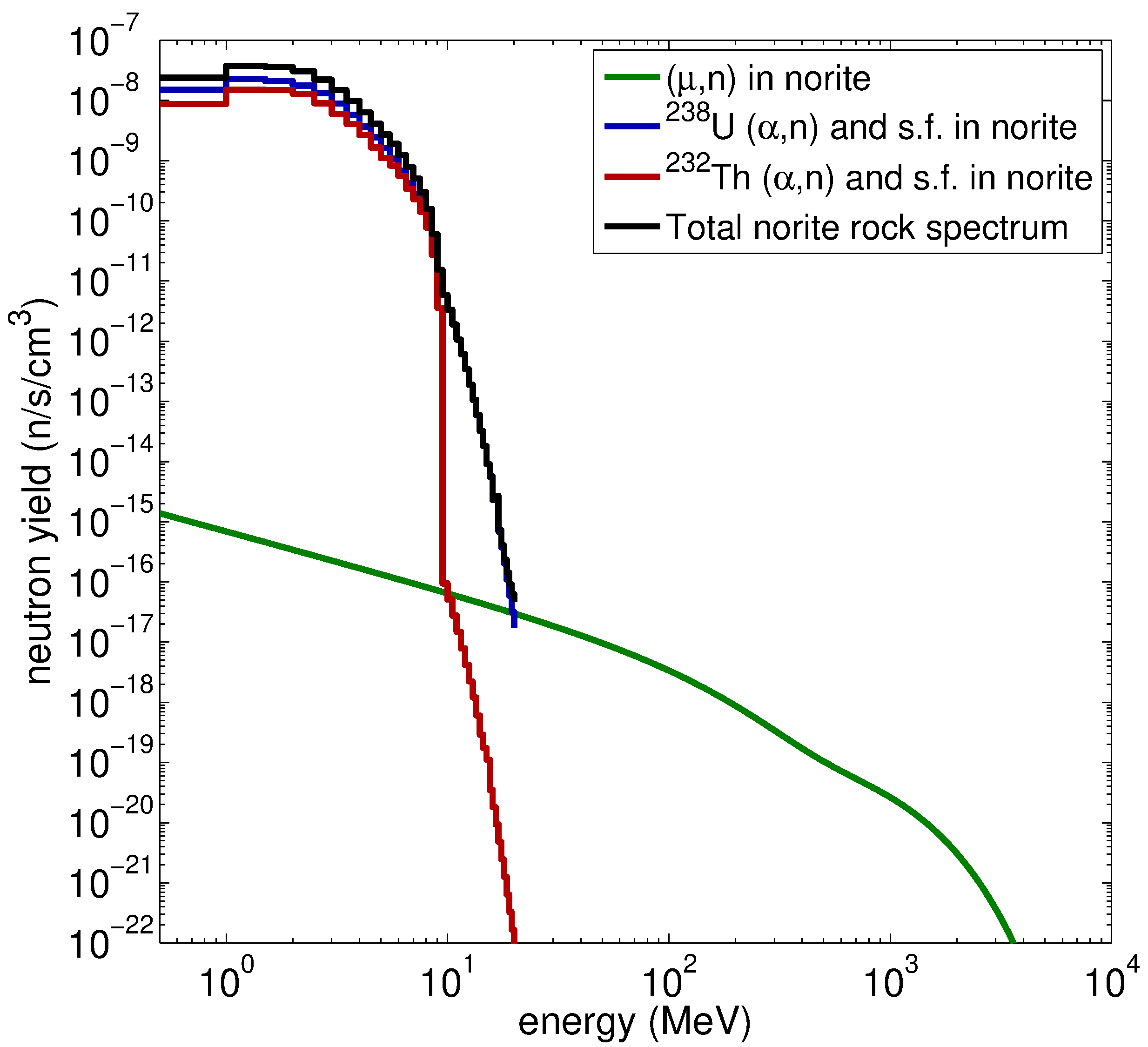}
\caption[Neutron energy spectra from ($\mu$,n) and from $^{238}$U and $^{232}$Th decays in norite]{The ($\mu$,n) neutron emission spectrum per cm$^3$ of norite rock, following parameterizations laid out in \cite{mei-06}, compared to that from spontaneous fission and ($\alpha$,n) neutrons from measured levels of $^{238}$U and $^{232}$Th contamination. The latter dominates the spectrum below 20 MeV in SNOLAB. The large drop in the $^{232}$Th spectrum at $\sim$10 MeV is due to the end-point of ($\alpha$,n) neutrons, leaving only the tail of the low-rate spontaneous fission neutrons.}
\label{fig:norite_n_spectrum}
\end{figure}

For each major component of the bubble chamber detector, out to and including the water shield, MCNP-PoliMi simulations were carried out treating the material in question as a neutron source, with energy spectrum given by Equation \ref{eq:mu_n_energy} and rate given by Equation \ref{eq:mu_n_rate_SNOLAB}. The ($\mu$,n) rate from the cavern walls (Figure \ref{fig:norite_n_spectrum})  is neglected here --- its treatment is relegated to Section \ref{sec:backgrounds_rock}. The ($\mu$,n) neutrons are assumed to be emitted preferentially in the downward direction due to the kinematics involved with collisions between high energy downward-traveling muons and nuclei at rest relative to the chamber. Table \ref{tab:cosmogenic_events} \begin{table}[t!]
\centering
\small{
\begin{tabular} {| l | c || c | c || c | c |}
\hline
\multicolumn{6}{| l |}{Cosmogenic ($\mu$,n) Neutrons} \\
\hline
Material & Data Set & \multicolumn{2}{| c ||}{Rate} & \multicolumn{2}{| c |}{Counts in Set} \\
 & & \multicolumn{2}{| c ||}{ \small{(10$^{-6}$ cts/kg$_\mathrm{CF_3I}$/day)}} & \multicolumn{2}{| c |}{\small{(10$^{-3}$ cts)}} \\
\cline{3-6}
 & & \small{($N = 1$)} & \small{($N > 1$)} & \small{($N =$ 1)} & \small{($N > 1$)} \\
\hline
304L Stainless Steel & DM-34$^\circ$C & $4.55^{+0.49}_{-0.48}$ & $1.58^{+0.15}_{-0.14}$ & $1.79^{+0.19}_{-0.19}$ & $0.62^{+0.06}_{-0.06}$ \\
(pressure vessel) & DM-37$^\circ$C & $4.86^{+0.51}_{-0.52}$ & $1.93^{+0.17}_{-0.17}$ & $0.43^{+0.05}_{-0.05}$ & $0.17^{+0.01}_{-0.02}$ \\
 & DM-40$^\circ$C & $5.13^{+0.54}_{-0.54}$ & $2.27^{+0.19}_{-0.19}$ & $0.36^{+0.04}_{-0.04}$ & $0.16^{+0.01}_{-0.01}$ \\
\hline
CF$_3$I & DM-34$^\circ$C & $3.73^{+0.40}_{-0.40}$ & $1.36^{+0.12}_{-0.12}$ & $1.47^{+0.16}_{-0.16}$ & $0.53^{+0.05}_{-0.05}$ \\
 & DM-37$^\circ$C & $4.00^{+0.42}_{-0.42}$ & $1.66^{+0.14}_{-0.14}$ & $0.35^{+0.04}_{-0.04}$ & $0.15^{+0.01}_{-0.01}$ \\
 & DM-40$^\circ$C & $4.21^{+0.44}_{-0.45}$ & $1.97^{+0.16}_{-0.16}$ & $0.30^{+0.03}_{-0.03}$ & $0.14^{+0.01}_{-0.01}$ \\
\hline
Propylene Glycol & DM-34$^\circ$C & $0.32^{+0.03}_{-0.03}$ & $0.11^{+0.01}_{-0.01}$ & $0.13^{+0.01}_{-0.01}$ & $0.04^{+0.004}_{-0.004}$ \\
(hydraulic fluid) & DM-37$^\circ$C & $0.35^{+0.04}_{-0.04}$ & $0.13^{+0.01}_{-0.01}$ & $0.03^{+0.003}_{-0.003}$ & $0.01^{+0.001}_{-0.001}$ \\
 & DM-40$^\circ$C & $0.37^{+0.04}_{-0.04}$ & $0.16^{+0.01}_{-0.01}$ & $0.03^{+0.003}_{-0.003}$ & $0.01^{+0.001}_{-0.001}$ \\
\hline
Water & DM-34$^\circ$C & $0.28^{+0.03}_{-0.03}$ & $0.10^{+0.01}_{-0.01}$ & $0.11^{+0.01}_{-0.01}$ & $0.04^{+0.004}_{-0.004}$ \\
(buffer) & DM-37$^\circ$C & $0.30^{+0.03}_{-0.03}$ & $0.12^{+0.01}_{-0.01}$ & $0.03^{+0.003}_{-0.003}$ & $0.01^{+0.001}_{-0.001}$ \\
 & DM-40$^\circ$C & $0.31^{+0.03}_{-0.03}$ & $0.14^{+0.01}_{-0.01}$ & $0.02^{+0.002}_{-0.002}$ & $0.01^{+0.001}_{-0.001}$ \\
\hline
\hline
Total & DM-34$^\circ$C & $8.94^{+0.57}_{-0.57}$ & $3.17^{+0.17}_{-0.17}$ & $3.52^{+0.22}_{-0.22}$ & $1.25^{+0.07}_{-0.07}$ \\
 & DM-37$^\circ$C & $9.56^{+0.60}_{-0.60}$ & $3.88^{+0.20}_{-0.20}$ & $0.85^{+0.05}_{-0.05}$ & $0.34^{+0.02}_{-0.02}$ \\
 & DM-40$^\circ$C & $10.09^{+0.63}_{-0.64}$ & $4.57^{+0.22}_{-0.22}$ & $0.72^{+0.05}_{-0.05}$ & $0.32^{+0.02}_{-0.02}$ \\
 \hline
\end{tabular} }
\caption[Cosmogenic background rate prediction]{The expected event rates and raw number of events expected from a cosmogenic neutron background in the DM-34$^\circ$C, DM-37$^\circ$C, and DM-40$^\circ$C data sets. The rates from the water shielding are not shown because they are an order of magnitude below those listed.}
\label{tab:cosmogenic_events}
\end{table} summarizes the expected number of events in the DM-34$^\circ$C, DM-37$^\circ$C, and DM-40$^\circ$C data sets from a cosmogenic ($\mu$,n) neutron background, assuming a flat bubble nucleation efficiency ($\eta_\mathrm{C,F}=0.46$) model (Figure \ref{fig:contour_a_eta}) and given a 79.86\% data cut efficiency on single-bubble events (Section \ref{sec:datasets_summary}). The uncertainty on the predictions is found by adding the 10\% uncertainty from the MCNP simulation in quadrature with the uncertainty on the threshold energy of each data set (Table \ref{tab:data_sets}). As expected, the nuclear recoil background rate imposed by cosmogenic neutrons on the chamber is negligible. Figure \ref{fig:summary} puts this background in context with all the other backgrounds calculated in this thesis.

\section{Rock-Generated Neutrons}
\label{sec:backgrounds_rock}

The 4 kg bubble chamber is located in the J-Drift of SNOLAB, already described as being within a deposit of norite rock. Norite has a higher density ($\rho = 2.894$ g/cm$^3$) and atomic mass ($A=24.77$) than standard rock, both of which serve to increase the potential ($\mu$,n) neutron rate originating in rock cavern walls --- a rate of $R_n = 6.23\times10^{-13}$ n/s/cm$^3$ is expected from Equation \ref{eq:mu_n_rate_SNOLAB}. However, in addition to the ($\mu$,n) neutron flux from the rock walls, there is also an expected rate originating from spontaneous fission and ($\alpha$,n) neutrons from the decays of the $^{238}$U and $^{232}$Th chains in the norite, with measured concentrations of 1.2 and 3.3 parts per million (ppm) by mass, respectively \citep{snolab-handbook}.

The ($\alpha$,n) and spontaneous fission neutron production rate and energy spectrum for the decays along the $^{238}$U and $^{232}$Th chains in norite were calculated using the SOURCES-4C code, a computer code system for calculating ($\alpha$,n), spontaneous fission, and delayed neutron sources and spectra \citep{wilson-05}. The code was supplied with the measured $^{238}$U, $^{232}$Th, and respective daughter isotope concentrations and the total composition of the material in question, assuming secular equilibrium along the decay chains and natural abundance of any ($\alpha$,n) target isotopes. From the SOURCES-4C calculation, the neutron yields and spectra from spontaneous fission and ($\alpha$,n) neutrons in $^{238}$U and $^{232}$Th was found to dominate the total spectrum of neutrons from norite out to 20 MeV (Figure \ref{fig:norite_n_spectrum}). Above 20 MeV, the ($\mu$,n) neutrons begin to contribute, however MCNP has a high-energy cutoff for neutrons above 20 MeV, so this high-energy tail is not simulated. The contribution from ($\mu$,n) neutrons in the norite rock could be studied further, but it is presumed to be negligible here.

The total fast neutron flux in SNOLAB has been estimated to be 4000 n/m$^2$/day \citep{snolab-handbook}. To determine the background event rate from such neutrons, a spherical shell 50 cm thick was defined in an MCNP-PoliMi simulation surrounding the detector geometry shown in Figure \ref{fig:mcnp_4kg} (left). This shell acted as a source of neutrons with randomly distributed origination points and an energy spectrum as in Figure \ref{fig:norite_n_spectrum}. Using a shell of norite as opposed to a spherical surface allowed for the moderation of the neutron spectrum by the cavern walls. The number of neutrons passing through the inner spherical surface of the shell was scaled to the total flux of 4000 n/m$^2$/day.

\begin{figure} [t!]
\centering
\includegraphics[scale=0.55]{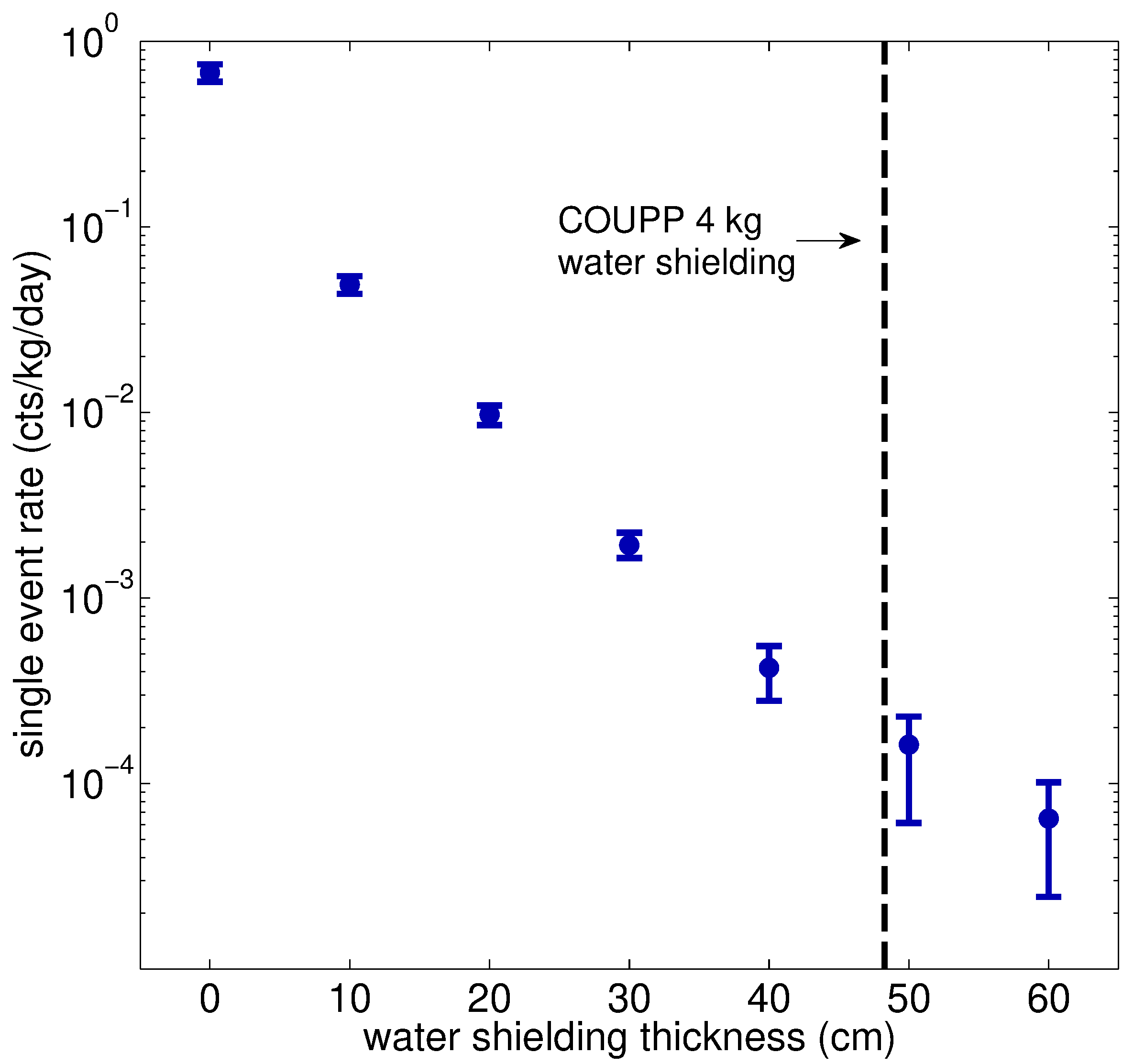}
\caption[Event rate from rock neutrons for various water shielding thicknesses]{The single-bubble event rate from the 4000 n/m$^2$/day flux of neutrons generated in the norite cavern walls for various thicknesses of water shielding. The water shielding installed around the 4 kg bubble chamber is $\sim$50 cm thick (48 cm of water + 2 cm of polyethylene).}
\label{fig:water_shield_thickness}
\end{figure}

Assuming a flat bubble nucleation efficiency ($\eta_\mathrm{C,F}=0.46$) and 79.86\% data cut efficiency on single-bubble events, the total expected event rate from neutrons generated in the norite rock cavern walls was generated for a variety of water shielding thicknesses (Figure \ref{fig:water_shield_thickness}).  From this, it was determined that the $\sim$50 cm water shielding in place was indeed adequate to reduce the expected number of events to acceptable levels. With the current water shield, the expected recoil-like event rates and predicted number of counts in each dark matter data set are shown in Table \ref{tab:rock_events}. \begin{table}[t!]
\centering
\small{
\begin{tabular} {| l | c || c | c || c | c |}
\hline
\multicolumn{6}{| l |}{Rock-Generated ($\alpha$,n) and Spontaneous Fission Neutrons} \\
\hline
Material & Data Set & \multicolumn{2}{| c ||}{Rate} & \multicolumn{2}{| c |}{Counts in Set} \\
 & & \multicolumn{2}{| c ||}{ \small{(10$^{-4}$ cts/kg$_\mathrm{CF_3I}$/day)}} & \multicolumn{2}{| c |}{\small{(10$^{-2}$ cts)}} \\
\cline{3-6}
 & & \small{($N = 1$)} & \small{($N > 1$)} & \small{($N = 1$)} & \small{($N > 1$)} \\
\hline
Norite & DM-34$^\circ$C & 1.26$^{+0.71}_{-0.57}$ & 0.25$^{+0.44}_{-0.16}$ & 4.96$^{+2.79}_{-2.24}$ & 0.98$^{+1.73}_{-0.63}$ \\
 & DM-37$^\circ$C & 1.26$^{+0.71}_{-0.57}$ & 0.50$^{+0.57}_{-0.32}$ & 1.11$^{+0.63}_{-0.50}$ & 0.44$^{0.50}_{-0.28}$ \\
 & DM-40$^\circ$C & 1.26$^{+0.71}_{-0.57}$ & 0.76$^{+0.58}_{-0.48}$ & 0.89$^{+0.50}_{-0.40}$ & 0.54$^{0.41}_{-0.34}$ \\
 \hline
\end{tabular} }
\caption[Rock-generated background rate prediction]{The expected event rates and raw number of events expected from an ($\alpha$,n) and spontaneous fission neutron background generated by $^{238}$U and $^{232}$Th contamination of the rock walls, for the DM-34$^\circ$C, DM-37$^\circ$C, and DM-40$^\circ$C data sets. The large error bars arise from the poor statistics of inadequately-sized MCNP simulations.}
\label{tab:rock_events}
\end{table}

\section{($\alpha$,n) Neutrons from Radon Dissolved in Liquids}
\label{sec:backgrounds_radon}

While detector installation at deep underground sites is particularly useful in limiting the cosmogenic neutron flux through the detector, one added problem arises in the increased density of radon gas in the atmosphere at low depths. Radon is soluble in water, and even more so in organic liquids like propylene glycol or refrigerants like CF$_3$I, and therefore any exposure of these liquids to the atmosphere, to other liquids containing dissolved radon, and to any $^{238}$U-contaminated materials (\emph{e.g.} the synthetic silica bell jar or the stainless steel) results in radon contamination of the system.

This affects the background rate of the bubble chamber in two ways. First, any radon dissolved in the CF$_3$I will act as an alpha source as the radon and its daughters decay. These events are only of concern when they occur within the CF$_3$I volume, due to the short range of alpha particles. As discussed, these alpha-generated events can be acoustically discriminated out of the data (Section \ref{sec:datasets_alpha}), and so this background can be eliminated. However, alphas can also initiate a neutron background through the ($\alpha$,n) reaction, and so any liquids in the detector containing dissolved radon will act as a weak ($\alpha$,n) neutron source, resulting in a recoil-like background, except in the case of the CF$_3$I where any ($\alpha$,n) neutrons are simultaneous with already-discriminated alpha events.

Efforts have been made to seal the liquids inside the chamber from the atmosphere, but there still is a consistent source of radon present in the detector, apparent from the steady 5.46 counts/kg/day rate of alpha events in the CF$_3$I volume (Section \ref{sec:datasets_alpharate}), nearly all of which come from $^{222}$Rn-$^{218}$Po-$^{214}$Po decays in the $^{238}$U decay chain (Section \ref{sec:efficiency_alphas})\footnote{A time correlation analysis confirms that the alpha event rate is almost entirely from the $^{222}$Rn-$^{218}$Po-$^{214}$Po decays \citep{behnke-08}. That analysis was repeated for the current data, with the same result \citep{lippincott-11a}.}. If one-third of the alpha events come from $^{222}$Rn particularly, then the number density of $^{222}$Rn in the CF$_3$I necessary to account for the total alpha rate is

\begin{equation}
\label{eq:Rn_density}
n_\mathrm{Rn} = \frac{R_\alpha}{3} \rho_\mathrm{CF_3I} \tau_\mathrm{Rn},
\end{equation}

\noindent with $^{222}$Rn lifetime $\tau_\mathrm{Rn} = 4.76\times10^5$ s and CF$_3$I density $\rho_\mathrm{CF_3I} = 1.99$ g/cm$^3$. From this, the number density of $^{222}$Rn necessary to account for the $R_\alpha = 5.46$ counts/kg/day alpha rate observed in the chamber is 0.02 atoms/cm$^3$, with equilibrium amounts of $^{218}$Po and $^{214}$Po. Whether or not the radon dissolved in the fluids can be explained by radon emanation from Suprasil synthetic fused silica bell jar and the stainless steel bellows (the only two materials in contact with the inner water/CF$_3$I volume) is discussed below.

\subsection{Emanation of Radon from Suprasil}
\label{sec:backgrounds_Rn_suprasil}

Measurements done by the EXO-200 collaboration found the $^{238}$U (and $^{232}$Th) content of the same Heraeus Suprasil synthetic silica used for the COUPP 4 kg bell jar to be 21$\pm$9 (59$\pm$14) parts per trillion (ppt) by mass \citep{leonard-08}. Assuming all the products of the $^{238}$U decay chain are in secular equilibrium, the number density of $^{222}$Rn atoms in the Suprasil is

\begin{equation}
\label{eq:suprasil_Rn_n}
n_\mathrm{Rn} = \frac{C_\mathrm{U} \rho N_A}{A_\mathrm{U}} \frac{\tau_\mathrm{Rn}}{\tau_\mathrm{U}},
\end{equation}

\noindent where $\rho = 2.2$ g/cm$^3$ is the density of Suprasil, $C_\mathrm{U}$ is the concentration by mass of $^{238}$U, $N_A$ is Avogadro's number, $A_\mathrm{U}$ is the atomic mass of $^{238}$U, and $\tau_\mathrm{U} = 2.05\times10^{17}$ s is the lifetime of $^{238}$U. With $C_\mathrm{U} = 21$ ppt, the number density of $^{222}$Rn atoms in Suprasil is $n_\mathrm{Rn} = 0.274$ atoms/cm$^3$.

The amount of $^{222}$Rn that emanates into the CF$_3$I volume can be roughly estimated. The number density of $^{222}$Rn atoms in the CF$_3$I is equal to the number density in the Suprasil times an emanation fraction. Based on a measurement of $^{222}$Rn emanation from the 8" Sudbury Neutrino Observatory (SNO) PMTs in vacuum\footnote{This measurement was also carried out for glass in water, with results within a factor of 3 of the values found for emanation in vacuum \citep{lee-94}.} \citep{liu-93}, the KamLAND experiment found that the emanation fraction from their 20" PMTs was around 0.004 \citep{gratta-98}. Assuming the synthetic silica of the 4 kg bell jar has a similar $^{222}$Rn diffusion length as the glass envelope of the SNO and KamLAND PMTs, this would imply that the number density of $^{222}$Rn in the CF$_3$I is only $1.10\times10^{-3}$ atoms/cm$^3$, far short of the total 0.02 atoms/cm$^{3}$ necessary to provide the 5.46 counts/kg/day rate of alpha events in the bulk of the CF$_3$I (one-third of which are from $^{222}$Rn decays).

Another estimate of the emanation fraction is found by considering the emanation from granular quartz, where the radon emanation fraction has been measured to be 0.046 \citep{sakoda-11}. Since the Suprasil bell jar is more crystalline in nature, this is probably an upper bound on the actual emanation fraction in our chamber. Still, using this upper bound to estimate the emanation out of the Suprasil gives approximately 0.013 atoms/cm$^3$ in the CF$_3$I, again short of the 0.02 atoms/cm$^3$ necessary to produce the measured background. In all, the amount of $^{222}$Rn measured from the alpha rate in the chamber is too high to originate from the Suprasil alone. Another contribution to the radon contamination of the system is necessary to explain the background rate.

\subsection{Emanation of Radon from Steel}
\label{sec:backgrounds_Rn_steel}

While the CF$_3$I volume is not in contact with any stainless steel, the water buffer above the CF$_3$I fills the stainless steel bellows. This bellows has 20 convolutions, and so the overall surface area of steel with the water buffer is substantial. Since $^{222}$Rn can also emanate from steel which can diffuse into the CF$_3$I volume, this is another potential source of $^{222}$Rn in the chamber.

For atomic mass $A$ and fluid density $\rho_\mathrm{fluid}$, the mole fraction of $^{222}$Rn in a fluid is given by

\begin{equation}
\label{eq:mole_fraction}
N_\mathrm{Rn} = n_\mathrm{Rn} \frac{A_\mathrm{fluid}}{\rho_\mathrm{fluid} N_A},
\end{equation}

\noindent which is also proportional to the square of its Hildebrand solubility parameter $\delta$

\begin{equation}
\label{eq:log_mole_fraction}
-\log{N} = k \delta^2 + \mathrm{constant},
\end{equation}

\noindent with $k=(4.79\pm0.58)\times10^{-3}$ (scaled for $\delta$ in units of $\mathrm{cal}^{1/2}/\mathrm{cm}^{3/2}$) and assuming similar solvent polarities between fluids \citep{kalmuk-07}. Therefore, for a given number density of $^{222}$Rn in the water buffer, the number density in CF$_3$I assuming chemical equilibrium is

\begin{equation}
\label{eq:steel_Rn_n}
n_\mathrm{Rn}^\mathrm{CF_3I} = n_\mathrm{Rn}^\mathrm{H_2O} \frac{A_\mathrm{H_2O}}{A_\mathrm{CF_3I}} \frac{\rho_\mathrm{CF_3I}}{\rho_\mathrm{H_2O}} 10^{k\left(\delta_\mathrm{H_2O}^2 - \delta_\mathrm{CF_3I}^2\right)}.
\end{equation}

\noindent The values of $\delta$ for CF$_3$I and water are $\delta_\mathrm{CF_3I} = 7.07 \: \mathrm{cal}^{1/2}/\mathrm{cm}^{3/2}$ \citep{lifke-96} and $\delta_\mathrm{H_2O} = 23.4 \: \mathrm{cal}^{1/2}/\mathrm{cm}^{3/2}$ \citep{hansen-67}, respectively.

The emanation of radon from stainless steel has been measured by the SNO collaboration to be about 4500 $^{222}$Rn atoms/m$^2$/day \citep{liu-90}. Since no specification was given for the radiopurity of the stainless steel used in the manufacturing of the bellows, this rate is assumed to apply to the bellows steel as well. Even without factoring in the bellows convolutions, the amount of surface area in contact with the water is on the order of 500 cm$^2$. This gives a number density of $^{222}$Rn in water of $n_\mathrm{Rn}^\mathrm{H_2O} = 0.5$ atoms/cm$^3$, and therefore $n_\mathrm{Rn}^\mathrm{CF_3I} \simeq 20$ atoms/cm$^3$ in the CF$_3$I. This is more than enough to account for the observed alpha rate in the chamber.

Therefore, a radon leak from the atmosphere into the inner volume is not necessary to explain the observed alpha event rate. However, when operating the COUPP 4 kg bubble chamber at a shallow site (the MINOS near detector hall at Fermilab), where the radon density in the atmosphere is lower, the alpha event rate was also lower, approximately 0.7 counts/kg/day \citep{behnke-11}. At first glance, this might imply a contribution to the alpha event rate from a radon leak into the inner water/CF$_3$I volume, the difference between the alpha rate at MINOS and at SNOLAB arising from the higher radon density in the atmosphere at SNOLAB. However, the bellows used for the Fermilab run was different than that at SNOLAB with possibly a different concentration of $^{238}$U (the precursor of emanated $^{222}$Rn), and so comparison between the two cases is difficult.

\subsection{Dissolution of Atmospheric Radon into Water}
\label{sec:backgrounds_radon_in_water}

The water inside the neutron shielding is, in fact, in contact with the atmosphere, and so the dissolution of $^{222}$Rn into water must be understood. In general, the amount of a gas that dissolves in a liquid is directly proportional to the partial pressure of the gas in equilibrium with the liquid, as given by Henry's law:

\begin{equation}
\label{eq:henrys_law}
n = k_H p,
\end{equation}

\noindent where $n$ is the concentration of the solute in the liquid, $p$ is the partial pressure of the solute in the gas, and $k_H$ is the Henry coefficient, which varies for each gas-liquid combination and is a function of the temperature $T$:

\begin{equation}
\label{eq:k_H_variation}
k_H(T) = k_H(T_0) \exp \left[C \left(\frac{1}{T} - \frac{1}{T_0}\right) \right].
\end{equation}

\noindent $T_0$ is a reference temperature where a value of $k_H$ is known and $C$ is a constant, again dependent on the gas-liquid combination. Values for the Henry coefficient $k_H$ and the temperature dependence constant $C$ for radon dissolved in water are readily available \citep{wilhelm-77}, with NIST-accepted values of $k_H^\mathrm{H_2O} = 9.3\times10^{-3}$ mol/kg/bar and $C = 2600$ K at $T_0 = 25^\circ$C.

The partial pressure of $^{222}$Rn in the atmosphere is found from the ideal gas law, $p = N k_B T / V$, where $k_B$ is the Boltzmann constant and N/V can be written in terms of the activity of $^{222}$Rn in the air as $N/V = A/V \times \tau_\mathrm{Rn}$ with $A/V$ the activity per unit volume in the atmosphere. Taken together with Equation \ref{eq:henrys_law}, this yields

\begin{equation}
\label{eq:henrys_law_Rn}
n_\mathrm{Rn} = \frac{A}{V} k_B k_H T \tau_\mathrm{Rn}.
\end{equation}

\noindent In SNOLAB, the amount of $^{222}$Rn in the atmosphere is $131.0\pm6.7$ Bq/m$^3$ \citep{snolab-handbook} and the average room temperature measured during the experiment was $T = 21.25^\circ$C. Therefore, the number density of $^{222}$Rn in the water of the neutron shielding is $n_\mathrm{Rn} = 15.9\pm0.8$ atoms/cm$^3$.

\subsection{Radon-Generated Neutron Rate}
\label{sec:backgrounds_radonrate}

Regardless of the source of $^{222}$Rn in the inner volume water/CF$_3$I system, a very good measure of its number density is found from the 5.46 counts/kg/day rate from the $^{222}$Rn-$^{218}$Po-$^{214}$Po alpha decays in the CF$_3$I. From this rate, the number density of $^{222}$Rn in CF$_3$I is found to be $2.00\times10^{-2}$ atoms/cm$^3$. The number densities of $^{218}$Po and $^{214}$Po can be found assuming equilibrium with $^{222}$Rn by multiplying by the life-time fraction (\emph{e.g.} $\tau_\mathrm{^{222}Rn} / \tau_\mathrm{^{218}Po}$). From Equation \ref{eq:steel_Rn_n}, the number density of $^{222}$Rn in the water buffer is then $4.52\times10^{-4}$ atoms/cm$^3$. In the water filling the neutron shielding tanks, it is taken to be 15.9 atoms/cm$^3$. That in the propylene glycol is handled separately (see Section \ref{sec:backgrounds_other}).

\begin{figure} [t!]
\centering
\includegraphics[scale=0.55]{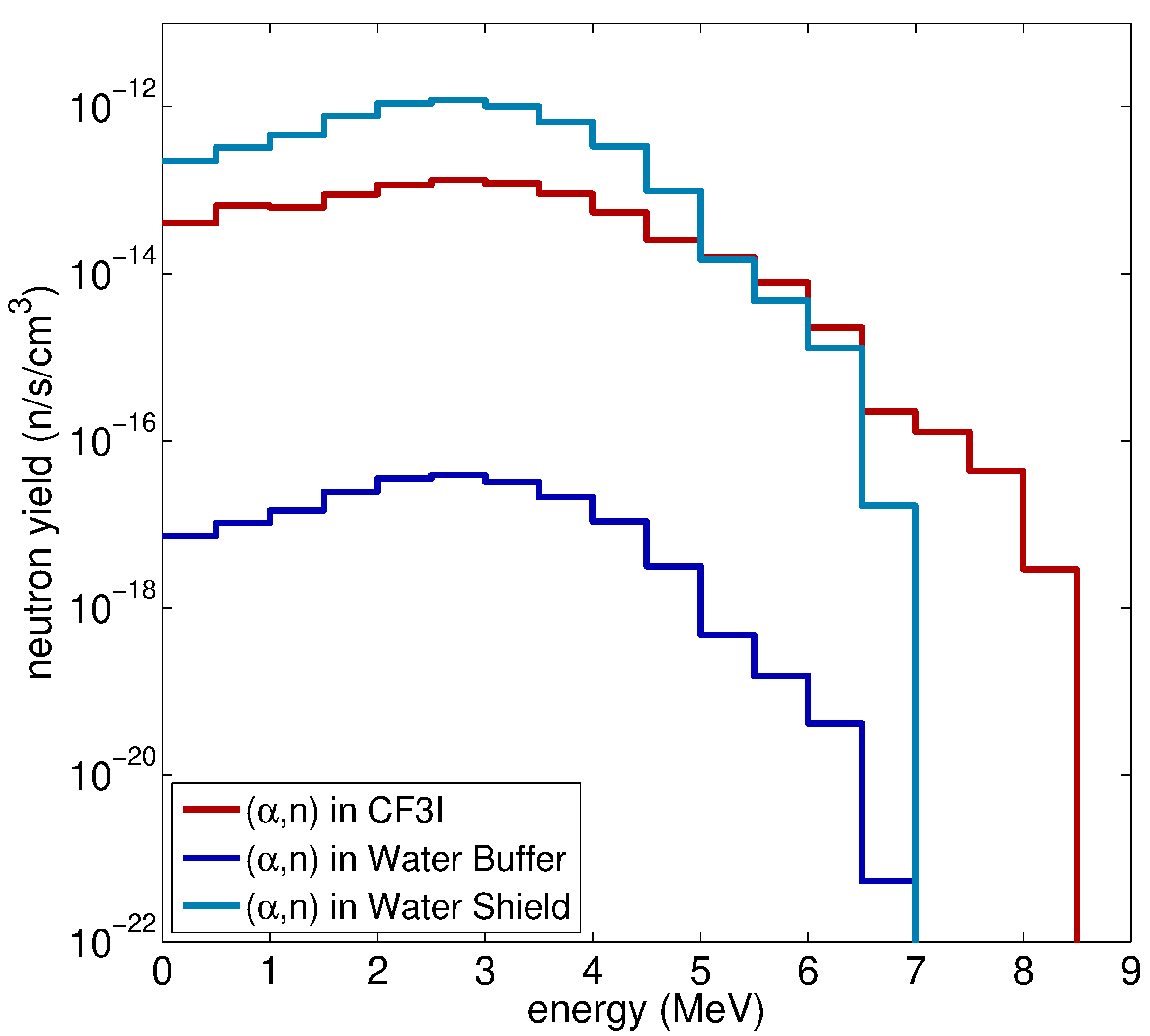}
\caption[($\alpha$,n) neutron emission energy spectra from $^{222}$Rn in detector fluids]{The ($\alpha$,n) neutron energy spectrum per cm$^3$ of fluid, given the number densities of $^{222}$Rn, $^{218}$Po, and $^{214}$Po dissolved in each.}
\label{fig:radon_n_spectrum}
\end{figure}

Each of these number densities is then used in a SOURCES-4C calculation to determine the energy spectrum and rate of neutrons generated via the ($\alpha$,n) reaction in each of these fluids (Figure \ref{fig:radon_n_spectrum},  also see note in Section \ref{sec:mcnp_sources_bugs}). For both water volumes, these spectra are used to define sources in an MCNP-PoliMi simulation, and the resulting event rates produced by the simulation is scaled by the rate of neutron production in each volume to produce predicted event rates and number of counts in the different dark matter search data sets, assuming a flat bubble nucleation efficiency ($\eta_\mathrm{C,F}=0.46$) and 79.86\% data cut efficiency on single-bubble events (Table \ref{tab:radon_events}). \begin{table}[t!]
\centering
\small{
\begin{tabular} {| l | c || c | c || c | c |}
\hline
\multicolumn{6}{| l |}{Radon-Generated ($\alpha$,n) Neutrons} \\
\hline
Material & Data Set & \multicolumn{2}{| c ||}{Rate} & \multicolumn{2}{| c |}{Counts in Set} \\
 & & \multicolumn{2}{| c ||}{ \small{(10$^{-7}$ cts/kg$_\mathrm{CF_3I}$/day)}} & \multicolumn{2}{| c |}{\small{(10$^{-5}$ cts)}} \\
\cline{3-6}
 & & \small{($N = 1$)} & \small{($N > 1$)} & \small{($N = 1$)} & \small{($N > 1$)} \\
\hline
Water (shield) & DM-34$^\circ$C & $1.35^{+0.25}_{-0.30}$ & $0.71^{+0.29}_{-0.20}$ & $5.33^{+0.97}_{-1.12}$ & $2.81^{+1.13}_{-0.78}$ \\
 & DM-37$^\circ$C & $1.66^{+0.30}_{-0.29}$ & $1.04^{+0.27}_{-0.25}$ & $1.47^{+0.26}_{-0.26}$ & $0.92^{+0.24}_{-0.22}$ \\
 & DM-40$^\circ$C & $1.83^{+0.31}_{-0.34}$ & $1.37^{+0.31}_{-0.31}$ & $1.30^{+0.22}_{-0.24}$ & $0.97^{+0.22}_{-0.22}$ \\
\hline
\end{tabular} }
\caption[Radon-generated background rate prediction]{The expected event rates and raw number of events expected from an ($\alpha$,n) neutron background generated by radon dissolved in the water shielding, for the DM-34$^\circ$C, DM-37$^\circ$C, and DM-40$^\circ$C data sets. The rates from the water buffer are not listed because they are more than 3 orders of magnitude below those listed.}
\label{tab:radon_events}
\end{table} Only the predictions for the water shielding are shown because the rate produced by the water buffer is 3 orders of magnitude below those from the water shielding. The radon in the CF$_3$I is ignored as an ($\alpha$,n) background because any ($\alpha$,n) neutrons are associated with simultaneously discriminated alpha events.

\section{Neutrons from Detector Components}
\label{sec:backgrounds_components}

The background event rates so far calculated have been small --- a total of $<$ 0.1 events per year is expected from their sum. This is anticipated because efforts have been made to minimize the neutron background from sources external to the bubble chamber, both cosmogenic neutrons and those generated by spontaneous fission and ($\alpha$,n) reactions in the surrounding rock, by operating at a deep site and using significant water shielding around the detector. However, a non-negligible background is produced internally from the $^{238}$U and $^{232}$Th contamination of the detector materials themselves. The reduction of this background requires meticulous radioactive screening of all detector components, a process which was not carried out prior to the experimental runs described in this thesis. In order to determine the level of the internal background in these results, each detector component was screened, resulting in a recommendation to replace two of the detector components that were found to lead to a measurable background event rate, limiting the exposure of a dedicated dark matter search.

To determine the present amount of $^{238}$U and $^{232}$Th in each detector material, a commercial low-background magnesium-cryostat high-purity germanium (HPGe) detector in a massive lead shielding 6 m.w.e. underground at the University of Chicago was used to count the total gamma flux at the different energy peaks associated with the decay of particular isotopes along the $^{238}$U and $^{232}$Th decay chains (Figures \ref{fig:U238_decay_chain} and \ref{fig:Th232_decay_chain}). An XIA Polaris digital spectrometer was used to collect data from a mass of the detector component positioned inside the lead shielding near the HPGe detector, in some instances using Marinelli beakers to maximize sensitivity. Gammas were counted at energies corresponding to the decay of $^{214}$Pb (295 keV and 352 keV) and $^{214}$Bi (609 keV, 1.120 MeV, and 1.764 MeV) in the $^{238}$U decay series and to the decay of $^{212}$Pb (238 keV), $^{208}$Th (583 keV and 2.614 MeV), and $^{228}$Ac (911 keV and 969 keV) in the $^{232}$Th decay series. These count rates were background subtracted using a reference run employing the HPGe detector with no material inserted. An MCNP-5 simulation was used to determine the efficiency for full-energy deposition in the germanium detector for each specific gamma energy. Using these efficiencies, the corrected rate of gamma emission at each energy was determined for each material.

For the lead-containing piezoelectric transducers (Section \ref{sec:backgrounds_piezos}), a separate measurement of the amount of $^{210}$Pb in the material was required, since the assumption could not be made that the isotopes below and including $^{210}$Pb are in equilibrium with the rest of the $^{238}$U decay chain. This enhancement in $^{210}$Pb concentration is characteristic of all lead-containing materials, arising from a contamination during the metallurgical extraction of lead. This $^{210}$Pb measurement required the counting of the 46 keV gamma from the $^{210}$Pb decay (Figure \ref{fig:U238_decay_chain}), which was not possible with the HPGe detector described above --- this $p$-type detector has a large dead layer separating source and Ge, completely moderating low-energy gammas. To count the 46 keV gamma line, an HPGe ``well" detector was used, also in a lead shielding 6 m.w.e. underground. This detector design bypasses the dead layer limitation (see Appendix \ref{ch:piezos}).

Given the measured count rate $R$ (in counts/s) of a particular spectral line from a gamma decay with nominal intensity $I$ (gammas/decay of the isotope) and detection efficiency $\varepsilon$ (counts/gamma), the concentration $C$ (g/g) of any isotope $X$ in secular equilibrium with the gamma emitter is

\begin{equation}
\label{eq:concentration}
C_X = \frac{R}{I \varepsilon} \frac{A_X}{N_A m} \tau_X \times \frac{1}{\mathrm{b.r.}},
\end{equation}

\noindent where $A_X$ is the atomic mass of $X$, $N_A$ is Avogadro's number, $m$ is the total sample mass being measured, $\tau_X$ is the mean lifetime of $X$, and b.r. is the branching ratio of the isotope being measured relative to $X$. Each gamma rate measurement performed provides a separate measurement $C_i \pm \sigma_i$ of the concentration of the parent isotope of a chain (see, \emph{e.g.}, Figure \ref{fig:Window_ppm}). The $\chi^2$ of the fit to $C_X$ given by $n$ such measurements is defined as

\begin{equation}
\label{eq:chi2_measurement}
\chi^2 \equiv \sum_{i=1}^n \frac{\left(C_i-\mu\right)^2}{\sigma_i^2},
\end{equation}

\noindent where $\mu$ is defined such that $\partial \chi^2 / \partial \mu = 0$, or

\begin{equation}
\label{eq:mu_measurement}
\mu = \frac{\sum_{i=1}^n C_i / \sigma_i^2}{\sum_{i=1}^n 1/\sigma_i^2}.
\end{equation}

\noindent Finding the value $\mu_e$ for which $\chi^2\left(\mu_e\right) = \chi^2\left(\mu\right) + 1$ gives the uncertainty on the fit $\mu$. In total, the best-fit concentration $C(X)$ for $n$ measurements with values $x_i \pm \sigma_i$ is then

\begin{equation}
\label{eq:concentration_fit}
C_X = \frac{\sum_{i=1}^n C_i / \sigma_i^2}{\sum_{i=1}^n 1/\sigma_i^2} \pm \frac{1}{\sqrt{{\sum_{i=1}^n 1/\sigma_i^2}}}.
\end{equation}

\noindent However, since an MCNP-5 simulation is involved in the counting rate, and additional 10\% uncertainty is added  in quadrature to the uncertainty in Equation \ref{eq:concentration_fit}.

\begin{figure} [t!]
\centering
\includegraphics[scale=0.55]{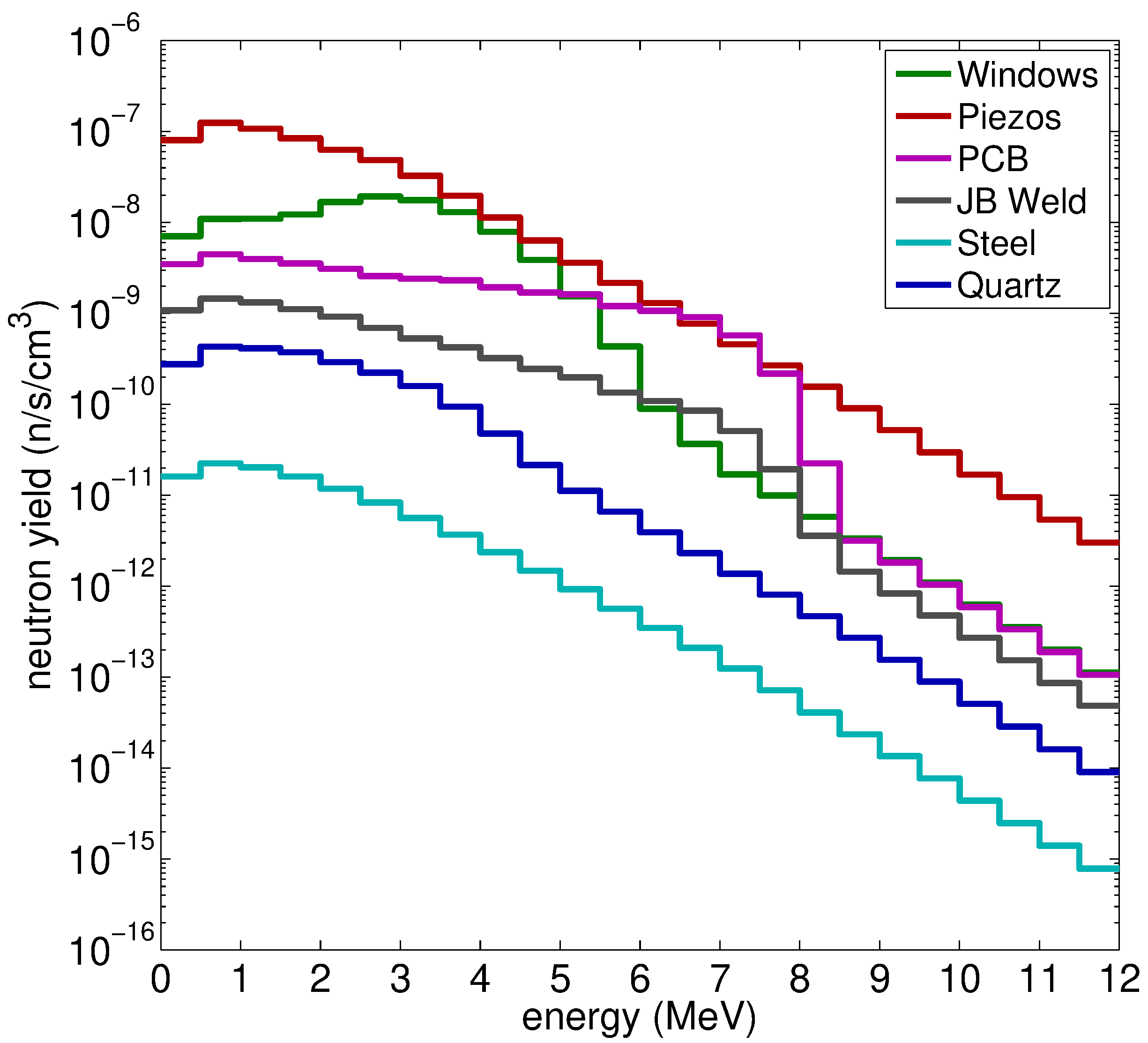}
\caption[Neutron energy spectra from $^{238}$U and $^{232}$Th in detector components]{The neutron emission energy spectra arising from the $^{238}$U and $^{232}$Th contamination of different detector components. The sum of the contributions from ($\alpha$,n) and spontaneous fission neutrons for both $^{238}$U and $^{232}$Th are shown for each component.}
\label{fig:UTh_n_spectrum}
\end{figure}

The concentration of $^{238}$U and $^{232}$Th in each material was determined from these measurements assuming secular equilibrium, except in the case of the lead-containing piezoelectric transducers, where the $^{210}$Pb concentration was specifically measured and the equilibrium was re-established from $^{210}$Pb to the bottom of the $^{238}$U decay chain. As with the rock neutrons, the ($\alpha$,n) and spontaneous fission neutron production rate and energy spectrum for each material were calculated using the SOURCES-4C code given the concentrations $C_\mathrm{U}$ and $C_\mathrm{Th}$ measured for each material (Figure \ref{fig:UTh_n_spectrum}).  These neutron spectra and yields were then used to define the sources in MCNP-PoliMi simulations, and a nuclear recoil event rate and livetime-scaled raw count prediction was generated for each material in the chamber that could act as an internal source of neutrons. Both the nuclear recoil event rate prediction and total number of counts expected in each dark matter search data set take into account a flat bubble nucleation efficiency ($\eta_\mathrm{C,F}$ = 0.46, Section \ref{sec:efficiency_fits}) and 79.86\% data cut efficiency (Section \ref{sec:datasets_summary}) on single-bubble events. The uncertainties in these predictions contain both a statistical uncertainty from the number of events simulated as well as a 10.5\% uncertainty from the MCNP-PoliMi simulation (Section \ref{sec:mcnp_uncertainty}), but do not include the 18\% uncertainty from the SOURCES-4C calculation \citep{charlton-98}.

\subsection{Borosilicate Glass Viewports}
\label{sec:backgrounds_windows}

Because the primary trigger for the COUPP experiment is the video trigger, a glass viewport capable of withstanding the pressure cycles and temperature fluctuations of a bubble chamber is necessary. In order to be properly pressure-rated, these windows were required to be quite large, with a total volume of 417 cm$^3$ of glass each. Borosilicate glass was used because it has very low coefficients of thermal expansion. The initial design of the 4 kg bubble chamber called for stereoscopic imaging using orthogonal cameras. For this design, a series of four windows were placed orthogonally around the steel pressure vessel. In actuality, only one viewport is used by both cameras, with a 20$^\circ$ separation angle, while the other viewports remain unused.

\begin{figure} [t!]
\begin{minipage}[b]{0.49\linewidth}
\centering
\includegraphics[scale=0.4]{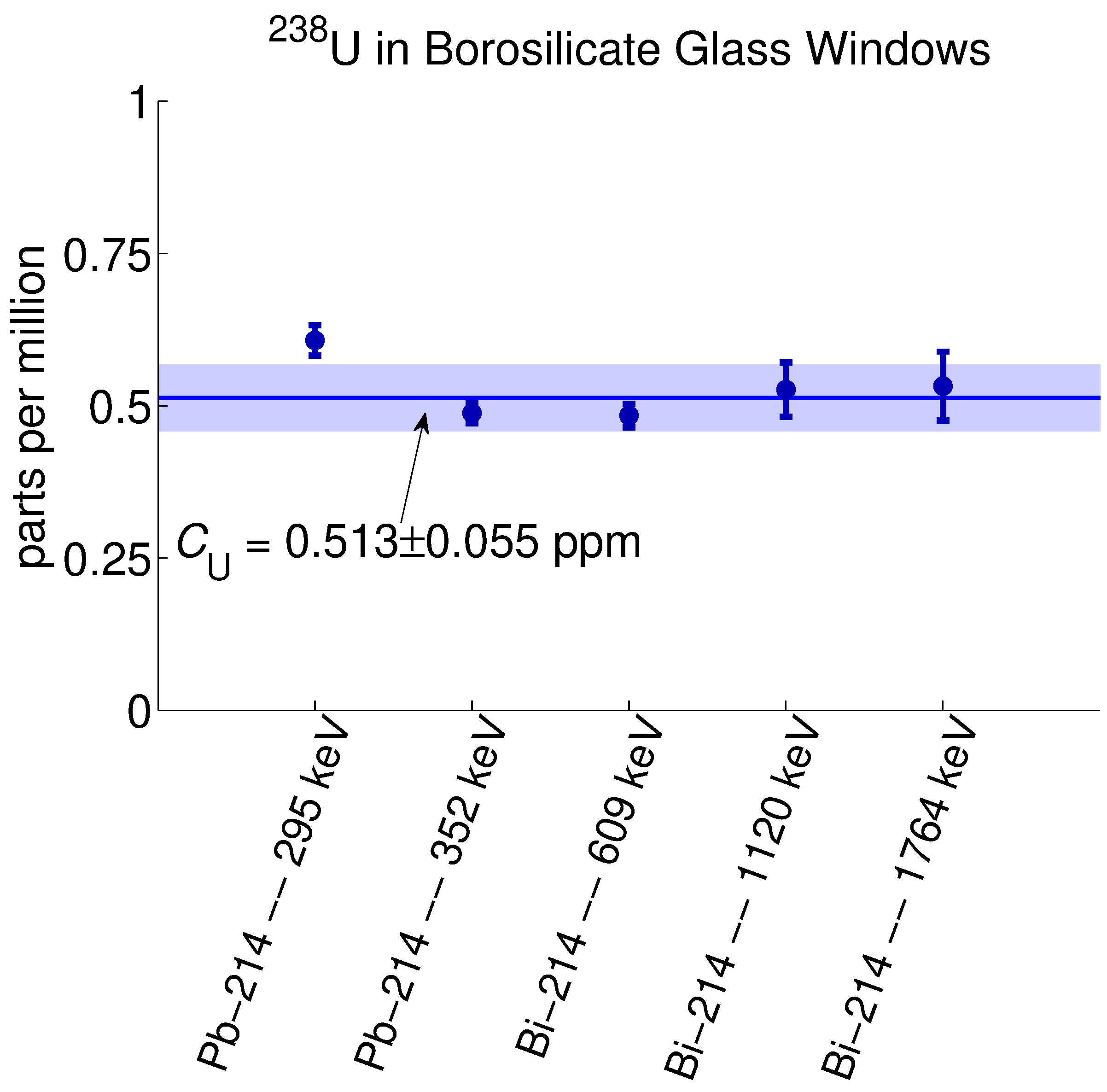}
\end{minipage}
\begin{minipage}[b]{0.51\linewidth}
\centering
\includegraphics[scale=0.4]{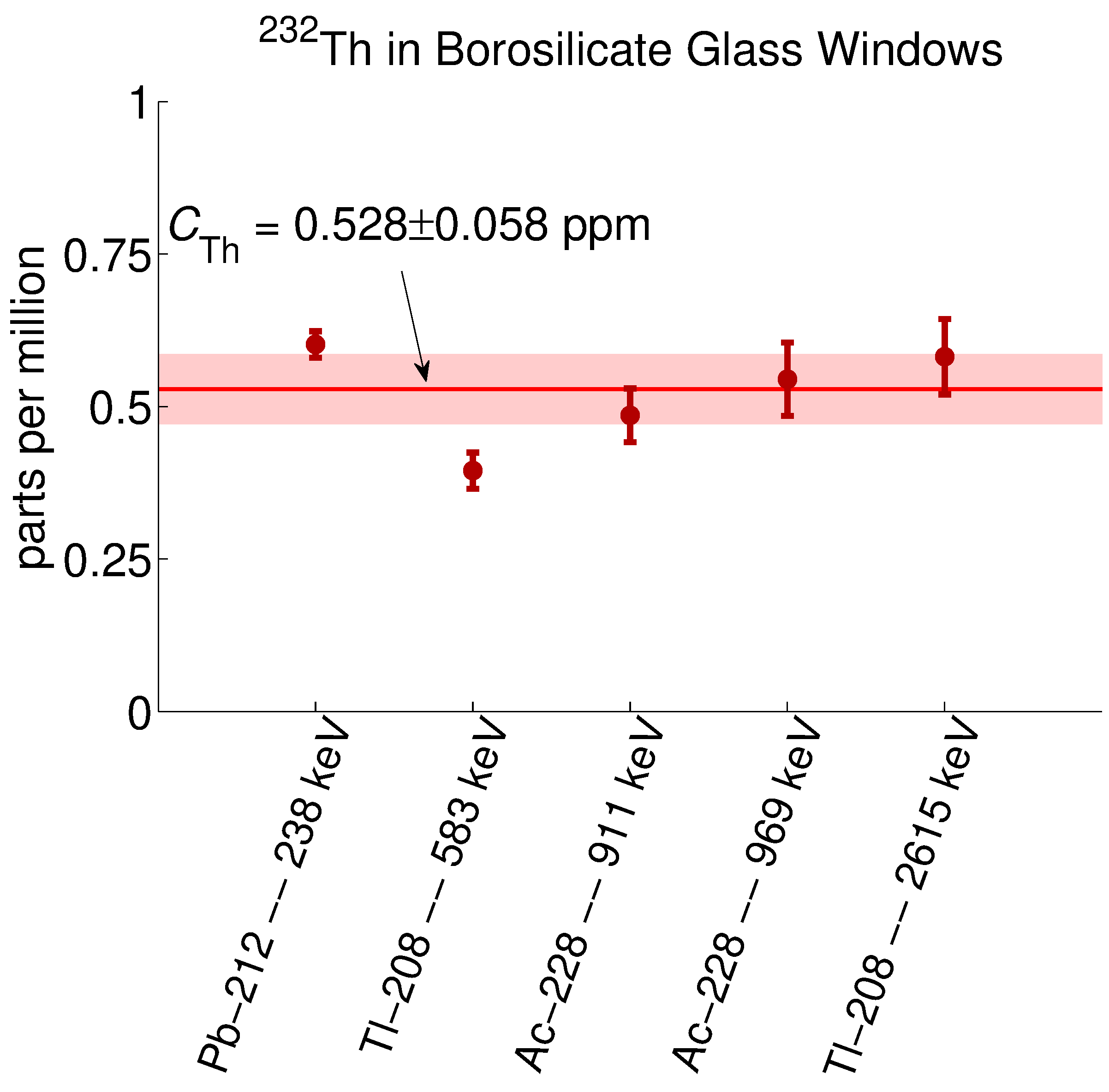}
\end{minipage}
\caption[Glass viewport $^{238}$U and $^{232}$Th concentration]{Composition (in ppm) of $^{238}$U and $^{232}$Th in the borosilicate glass of the viewports. Each gamma energy peak implies an amount of parent isotope, suggesting all isotopes in the decay chain are in equilibrium.}
\label{fig:Window_ppm}
\end{figure}

One window was counted for 2.36 days using the HPGe detector as described above. The best-fit for the $^{238}$U and $^{232}$Th concentrations were found to be 0.513$\pm$0.055 ppm and 0.528$\pm$0.058 ppm, respectively (Figure \ref{fig:Window_ppm}).  Using the SOURCES-4C code, the neutron rates and spectra from the ($\alpha$,n) and spontaneous fission reactions in the glass were calculated, giving a total neutron rate of 1.22$\times$10$^{-7}$ n/s/cm$^3$ and spectrum as in Figure \ref{fig:UTh_n_spectrum}. This rate is dominated by the ($\alpha$,n) neutrons, by nature of the significant fraction of ideal ($\alpha$,n) emitters (primarily light elements such as $^{10}$B, $^{11}$B, $^{23}$Na, $^{27}$Al, $^{29}$Si, and $^{30}$Si) in the composition of the borosilicate glass --- assumed to be a typical 81\% SiO$_2$, 13\% B$_2$O$_3$, 4\% Na$_2$O, and 2\% Al$_2$O$_3$. Because of the large total volume of glass present in the chamber with these relatively high rates of neutron production, the MCNP-PoliMi simulations predicted very high rates of recoil-like events in the chamber from the windows with these concentrations of $^{238}$U and $^{232}$Th (Table \ref{tab:UTh_events}).

\begin{table}[t!]
\centering
\small{
\begin{tabular} {| l | c || c | c || c | c |}
\hline
\multicolumn{6}{| l |}{$^{238}$U and $^{232}$Th ($\alpha$,n) and Spontaneous Fission Neutrons} \\
\hline
Material & Data Set & \multicolumn{2}{| c ||}{Rate} & \multicolumn{2}{| c |}{Counts in Set} \\
 & & \multicolumn{2}{| c ||}{ \small{(10$^{-3}$ cts/kg$_\mathrm{CF_3I}$/day)}} & \multicolumn{2}{| c |}{\small{(cts)}} \\
\cline{3-6}
 & & \small{($N = 1$)} & \small{($N > 1$)} & \small{($N = 1$)} & \small{($N > 1$)} \\
\hline
Borosilicate Glass & DM-34$^\circ$C & $5.56^{+0.44}_{-0.44}$ & $2.12^{+0.14}_{-0.15}$ & $2.23^{+0.18}_{-0.17}$ & $0.83^{+0.05}_{-0.06}$ \\
(viewports) & DM-37$^\circ$C & $6.14^{+0.48}_{-0.47}$ & $2.64^{+0.16}_{-0.16}$ & $0.54^{+0.04}_{-0.04}$ & $0.23^{+0.01}_{-0.01}$ \\
 & DM-40$^\circ$C & $6.53^{+0.50}_{-0.50}$ & $3.16^{+0.19}_{-0.19}$ & $0.46^{+0.04}_{-0.04}$ & $0.22^{+0.01}_{-0.01}$ \\
\hline
PZT & DM-34$^\circ$C & $2.31^{+0.19}_{-0.19}$ & $0.88^{+0.06}_{-0.06}$ & $0.91^{+0.08}_{-0.07}$ & $0.35^{+0.02}_{-0.02}$ \\
(piezos) & DM-37$^\circ$C & $2.52^{+0.20}_{-0.20}$ & $1.11^{+0.07}_{-0.07}$ & $0.22^{+0.02}_{-0.02}$ & $0.10^{+0.01}_{-0.01}$ \\
 & DM-40$^\circ$C & $2.69^{+0.22}_{-0.22}$ & $1.34^{+0.08}_{-0.08}$ & $0.19^{+0.02}_{-0.02}$ & $0.10^{+0.01}_{-0.01}$ \\
\hline
PCB & DM-34$^\circ$C & $0.127^{+0.008}_{-0.008}$ & $0.047^{+0.003}_{-0.003}$ & $0.050^{+0.003}_{-0.003}$ & $0.019^{+0.001}_{-0.001}$ \\
(preamps) & DM-37$^\circ$C & $0.138^{+0.009}_{-0.009}$ & $0.059^{+0.003}_{-0.003}$ & $0.012^{+0.001}_{-0.001}$ & $0.005^{+0.0003}_{-0.0003}$ \\
 & DM-40$^\circ$C & $0.146^{+0.009}_{-0.009}$ & $0.070^{+0.003}_{-0.003}$ & $0.010^{+0.001}_{-0.001}$ & $0.005^{+0.0002}_{-0.0002}$ \\
\hline
304L Stainless Steel & DM-34$^\circ$C & $0.093^{+0.007}_{-0.007}$ & $0.035^{+0.002}_{-0.002}$ & $0.037^{+0.003}_{-0.003}$ & $0.014^{+0.001}_{-0.001}$ \\
(pressure vessel) & DM-37$^\circ$C & $0.102^{+0.008}_{-0.008}$ & $0.043^{+0.003}_{-0.003}$ & $0.009^{+0.001}_{-0.001}$ & $0.004^{+0.0002}_{-0.0002}$ \\
 & DM-40$^\circ$C & $0.108^{+0.008}_{-0.008}$ & $0.052^{+0.003}_{-0.003}$ & $0.008^{+0.001}_{-0.001}$ & $0.004^{+0.0002}_{-0.0002}$ \\
\hline
JB Weld & DM-34$^\circ$C & $0.084^{+0.007}_{-0.007}$ & $0.032^{+0.002}_{-0.002}$ & $0.033^{+0.003}_{-0.003}$ & $0.013^{+0.001}_{-0.001}$ \\
(epoxy) & DM-37$^\circ$C & $0.090^{+0.007}_{-0.007}$ & $0.040^{+0.003}_{-0.003}$ & $0.008^{+0.001}_{-0.001}$ & $0.004^{+0.0002}_{-0.0002}$ \\
 & DM-40$^\circ$C & $0.096^{+0.008}_{-0.008}$ & $0.048^{+0.003}_{-0.003}$ & $0.007^{+0.001}_{-0.001}$ & $0.003^{+0.0002}_{-0.0002}$ \\
\hline
Quartz & DM-34$^\circ$C & $0.037^{+0.003}_{-0.003}$ & $0.014^{+0.001}_{-0.001}$ & $0.015^{+0.001}_{-0.001}$ & $0.005^{+0.0003}_{-0.0003}$ \\
(bell jar flange) & DM-37$^\circ$C & $0.040^{+0.003}_{-0.003}$ & $0.017^{+0.001}_{-0.001}$ & $0.004^{+0.0002}_{-0.0002}$ & $0.002^{+0.0001}_{-0.0001}$ \\
 & DM-40$^\circ$C & $0.043^{+0.003}_{-0.003}$ & $0.021^{+0.001}_{-0.001}$ & $0.003^{+0.0002}_{-0.0002}$ & $0.002^{+0.0001}_{-0.0001}$ \\
\hline
\hline
Total & DM-34$^\circ$C & $8.32^{+0.48}_{-0.48}$ & $3.13^{+0.15}_{-0.16}$ & $3.28^{+0.19}_{-0.19}$ & $1.23^{+0.06}_{-0.06}$ \\
 & DM-37$^\circ$C & $9.04^{+0.52}_{-0.52}$ & $3.91^{+0.18}_{-0.18}$ & $0.80^{+0.05}_{-0.05}$ & $0.35^{+0.02}_{-0.02}$ \\
 & DM-40$^\circ$C & $9.62^{+0.55}_{-0.55}$ & $4.69^{+0.21}_{-0.21}$ & $0.68^{+0.04}_{-0.04}$ & $0.33^{+0.01}_{-0.01}$ \\
 \hline
\end{tabular} }
\caption[Detector component background rate prediction]{The expected event rates and raw number of events expected from the $^{238}$U and $^{232}$Th contamination of detector components, for the DM-34$^\circ$C, DM-37$^\circ$C, and DM-40$^\circ$C data sets.}
\label{tab:UTh_events}
\end{table}

The expected recoil-like event rate generated by the windows is unacceptable for a competitive dark matter experiment. For an upcoming data run using the COUPP 4 kg bubble chamber at SNOLAB, the four windows are being replaced by one Sico SQ1 synthetic silica window (similar to the material used for the 4 kg bell jar) and three blind steel flanges. While the radiopurity of the new window is three to four orders of magnitude higher, it loses in thermal expansion properties to the borosilicate glass. For this reason, a very large steel retaining flange is necessary to encase the slightly larger synthetic silica window (523 cm$^3$ volume as opposed to the 417 cm$^3$ volume for each of the older windows) to withstand the temperature and pressure fluctuations of the bubble chamber. If the radiopurity of this window is of the same order as the synthetic silica bell jar (21 ppt $^{238}$U and 59 ppt $^{232}$Th), then the recoil-like event rate at a 15.8 keV threshold would be reduced to $\sim$$1\times10^{-7}$ counts/kg/day, comparable to those produced by other small but presumably irreducible backgrounds. Screening of this new window is in progress at the time of this writing.

\subsection{Piezoelectric Transducers}
\label{sec:backgrounds_piezos}

\begin{figure} [t!]
\begin{minipage}[b]{0.49\linewidth}
\centering
\includegraphics[scale=0.4]{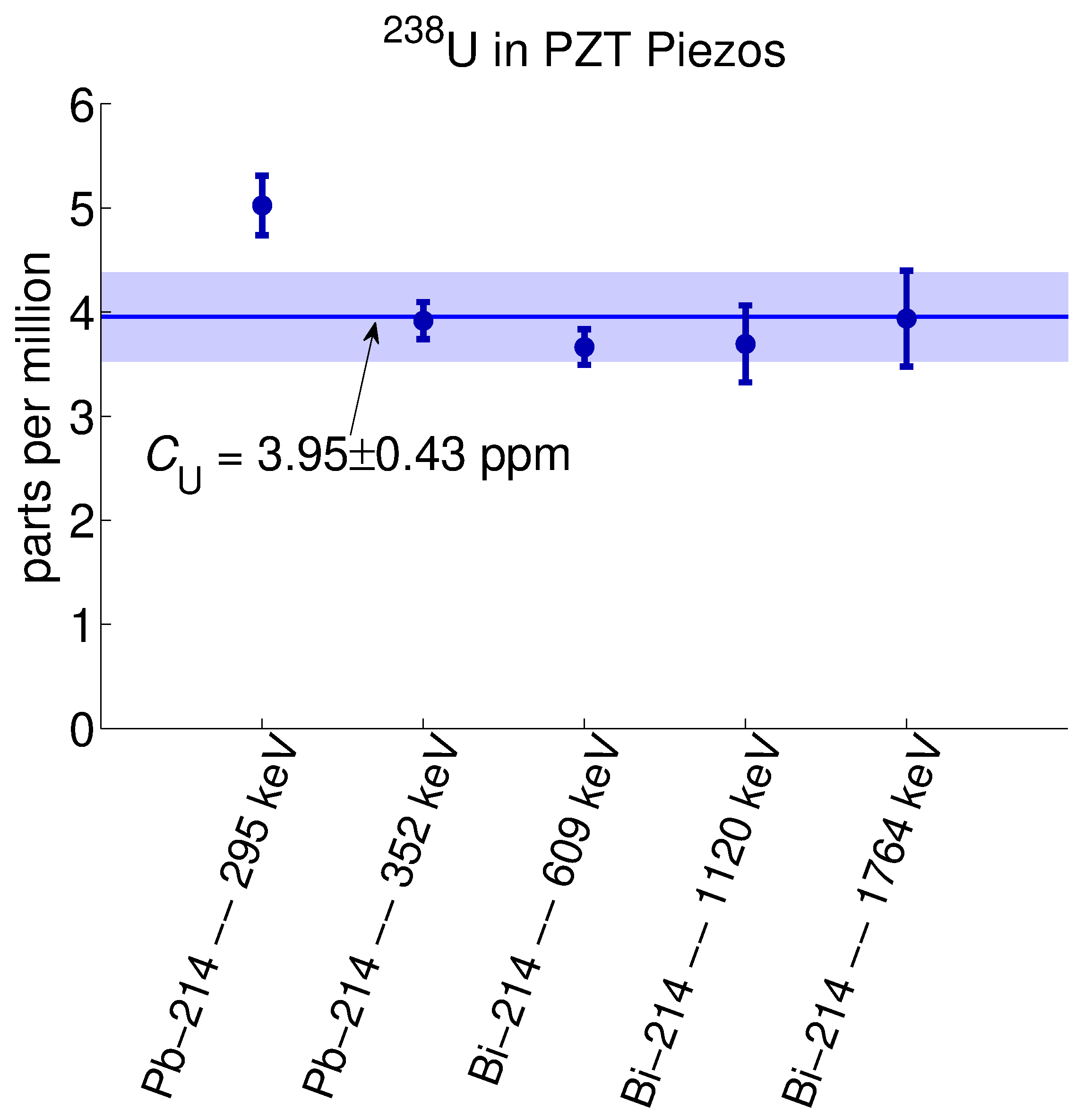}
\end{minipage}
\begin{minipage}[b]{0.51\linewidth}
\centering
\includegraphics[scale=0.4]{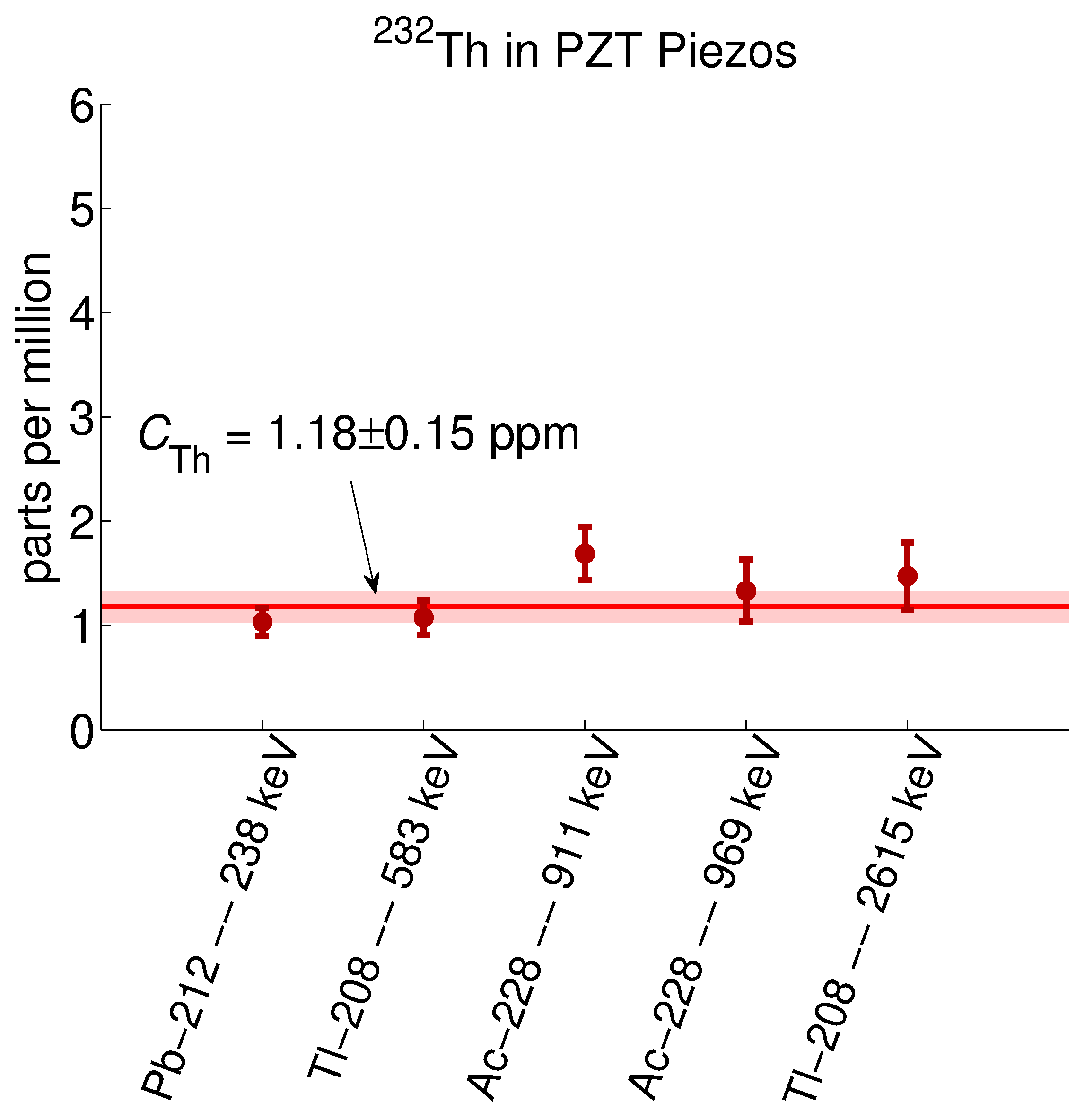}
\end{minipage}
\caption[PZT $^{238}$U and $^{232}$Th concentration]{Composition (in ppm) of $^{238}$U and $^{232}$Th in the PZT piezoelectric transducers. Each gamma energy peak implies an amount of parent isotope, suggesting all isotopes in the decay chain are in equilibrium.}
\label{fig:PZT_ppm}
\end{figure}

As described in Section \ref{sec:coupp_sensors}, piezoelectric transducers were used to provide an acoustic measurement of the bubble formation allowing for event-by-event acoustic discrimination of alphas. The piezoelectric material used to create transducers are typically ceramic in nature, and in the case of the COUPP 4 kg bubble chamber results reported in this thesis, lead zirconate titanate (PZT for short, Pb[Zr$_x$Ti$_{1-x}$]O$_3$, $0 \leq x \leq 1$) was used ($x=0.5$ assumed here). This material is especially problematic because it contains relatively large amounts of $^{210}$Pb, $^{238}$U, and $^{232}$Th embedded within a low-$Z$ (and therefore high neutron yield) environment of the ceramic. The small amounts of $^{17}$O and $^{18}$O are an ideal ($\alpha$,n) target, as are the titanium isotopes (see Section \ref{sec:mcnp_sources_bugs}). The choice of salts used in the creation of the PZT could not be screened prior to its usage (these being commercial piezos from Ferroperm), which lead to unacceptable levels of radioactivity in the piezos (Figure \ref{fig:PZT_ppm}).  Mostly through the spontaneous fission channel, but with some contribution from the ($\alpha$,n) reactions, these piezos became a significant source of neutrons contributing to the overall background.

A 67 g sample of PZT was counted for 4.3 hours using the low-background HPGe detector. The best-fit for the $^{238}$U and $^{232}$Th concentrations were found to be 3.95$\pm$0.43 ppm and 1.18$\pm$0.15 ppm, respectively. The HPGe ``well" detector was used to count the 46 keV gamma from $^{210}$Pb to give the enhancement in $^{210}$Pb concentration above that expected from equilibrium with $^{238}$U. The measured amount of $^{210}$Pb was 161.2 Bq/kg$_\mathrm{PZT}$. The SOURCES-4C code was used to calculate a total neutron yield of 5.88$\times$10$^{-7}$ n/s/cm$^3$ from the commercial piezos, using the concentrations of $^{238}$U, $^{232}$Th, and $^{210}$Pb measured. There were 8 piezos mounted relatively close to the CF$_3$I (Figure \ref{fig:coupp_pictures}, right), each with a volume of 1.7 cm$^3$, resulting in a total mass of 106 g of PZT used in this experiment. In total, this resulted in a non-negligible recoil-like event rate in the chamber originating from the piezos (Table \ref{tab:UTh_events}). It is apparent that better screening of the lead oxide salts used to manufacture the piezos is necessary. For the upcoming data run using the COUPP 4 kg bubble chamber at SNOLAB, the piezos have been replaced with custom-made radiopure versions. The preparation of these radiopure piezos is briefly described in Appendix \ref{ch:piezos}.

\subsection{PCB in Piezo Preamplifiers}
\label{sec:backgrounds_pcb}

Each piezo has an on-board preamplifier made of standard printed circuit board (PCB) material. PCB is typically made of plastics with encased fibrous glass, and is not particularly low in $^{238}$U and $^{232}$Th. For calculations here, the composition of PCB is taken to be SiO$_2$ and some plastic material C$_{13}$H$_{12.5}$O$_{8}$ (composition scaled so the abundances of C, H, and O stoichiometrically match 1 part SiO$_2$). Each of the 8 PCB preamps represents approximately 1.3 cm$^3$ of material, resulting in a total mass of 19.2 g of PCB used.

\begin{figure} [t!]
\begin{minipage}[b]{0.49\linewidth}
\centering
\includegraphics[scale=0.4]{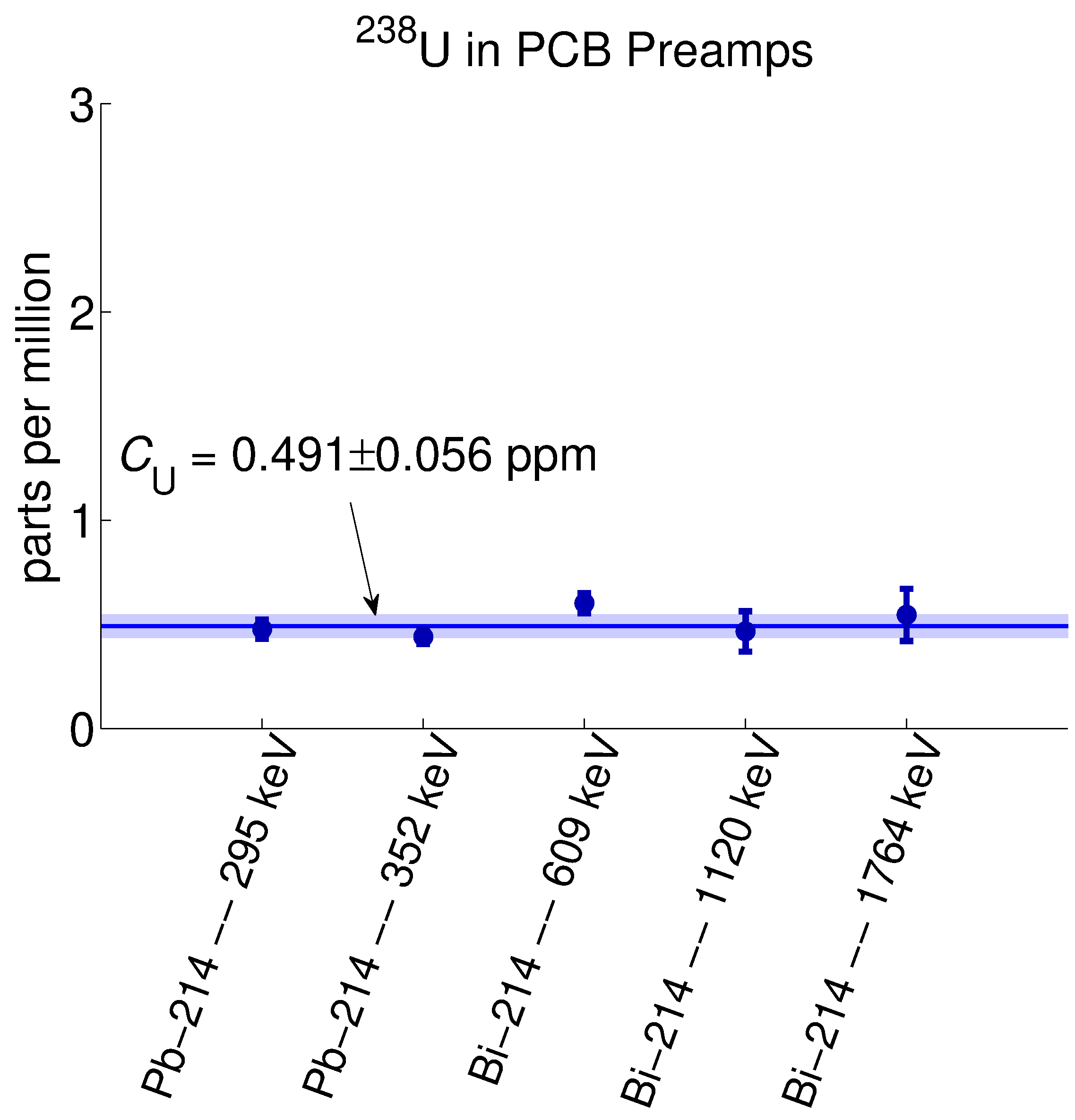}
\end{minipage}
\begin{minipage}[b]{0.51\linewidth}
\centering
\includegraphics[scale=0.4]{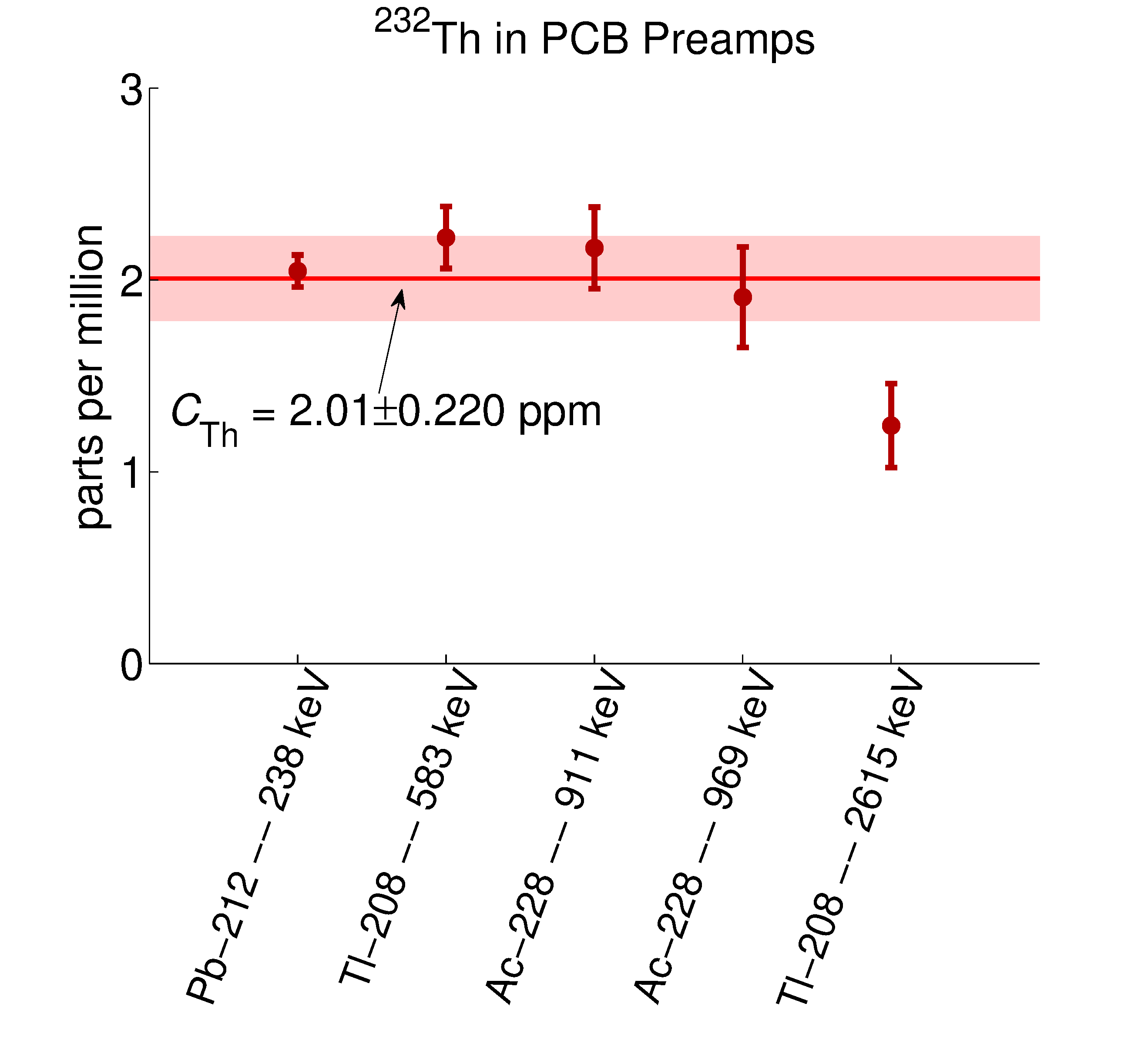}
\end{minipage}
\caption[PCB $^{238}$U and $^{232}$Th concentration]{Composition (in ppm) of $^{238}$U and $^{232}$Th in the PCB piezo preamps. Each gamma energy peak implies an amount of parent isotope, suggesting all isotopes in the decay chain are in equilibrium.}
\label{fig:PCB_ppm}
\end{figure}

Measurements of a 42.8 g sample of PCB (18 boards taken together) were performed with the HPGe detector for 19.2 hours, yielding 0.491$\pm$0.056 ppm and 2.01$\pm$0.22 ppm of $^{238}$U and $^{232}$Th, respectively (Figure \ref{fig:PCB_ppm}).  Given these concentrations in a SOURCES-4C calculation, the total neutron yield was $3.51\times10^{-8}$ n/s/cm$^3$ in the PCB. With a total volume of approximately 10.4 cm$^3$ used, this leads to a recoil-like event rate only on the order of 10$^{-4}$ counts/kg/day, making it less important than either the viewports or the PZT in the piezos (Table \ref{tab:UTh_events}). Still, while upgrading the piezos, the PCB boards are being replaced with substantially more radiopure CuFlon circuit boards. Although CuFlon is a teflon-based material (rich in $^{19}$F, which is an ideal ($\alpha$,n) target), the measured $^{238}$U and $^{232}$Th composition of the boards is three orders of magnitude lower than the PCB, reducing the expected rate to $\sim$10$^{-5}$ counts/kg/day.

\subsection{J-B Weld Epoxy}
\label{sec:backgrounds_jbweld}

Each of the 8 piezos were fastened to the neck of the bell jar using J-B Weld, an iron-containing Bisphenol A Diglycidyl Ether (BADGE) epoxy, with some amounts of Novolac epoxy resin, and 2,4,6-tris(dimethylaminomethyl) phenol. Although the specific chemical composition of J-B Weld is proprietary, it was estimated to be similar to other BADGE epoxies, where the resin has an approximate chemical composition of H$_{3.9}$C$_{4.3}$O$_{0.8}$N$_{0.05}$, and there is some amount of CaCO$_3$ filler along with the iron (with composition 1:2:5, respectively). The density of J-B Weld was measured to be 2.85 g/cm$^3$ and it was approximated that about 3.5 cm$^3$ of J-B Weld was used per piezo. There are copious amounts of potential ($\alpha$,n) target nuclei in J-B Weld, and so it can also serve as a weak neutron source for the chamber.

\begin{figure} [t!]
\begin{minipage}[b]{0.49\linewidth}
\centering
\includegraphics[scale=0.4]{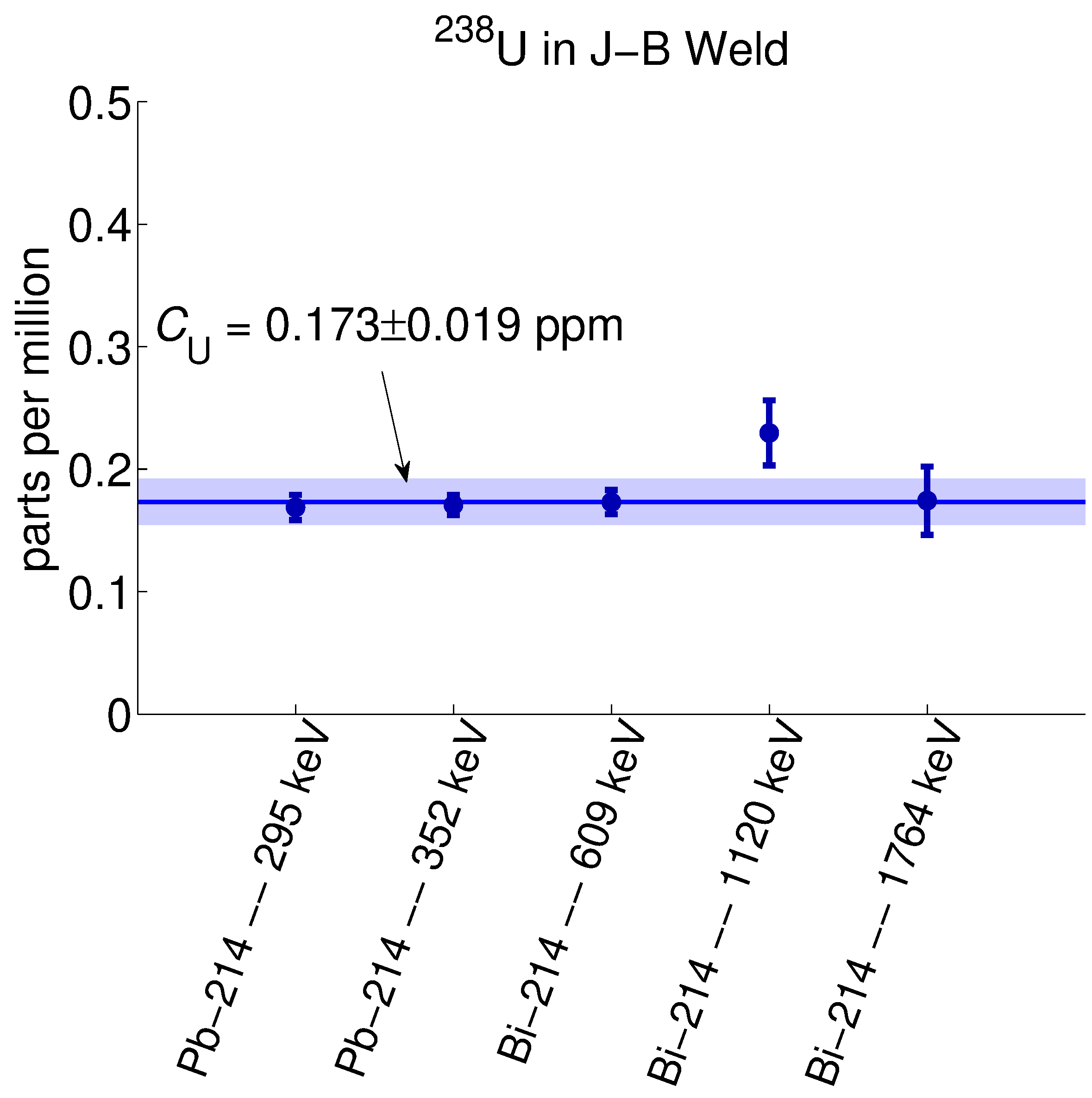}
\end{minipage}
\begin{minipage}[b]{0.51\linewidth}
\centering
\includegraphics[scale=0.4]{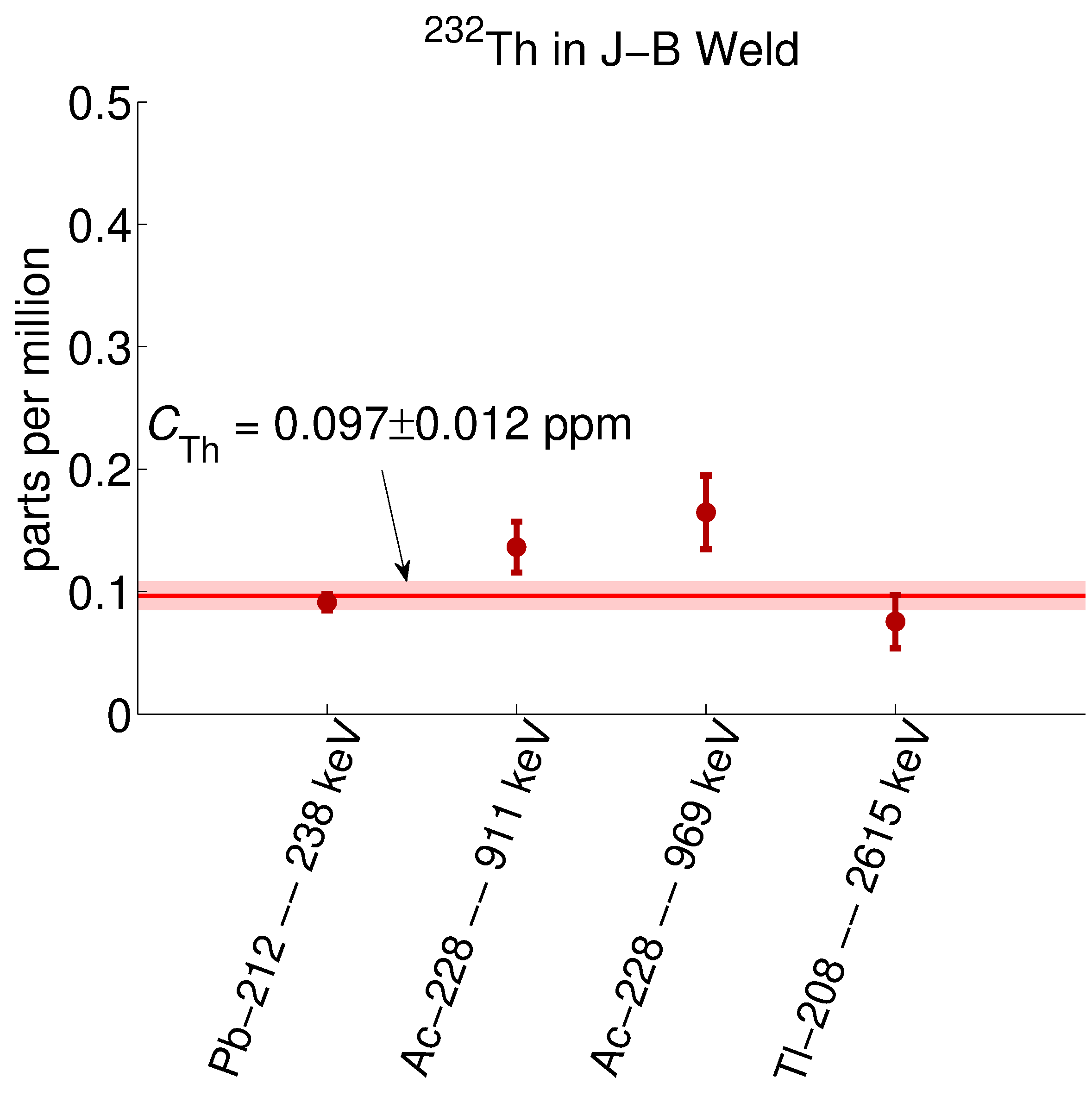}
\end{minipage}
\caption[J-B Weld $^{238}$U and $^{232}$Th concentration]{Composition (in ppm) of $^{238}$U and $^{232}$Th in the J-B Weld epoxy. Each gamma energy peak implies an amount of parent isotope, suggesting all isotopes in the decay chain are in equilibrium.}
\label{fig:JBWeld_ppm}
\end{figure}

A 28 g sample of J-B Weld was counted for 3.77 days with the HPGe detector. The best-fit for the $^{238}$U and $^{232}$Th concentrations in J-B Weld were found to be 0.173$\pm$0.019 ppm and 0.097$\pm$0.012 ppm, respectively (Figure \ref{fig:JBWeld_ppm}).  Given these concentrations in a SOURCES-4C calculation, the total neutron yield was $8.71\times10^{-9}$ n/s/cm$^3$ in the J-B Weld. With a total volume of approximately 28 cm$^3$ used, this leads to a recoil-like event rate of $\sim$10$^{-4}$ counts/kg/day (Table \ref{tab:UTh_events}). Although this rate is low, care should be taken when using epoxies in attaching or encapsulating piezos in the future.

\subsection{Stainless Steel Pressure Vessel}
\label{sec:backgrounds_steel}

The type 304L stainless steel pressure vessel (68\% Fe, 18\% Cr, 12\% Ni, and 2\% Mn), while having a typically low concentration of $^{238}$U and $^{232}$Th and an even lower concentration of potential ($\alpha$,n) targets (trace amounts of $^{13}$C, $^{29}$Si, $^{30}$Si, and $^{31}$P were assumed), can still contribute to the overall background of the chamber due to the sheer size of the vessel itself --- approximately $6\times10^4$ cm$^3$ of steel (about 600 kg) near the CF$_3$I volume was taken into account. The $^{238}$U and $^{232}$Th concentrations of the steel used were not specifically measured in this experiment, but assumptions that both are $\sim$1 ppb are well-motivated \citep{ilias-12}. From these assumptions, a total neutron production rate of $1.11\times10^{-10}$ n/s/cm$^3$ was calculated with SOURCES-4C, dominated by the spontaneous fission of $^{238}$U (Table \ref{tab:UTh_events}). While in the current experiment, this does not represent a significant background (it is about two orders of magnitude below the total background), once the windows and piezos are replaced with more radiopure versions, $^{238}$U spontaneous fission in the stainless steel pressure vessel will represent the most significant background for the 4 kg chamber.

\subsection{Quartz Flange}
\label{sec:backgrounds_quartz}

Finally, while the bell jar for the bubble chamber is made of radiopure Suprasil synthetic fused silica quartz with a concentration of $^{238}$U and $^{232}$Th of 21$\pm$9 and 59$\pm$14 ppt, respectively \citep{leonard-08}, the flange on the top of the bell jar was still made of the standard quartz used for the bell jar of a previous version of the COUPP experiment (the 2 kg bubble chamber). This flange had a total volume of about 193 cm$^3$. The concentrations of $^{238}$U and $^{232}$Th in standard quartz were measured by the HPGe detector to be 42$\pm$4 and $<$ 30 ppb, respectively\footnote{For a conservative limit, the concentration of $^{232}$Th was taken to be $\sim$30 ppb.} \citep{szydagis-11}. For these concentrations in SiO$_2$, a total neutron production rate of $2.36\times10^{-9}$ n/s/cm$^3$ was calculated with SOURCES-4C (Table \ref{tab:UTh_events}). The contribution to the overall background from the quartz flange is small, much below the rate expected from the steel pressure vessel.

\subsection{Other Negligible Components}
\label{sec:backgrounds_other}

The propylene glycol hydraulic fluid filling the stainless steel pressure vessel also has radioactive contamination of $^{238}$U and $^{232}$Th. After the bubble chamber runs at SNOLAB described in this thesis were completed, a 1.04 kg sample of the glycol was counted using an HPGe detector in SNOLAB for a total of 6.86 days. The $^{238}$U concentration was inferred (assuming secular equilibrium) from the $^{226}$Ra gamma spectrum to be 31$\pm$0.42 ppt. While radon seepage into the glycol from the atmosphere or from its emanation from the steel pressure vessel and bellows could cause this equilibrium to be broken (Section \ref{sec:backgrounds_radon}), the assumption is made that the measurement performed implies an equilibrium of all components along the $^{238}$U decay chain. The measured composition of $^{232}$Th was below background. The neutron production rate using SOURCES-4C was found to be $6.38\times10^{-13}$ n/s/cm$^3$, which leads to a recoil-like event rate of only $\sim$10$^{-5}$ counts/kg/day, below even the contribution from the quartz flange.

In addition to those listed, many other components of the COUPP 4 kg bubble chamber were screened for contamination of $^{238}$U and $^{232}$Th, including the cameras, lenses, Setra pressure sensors, epoxies, RTVs, backing piezo metal powder, light reflector, and other materials used in very small amounts in piezo construction. In all, no other material in the experiment was deemed to have a radioactive contamination, mass, or composition sufficient to provide a non-negligible nucleation rate in this chamber.

\section{Total Neutron Background Expectations}
\label{sec:backgrounds_total}

In all, the neutron backgrounds in these results of the COUPP 4 kg bubble chamber operated at SNOLAB are dominated by two sources: the borosilicate glass viewports and the PZT of the piezoelectric transducers. By replacing both components with more radiopure versions, the background recoil-like events can be reduced to $<$ 1 event per year total. All recoil-like background rates from all neutron sources in this chapter are listed in Table \ref{tab:total_rates}, in descending order of contribution. Figure \ref{fig:summary} depicts the most dominant of these backgrounds with respect to the dark matter search data taken.

\begin{table}[p!]
\centering
\begin{minipage}{\textwidth}
\small{
\begin{tabular} {| l | c | c || c | c | c |}
\hline
Material & Source & Concentration & Rate in & Rate in & Rate in \\
 & & & DM-34$^\circ$C & DM-37$^\circ$C & DM-40$^\circ$C \\
\hline
\multicolumn{3}{| c }{} & \multicolumn{3}{ c |}{(10$^{-3}$ cts/kg$_\mathrm{CF_3I}$/day)} \\
\hline
Glass \footnotesize{(viewports)} & $^{238}$U ($\alpha$,n) & 0.513$\pm$0.055 ppm & $3.87^{+0.42}_{-0.42}$ & $4.20^{+0.45}_{-0.45}$ & $4.46^{+0.47}_{-0.47}$ \\
PZT \footnotesize{(piezos)} & $^{238}$U s.f. & 3.95$\pm$0.43 ppm & $1.64^{+0.18}_{-0.18}$ & $1.79^{+0.19}_{-0.19}$ & $1.90^{+0.20}_{-0.20}$ \\
Glass \footnotesize{(viewports)} & $^{232}$Th ($\alpha$,n) & 0.528$\pm$0.058 ppm & $1.30^{+0.14}_{-0.14}$ & $1.41^{+0.15}_{-0.15}$ & $1.50^{+0.16}_{-0.16}$ \\
\hline
\multicolumn{3}{| c }{} & \multicolumn{3}{ c |}{(10$^{-4}$ cts/kg$_\mathrm{CF_3I}$/day)} \\
\hline
PZT \footnotesize{(piezos)} & $^{238}$U ($\alpha$,n) & 3.95$\pm$0.43 ppm & $6.04^{+0.66}_{-0.66}$ & $6.66^{+0.71}_{-0.71}$ & $7.14^{+0.75}_{-0.75}$ \\
Glass \footnotesize{(viewports)} & $^{238}$U s.f. & 0.513$\pm$0.055 ppm & $4.87^{+0.52}_{-0.53}$ & $5.29^{+0.56}_{-0.56}$ & $5.63^{+0.59}_{-0.60}$ \\
Norite \footnotesize{(rock walls)} & $^{238}$U/$^{232}$Th & 1.2/3.3 ppt & $1.26^{+0.71}_{-0.57}$ & $1.26^{+0.71}_{-0.57}$ & $1.26^{+0.71}_{-0.57}$ \\
Steel \footnotesize{(pressure vessel)} & $^{238}$U s.f. & $\sim$1 ppb & $0.92^{+0.10}_{-0.10}$ & $1.00^{+0.11}_{-0.11}$ & $1.07^{+0.11}_{-0.11}$ \\
\hline
\multicolumn{3}{| c }{} & \multicolumn{3}{ c |}{(10$^{-5}$ cts/kg$_\mathrm{CF_3I}$/day)} \\
\hline
PZT \footnotesize{(piezos)} & $^{232}$Th ($\alpha$,n) & 1.18$\pm$0.15 ppm & $6.33^{+0.69}_{-0.69}$ & $6.98^{+0.75}_{-0.74}$ & $7.51^{+0.80}_{-0.80}$ \\
J-B Weld \footnotesize{(epoxy)} & $^{238}$U s.f. & 0.173$\pm$0.019 ppm & $5.83^{+0.63}_{-0.63}$ & $6.34^{+0.67}_{-0.68}$ & $6.74^{+0.72}_{-0.71}$ \\
PCB \footnotesize{(preamps)} & $^{232}$Th ($\alpha$,n) & 2.01$\pm$0.22 ppm & $4.89^{+0.53}_{-0.53}$ & $5.32^{+0.56}_{-0.57}$ & $5.63^{+0.60}_{-0.60}$ \\
PCB \footnotesize{(preamps)} & $^{238}$U s.f. & 0.491$\pm$0.056 ppm & $4.41^{+0.48}_{-0.48}$ & $4.80^{+0.51}_{-0.51}$ & $5.11^{+0.54}_{-0.54}$ \\
PCB \footnotesize{(preamps)} & $^{238}$U ($\alpha$,n) & 0.491$\pm$0.056 ppm & $3.36^{+0.36}_{-0.36}$ & $3.63^{+0.39}_{-0.39}$ & $3.87^{+0.41}_{-0.41}$ \\
J-B Weld \footnotesize{(epoxy)} & $^{238}$U ($\alpha$,n) & 0.173$\pm$0.019 ppm & $2.11^{+0.23}_{-0.23}$ & $2.26^{+0.24}_{-0.24}$ & $2.38^{+0.25}_{-0.25}$ \\
Quartz \footnotesize{(flange)} & $^{238}$U s.f. & 42$\pm$4 ppb & $1.92^{+0.21}_{-0.21}$ & $2.09^{+0.22}_{-0.22}$ & $2.21^{+0.23}_{-0.23}$ \\
Quartz \footnotesize{(flange)} & $^{238}$U ($\alpha$,n) & 42$\pm$4 ppb & $1.40^{+0.15}_{-0.15}$ & $1.52^{+0.16}_{-0.16}$ & $1.64^{+0.17}_{-0.18}$ \\
\hline
\multicolumn{3}{| c }{} & \multicolumn{3}{ c |}{(10$^{-6}$ cts/kg$_\mathrm{CF_3I}$/day)} \\
\hline
Steel \footnotesize{(pressure vessel)} & ($\mu$,n) &  & $4.55^{+0.49}_{-0.49}$ & $4.87^{+0.52}_{-0.52}$ & $5.12^{+0.54}_{-0.54}$ \\
Glycol \footnotesize{(hydraulic fluid)} & $^{238}$U total & 31$\pm$0.42 ppt & $4.40^{+0.48}_{-0.47}$ & $4.76^{+0.51}_{-0.51}$ & $5.06^{+0.53}_{-0.53}$ \\
J-B Weld \footnotesize{(epoxy)} & $^{232}$Th ($\alpha$,n) & 0.097$\pm$0.012 ppm & $4.05^{+0.43}_{-0.43}$ & $4.31^{+0.46}_{-0.46}$ & $4.54^{+0.48}_{-0.48}$ \\
Quartz \footnotesize{(flange)} & $^{232}$Th ($\alpha$,n) & $\sim$30 ppb\footnote{Rate was below background, so concentration is conservatively taken as the upper limit value.} & $3.88^{+0.42}_{-0.42}$ & $4.22^{+0.45}_{-0.45}$ & $4.54^{+0.48}_{-0.48}$ \\
CF$_3$I \footnotesize{(target)} & ($\mu$,n) &  & $3.74^{+0.40}_{-0.40}$ & $3.99^{+0.42}_{-0.42}$ & $4.21^{+0.44}_{-0.44}$ \\
Steel \footnotesize{(pressure vessel)} & $^{238}$U ($\alpha$,n) & $\sim$1 ppb & $0.95^{+0.10}_{-0.10}$ & $1.04^{+0.11}_{-0.11}$ & $1.11^{+0.12}_{-0.12}$ \\
\hline
\multicolumn{3}{| c }{} & \multicolumn{3}{ c |}{(10$^{-7}$ cts/kg$_\mathrm{CF_3I}$/day)} \\
\hline
Steel \footnotesize{(pressure vessel)} & $^{232}$Th ($\alpha$,n) & $\sim$1 ppb & $3.88^{+0.42}_{-0.42}$ & $4.26^{+0.45}_{-0.46}$ & $4.54^{+0.48}_{-0.48}$ \\
Glycol \footnotesize{(hydraulic fluid)} & ($\mu$,n) &  & $3.16^{+0.34}_{-0.34}$ & $3.44^{+0.37}_{-0.37}$ & $3.69^{+0.39}_{-0.39}$ \\
Water \footnotesize{(buffer fluid)} & ($\mu$,n) &  & $2.84^{+0.30}_{-0.30}$ & $2.99^{+0.32}_{-0.32}$ & $3.12^{+0.33}_{-0.33}$ \\
Water \footnotesize{(shielding)} & $^{222}$Rn ($\alpha$,n) & 15.9 atoms/cm$^3$ & $1.35^{+0.25}_{-0.30}$ & $1.66^{+0.30}_{-0.29}$ & $1.83^{+0.31}_{-0.34}$ \\
\hline
\multicolumn{3}{| c }{} & \multicolumn{3}{ c |}{(10$^{-8}$ cts/kg$_\mathrm{CF_3I}$/day)} \\
\hline
Water \footnotesize{(shielding)} & ($\mu$,n) &  & $5.23^{+0.57}_{-0.56}$ & $5.54^{+0.61}_{-0.59}$ & $5.92^{+0.63}_{-0.65}$ \\
\hline
\hline
\multicolumn{3}{| c }{} & \multicolumn{3}{ c |}{(10$^{-3}$ cts/kg$_\mathrm{CF_3I}$/day)} \\
\hline
Total & & & $8.33^{+0.48}_{-0.48}$ & $9.05^{+0.52}_{-0.52}$ & $9.63^{+0.55}_{-0.55}$ \\
 \hline
\end{tabular} }
\end{minipage}
\caption[Total neutron background predictions]{The predicted neutron background rates from each source, for the DM-34$^\circ$C, DM-37$^\circ$C, and DM-40$^\circ$C data sets.}
\label{tab:total_rates}
\end{table}

\singlespacing
\chapter{Gamma Background Characterization}
\label{ch:gammas}
\doublespacing

As discussed in Chapter \ref{ch:bubblechambers}, the most important feature separating the bubble chamber from other WIMP direct detection techniques is its ability to become essentially insensitive to gamma-induced events, more than any other present approach. This is accomplished by the stopping power threshold requirement represented in Figure \ref{fig:E_vs_dEdx} --- by proper tuning of the degree of superheat of the chamber, only particles with large stopping powers will nucleate bubbles. This includes nuclear recoils from alpha decays (Section \ref{sec:datasets_alpha}), neutron backgrounds (Chapter \ref{ch:backgrounds}), the rare photonuclear interaction (Section \ref{sec:gammas_highenergy}), and WIMPs. To first order, muons, gammas, x-rays, electrons, \emph{etc.} are excluded from the list of backgrounds, because they fail to cross the stopping power threshold. However, this exclusion is not perfect, especially at low thresholds (high degrees of superheat), and so a proper study of the most dominant of these minimum-ionizing backgrounds (\emph{i.e.} gammas) is well-motivated.

To empirically determine the rate of gamma-induced events in the CF$_3$I, two pieces of information are required: the level of rejection of gamma events in the chamber for a specific threshold and the gamma flux and spectrum incident on the active volume of the bubble chamber. The former is accomplished by performing gamma calibrations using very high-activity gamma sources to nucleate bubbles in the chamber, and comparing the observed rate to the expected rate from MCNP-PoliMi simulations (Section \ref{sec:gammas_calibrations}). The latter is accomplished by actively measuring the gamma rate at the location of the bubble chamber with a low-background NaI detector (Section \ref{sec:gammas_measurement}).

\section{Gamma Calibrations}
\label{sec:gammas_calibrations}

Three different gamma calibrations were performed with the COUPP 4 kg chamber at SNOLAB. In January 2011, a provisional relatively-low intensity gamma source ($\sim$5 $\mu$Ci total) was created by combining several weaker gamma sources together --- a $\sim$7.6 nCi $^{22}$Na source, a $\sim$81 nCi $^{60}$Co source, a $\sim$4.2 $\mu$Ci $^{137}$Cs source, and a $\sim$0.76 $\mu$Ci $^{152}$Eu source. Due to the complications of using such an amalgamation of sources, single-isotope sources were obtained (a 1.0$\pm$0.15 mCi $^{133}$Ba source and a 100$\pm$15 $\mu$Ci $^{60}$Co source, both with a fabrication date of May 2011), which had the added benefit of having higher source strengths. These sources were used for gamma calibrations that took place from July to August 2011, and again in October 2011. All gamma sources used for calibrations were positioned 140 cm below the top of the water shielding, just below the center of the viewports, in very close proximity to the outside of the pressure vessel.

The results of these runs are highlighted in Tables \ref{tab:counts} and \ref{tab:rates}. Of note in these measurements is the apparent elevated multi-bubble count rates observed --- a indication of an additional neutron background introduced by the gamma sources themselves. The statistical significance of the multi-bubble count rate is not great, however, and there are no gamma energies high enough in either $^{133}$Ba or $^{60}$Co that could initiate a ($\gamma$,n) reaction leading to a neutron yield. If this problem persists in future gamma calibrations, a more in-depth analysis will be required. Here, it is assumed that the contribution to the single-bubble event population from any neutrons from these sources is negligible, an assumption which leans to the conservative side of gamma rejection predictions. The calibrations at high pressures ($>$ 30 psia) will not be used, leaving eight gamma calibrations taken at thresholds near those used in the dark matter search.

The gamma rejection at each threshold is found by taking the 1$\sigma$ upper limit \citep{feldman-98} to the observed number of single-bubble events (for a conservative gamma rejection factor limit) in each gamma calibration (Table \ref{tab:counts}) and comparing the observed bubble nucleation rate with the gamma interaction rate expected from a corresponding MCNP-PoliMi simulation of the gamma calibration \citep{behnke-08}. The MCNP-PoliMi expected number of gamma interactions comes from taking the number of gammas that deposited energy via Compton or photoelectric scattering in the CF$_3$I volume above the energy threshold $E_c$, ignoring the stopping power threshold, but including the data cut efficiency of 79.86\% (Section \ref{sec:datasets_summary}). Individual gammas in the simulation that had more than one interaction depositing energy above $E_c$ were treated as if they nucleated only a single-bubble event (since, in reality, no gamma-generated event will have multiple bubbles). This total number of gamma interactions above threshold is then scaled to the exposure of the comparison gamma calibration. The ratio of the observed single-bubble events in the gamma calibration data to the expected number of gamma interactions is defined as the gamma rejection factor, which is seen to be gamma energy- and threshold-dependent. The results of each gamma calibration are shown in Table \ref{tab:gamma_rejection}. \begin{table}[t!]
\centering
\small{
\begin{tabular} {| c | c | c | c | c | c |}
\hline
Data Set & Threshold & Exposure & Counts & MCNP & Rejection \\
 & \small{(keV)} & \small{(kg-days)} & \small{(68\% C.L.)} & prediction & factor \\
\hline
Mix-62"-41$^\circ$C-30psia & 7.22$^{+0.58}_{-0.53}$ & 7.09 & $\lesssim$22.82 & $1.54\times10^{8}$ & $\sim$$1.5\times10^{-7}$ \\
Mix-62"-40$^\circ$C-30psia & 8.00$^{+0.65}_{-0.59}$ & 4.11 & $\lesssim$5.30 & $8.90\times10^{7}$ & $\lesssim$$6.0\times10^{-8}$ \\
Mix-62"-37$^\circ$C-30psia & 11.27$^{+0.97}_{-0.88}$ & 7.49 & $\lesssim$1.29 & $1.62\times10^{8}$ & $\lesssim$$8.0\times10^{-9}$\\
\hline
Co60-55"-40$^\circ$C-30psia & 7.96$^{+0.65}_{-0.59}$ & 0.30 & $\lesssim$5.30 & $1.16\times10^{8}$ & $\lesssim$$3.4\times10^{-8}$ \\
Co60-55"-34$^\circ$C-30psia & 15.77$^{+1.44}_{-1.30}$ & 15.49 & $\lesssim$2.75 & $8.02\times10^{9}$ & $\lesssim$$3.4\times10^{-10}$ \\
\hline
Ba133-55"-40$^\circ$C-30psia & 7.99$^{+0.65}_{-0.59}$ & 0.97 & $\lesssim$31.38 & $2.22\times10^{9}$ & $\sim$$1.4\times10^{-8}$ \\
Ba133-55"-37$^\circ$C-30psia & 11.19$^{+0.96}_{-0.87}$ & 11.98 & $\lesssim$4.25 & $2.74\times10^{10}$ & $\lesssim$$1.6\times10^{-10}$ \\
Ba133-55"-34$^\circ$C-30psia & 15.84$^{+1.46}_{-1.31}$ & 25.40 & $\lesssim$1.29 & $5.79\times10^{10}$ & $\lesssim$$2.2\times10^{-11}$ \\
\hline
\end{tabular} }
\caption[Gamma rejection for different thresholds]{The gamma event rejection factors measured with different calibration sources at different thresholds. Only upper limits on the rejection factor could be placed for such small count sizes in the gamma calibrations. The rejection found from the 7.2 keV mixed-source calibration and the 8.0 keV $^{133}$Ba calibration are taken as actual values, not limits. The uncertainty in the rejection factor is dominated by the $\sim$10\% uncertainty from the MCNP prediction.}
\label{tab:gamma_rejection}
\end{table}

\begin{figure} [t!]
\centering
\includegraphics[scale=0.46]{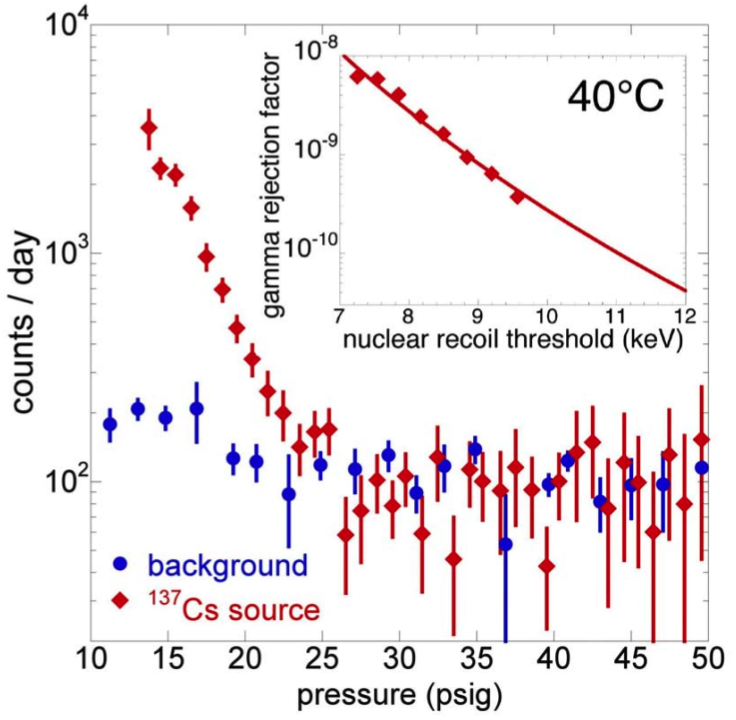}
\caption[Gamma rejection factor at different thresholds]{The $^{137}$Cs gamma calibration results from the COUPP 2 kg bubble chamber. The gamma rejection factor at various thresholds is shown in the inset. Figure from \cite{behnke-08} reprinted with permission from AAAS.}
\label{fig:gamma_rejection}
\end{figure}

Because of the low number of single-bubble events available in most calibration sets, only two (the 7.2 keV threshold mixed-source calibration and the 8.0 keV threshold $^{133}$Ba calibration) are taken as being indicative of the actual rejection factor at those thresholds. The rest are no better than upper limits. However, the gamma calibrations from the COUPP 2 kg chamber (the predecessor to the 4 kg chamber, using the same pressure vessel but a smaller inner volume, \citep{behnke-08}) featured much more substantial gamma exposure ($^{137}$Cs sources up to 13 mCi), from which a definite falling exponential trend is seen in the rejection factor (Figure \ref{fig:gamma_rejection}).  With the 2 kg chamber, the same calibrations were performed with three different intensity $^{137}$Cs sources (separated by a decade in intensity). Each calibration provided the same power dependence on $E_c$. Using the rejection factor found for the 8.0 keV threshold $^{133}$Ba calibration in Table \ref{tab:gamma_rejection} and scaling this to higher thresholds by the trend seen in Figure \ref{fig:gamma_rejection} (inset), the gamma rejection factor $f_r$ as a function of threshold energy $E_c$ (in keV) is found to be

\begin{equation}
\label{eq:rejection_factor}
f_r(E_c) = 1.29\times10^{-4} e^{-1.14 E_c}.
\end{equation}

The variation of the rejection factor with threshold is expected --- at higher thresholds, there is a substantially smaller possibility for gamma interactions to deposit their energy in a small enough region to nucleate a bubble (see Figure \ref{fig:E_vs_dEdx}). The potential variation in gamma rejection with gamma energy when comparing the rejection factors for $^{133}$Ba and $^{60}$Co (Table \ref{tab:gamma_rejection}) is notable. \cite{peyrou-67} has wondered if the dominant contribution to gamma sensitivity in a bubble chamber arises from sensitivity to Auger electrons following photoelectric interactions. A lower energy gamma source (\emph{e.g.} $^{133}$Ba) would be dominated by photoelectric scattering, while a higher energy source (\emph{e.g.} $^{60}$Co) would have a larger contribution from Compton scattering. The data in Table \ref{tab:gamma_rejection} seem to contradict the hypothesis of Auger electrons leading to gamma sensitivity, as the rejection is in fact \emph{better} for the lower energy source.

\section{Gamma Flux Measurement}
\label{sec:gammas_measurement}

\begin{figure} [t!]
\begin{minipage}[b]{0.5\linewidth}
\centering
\includegraphics[scale=0.42]{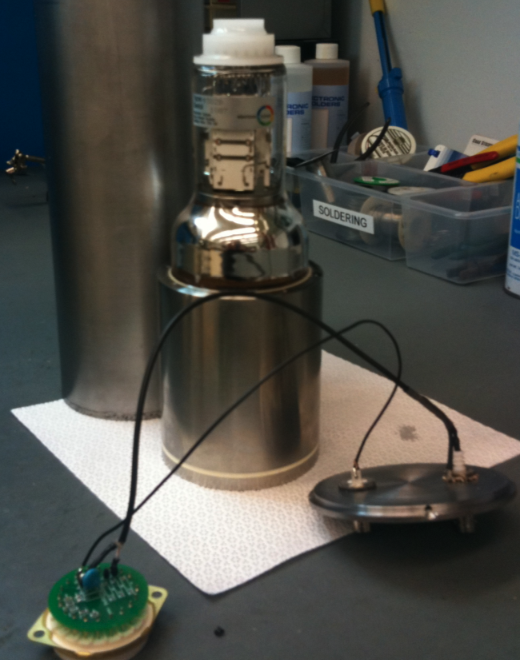}
\end{minipage}
\begin{minipage}[b]{0.5\linewidth}
\centering
\includegraphics[scale=0.513]{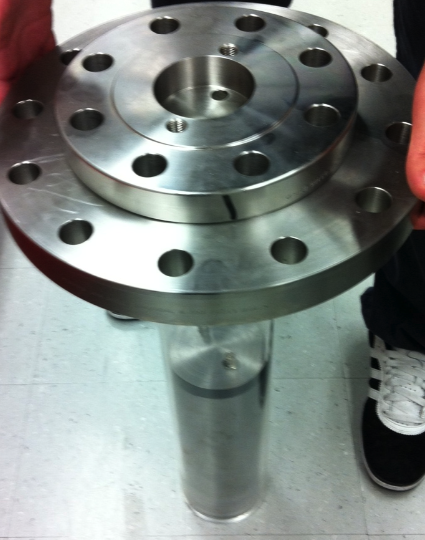}
\end{minipage}
\caption[Thallium-doped NaI detector used for gamma flux measurement]{The NaI[Tl] detector with mounted PMT next to the steel light-shielding can (left). The detector-containing steel can sits at the bottom of an acrylic jar held in place below a stainless steel top flange (right), which replaces the top flange attached to the inner volume assembly of the bubble chamber.}
\label{fig:nai_pictures}
\end{figure}

In order to determine the total environmental gamma flux reaching the CF$_3$I volume and what its energy spectrum is, a low-background thallium-doped NaI detector was placed at the approximate location of the CF$_3$I volume. This detector consisted of a 1.78 kg cylindrical NaI[Tl] scintillator crystal grown by Alpha Spectra, Inc. using low-background techniques. The crystal is surrounded by a thin teflon reflector and encased within a steel holder, specifically counted for the absence of $^{60}$Co, with Gore-Tex (a sponge-like form of teflon) and oxygen-free high thermal conductivity (OFHC) copper ancillary parts and a viewport made of Suprasil synthetic fused silica. To record the scintillation light from the NaI[Tl], an ultra-low background PMT manufactured by Electron Tubes Enterprises (model number 9302B) was attached to the Suprasil window using optical room temperature vulcanizing (RTV) silicone rubber (Figure \ref{fig:nai_pictures}).  This same NaI[Tl] crystal and PMT was used as part of an anti-Compton veto for the CoGeNT experiment \citep{aalseth-08}.

\begin{figure} [t!]
\centering
\includegraphics[scale=0.45]{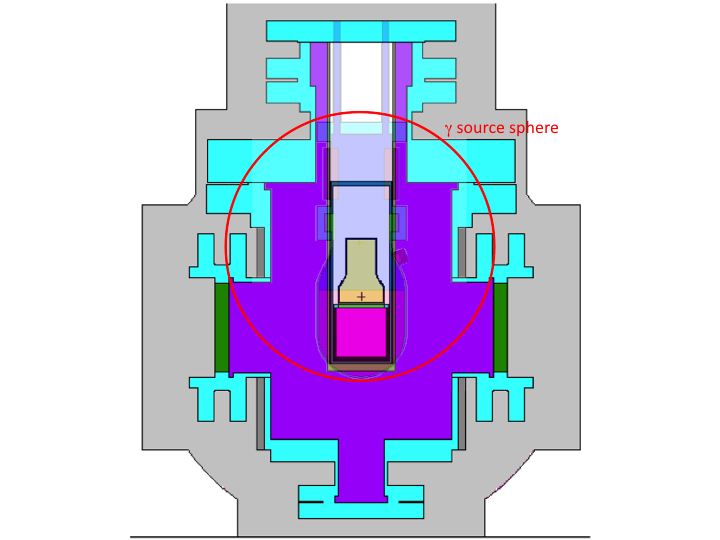}
\caption[Location of NaI detector in pressure vessel]{The location of the NaI detector in the pressure vessel in the MCNP simulations overlaying the position of the CF$_3$I volume that the NaI detector replaced (compare to Figure \ref{fig:mcnp_4kg}). The position of the NaI crystal (pink) is approximately the same as that of the CF$_3$I volume. The sphere used as a gamma source is shown in red.}
\label{fig:mcnp_nai}
\end{figure}

A light-shielding can was fashioned from a 10.16 cm OD, 9.56 cm ID steel pipe, also screened for absence of $^{60}$Co, with an end-cap welded on using non-radioactive lanthanated welding tips and a removable lid with SHV and BNC feedthroughs for connection to the PMT. The NaI detector with attached PMT was enclosed in this can and held stable by OFHC copper rods and a layer of Gore-Tex padding. This entire enclosure was placed within a glycol-tight acrylic bottle, which was itself lowered into the glycol hydraulic fluid bath inside the 4 kg steel pressure vessel, once the inner volume assembly of the bubble chamber had been removed. The position of the NaI[Tl] crystal was almost exactly at the center of where the CF$_3$I volume was located (Figure \ref{fig:mcnp_nai}).  A Spectrum Techniques multi-channel analyzer (MCA) was used as a high voltage supply to bias the PMT and as a data acquisition unit. The PMT was biased to +750 V, and the signal was passed through the preamp input of the MCA, operating in pulse height analysis (PHA) mode.

Although the NaI detector was manufactured using low-background techniques using radioclean materials, there is still an inherent background in the detector coming from the low levels of radioactivity intrinsic to the device. Some of these backgrounds come from the gamma emission from elements along the $^{238}$U and $^{232}$Th decay chains (Figures \ref{fig:U238_decay_chain} and \ref{fig:Th232_decay_chain}) from the detector components, while some of them come from the alpha decays along these chains from internal contamination of the NaI crystal itself \citep{ichihara-03}. To measure these intrinsic backgrounds, the NaI detector, complete with the PMT and light-shielding can, was installed within a lead shielding (the innermost layer being low-background, $\sim$14 Bq/kg in $^{210}$Pb) and run for 1 week at 330 m.w.e. underground in a pumping station of Chicago's Tunnel and Reservoir Plan (TARP). The intrinsic backgrounds measured at TARP below the highest gamma energy in the $^{238}$U or $^{232}$Th chains (2615 keV from $^{208}$Tl, referred to as the thallium cutoff) were two orders of magnitude below those measured at SNOLAB, so no intrinsic background subtraction was required for any measurement described below. Those backgrounds measured above the thallium cutoff were dominated by alphas in the NaI crystal (peaks at $\sim$3.5 MeV and $\sim$5.5 MeV \citep{ichihara-03}) and muons from the cosmic ray flux still prominent at 330 m.w.e.

Initially, a Spectrum Techniques UCS-20 MCA was used as the data acquisition box. However, this particular UCS-20 MCA had a readout error for low-rate ($<$ 8 cts/s) counting --- only $\sim$10\% of counts were being recorded if below the 8 Hz rate. This had no effect for most data, as the count rate was high enough. However, when taking the intrinsic background data or when taking high energy gamma data (Section \ref{sec:gammas_highenergy}), this defect effectively reduced the live-time of the detector by an order of magnitude. The MCA was eventually replaced with a newer UCS-30 model, which eliminated this problem. As a cross-check, the rates from the UCS-30 were compared with those from the UCS-20 for equivalent high-rate measurements, with very good agreement found.

\subsection{NaI Detector Efficiency}
\label{sec:gammas_nai_efficiency}

\begin{figure} [t!]
\centering
\includegraphics[scale=0.55]{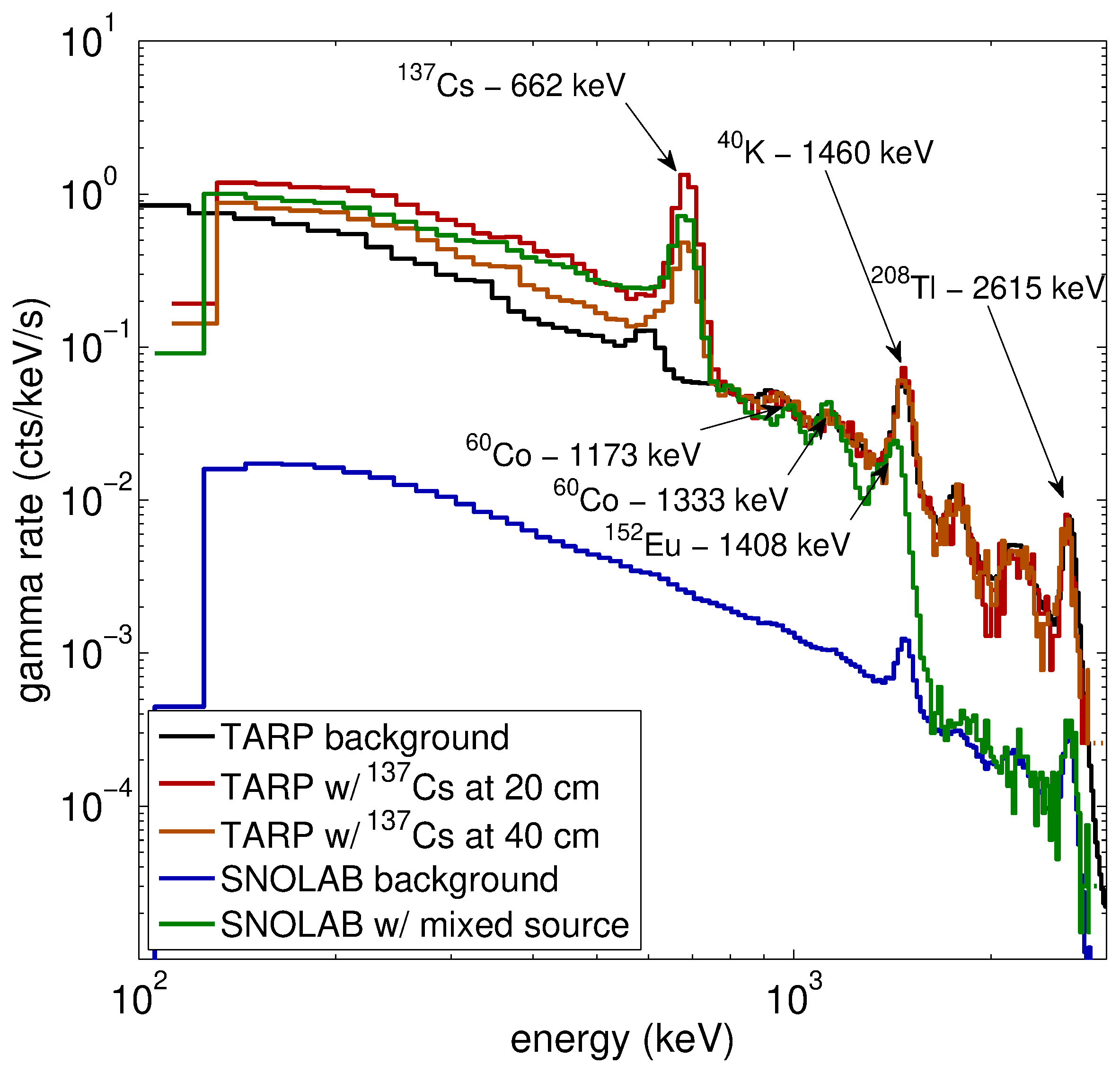}
\caption[Gamma measurements used for NaI efficiency analysis]{The environmental gamma rate measured by the NaI detector at both TARP and SNOLAB, compared with the calibrations using a $\sim$0.84 $\mu$Ci $^{137}$Cs source and a $\sim$5 $\mu$Ci mixed source, respectively.}
\label{fig:nai_efficiency}
\end{figure}

While measuring the intrinsic backgrounds of the NaI detector in TARP in September 2011, gamma calibration runs were also performed. The environmental gamma flux in TARP was measured for 2.0 days with the detector outside of the lead shielding. A $\sim$0.84 $\mu$Ci $^{137}$Cs source (measured as $\sim$1 $\mu$Ci in January 2004) was then placed at 20 cm from the outside diameter of the light-shielding can and then again at 40 cm, where data were taken for 200.6 s and 201.2 s, respectively. Subtracting the environmental gamma rate in energy bins below the $^{137}$Cs gamma energy (661.7 keV) and comparing the observed spectrum with that expected from an MCNP-5 simulation, the detector efficiency was found to be 77.8$\pm$4.7\% and 83.8$\pm$10.2\%, respectively. A similar calibration was performed while the NaI detector was installed in the pressure vessel at SNOLAB. Here, the gamma source mixture described above ($^{22}$Na/$^{60}$Co/$^{137}$Cs/$^{152}$Eu) was positioned 140 cm below the top of the water shielding, where data were recorded for 42.3 min. Using the observed rate from the entire spectrum background subtracted (from a 1.2 day background exposure), the detector efficiency was found to be 90.1$\pm$9.5\%. Figure \ref{fig:nai_efficiency}  shows the measured environmental rate of gammas at both TARP and SNOLAB, along with the comparison calibration measurements. Given that the detector itself has an energy threshold of $\sim$100 keV\footnote{Environmental gammas with energies lower than this do not get through the steel cans, in any case.}, below which a considerable contribution to the overall count rate is expected from the MCNP simulations, it is not surprising that the simulations slightly over-predict the observed count rate. The calibrations performed provide evidence that the detector and DAQ triggering efficiency is therefore consistent with 100\%.

\subsection{Unfolding of the Gamma Spectrum Measurement}
\label{sec:gammas_unfolding}

The background gamma measurement with the NaI detector in the pressure vessel (Figure \ref{fig:nai_efficiency}, blue) provides rate measurements $R_i$ for specific gamma energy bin $i$ (with $n$ total energy bins). These rate measurements are indicative of the actual gamma flux $\Phi_i$ in the environment if a detector response matrix $\mathbb{R}$ (size $n \times n$) is incorporated:

\begin{equation}
\label{eq:unfolding}
R = \mathbb{R} \Phi.
\end{equation}

\noindent This response matrix is determined via MCNP-5 simulations in the following manner. The gamma source for the simulation is defined on the smallest sphere that encompasses both the bubble chamber inner volume and the NaI detector setup (Figure \ref{fig:mcnp_nai}, red circle). From that sphere, gammas are initiated in an inward direction (any isotropic orientation that is initially towards the inside of the sphere) that are entirely composed of energies uniformly sampled within one specific energy bin $i$ at a time. This is repeated $n$ times, for all energy bins. For each simulation, the pulse height tally (tally 8 in MCNP-5) in the NaI cell of the geometry (Figure \ref{fig:mcnp_nai}, pink) is recorded for all energy bins (this tally has length $n$). This gives the user a count rate observed by the NaI detector in each energy bin, for incoming gammas initiated with energy in bin $i$. Column $i$ in the response matrix $\mathbb{R}$ is then equal to the tally result generated by gammas with energy in bin $i$.

\begin{figure} [t!]
\centering
\includegraphics[scale=0.55]{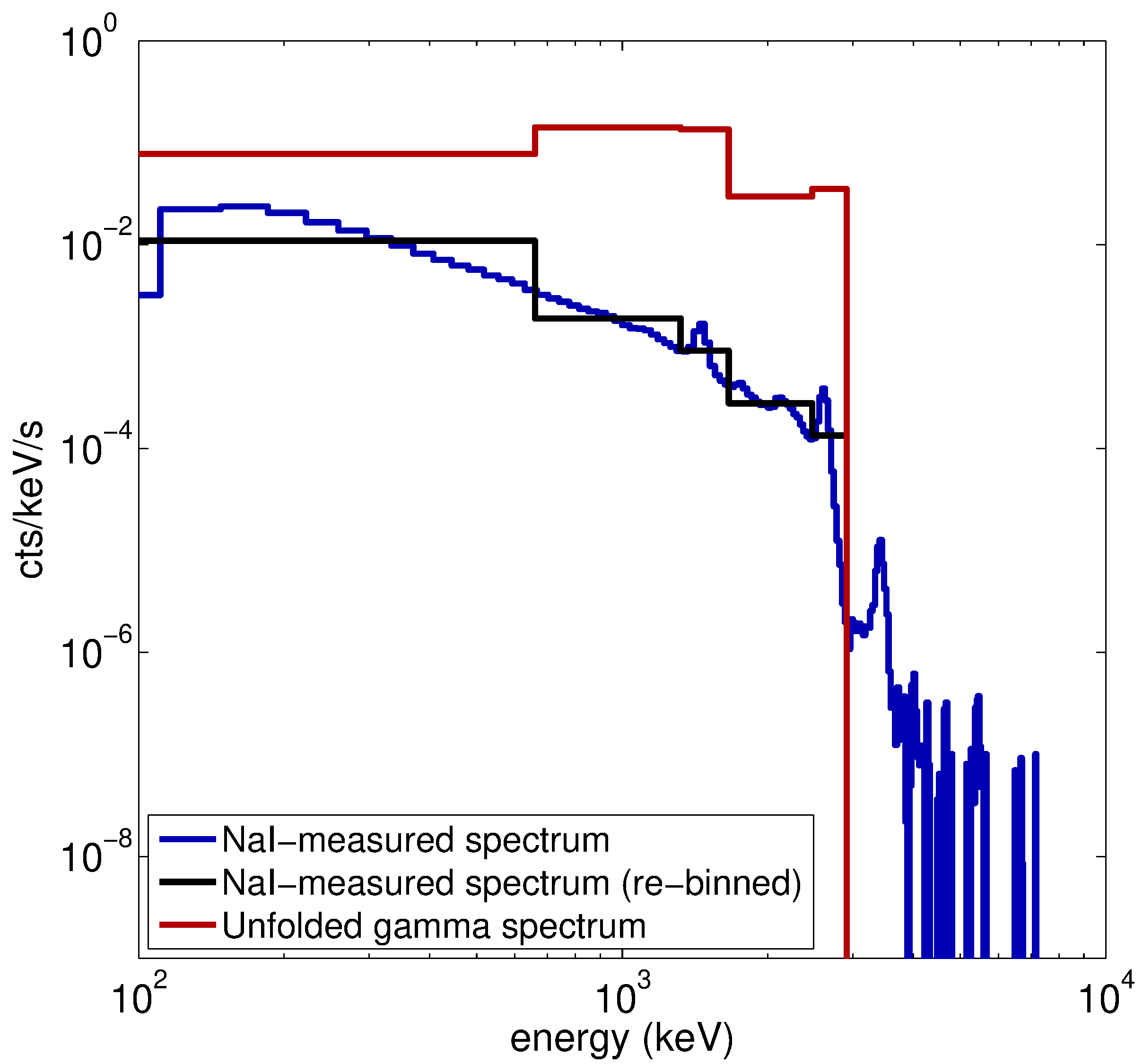}
\caption[Gamma spectrum measured at location of CF$_3$I volume]{The gamma rate and spectrum measured by the NaI detector at the location of the CF$_3$I volume (blue). The spectrum was reshaped into the chosen energy bins for unfolding (black). The calculated unfolded spectrum $\Phi$ is shown in red. Counts above the thallium cutoff (2615 keV) are taken to be NaI-internal alphas.}
\label{fig:gamma_spectrum}
\end{figure}

For this analysis, energy bins were chosen with the following lower limits: 100 keV, 660 keV, 1320 keV, 1660 keV, 2470 keV, and 2910 keV (Figure \ref{fig:gamma_spectrum}), and upper limit of 10 MeV.  From the MCNP-5 simulations with this choice of energy bins, the detector response matrix is found to be

\begin{equation}
\label{eq:response_matrix}
\mathbb{R} = \begin{bmatrix}
2.36 & 2.80 & 2.85 & 2.94 & 3.03 & 3.45 \\
0 & 0.815 & 0.673 & 0.508 & 0.375 & 0.226 \\
0 & 0 & 0.482 & 0.256 & 0.168 & 0.0697 \\
0 & 0 & 0 & 0.593 & 0.551 & 0.144 \\
0 & 0 & 0 & 0 & 0.382 & 0.0911 \\
0 & 0 & 0 & 0 & 0 & 1.23 \\
\end{bmatrix} \times 10^{-2}.
\end{equation}

\noindent Because the response matrix is diagonal, and therefore invertible, the solution to Equation \ref{eq:unfolding} is exactly solvable. This is not often the case with such measurements --- typically a complicated unfolding procedure must be undertaken by fitting the observed spectrum with the predicted one by a least-squares non-negative fit or similar procedure \citep{doroshenko-77}. In this case, solving for $\Phi$ simply requires left multiplying Equation \ref{eq:unfolding} by the response matrix $\mathbb{R}$, which yields

\begin{equation}
\label{eq:gamma_spectrum}
\Phi =  \begin{pmatrix}
68.44 \\
147.2 \\
72.54 \\
37.90 \\
24.46 \\
0 \\
\end{pmatrix} \gamma / \mathrm{m}^2 / \mathrm{s},
\end{equation}

\noindent for the flux of environmental gammas reaching the CF$_3$I volume in the energy bins defined above.

\subsection{Gamma-Induced Event Rates}
\label{sec:gamma_events}

The gamma spectrum $\Phi$ calculated here matches expectations: there are contributions mostly in the energy bins containing the $^{40}$K 1460 keV line (there is a significant amount of $^{40}$K at SNOLAB and in the borosilicate glass viewports) and the bin below this (bins 2 and 3), the latter of which contains many of the gammas in the $^{238}$U and $^{232}$Th chains (see Figures \ref{fig:U238_decay_chain} and \ref{fig:Th232_decay_chain}). The peak associated with the thallium cutoff at 2615 keV is also reproduced. Using the spectrum $\Phi$ as a source of gammas originating from the same source sphere used for determining the response matrix $\mathbb{R}$ (Figure \ref{fig:mcnp_nai}, red), an MNCP-PoliMi gamma interaction rate prediction was generated for this spectrum of environmental gammas incident on the CF$_3$I volume in the bubble chamber. Using the values for the rejection factor calculated in Equation \ref{eq:rejection_factor}, the expected bubble nucleation rate and predicted number of counts for each dark matter data set originating from the environmental gamma background are shown in Table \ref{tab:gamma_events}. Figure \ref{fig:summary} shows the predicted gamma event rate with the COUPP 4 kg bubble chamber at different operating recoil thresholds. At recoil thresholds that are typically used, limitations arising from a gamma background are currently far below those from most neutron backgrounds, and so the gamma background will not be a concern for this chamber.

\begin{table}[t!]
\centering
\small{
\begin{tabular} {| c | c || c | c |}
\hline
\multicolumn{4}{| l |}{Environmental Gammas} \\
\hline
Data Set & Rejection & Rate & Counts in Set \\
 & & \small{(cts/kg/day)} & \small{(cts)} \\
\hline
DM-34$^\circ$C & $\sim$$1.8\times10^{-12}$ & $\left(6.14^{+0.64}_{-0.64}\right)\times10^{-7}$ & $\left(1.62^{+0.18}_{-0.18}\right)\times10^{-4}$ \\
DM-37$^\circ$C & $\sim$$3.5\times10^{-10}$ & $\left(1.21^{+0.13}_{-0.13}\right)\times10^{-4}$ & $\left(4.19^{+0.45}_{-0.46}\right)\times10^{-5}$ \\
DM-40$^\circ$C & $\sim$$1.4\times10^{-8}$ & $\left(4.74^{+0.50}_{-0.50}\right)\times10^{-3}$ & $\left(3.72^{+0.40}_{-0.40}\right)\times10^{-5}$ \\
\hline
\end{tabular} }
\caption[Environmental gamma background rate prediction]{The expected single-bubble event rates and raw number of events expected from environmental gammas at SNOLAB, derived from a measurement with a NaI detector in position at the CF$_3$I location, for the DM-34$^\circ$C, DM-37$^\circ$C, and DM-40$^\circ$C data sets.}
\label{tab:gamma_events}
\end{table}

\section{High Energy Gammas}
\label{sec:gammas_highenergy}

A measurable very high energy ($\gtrsim 10$ MeV) gamma flux incident on the bubble chamber could initiate photonuclear ($\gamma$,$x$) interactions ($x$ denoting any number of disintegration products) in the CF$_3$I leading to a nuclear recoil background which has not to this point been considered. While the sources that can be listed for such gammas are scarce, this background would be a significant hurdle for the bubble chamber due to the unacceptable cost and effort require to shield against it, and so a determination of the rate of photonuclear events in the CF$_3$I must be made. To measure the high energy (10 MeV $< E_\gamma < $ 50 MeV) gamma flux in the J-Drift at SNOLAB, the NaI detector described above was used with a widened energy range for detection --- accomplished by lowering the bias on the PMT (or equivalently by attenuating the signal from the PMT). Given that high energy gammas are not substantially moderated by the water shield or the detector materials themselves, this measurement was not required to be made with the detector in place within the pressure vessel.

\begin{figure} [t!]
\centering
\includegraphics[scale=0.55]{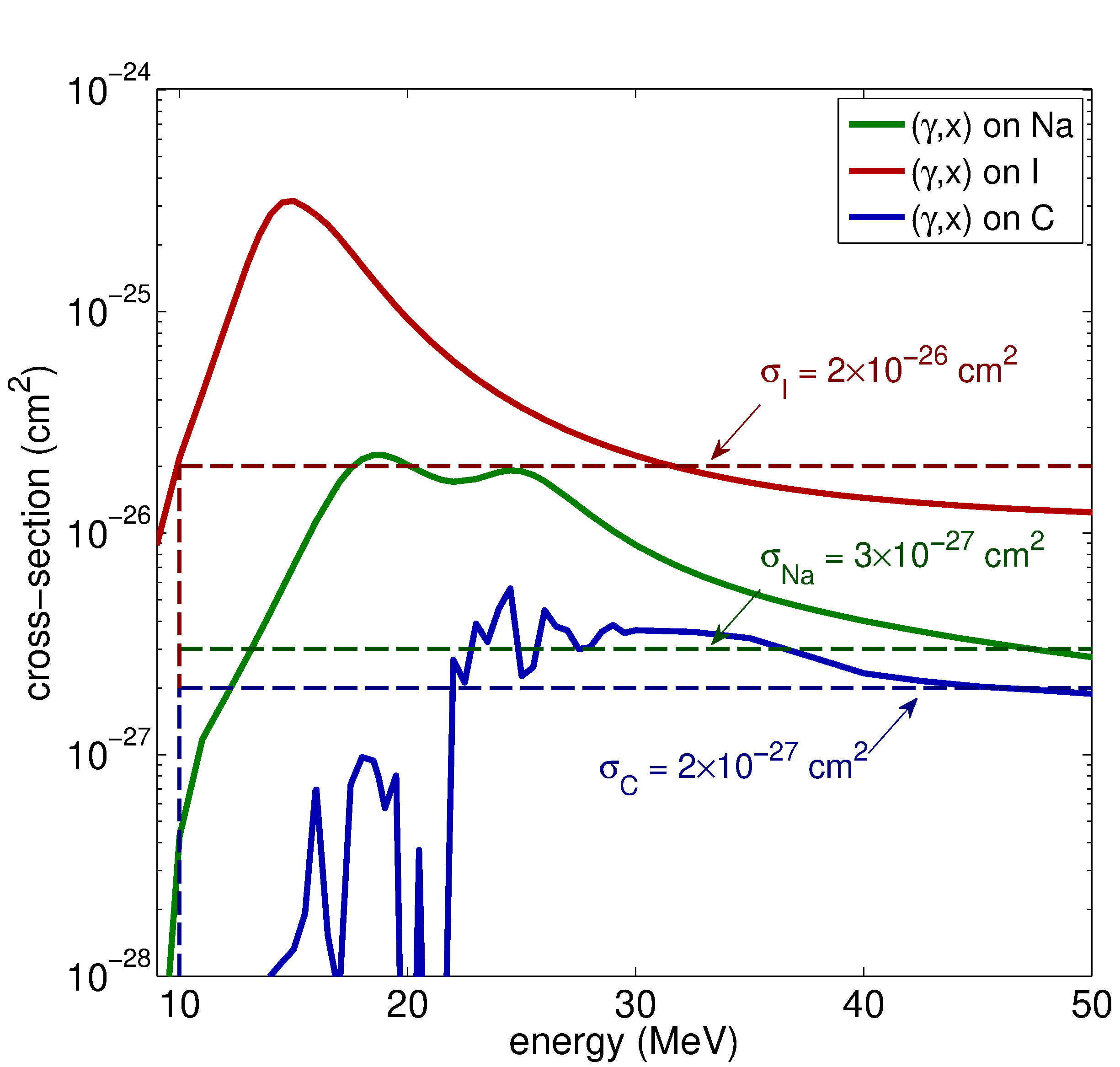}
\caption[($\gamma$,$x$) cross-sections on Na and I]{The ($\gamma$,$x$) cross sections on Na and I \citep{chadwick-11}. For simplicity, a flat cross-section for each is used, shown as the dashed line.}
\label{fig:gx_on_nai}
\end{figure}

The flux of high energy gammas $\Phi_\mathrm{h.e.}$ (in $\gamma$/cm$^2$/day) can be approximated from the NaI measurements by

\begin{equation}
\label{eq:Phi_he_gamma}
\Phi_\mathrm{h.e.} \simeq \frac{R}{n_\mathrm{Na} \sigma_\mathrm{Na} + n_\mathrm{I} \sigma_\mathrm{I}},
\end{equation}

\noindent where $R$ is the measured rate of gammas above the ($\gamma$,$x$) threshold on NaI (in cts/s/cm$^3$), $n_i$ is the number density of material $i$ in the detector, and $\sigma_i$ is the ($\gamma$,$x$) cross section. For the present calculation, the cross sections are approximated as step functions, with conservative values chosen for the magnitudes, and a ($\gamma$,$x$) threshold that is $\sim$10 MeV for all nuclei (Figure \ref{fig:gx_on_nai}). 

NaI has a density of 3.67 g/cm$^3$, and so the number densities are $n_\mathrm{Na,I} = 1.47\times10^{22}$ atoms/cm$^3$, and the cross sections are as seen in Figure \ref{fig:gx_on_nai}. For $t$ days of data with no observed counts above 10 MeV (with a 90\% confidence level upper-limit of 2.44 \citep{feldman-98}) using the 485 cm$^3$ detector, a limit is set on the high energy gamma flux of $\Phi_\mathrm{h.e.} \lesssim 14.88 t^{-1}$ $\gamma$/cm$^2$/day. Applying this flux limit to the CF$_3$I (with density 1.99 g/cm$^3$), using the cross sections shown in Figure \ref{fig:gx_on_nai} (and assuming a similar cross section for F as for Na), the limit on the count rate in CF$_3$I is then

\begin{equation}
\label{eq:R_he_CF3I}
R \lesssim \frac{1.42}{t} \: \mathrm{cts}/\mathrm{kg}/\mathrm{day}.
\end{equation}

\noindent At the time of this writing, 35 days of data has been taken with the NaI detector with no observed gammas above 10 MeV. Therefore, the limit on the photonuclear background is $R \lesssim 0.04$ cts/kg/day. As more data is collected with no observed counts, this limit will be improved upon.

\singlespacing
\chapter{Dark Matter Limits from the 4 kg Bubble Chamber}
\label{ch:limits}
\doublespacing

This thesis reports the WIMP direct detection results from the COUPP 4 kg bubble chamber operated from September 2010 to August 2011 in the J-Drift of the SNOLAB underground laboratory, at a depth of 6000 m.w.e. The bubble chamber employs superheated CF$_3$I as a WIMP detector, which as a target fluid gives the ability to set limits on both the spin-independent WIMP-nucleon interactions (due to the high mass of target $^{127}$I and, in some part, to $^{19}$F as a low-mass WIMP target, see Figure \ref{fig:SI_WIMP_recoil_CF3I}) as well as the spin-dependent WIMP-proton interactions (due to the abundance of the odd-proton target $^{19}$F, see Figure \ref{fig:SD_WIMP_recoil_CF3I}). The dual threshold nature of the bubble chamber makes it an ideal technology for low-signal searches because the prominent gamma background that plagues many such searches is eliminated by imposing a stopping power threshold (Chapter \ref{ch:gammas}). The next most significant background, events generated by alpha decays, can also be rejected with use of acoustic discrimination (Section \ref{sec:datasets_alpha}). What remains is to understand and eliminate the low levels of background generated by neutrons (Chapter \ref{ch:backgrounds}). With proper preparation of a bubble chamber operated at a deep underground site (to reduce cosmogenic neutron backgrounds) with a significant low-$Z$ shielding (to reduce rock-generated neutron backgrounds), and with extensive screening of the radioactive contamination of detector components, it is quite feasible that a zero-background experimental search for WIMP dark matter can be performed.

The experimental results reported here did, however, have an identifiable neutron background, almost exclusively from the borosilicate glass viewports in the pressure vessel (Section \ref{sec:backgrounds_windows}) and the piezoelectric transducers necessary for acoustic alpha discrimination (Section \ref{sec:backgrounds_piezos}). Proper screening of replacement materials for these parts in the next data run of the 4 kg bubble chamber has been carried out (see Appendix \ref{ch:piezos}), and data-taking is set to begin in March 2012 with this chamber at SNOLAB. The current data consists of 3 data sets collected at 3 bubble nucleation thresholds --- $8.00^{+0.67}_{-0.61}$, $11.21^{+0.97}_{-0.87}$, and $15.83^{+1.46}_{-1.31}$ keV --- for exposures of 70.98, 88.41, and 393.64 kg-days, respectively. As summarized in Table \ref{tab:counts}, this analysis resulted in a total of 20 recoil-like events in these data sets combined, with 3 multi-bubble events --- the latter being an incontrovertible signature of a neutron background in the recoil-like events (only neutrons have the ability to multiple-scatter in the detector volume). Before using these recoil-like events to set dark matter limits with this data, they need to be analyzed is closer detail. Specific information for each of the 20 recoil-like events, as well as several other events of interest, is shown in Table \ref{tab:interesting_events}.

\begin{table}[p!]
\centering
\small{
\begin{tabular} {| c | c | c | c | c | c | c | c | c |}
\hline
\multicolumn{9}{| l |}{DM-34$^\circ$C Data} \\
\hline
\multicolumn{9}{| c |}{Recoil-Like Events} \\
\hline
Run & Event & Date & Time & $T$ ($^\circ$C) & $P$ (psia) & $AP$ & $\Delta t$ (min) & Details \\
\hline
20110320\_0 & 135 & 03/20/11 & 21:22:21 & 33.47 & 30.43 & 1.13 & 14.7 & \\
20110420\_0 & 146 & 04/21/11 & 14:15:09 & 33.48 & 30.61 & 0.95 & 14.9 & \\
20110421\_0 & 0 & 04/21/11 & 22:17:06 & 33.47 & 30.74 & 1.14 & 89.8 & \\
20110424\_0 & 51 & 04/24/11 & 11:56:28 & 33.47 & 30.67 & 0.99 & 54.0 & \\
20110508\_0 & 107 & 05/09/11 & 00:03:49 & 33.48 & 30.63 & 0.94 & 13.3 & \\
20110523\_0 & 168 & 05/24/11 & 21:59:01 & 33.49 & 30.80 & 1.04 & 191.0 & \\
20110526\_0 & 11 & 05/26/11 & 08:06:04 & 33.48 & 30.75 & 0.98 & 15.6 & \\
20110604\_0 & 107 & 06/04/11 & 23:39:30 & 33.48 & 30.61 & 0.97 & 40.0 & \\
\hline
\multicolumn{9}{| c |}{Multi-Bubble Events} \\
\hline
20110512\_0 & 105 & 05/12/11 & 23:24:26 & 33.49 & 30.61 & 2.18 & 205.2 & $N$=2 \\
\hline
\hline
\multicolumn{9}{| l |}{DM-37$^\circ$C Data} \\
\hline
\multicolumn{9}{| c |}{Recoil-Like Events} \\
\hline
Run & Event & Date & Time & $T$ ($^\circ$C) & $P$ (psia) & $AP$ & $\Delta t$ (min) & Details \\
\hline
20101214\_0 & 171 & 12/14/10 & 20:29:16 & 36.23 & 30.60 & 1.18 & 40.4 & \\
20101226\_0 & 29 & 12/26/10 & 09:49:53 & 36.25 & 30.97 & 0.90 & 15.4 & \\
20101228\_0 & 160 & 12/29/10 & 08:48:48 & 36.25 & 30.85 & 1.26 & 1.6 & \\
20101228\_0 & 161 & 12/29/10 & 08:51:06 & 36.25 & 30.64 & 1.25 & 2.3 & \\
20101229\_0 & 189 & 12/30/10 & 10:28:20 & 36.24 & 30.88 & 0.88 & 6.6 & \\
20110103\_0 & 72 & 01/03/11 & 09:34:45 & 36.24 & 30.81 & 1.15 & 37.6 & \\
\hline
\multicolumn{9}{| c |}{Multi-Bubble Events} \\
\hline
20101211\_0 & 55 & 12/11/10 & 22:38:52 & 36.24 & 30.77 & 3.38 & 10.5 & $N$=3 \\
20101230\_0 & 175 & 12/31/10 & 10:44:44 & 36.24 & 30.49 & 3.92 & 10.7 & $N$=3 \\
\hline
\multicolumn{9}{| c |}{Other Interesting Events} \\
\hline
20101228\_0 & 73 & 12/28/10 & 21:52:54 & 36.23 & 30.62 & 1.30 & 62.1 & cut by $AP$ \\
20101231\_0 & 125 & 01/01/11 & 09:21:04 & 36.24 & 30.47 & 1.32 & 7.8 & cut by $AP$ \\
\hline
\hline
\multicolumn{9}{| l |}{DM-40$^\circ$C Data} \\
\hline
\multicolumn{9}{| c |}{Recoil-Like Events} \\
\hline
Run & Event & Date & Time & $T$ ($^\circ$C) & $P$ (psia) & $AP$ & $\Delta t$ (min) & Details \\
\hline
20101112\_0 & 92 & 11/12/10 & 23:46:04 & 39.01 & 31.13 & 0.97 & 2.1 & \\
20101112\_0 & 118 & 11/13/10 & 02:12:43 & 38.99 & 31.05 & 1.22 & 2.7 & \\
20101112\_0 & 154 & 11/13/10 & 05:57:50 & 39.01 & 30.97 & 1.03 & 1.5 & \\
20101126\_0 & 56 & 11/27/10 & 01:30:51 & 39.02 & 30.34 & 1.06 & 107.1 & \\
20101130\_0 & 157 & 12/01/10 & 02:42:07 & 39.04 & 30.51 & 0.97 & 8.4 & \\
20101201\_0 & 88 & 12/02/10 & 02:43:59 & 39.05 & 30.47 & 0.97 & 37.6 & \\
\hline
\multicolumn{9}{| c |}{Other Interesting Events} \\
\hline
20101106\_0 & 169 & 11/07/10 & 10:08:22 & 39.03 & 30.99 & 1.35 & 6.2 & cut by $AP$ \\
20101106\_0 & 196 & 11/07/10 & 12:57:55 & 39.02 & 30.91 & 1.36 & 6.6 & cut by $AP$ \\
20101112\_0 & 94 & 11/12/10 & 23:49:26 & 38.99 & 30.98 & 1.47 & 3.4 & cut by $AP$ \\
20101128\_0 & 82 & 11/28/10 & 18:37:02 & 39.02 & 30.73 & 1.30 & 1.5 & cut by $AP$ \\
\hline
\end{tabular} }
\caption[Recoil-like events in the dark matter search data]{Details of events of interest in the dark matter search data.}
\label{tab:interesting_events}
\end{table}

\section{Relaxed Recoil-Like $AP$ Cut}
\label{sec:limits_relaxedAP}

\begin{figure} [t!]
\centering
\includegraphics[scale=0.55]{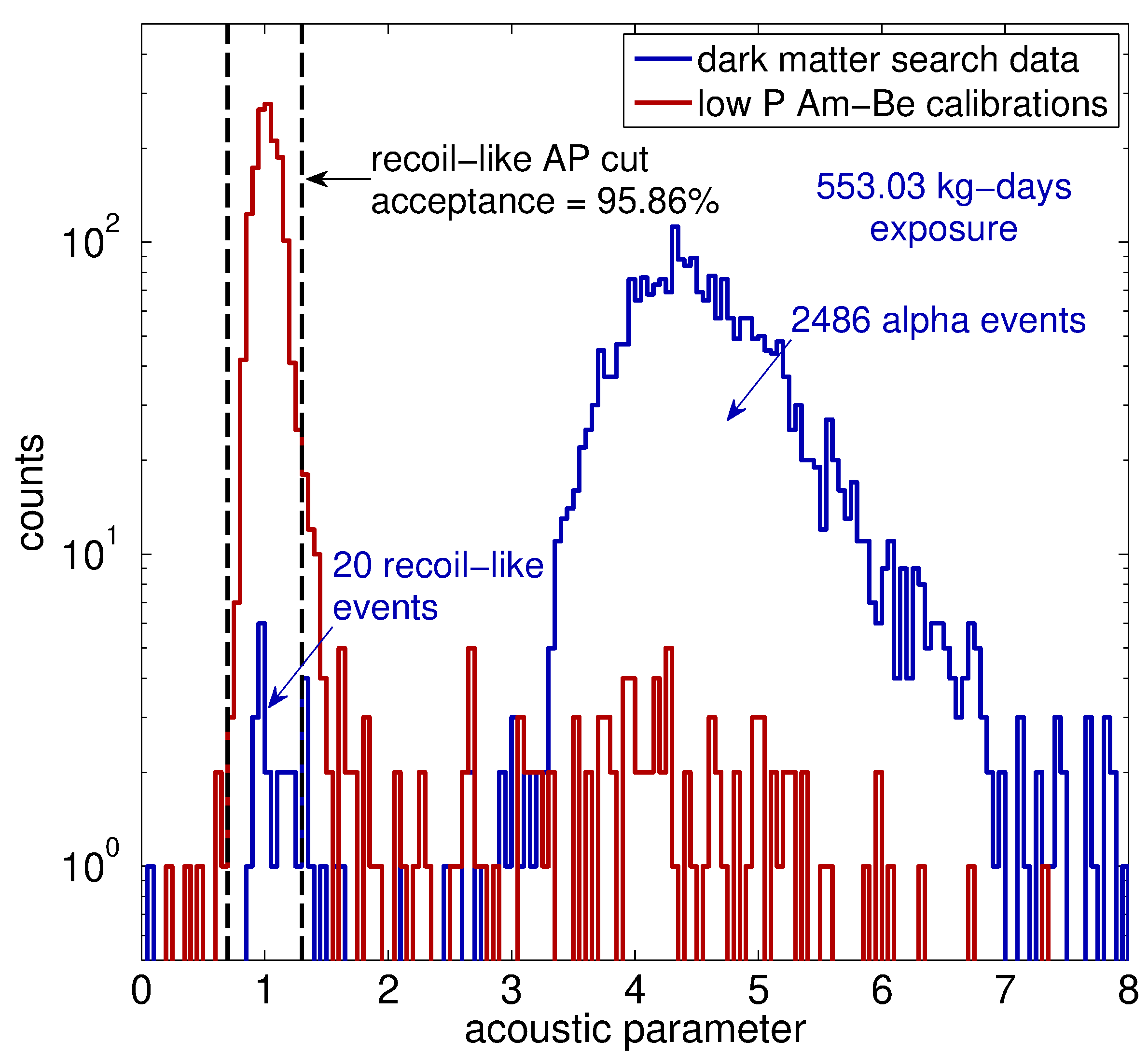}
\caption[$AP$ (log) of dark matter search and neutron calibration data]{The distribution of $AP$ in the alpha-dominated dark matter search data compared to the low-pressure Am-Be neutron calibration data. Problematic high-$AP$ events in the dark matter search data are evident just above the upper limit of the applied $AP$ cut.}
\label{fig:AP_histogram}
\end{figure}

In the bubble chamber, the piezos are used to record the ultrasonic acoustic emission of bubble formation, information that can be used to discriminate recoil-like events (those generated from single neutrons and, presumably, WIMPs) from alpha decay events. An acoustic parameter ($AP$) is defined from the frequency-weighted acoustic power density integral as in Equation \ref{eq:AP}. Comparing the $AP$ histogram of events from neutron calibration runs to that from the alpha-dominated dark matter search data (Figure \ref{fig:AP_histogram})  illustrates the power of using the $AP$ to discriminate neutrons from alphas. The $AP$ boundary set in the recoil-like $AP$ cut defined in Section \ref{sec:datasets_recoilAP} is $0.7 \leq AP \leq 1.3$. This limit has been fixed since the engineering run at the MINOS near detector hall at Fermilab \citep{behnke-11}, and was based on the distribution of neutron event from similar Am-Be calibrations performed there. By relaxing the limits on the $AP$ cut, very little neutron acceptance is gained, and so these limit values seem well-justified. For the events in the dark matter search data at SNOLAB, however, there was a significant population of events that were rejected from the recoil-like WIMP candidate set that would have been included by a very small increase to the upper limit of the $AP$ cut. The question then arises: should these high-$AP$ events be included in the WIMP candidate set?

It is expected that all WIMP recoils in the detector are indistinguishable from nuclear recoils --- the recoiling nucleus is indifferent to the identity of the particle scattering off of it, and the superheated fluid impartial to the nuclear species of the recoil so long as it is above threshold. This being the case, the population of recoil-like WIMP candidate events should closely follow the distribution peaked around $AP=1$ in the neutron calibration data. The slight increase in population above $AP=1.3$ does not match that expected from the neutron simulations, which suggests an origin not related to a recoil-like background.

\section{$\Delta t$ Cut}
\label{sec:limits_delta_t}

\begin{figure} [t!]
\centering
\includegraphics[scale=0.55]{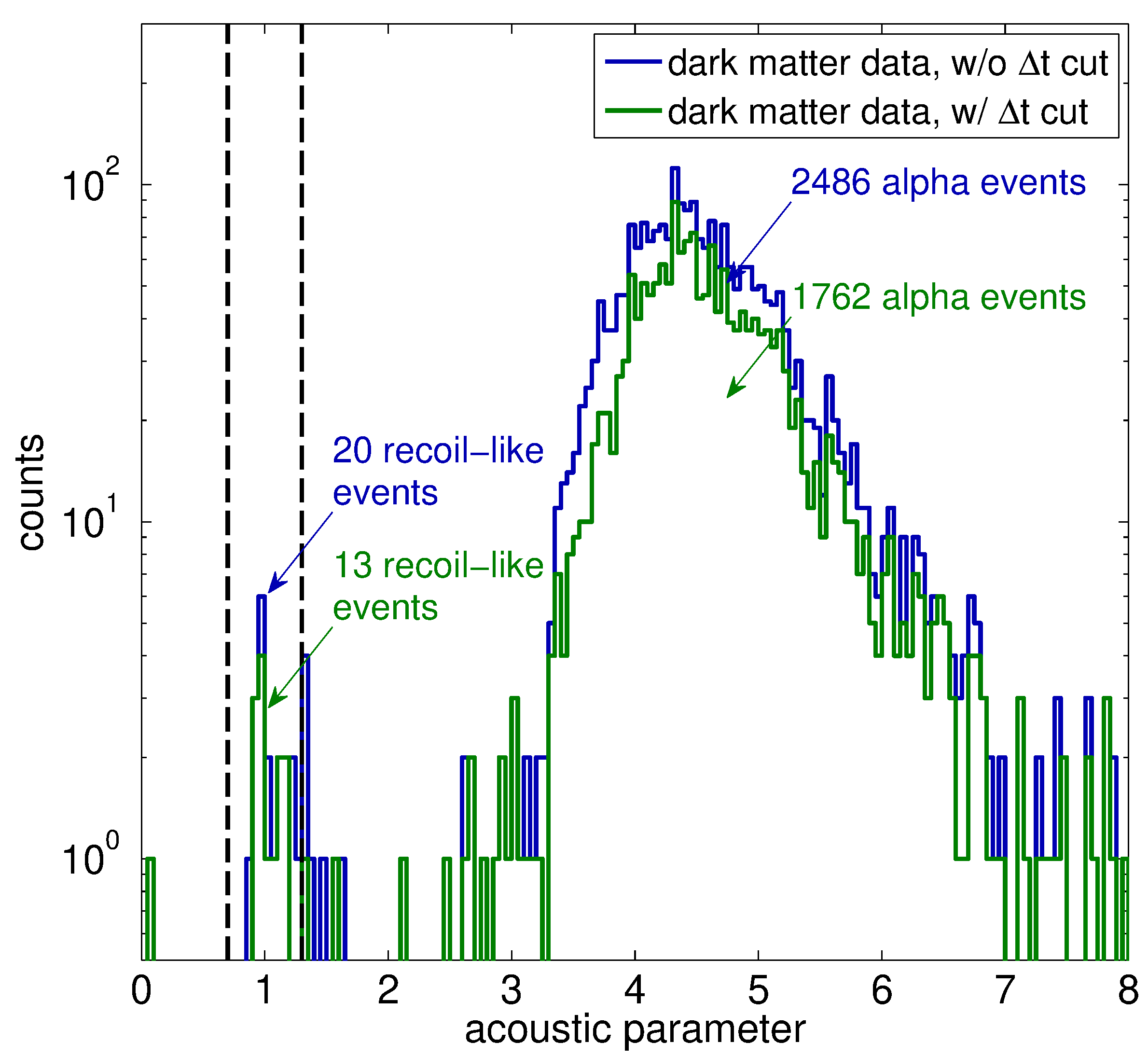}
\caption[$AP$ (log) of dark matter search data with and without $\Delta t$ cut]{The dark matter search data before and after applying a $\Delta t$ cut requiring that each bubble is preceded by a timeout-trigger. The un-recoil-like distribution around $AP=1$ is eliminated, suggesting problematic events comprise the high $AP$ data.}
\label{fig:AP_histogram_delta_t}
\end{figure}

Closely analyzing the events of interest in the dark matter search data found in Table \ref{tab:interesting_events}, one becomes aware of an evident time-clustering of recoil-like events that is completely adverse to being characteristic of a WIMP- or neutron-like\footnote{The possibilities of beta-delayed neutron emission following spontaneous fission in detector materials was examined and found insufficient to explain the time-clustering of events.} signal. In fact, by looking at the time between a particular event and the previous bubble-containing event $\Delta t$, it is apparent that several of the recoil-like events occurred immediately after (\emph{i.e.} as the following event to) another bubble event. Requiring that each event in the dark matter search data is preceded by a timeout-trigger (which occurs after a 500 s uneventful expansion), a cut on $\Delta t$ can be established which requires $\Delta t > 530$ s $= 8.83$ min (the additional 30 s is from the minimum expansion time requirement, Section \ref{sec:datasets_expansion}). Imposing this cut on the dark matter search data (Figure \ref{fig:AP_histogram_delta_t})  leaves the DM-34$^\circ$C data as is, reduces the recoil-like events in the DM-37$^\circ$C data from 6 to 3 (also eliminating 1 of 2 high $AP$ events), and reduces the recoil-like events in the DM-40$^\circ$C from 6 to 2 (eliminating all 4 high $AP$ events). However, it also reduces the effective live time by about 20\% \citep{lippincott-11b}.

The physics behind the time-clustering of events is unknown at present, but its importance is quite apparent. With a $\Delta t$ cut, the unexpected distribution from high $AP$ ($>$ 1.3) recoils is eliminated. Also eliminated are time-clustered events in the dark matter search data that do not represent a WIMP-like signal. As noted in \cite{lippincott-11a}, of the 8 recoil-like events in the DM-40$^\circ$C data that are cut by the $\Delta t$ cut, 7 of them are preceded by an event at the wall-surface interface (that is, on the ring around the top of the CF$_3$I volume, adjacent to the bell jar) --- while only 6\% of all total events occur on this ring. No hypothesis for the cause of the time-clustering events is available at this time, but it is notable that the phenomenon only seems to occur at low thresholds (high temperatures).

\section{COUPP 4 kg Data Summary}
\label{sec:limits_summary}

\begin{figure} [p!]
\centering
\includegraphics[scale=0.73]{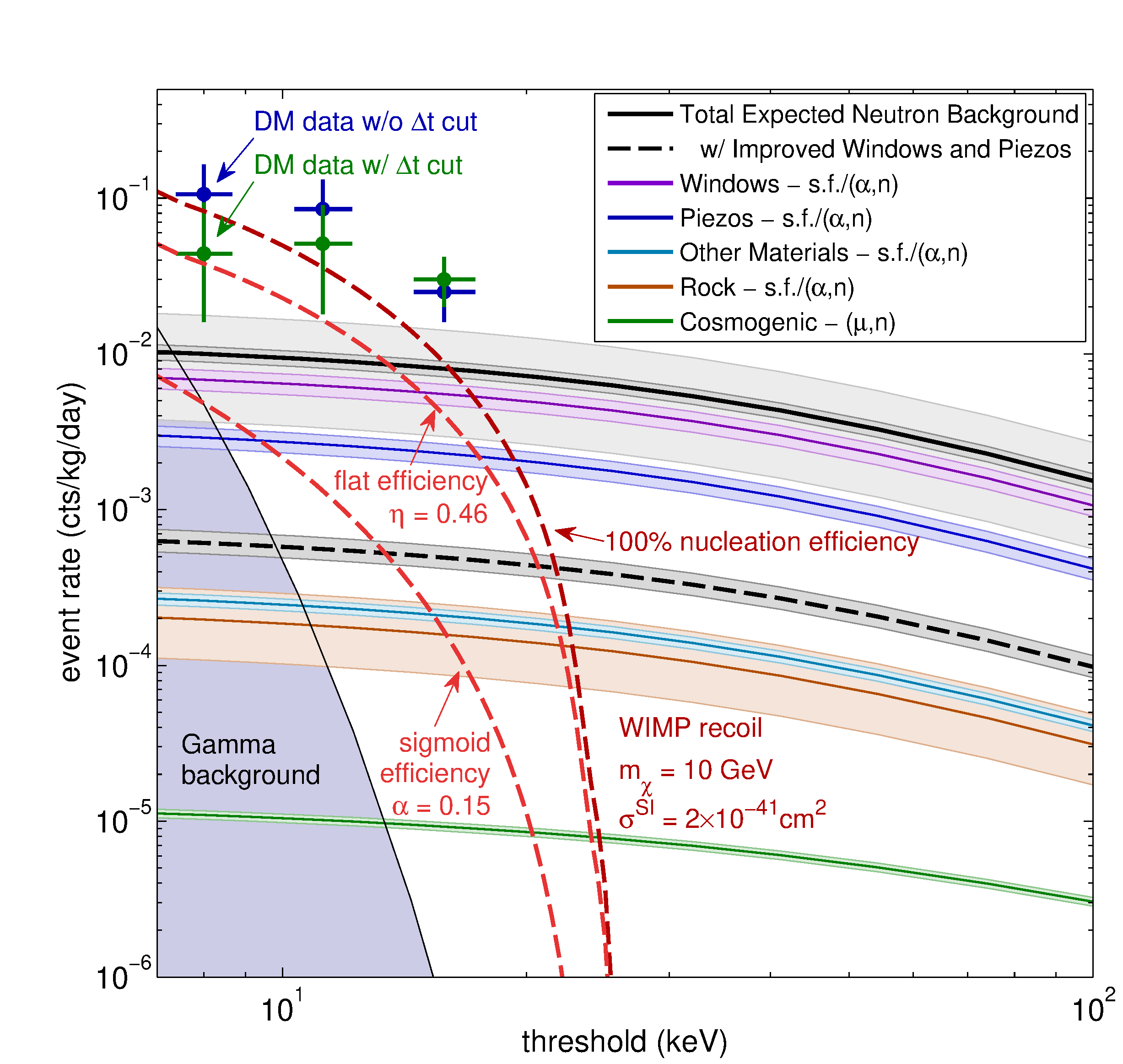}
\caption[Comprehensive view of dark matter search data as they relate to expected backgrounds]{The WIMP direct detection results from the first dark matter search with the COUPP 4 kg bubble chamber at SNOLAB. Data were collected at 3 recoil thresholds, where 20 excess single-bubble events were observed (13 if a cut on the time between events is applied, see Section \ref{sec:limits_delta_t}). The expected neutron backgrounds are shown with 1$\sigma$ error bars, primarily from Monte Carlo uncertainties. The wide (light gray) band around the total expected neutron background prediction is generated from scaling the observed multi-bubble rate to an expected single-bubble rate (see text). The 1$\sigma$ upper limit gamma background prediction is based on a measurement of the gamma flux at the location of the CF$_3$I volume (Section \ref{sec:gammas_measurement}) and the gamma rejection factor determined from gamma source calibrations (Section \ref{sec:gammas_calibrations}). Predictions incorporate a data cut efficiency (Chapter \ref{ch:datasets}) and bubble nucleation efficiency (Chapter \ref{ch:efficiency}). The single-bubble event rate expected from a representative WIMP candidate fit to the most recent DAMA, CoGeNT, and CRESST results \citep{hooper-11} is shown under different nucleation efficiency models (Section \ref{sec:efficiency_fits}).}
\label{fig:summary}
\end{figure}

For the results from this WIMP direct detection experiment, nearly every conceivable avenue has been explored to understand the 20 (or 13, employing the ${\Delta}t$ cut) dark matter candidate events observed. These efforts are summarized in Figure \ref{fig:summary}. All background characterization attempted relies heavily on Monte Carlo techniques to provide event rate predictions for the bubble chamber. This being the case, the uncertainties expected in any of these predictions is quite large. In an attempt to acknowledge these uncertainties, large errors of approximately 10\% were assigned to all MCNP predictions (see Section \ref{sec:mcnp_uncertainty}). An additional 18\% uncertainty was added to any prediction requiring the use of SOURCES-4C \citep{charlton-98}. As these add in quadrature, the resulting total uncertainty to any background prediction can fail to fully express the true possible bias. Background predictions in Figure \ref{fig:summary} should therefore be taken as indicative --- the total observed rates may well be explained by these, and only future runs following upgrades will be able to illuminate their origin.

An independent projection for the total single-bubble neutron background rate in the chamber is calculable by scaling the number of multi-bubble events observed in the detector in all three data sets by the experimental or MCNP-simulated ratio of single-bubble events to multi-bubble events. Table \ref{tab:rates} gives the results of the neutron calibrations performed at many different nuclear recoil thresholds for a variety of bubble multiplicities. At thresholds comparable to those of the dark matter search data sets, the experimentally-observed ratio of single-bubble events to multi-bubble events is approximately 2 ($\sim$1.7 for 8.0 keV and $\sim$2.7 for 15.8 keV calibrations). A similar result is produced by the MCNP-PoliMi simulations --- which is not a coincidence, since these simulations are scaled by the best-fit bubble nucleation efficiency. Using a 2:1 ratio of single-bubble events to multi-bubble events and the fact that there were 3 observed multi-bubble events, 6 total single-bubble events are expected from a neutron background. In the total exposure of 553.0 kg-days, an event rate of around $1\times10^{-2}$ counts/kg/day is therefore expected from this approach, in excellent agreement with that found from the calculations of Chapter \ref{ch:backgrounds} and shown in Figure \ref{fig:summary}. Being based on only 3 counts, the uncertainty in this calculation is quite large, shown as the light gray band on the total expected neutron background prediction in Figure \ref{fig:summary}.

Most central to the results of this thesis has been characterizing the backgrounds expected in the dark matter search data, arising mostly from neutrons generated from radioactive contamination of the borosilicate glass viewports and the PZT piezoelectric transducers (Sections \ref{sec:backgrounds_windows} and \ref{sec:backgrounds_piezos}). With proper screening of a replacement window and piezos (Appendix \ref{ch:piezos}), the total background expected from neutron sources in the bubble chamber should drop by over an order of magnitude (Figure \ref{fig:summary}, black dashed). At this level, the contributions from the spontaneous fission of $^{238}$U and $^{232}$Th in the stainless steel pressure vessel (Section \ref{sec:backgrounds_steel}) and from rock-generated neutrons punching through the $\sim$50 cm water shielding (Section \ref{sec:backgrounds_rock}) will dominate. Without a significant redesign of the 4 kg bubble chamber, these cannot be reduced further. The gamma background expected is based on measurements made of the gamma flux at the site of the CF$_3$I volume, using a NaI[Tl] scintillator detector (Chapter \ref{ch:gammas}). The rate at which the sensitivity to gammas falls off is based upon gamma source calibrations of both this chamber (Section \ref{sec:gammas_calibrations}) and with the older COUPP 2 kg bubble chamber \citep{behnke-08}. In total, the backgrounds of the COUPP 4 kg bubble chamber are significant at present, but with the planned improvements to the windows and piezos, the WIMP sensitivity increases drastically. With the reduced backgrounds expected in the upcoming data run, COUPP may begin to see the signature from a light WIMP ($m_\chi \sim $10 GeV, $\sigma^\mathrm{SI} \sim 2\times10^{-41}$ cm$^2$), the current best-fit to the combined results of DAMA, CoGeNT, and CRESST \citep{hooper-11}, or to firmly exclude such a possibility.

\section{COUPP 4 kg Dark Matter Limits}
\label{sec:limits_limits}

\begin{figure} [p!]
\centering
\includegraphics[scale=0.8]{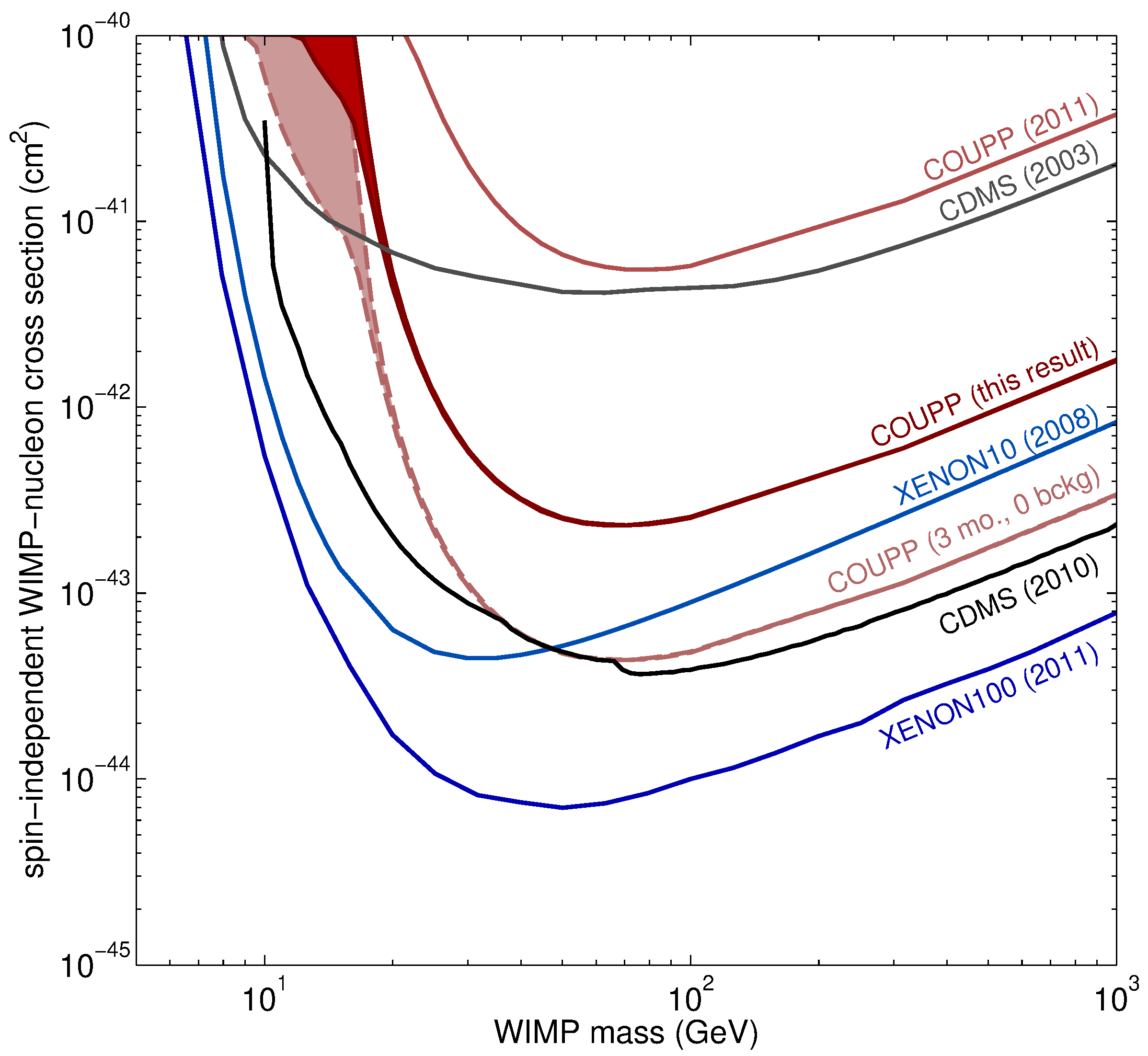}
\caption[Spin-independent WIMP-nucleon cross section limit]{The 90\% confidence level limit on the spin-independent WIMP-nucleon scattering cross section from this result is shown in red. This is based on 20 WIMP recoil-like events in 553.0 kg-days of exposure. The spread on the limits at low mass arises from the choice of bubble nucleation efficiency used (see text and Section \ref{sec:efficiency_fits}). The recent result from the COUPP 4 kg bubble chamber operating in the MINOS near detector hall in Fermilab \citep{behnke-11} is shown in pink, and a projection for this bubble chamber operating for three months background free at SNOLAB is also shown. Direct detection limits from CMDS \citep{akerib-03,ahmed-10}, XENON10 \citep{angle-08}, and XENON100 \citep{aprile-11} are represented in black and blue.}
\label{fig:SI_limit}
\end{figure}

\begin{figure} [p!]
\centering
\includegraphics[scale=0.8]{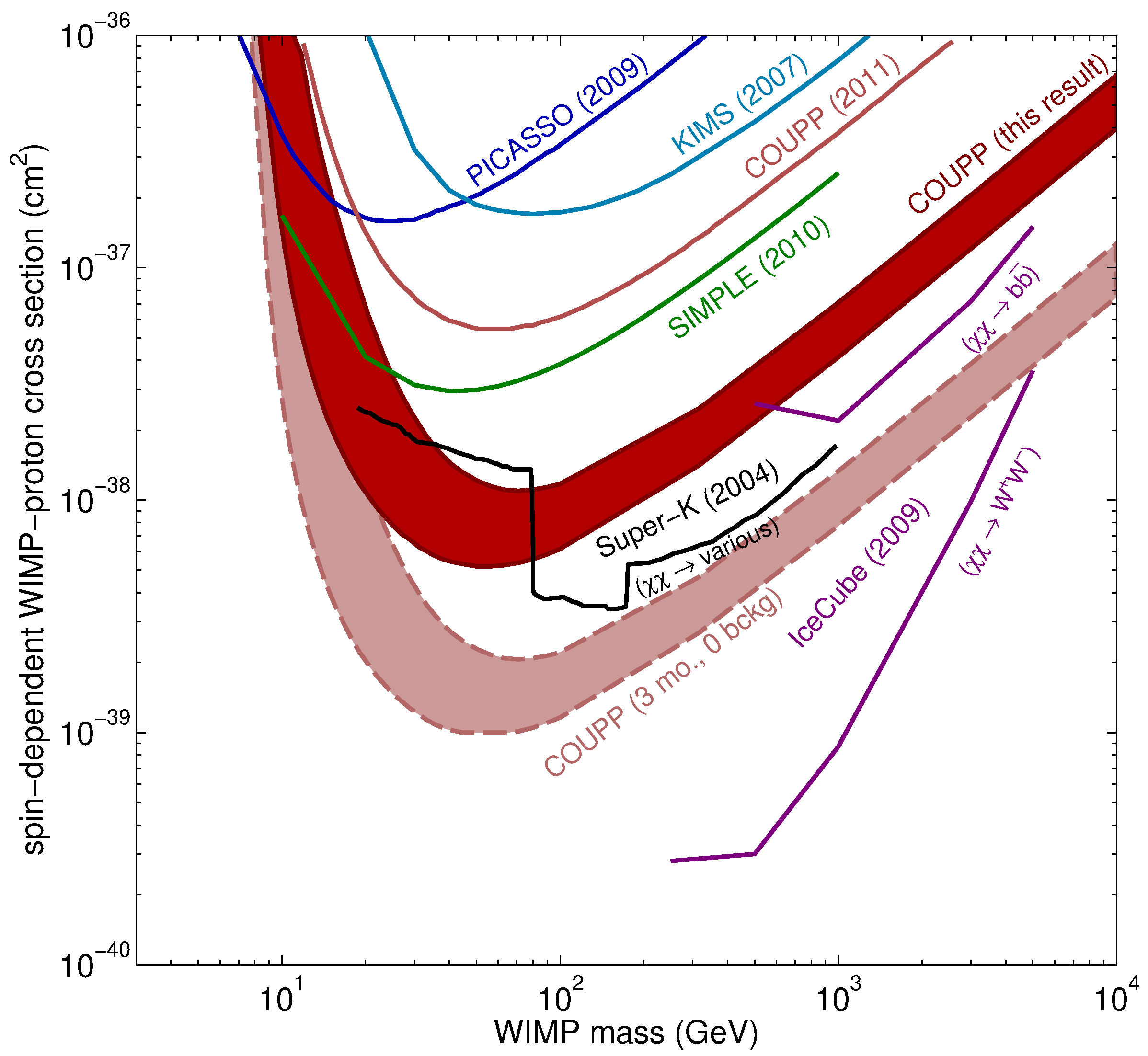}
\caption[Spin-dependent WIMP-proton cross section limit]{The 90\% confidence level limit on the spin-dependent WIMP-proton scattering cross section from this result is shown in red. This is based on 20 WIMP recoil-like events in 553.0 kg-days of exposure. The spread on the limits arises from the choice of bubble nucleation efficiency used (see text and Section \ref{sec:efficiency_fits}). The recent result from the COUPP 4 kg bubble chamber operating in the MINOS near detector hall in Fermilab \citep{behnke-11} is shown in pink, and a projection for this bubble chamber operating for three months background free at SNOLAB is also shown. Direct detection limits from PICASSO \citep{archambault-09}, KIMS \citep{lee-07}, and SIMPLE \citep{felizardo-10} are represented in blue and green. Limits on neutralino annihilation in the Sun from IceCube \citep{abbasi-09} and Super-Kamiokande \citep{desai-04} and also depicted. These indirect detection limits have additional dependence on the branching fractions of the annihilation products.}
\label{fig:SD_limit}
\end{figure}

The COUPP 4 kg bubble chamber employed a 4.048 kg superheated CF$_3$I target volume to search for WIMP dark matter. Being high-mass, the $^{127}$I nuclei make ideal spin-independent WIMP-nucleon scattering targets. At low WIMP mass, the $^{19}$F nuclei also begin to contribute (Figure \ref{fig:SI_WIMP_recoil_CF3I}). For spin-dependent WIMP-proton scattering, the $^{19}$F with its spin-uncoupled proton is the dominant target (Figure \ref{fig:SD_WIMP_recoil_CF3I}). Taking the 90\% confidence level upper limit to the number of observed events, and comparing the corresponding rates to those expected from a WIMP signal (Equation \ref{eq:dR_dER_fin}, integrated) with mass and cross section parameters, limits are obtained on the largest elastic cross section allowed by the data (Figures \ref{fig:SI_limit} and \ref{fig:SD_limit}).   These limits are calculated using the standard halo parameterization \citep{lewin-96}: $\rho_\chi = 0.3$ GeV/cm$^3$, $v_\mathrm{esc} = 544$ km/s, $v_E = 244$ km/s, and $v_0 = 230$ km/s. The Helm formalism \citep{helm-56} is used for the spin-independent nuclear form factor (Equation \ref{eq:F_SI}), and the DKVS formalism \citep{cannoni-11b} is used for the spin-dependent form factors (see Section \ref{sec:directdetection_crosssections}). The width on the exclusion curves arises from the choice of the bubble nucleation efficiency model --- the upper limit on the curve is from the sigmoid model (Equation \ref{eq:P_PICASSO}) with best-fit parameter $\alpha = 0.15$ (Figure \ref{fig:contour_a_alpha}), while the lower limit is from the flat efficiency model (Equation \ref{eq:P_flat}) with best-fit parameter $\eta = 0.46$ (Figure \ref{fig:contour_a_eta}). Because the spin-independent WIMP-nucleon scattering limit is determined by the 100\% efficient $^{127}$I recoil (Chapter \ref{ch:efficiency}), no spread is found in the high-mass region of the exclusion plot. Given the systematic uncertainties in the background predictions, no background subtraction has been attempted in producing dark matter limits with these results.

In the spin-dependent sector, this result sets a new world-leading limit on the WIMP-proton scattering cross section. Following the identification reported in this thesis that certain components of the detector act as weak neutron sources limiting these results, a new run of the COUPP 4 kg bubble chamber is about to begin in SNOLAB with more radioclean detector materials. According to this analysis, the background-generated event rate expected in the new version of this chamber will be reduced by over an order of magnitude with respect to the current results (Figure \ref{fig:summary}). A dark matter search of over three months with no expected backgrounds is now possible. This would improve upon the world-leading limits already set in the spin-dependent case, and would make COUPP comparable to CDMS in the spin-independent case. As more data is collected without a measurable neutron background, these limits will be improved.

\appendix

\singlespacing
\chapter{CF$_3$I Pressure Gauge Offset}
\label{ch:P1offset}
\doublespacing

As described in Section \ref{sec:coupp_sensors}, the COUPP 4 kg bubble chamber makes use of five pressure sensors: two Setra GCT-225 ultra high purity pressure transducers (PT1 measuring the pressure of the inner water/CF$_3$I volume, and PT2 measuring the pressure of the glycol inside the steel pressure vessel), two Omegadyne PX309 pressure transducers (PT4 measuring the pressure of the glycol in the hydraulic cart, and PT5 measuring the ambient air pressure of the room), and one Dytran 2005V fast pressure transducer also measuring the pressure of the water/CF$_3$I volume (PT3). Here, the reading of transducer PT$n$ will be labeled $PT_n$ (\emph{measured} pressure), as opposed to the actual value of the pressure being measured, which is labeled $P_n$ (\emph{actual} pressure). For instance, the transducer attached to the water/CF$_3$I volume (PT1) reads out a value of $PT_1$, pointing to an actual pressure $P_1$, where $P_1 = PT_1+\textrm{cal}_1$, for some calibration value.

Initially, the pressure transducers were all calibrated and the mean of the $PT_1$ and $PT_2$ measurements were the same ($P_1$ and $P_2$ were equal, and the values of $PT_1$ and $PT_2$ reflected the true pressures $P_1$ and $P_2$). After a few weeks of running, some events of unknown nature (presumably bubbles in the water volume near PT1) caused the baseline of PT1 to drastically change, which introduced a large offset between $PT_1$ and $PT_2$ in the form of a change to the calibration value for PT1. This offset is presumed to apply only to the response of PT1, not in an overall pressure differential between $P_1$ and $P_2$, because the value of $P_1$ is quite clearly not in the range of 160 psia and the events that caused this shift were unexceptional in nature.

\section{$PT_1$ Offset Step Classification}
\label{P1offset_steps}

Over the course of the data and commissioning runs of the 4 kg chamber at SNOLAB, a number of smaller $PT_1$ offset steps occurred (data between October 10, 2010 and June 28, 2011 is shown below in Figure \ref{fig:PT1_offset_corrected}, with a number of these steps shown in black). While some of these steps correspond to response changes in either PT1 or PT2 (which would not change the relationship between the actual pressure values $P_1$ and $P_2$, only $PT_1$ and $PT_2$), some of them may in fact correspond to a change in the bellows force, which would offset $P_1$ from $P_2$. The steps corresponding to response changes will subsequently be referred to as Type 1 steps, whereas the steps corresponding to actual pressure changes between $P_1$ and $P_2$ will be referred to as Type 2 steps.

In order to classify each $PT_1$ offset step as either a Type 1 or Type 2 step, two other parameters that scale with $P_1$ were studied: the baseline of the Dytran mounted on the pressure vessel and the acoustic parameter measured by the three active piezoelectric transducers epoxied to the quartz jar. By looking for steps in the Dytran baseline and the acoustic parameter across the $PT_1$ offset steps in question, one can determine whether the value of $P_1$ itself has actually changed across that step\footnotemark. If it has, the step is defined as a Type 2 step.

\footnotetext{Only events with one bubble and for which the acoustic parameter is greater than 0 and less than 2 are considered here.}

\begin{figure} [t!]
\centering
\includegraphics[scale=0.6]{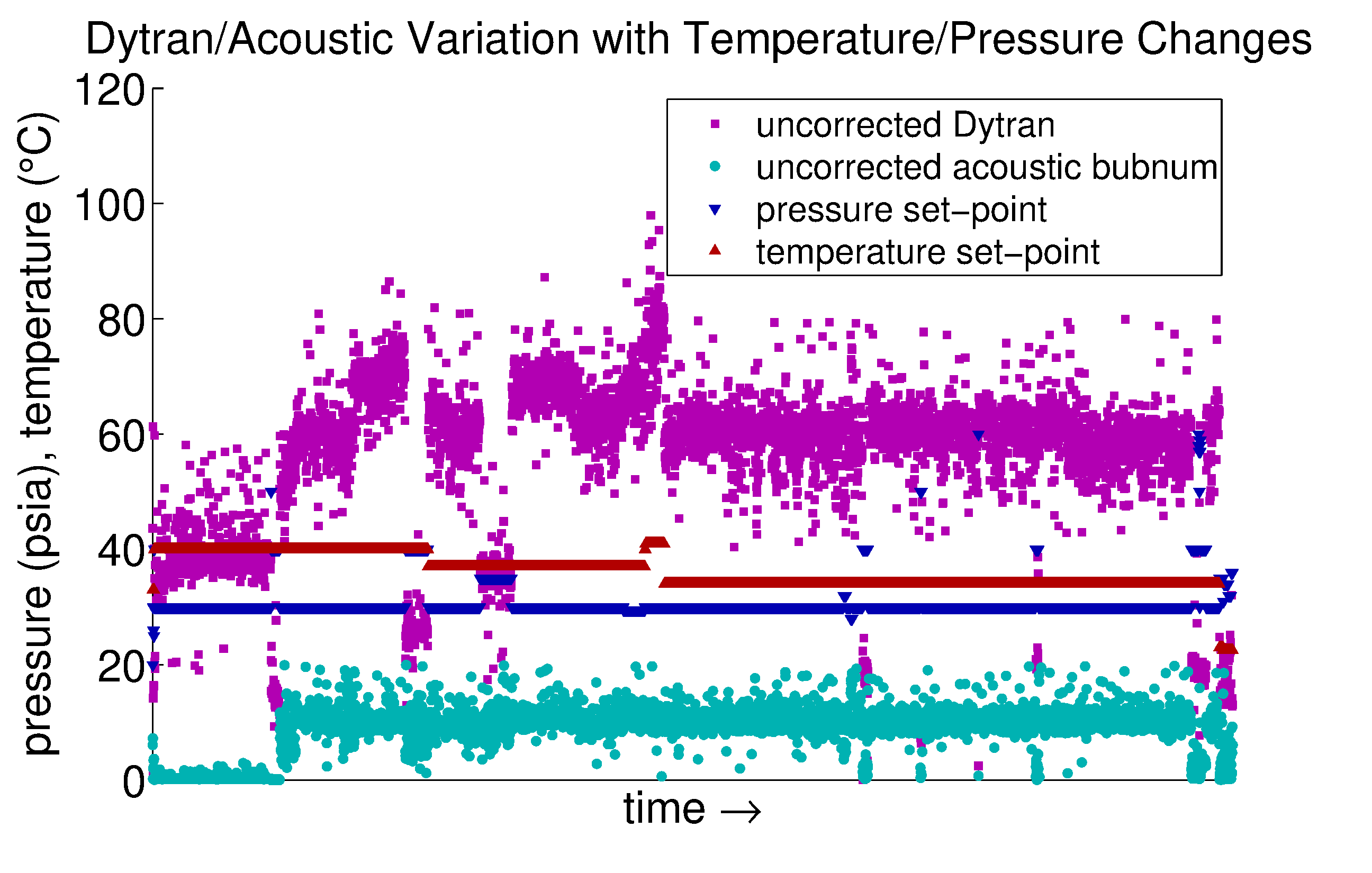}
\caption[Dytran/acoustic variation with temperature/pressure changes]{Data from October 10, 2010 to June 28, 2011, showing temperature (red) and pressure (blue) set-points as well as the values of the Dytran baseline (magenta) and acoustic parameter (cyan). A clear temperature and pressure dependence can be seen in the value of the Dytran, while only a pressure dependence is evident in the value of the acoustic parameter. The values of the Dytran baseline and acoustic parameter are scaled to fit within the range of the plot, and therefore have arbitrary units on the y-axis.}
\label{fig:PT1_offset_uncorrected}
\end{figure}

In using the Dytran baseline and acoustic parameter as variables to classify $PT_1$ offset steps, it is important to work with a consistent data set --- one in which these parameters are expected to be comparable across each of these steps. The Dytran baseline varies with changes to the temperature or pressure set-points of the chamber, while the acoustic parameter only varies with changes to the pressure (scaled values of the Dytran baseline and acoustic parameter are shown in Figure \ref{fig:PT1_offset_uncorrected} along with changes to the temperature and pressure set-points).  In order to remove the dependence on the temperature and pressure set-points of the chamber, corrected Dytran baseline and acoustic parameter variables were created which scale the original values of these parameters with changes to the temperature and pressure.

\section{Dytran and Piezo Corrections}
\label{P1offset_fits}

\begin{figure} [t!]
\begin{minipage}[b]{0.5\linewidth}
\centering
\includegraphics[trim=1cm 0cm 1cm 0cm, clip=true, scale=0.4]{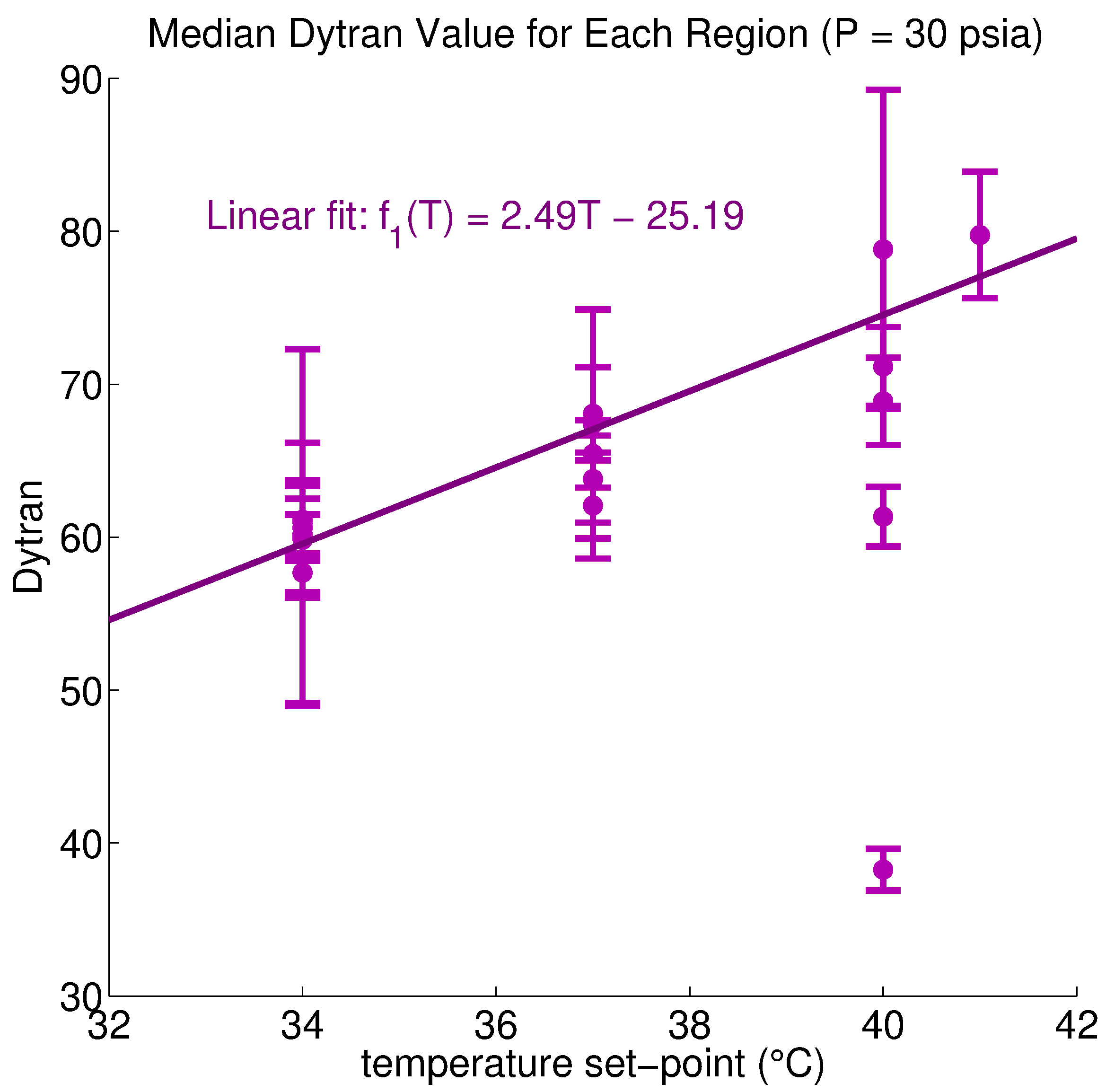}
\end{minipage}
\begin{minipage}[b]{0.5\linewidth}
\centering
\includegraphics[trim=0.5cm 0cm 1cm 0cm, clip=true, scale=0.4]{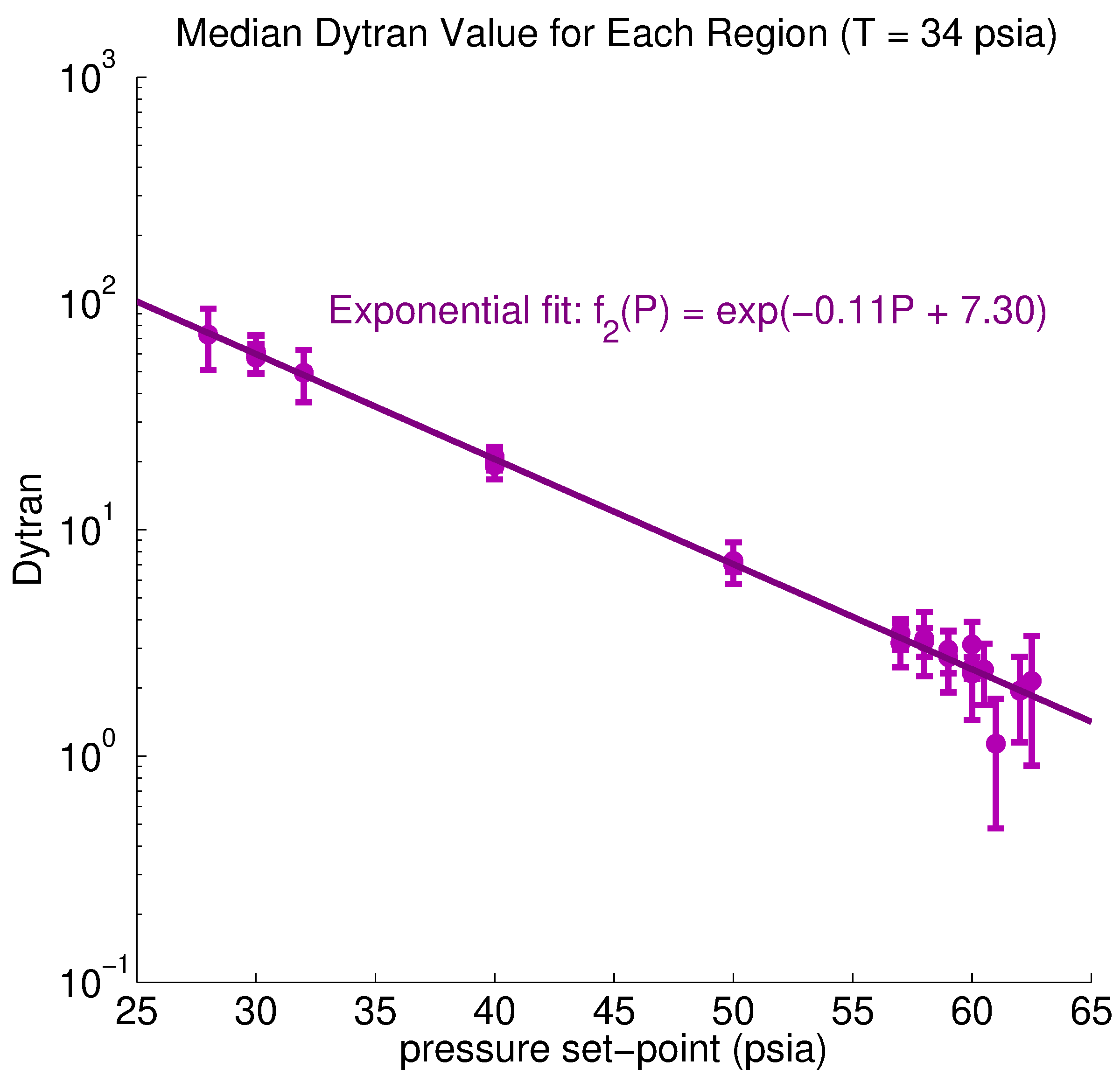}
\end{minipage}
\caption[Median Dytran baseline with variation in temperature and pressure]{Median values of the Dytran baseline varying over different temperature set-points (and a constant pressure set-point of 34 psia) and over different pressure set-points (and a constant temperature set-point of 30$^{\circ}$C) along with their best fits ($f_1(T)$ and $f_2(P)$, respectively).}
\label{fig:dytran_variation}
\end{figure}

To determine this correction, a number of regions was defined across the data where the temperature and pressure set-points were held constant. The median of the Dytran baseline and the acoustic parameter were recorded for each of these regions, and fits were made of the temperature- and pressure-dependence. The median Dytran values for each region with a pressure set-point of 30 psia (the most common pressure set-point) were plotted against the temperature set-point of the region and fit to a temperature-dependent Dytran correction function (Figure \ref{fig:dytran_variation}, left).  The pressure-dependence of the Dytran was determined in a similar fashion by plotting the median Dytran values with a temperature set-point of 34$^{\circ}$C (the most common temperature set-point) against the pressure set-point of the region (Figure \ref{fig:dytran_variation}, right). The resulting Dytran correction term was found to be

\begin{equation}
C_D(T,P)=2.49T + e^{-0.11P+7.33} + \mathrm{constant}.
\label{eq:dytran_correction}
\end{equation}

\begin{figure} [h!]
\centering $
\begin{array}{c}
\includegraphics[scale=0.4]{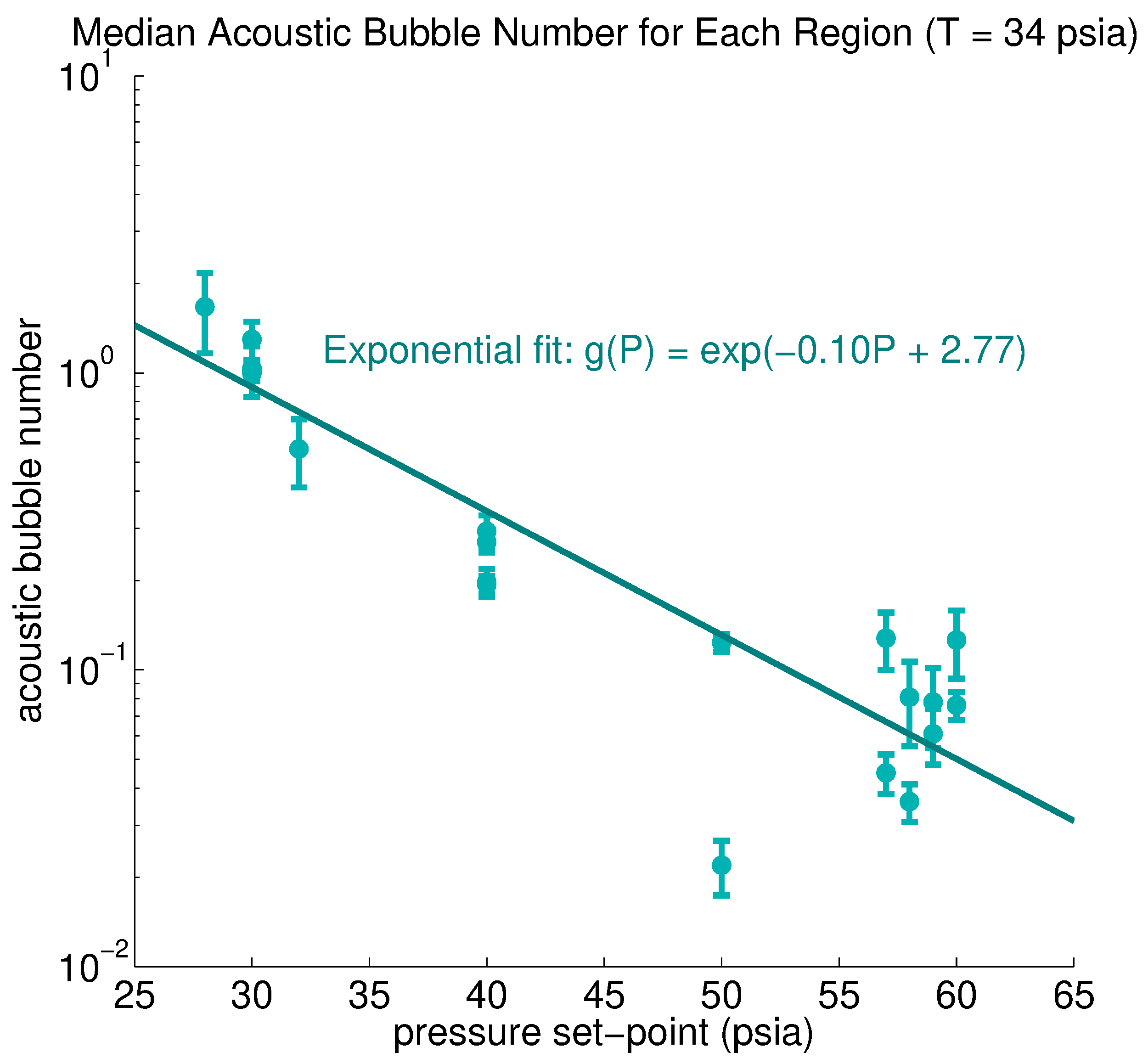}
\end{array} $
\caption[Median acoustic parameter with variation in pressure]{Median values of the acoustic parameter varying over different pressure set-points (and a constant temperature set-point of 30$^{\circ}$C) along with its best fit ($g(P)$).}
\label{fig:acoustic_variation}
\vspace{20pt}
\end{figure}

As stated earlier, the median acoustic parameter values showed no temperature-dependence. The pressure-dependence was determined by plotting the median acoustic parameter of regions with a temperature set-point of 34$^{\circ}$C against the pressure set-point of the region (Figure \ref{fig:acoustic_variation})  and was found to be

\begin{equation}
C_{AP}(P)=e^{-0.10P+2.78} + \mathrm{constant}.
\label{eq:acoustic_correction}
\end{equation}

\begin{figure} [t!]
\centering
\includegraphics[scale=0.6]{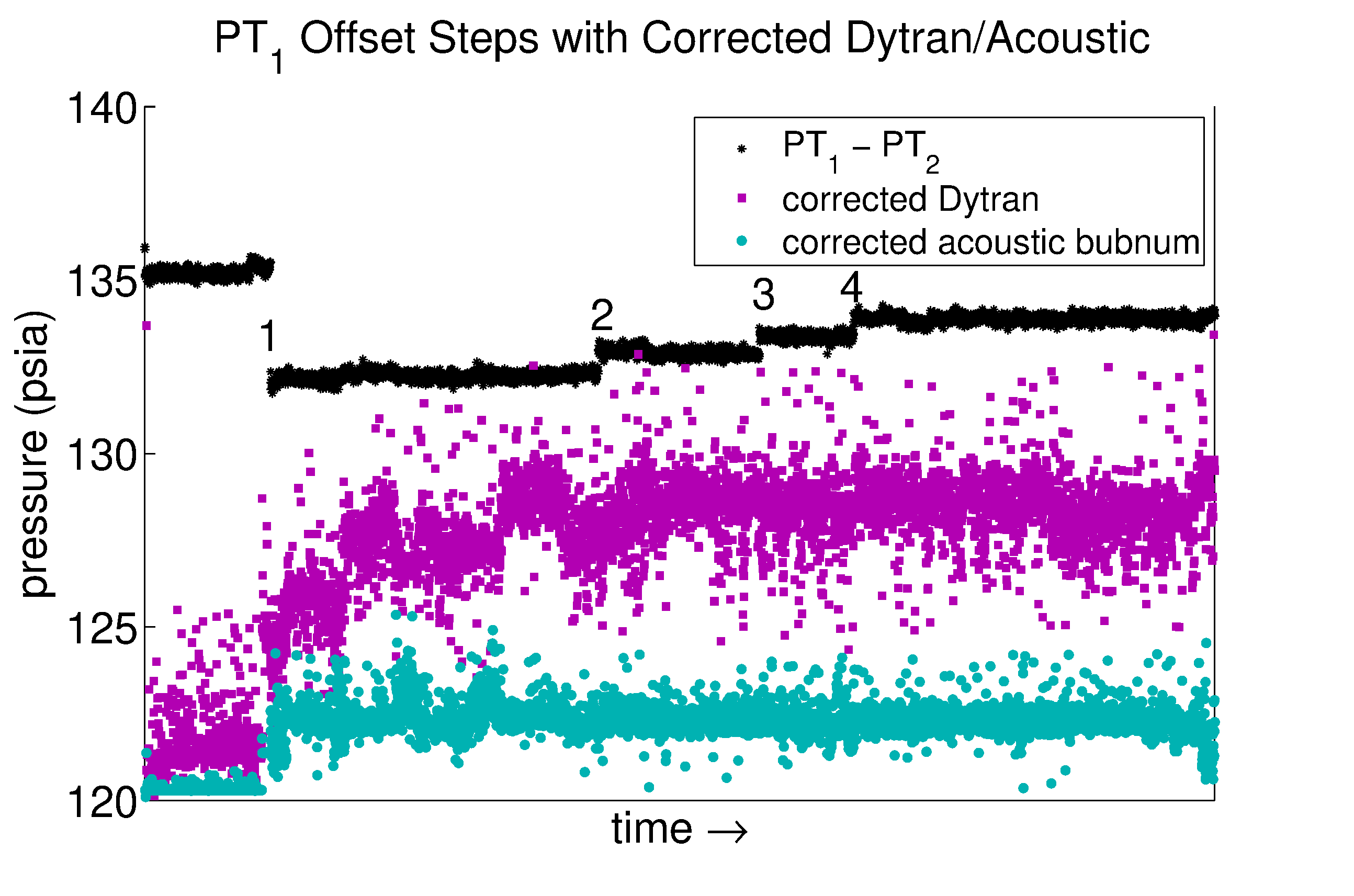}
\caption[$PT_1$ offset steps with uncorrected Dytran]{$PT_1-PT_2$ values (black) from October 10, 2010 to June 28, 2011, for events with $\mathtt{nbub}=1$ and uncorrected acoustic parameter between 0 and 2, showing various offset steps, labeled 1 through 4. Also shown is the corrected Dytran baseline (magenta) and acoustic parameter (cyan), scaled to fit on the plot, and therefore with arbitrary units on the y-axis. Discrete jumps in the Dytran baseline and acoustic parameter corresponding to a $PT_1$ offset step suggests a pressure differential across that step, resulting in the step being labeled as Type 2.}
\label{fig:PT1_offset_corrected}
\end{figure}

\noindent The constants in Equations \ref{eq:dytran_correction} and \ref{eq:acoustic_correction} are unimportant, since the corrected Dytran baseline and acoustic parameter are ultimately scaled to fit on Figure \ref{fig:PT1_offset_corrected}.

The corrected Dytran baseline ($D_{\mathrm{cor}}^i$) for the $i^{\mathrm{th}}$ event with temperature $T_i$ and pressure $P_i$ was calculated by subtracting the Dytran correction from Equation \ref{eq:dytran_correction} from the original Dytran baseline value ($D_0^i$) for that event:

\begin{equation}
D_{\mathrm{cor}}^i = D_0^i - C_D(T_i,P_i).
\label{eq:corrected_dytran}
\end{equation}

\noindent Similarly, the corrected acoustic parameter ($AP_{\mathrm{cor}}^i$) for each event with pressure $P_i$ was found by subtracting the acoustic parameter correction from Equation \ref{eq:acoustic_correction} from the original acoustic parameter ($AP_0^i$) for that event:

\begin{equation}
AP_{\mathrm{cor}}^i = AP_0^i - C_{AP}(P_i).
\label{eq:corrected_acoustic}
\end{equation}

The corrected Dytran baseline and acoustic parameter values were plotted along with the difference between $PT_1$ and $PT_2$ for each event with $\mathtt{nbub}=1$ and uncorrected acoustic parameter between 0 and 2 (Figure \ref{fig:PT1_offset_corrected}).  The $PT_1$ offset steps in question are labeled 1 through 4 on the plot. Since the Dytran and the piezoelectric transducers producing the acoustic parameter are sensitive to pressure changes in the water/CF$_3$I volume ($P_1$), whether an offset step is Type 1 or Type 2 is determined by whether the corrected Dytran baseline and acoustic parameter has a sharp corresponding offset at the same time (i.e. whether $P_1$ changes at that step). By analyzing Figure \ref{fig:PT1_offset_corrected}, it is obvious that $PT_1$ offset step 1 resulted in a pressure change in $P_1$, and is thus a Type 2 step. Within error, the others seem to correspond only to a response change in PT1, and are thus Type 1 steps.

\section{$PT_1$ Correction}
\label{P1offset_correction}

That $PT_1$ offset step 1 is a Type 2 step comes as no surprise. This step corresponds to runs 20101018\_9 and 20101019\_0, during which the bellows became fully extended due to an error in the DAQ system that failed to curtail the boiling of the chamber when the nitrogen bottle used by the pressure cart ran out overnight. The overextension of the bellows could have affected its spring constant, resulting in a change in the pressure differential between the inside and outside volumes (that is, between $P_1$ and $P_2$). Attributing this entire $PT_1$ offset step to a real pressure differential, one can assume that while $PT_1$ and $PT_2$ should have been equal before this $PT_1$ offset, after the offset $PT_1$ should be less than $PT_2$ by the total amount of the offset, or

\begin{equation}
PT_1 = \left\{ \begin{array}{ll}
PT_2 & \mbox{ if run $<$ 20101019\_0} \\
PT_2 - 2.84\:\mathrm{psia} & \mbox{ if run $\geq$ 20101019\_0.} \\
\end{array}
\right. \\
\label{eq:PT1_correction}
\end{equation}

\noindent Therefore, in lieu of using PT1 as the gauge for the inner CF$_3$I/water volume pressure reading, PT2 will be used together with the scaling given here.

\singlespacing
\chapter{Simulation Packages}
\label{ch:mcnp}
\doublespacing

Three different computer codes were used for this thesis. The SRIM package \citep{ziegler-96} is a Monte Carlo simulation package developed to calculate the interactions of ions with matter. This was used to determine the stopping power and projected ranges of different nuclear recoils in the CF$_3$I volume. The SOURCES-4C code \citep{wilson-05} is a computer code system for calculating ($\alpha$,n), spontaneous fission, and delayed neutron sources and spectra. This was used to determine the neutron yields and energy spectra given different radioactive concentrations in materials in and around the detector. Finally, the Monte Carlo N-Particle (MCNP) transport code \citep{sweezy-05a} is a Monte Carlo software package to simulate nuclear processes. This simulation package was used extensively to calculate the expected event rates from neutron and gamma calibrations and those arising from neutron backgrounds calculated using SOURCES-4C. When necessary, discussion of each follows.

\section{MCNP}
\label{sec:mcnp_mcnp}

The MCNP transport code is a Monte Carlo software package for simulating transport of gamma, neutron, and electron radiation in materials, developed by Los Alamos National Laboratory and distributed in the United States by the Radiation Safety Information Computational Center at the Oak Ridge National Laboratory \citep{sweezy-05a}. It is designed to simulate particle interactions involving neutrons, gammas, and electrons, including secondary gammas resulting from neutron interactions and secondary electrons resulting from gamma interactions. The standard version of MCNP operates in such a way to give a correct answer when averaged over a large number of particle histories, but its physics deliberately deviates from reality on an event-by-event basis in order to gain an advantage in efficiency \citep{pozzi-03}. When information from individual histories is desired, a more precise code describing each interaction is necessary. The MCNP-PoliMi (MCNP Politecnico di Milano) code was designed out of the standard MCNP-4c (version 4c) package in order to provide this event-by-event resolution for simulations \citep{padovani-02}. When this specificity is not required, standard MCNP-5 (version 5) was used.

When employing MCNP, the user first specifies a source for neutrons or photons, including the source spectrum and the location, orientation, and geometry of the emission. The source particles traverse some user-defined geometry filled with volumes of different materials representing the experimental setup until they arrive at a designated target volume or surface. The physics experienced by the source particles as they interact with their environment is also tunable (\emph{e.g.} whether Bremsstrahlung is included or ignored when transporting gammas). If MCNP-5 is used, a tally is calculated at the target volume or surface, recording summary information such as the particle current, flux, or energy deposition, depending on which tally is used. In MCNP-PoliMi, information for each interaction in a target volume is recorded, including the type of interaction (\emph{e.g.} elastic scatter, capture, \emph{etc.}), the source particle and target particle involved in the interaction, the energy deposited in and location of the interaction, and so on.

In order to test the reliability of using MCNP-PoliMi in simulations, comparisons with another simulation package were made. For the neutron calibration study described in Chapter \ref{ch:efficiency}, a parallel analysis was performed by E. Vazquez-Jauregui using the GEANT-4 simulation package \citep{agostinelli-03}, in lieu of using MCNP-PoliMi. The predicted rates generated by GEANT-4 matched those from MCNP-PoliMi within error, and therefore the use of MCNP-PoliMi throughout this work is presumed to be well-motivated.

\subsection{A Note on MCNP Uncertainty}
\label{sec:mcnp_uncertainty}

Several factors contribute to the overall systematic error (accuracy) of the predictions generated by MCNP, including the physics and mathematical models used, uncertainties in the cross sections and nuclear/atomic data, \emph{etc.} \citep{shultis-faw-06}. It is assumed that any statistical error (precision) is made negligible by iterating over a large number of histories. In this work, the systematic error is separated into three categories. Geometrical error in any prediction is generated by the accuracy of the MCNP geometry as it represents the real physical components of the detector --- be it in their location and dimensions or in their physical composition or density, impurity concentration, \emph{etc}. To determine the geometrical error, the dimensions of the 4 kg geometry were varied to maximize and minimize the amount of moderator between source and target, within the tolerances allowed by the manufacturer of each part of the chamber. From these, the geometrical error was found to be a $\sim$2\% effect. Experimental error in any prediction is generated by different uncertainties arising from the fact that physical sources and detectors are being dealt with (\emph{e.g.} the uncertainty in the source strength of a calibration source will propagate its way through to uncertainties in the prediction generated by MCNP). This error was similarly found to be small. Finally, an inherent uncertainty arises simply due to the fact that a Monte Carlo simulation is being used, which has any number of immeasurable inaccuracies built in that keeps it from faithfully representing reality. In comparison to the first two, this uncertainty dominates.

To determine the inherent error of the MCNP to match physical observations, measurements were made of the yield from five different neutron sources of known activities using two $^{3}$He neutron proportional counters in a total of 11 configurations \citep{collar-07}. For a given measurement, an MCNP simulation was used to determine the efficiency of the detector. Therefore, comparing the nominal yield of the neutron source with the predicted yield (based on simulated efficiencies and measurements) provides an estimation of the uncertainty arising from MCNP. In total, this uncertainty was found to be 10.5\%. The results of the measurements are shown in Table \ref{tab:mcnp_test}.

\begin{table}[t!]
\centering
\begin{tabular} {| l | l | c | c |}
\hline
Configuration & Source & Nominal Yield & Measured Yield \\
 & & \small{(n/s)} & \small{(n/s)} \\
\hline
\hline
\multicolumn{4}{| l |}{Small $^3$He Counter} \\
\hline
Bare & Am-Be(30 mCi $\alpha$) \#1 & $8.55\times10^4$ & $6.95\times10^4$ \\
Bare & Am-Be(3 mCi $\alpha$) & $8.55\times10^3$ & $6.51\times10^3$ \\
Bare & Am-Be(30 mCi $\alpha$) \#2 & $8.55\times10^4$ & $7.09\times10^4$ \\
Cd-wrapped & Am-Be(30 mCi $\alpha$) \#1 & $8.55\times10^4$ & $8.51\times10^4$ \\
Cd-wrapped & Am-Be(30 mCi $\alpha$) \#2 & $8.55\times10^4$ & $8.43\times10^4$ \\
Cd-wrapped & Pu-Be (1 Ci $\alpha$) & $2.85\times10^6$ & $2.78\times10^6$ \\
Cd-wrapped & $^{252}$Cf & $1.00\times10^7$ & $8.43\times10^6$ \\
\hline
\hline
\multicolumn{4}{| l |}{Large $^3$He Counter} \\
\hline
Bare & Am-Be(30 mCi $\alpha$) \#1 & $8.55\times10^4$ & $8.92^{+3.36}_{-1.69}\times10^4$ \\
Bare & Am-Be(3 mCi $\alpha$) & $8.55\times10^3$ & $7.91^{+2.99}_{-1.56}\times10^3$ \\
Bare & Am-Be(30 mCi $\alpha$) \#2 & $8.55\times10^4$ & $9.35^{+3.54}_{-1.76}\times10^4$ \\
Bare & $^{252}$Cf & $1.00\times10^7$ & $8.54^{+3.23}_{-1.61}\times10^6$ \\
\hline
\end{tabular}
\caption[Known-activity source measurements used to test MCNP predictions]{Comparisons of the nominal yield from known sources and the measured yield, after an MCNP detector efficiency simulation. Measurements were taken with both a small and a large $^3$He counter, with a second set using the small counter with a Cd-wrapping. Data from \cite{collar-07}.}
\label{tab:mcnp_test}
\end{table}

\subsection{MCNP 4 kg Bubble Chamber Geometry}
\label{sec:mcnp_geometry}

\begin{figure} [t!]
\begin{minipage}[b]{0.46\linewidth}
\centering
\includegraphics[scale=0.5]{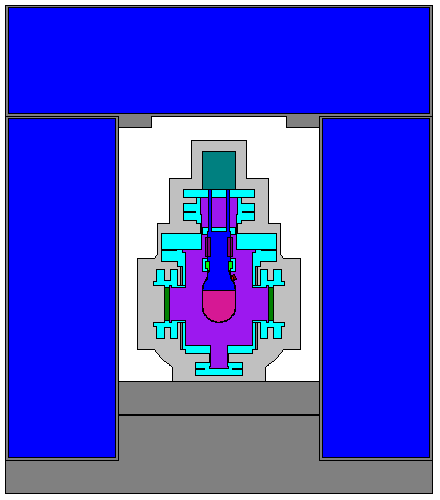}
\end{minipage}
\begin{minipage}[b]{0.54\linewidth}
\centering
\includegraphics[scale=0.42]{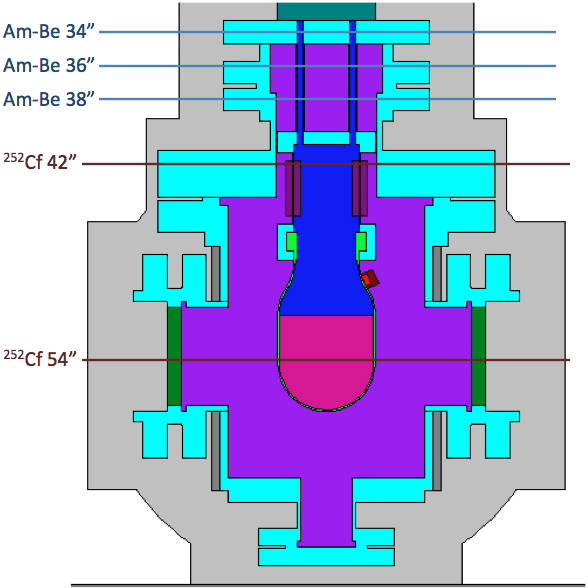}
\end{minipage}
\caption[MCNP geometry of the 4 kg chamber]{The $x$-$z$ projection of the MCNP geometry for the 4 kg bubble chamber, showing the entire geometry (left) and a zoomed-in view (right) (compare to Figure \ref{fig:coupp_pictures}). The approximate height of the calibration sources in different configurations is also shown, although their actual position is off the $y=0$ plane, and thus not visible in this projection --- for calibrations, the sources are inserted through a sealable source hole in the top panel of the water shield, 10"$\times$10" from the center of the top panel.}
\label{fig:mcnp_4kg}
\end{figure}

The COUPP 4 kg bubble chamber geometry used in MCNP consists of accurate representations of the inner volume, the pressure vessel, the thermal shielding, the base pedestal, and the water shielding (Figure \ref{fig:mcnp_4kg}).  Certain features unimportant to the propagation of source particles are not accurately represented, including steel bolt-heads and nuts, the specifics of the plumbing above the top flange, and the thin steel legs. However, the source encasings shown in Figure \ref{fig:Cf_AmBe_sources} (not shown in Figure \ref{fig:mcnp_4kg} because the source locations are off the $y=0$ plane for the $x$-$z$ projection), the geometry of the piezoelectric transducers (Figure \ref{fig:mcnp_4kg}, red), and other specifics that would noticeably change the predictions from sources are included with thorough detail.

\subsection{MCNP Libraries}
\label{sec:mcnp_libraries}

For each isotope in every material of the MCNP geometry, the cross section library to be used in the simulation was required. When available, the ENDF/B-6.0 libraries \citep{rose-91} were used in this analysis. Exceptions include the following elements (and data library used): argon (ENDL-92), chromium (RMCCS), manganese (RMCCS), iron (RMCCS), nickel (RMCCS), and lead (RMCCS).

\subsection{Known MCNP Bugs}
\label{sec:mcnp_bugs}

Initially, use of the MCNP-PoliMi code proved problematic due to the fact that certain elastic scatter interactions written to the dump log file were duplicated. This was confirmed by E. Padovani, one of its authors, as a printing error arising in an early version of MCNP-PoliMi that had been fixed in an updated version. With a newer version of MCNP-PoliMi, this problem was eliminated.

\section{SOURCES-4C}
\label{sec:mcnp_sources}

The SOURCES-4C code is a computer code system for calculating ($\alpha$,n), spontaneous fission, and delayed neutron sources and spectra \citep{wilson-05}. The user specifies the material composition, including radioactive source number densities and ($\alpha$,n) target nuclei atom fractions, and the total rate and energy spectrum of neutrons arising from both ($\alpha$,n) and spontaneous fission is produced (along with the delayed neutron rate and spectrum, but these were not useful in this analysis). This code was used whenever a neutron source arising from radioactive contamination of different detector materials (and those surrounding the detector) was needed for an MCNP-PoliMi simulation. Benchmarking of the SOURCES-3A code \citep{charlton-98} has produced very good agreement between SOURCES calculations and experimental results for a variety of test cases, with a quoted accuracy of 18\%.

\subsection{Known SOURCES-4C Bugs}
\label{sec:mcnp_sources_bugs}

Several bugs were noted in the SOURCES-4C code during analysis. First, the maximum alpha energy allowed by the calculation is 6.5 MeV due to cross section library limitations. However, in the decay data library (tape5), several isotopes have alpha energies exceeding this upper energy limit. If these isotopes are used in a calculation, an error is produced and the calculation fails. As a work-around, the tape5 library was altered to ensure that all alpha energies were below this 6.5 MeV limit --- when an energy exceeded 6.5 MeV, the energy was set to 6.5 MeV instead. The effect of this work-around is to make the resultant neutron spectrum slightly softer (since higher energy alphas are discounted), but a better solution to this problem is not known.

Further, the use of certain ($\alpha$,n) target isotopes made the calculation fail for unknown reasons. The one determined to fail here was $^{19}$F, useful in calculating the neutron rates from $^{222}$Rn dissolved in the CF$_3$I volume. As a work-around, any instance of a $^{19}$F target was replaced by an equivalent atom fraction of $^{17}$O, the nearest neighbor to $^{19}$F with similar $A/Z$. A similar treatment was done for the titanium isotopes in the PZT of the piezos (Section \ref{sec:backgrounds_piezos}). SOURCES-4C does not have $^{48}$Ti as an ($\alpha$,n) target, although it will certainly contribute somewhat, especially in materials rich in titanium. For this, replacement by $^{37}$Cl was made.

\begin{figure} [t!]
\centering
\includegraphics[scale=0.55]{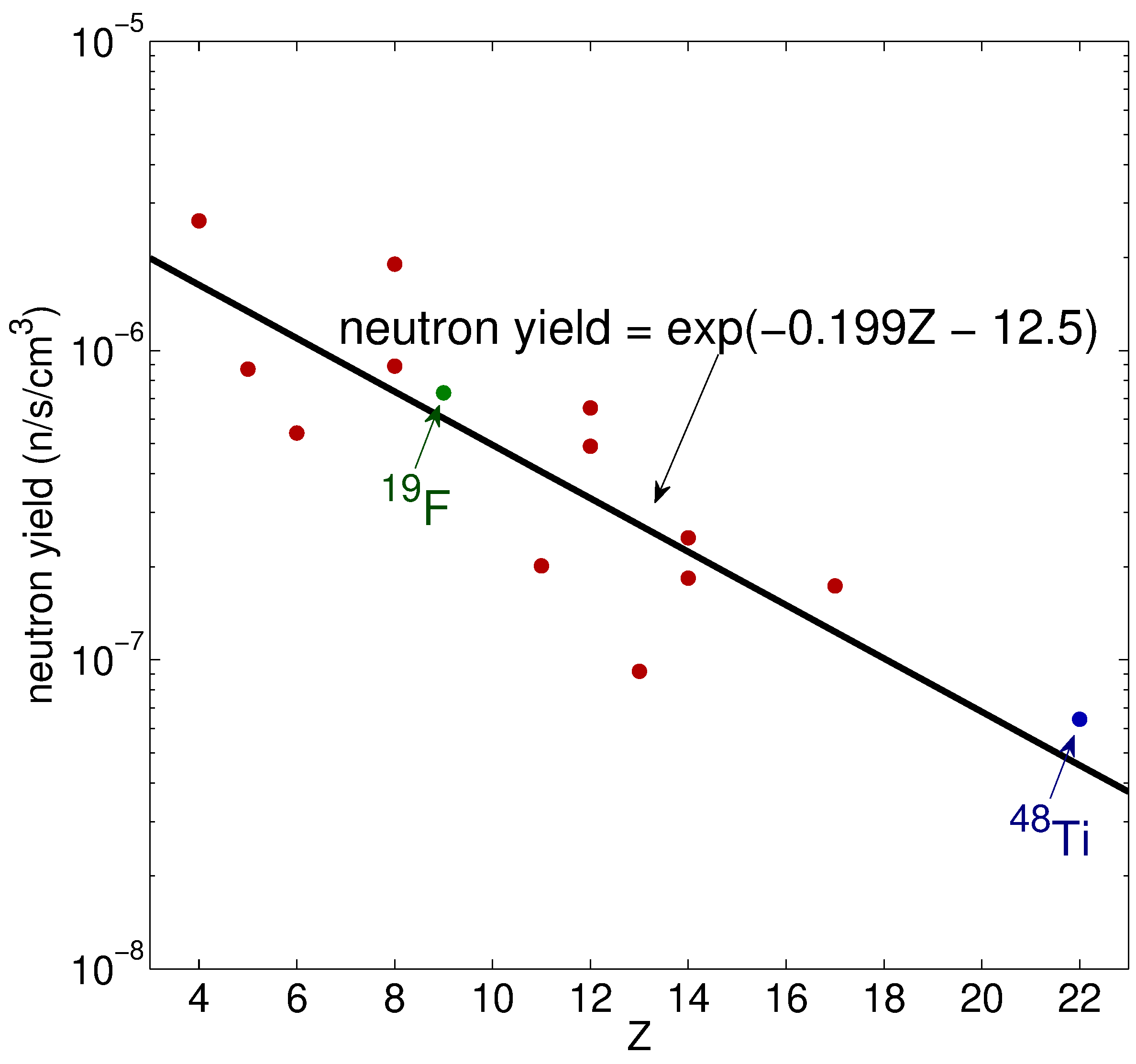}
\caption[Neutron yield per ($\alpha$,n) target nucleus in SOURCES-4C]{The neutron yield calculated by SOURCES-4C of each ($\alpha$,n) target material with $A/Z>2$. The use of neighboring nuclei in place of those missing from the SOURCES-4C data libraries is well-motivated, if proper scaling given by the trend line is followed.}
\label{fig:SOURCES_target_practice}
\end{figure}

The effect of replacing one isotope with another as an ($\alpha$,n) target has been estimated as follows. The neutron yield of a material with a constant composition of a particular ($\alpha$,n) target with $A/Z>2$ built in to SOURCES-4C was calculated, using a fixed abundance of the $^{232}$Th decay chain in equilibrium. This was repeated for each ($\alpha$,n) target isotope available with $A/Z>2$. An overall exponential trend was observed between $Z$ and the calculated yield (Figure \ref{fig:SOURCES_target_practice}).  A similar trend between target $Z$ and ($\alpha$,n) neutron yield has been reported for both 6 MeV and 8 MeV alphas \citep{heaton-89}. The trend shown in Figure \ref{fig:SOURCES_target_practice} fits in between those produce for 6 MeV and 8 MeV alphas, which is anticipated since it was produced for the whole spectrum of $^{232}$Th alpha energies. By using $^{17}$O in place of $^{19}$F or by using $^{37}$Cl in place of $^{48}$Ti, and scaling the atom fraction by the neutron yield fit, the values shown in green and blue (respectively) are produced. Both represent believable scalings, and so this technique will be used for ($\alpha$,n) targets either not functioning with or not included in the SOURCES-4C code.

\singlespacing
\chapter{Preparation of Radiopure Piezoelectric Transducers}
\label{ch:piezos}
\doublespacing

\begin{figure} [t!]
\centering
\includegraphics[scale=0.55]{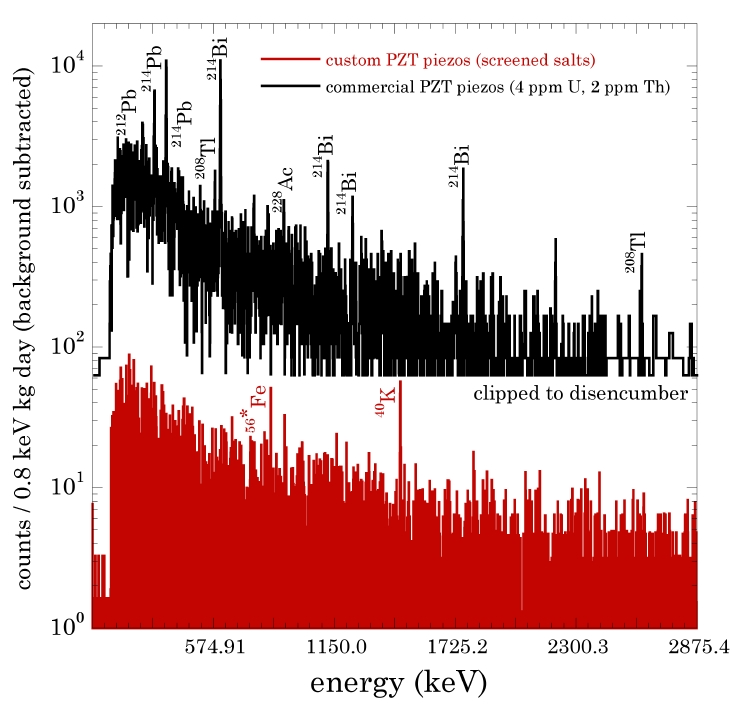}
\caption[Comparison between $^{238}$U and $^{232}$Th content in piezos]{Comparison between the $^{238}$U and $^{232}$Th activity of commercial Ferroperm PZT and second generation radioclean PZT manufactured by S. Priya's group  at Virginia Tech. The achived reduction in $^{238}$U content is a factor $\sim$400 and $\sim$60 for the $^{232}$Th chain. Figure courtesy J.I. Collar.}
\label{fig:piezo_U_Th}
\end{figure}

\begin{figure} [t!]
\centering
\includegraphics[scale=0.54]{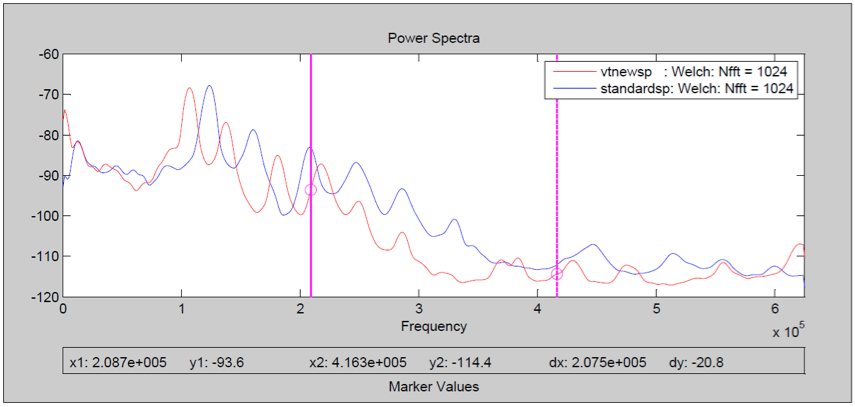}
\caption[Comparison of acoustic sensitivity of first and second generation piezos]{Comparison of acoustic sensitivity of first generation (blue) and second generation (red) piezos. Similar response is seen in all frequency bands from 0 to 600 kHz. Figure courtesy I. Levine and E. Behnke.}
\label{fig:piezo_response}
\end{figure}

As discussed in Section \ref{sec:backgrounds_piezos}, commercial PZT piezoelectric transducers manufactured by Ferroperm Piezoceramics A/S were used to allow for an efficient acoustic alpha-recoil discrimination in the physics runs described in this thesis. These ``first-generation" piezo ceramics were unfortunately found to contain a concentration in alpha emitters from the $^{238}$U and $^{232}$Th chains sufficient to generate a significant neutron background via the ($\alpha$,n) reaction (concentrations of 3.95$\pm$0.43 ppm in $^{238}$U and 1.18$\pm$0.15 ppm in $^{232}$Th were measured). A campaign of screening component salts led to the selection of raw materials with much lower $^{238}$U and $^{232}$Th content. This screening was performed using a low-background HPGe detector at the Laboratory for Astrophysics and Space Research (LASR) underground laboratory (at a depth of 6 m.w.e.) at the University of Chicago. Custom PZT piezo manufacture using these salts, with additional precautions to avoid any contamination, was performed at Virginia Tech by COUPP collaborators. These ``second generation" radioclean piezos were confirmed to have both a much lower content in alpha emitters (9.77$\pm$1.68 ppb $^{238}$U and 20.95$\pm$4.78 ppb $^{232}$Th, see Figure \ref{fig:piezo_U_Th})  and a similar acoustic detection performance (Figure \ref{fig:piezo_response})  to commercial units. Their installation in the 4 kg chamber is ongoing at the time of this writing. As can be seen from Figures \ref{fig:SI_limit} and \ref{fig:SD_limit}, they are expected to allow for an improvement of almost an order of magnitude on present WIMP limits. We expect to have exhausted the physics reach of the 4 kg chamber before these new piezos can impose any limitations.

\begin{figure} [t!]
\centering
\includegraphics[scale=0.62]{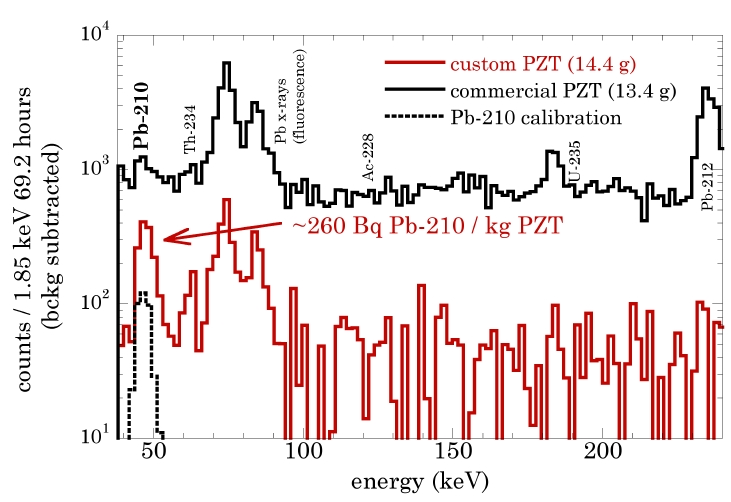}
\caption[Comparison between $^{210}$Pb content in piezos]{Comparable $^{210}$Pb content of commercial and second generation radioclean PZT, as derived from the 46 keV gamma emission detected with a dedicated $p$-type low-background ``well" HPGe detector at the underground 6 m.w.e. LASR laboratory of the University of Chicago. This detector type and geometry is necessary to identify this low-energy emission with a significant efficiency. The simulations of the detection efficiency used to obtain the $^{210}$Pb concentration were sanctioned via calibrations using a liquid $^{210}$Pb source of known strength (0.55$\pm$0.12 counts/s expected from simulation, 0.63 count/s observed, calibration not to scale in this figure). Figure courtesy J.I. Collar.}
\label{fig:piezo_Pb}
\end{figure}

These improved piezos however still present a similar content in $^{210}$Pb to the commercial units (Figure \ref{fig:piezo_Pb})  --- a $^{210}$Pb concentration of 161.2 Bq/kg$_\mathrm{PZT}$ was measured for the first generation piezos, while $\sim$260 Bq/kg$_\mathrm{PZT}$ was measured for the second generation piezos. As discussed in Section \ref{sec:backgrounds_components}, $^{210}$Pb is typically found in out-of-equilibrium enhanced concentrations in any lead-containing materials. This is due to a chemical affinity of the naturally-occurring ($^{238}$U chain) $^{210}$Pb in any materials (e.g., coke, limestone, \emph{etc.}) coming in contact with the lead ore during its sintering \citep{peden-98}, leading to its absorption and preferential concentration in the molten lead. In the case of PZT, this contamination is introduced via the PbO (``yellow" lead oxide) or Pb$_3$O$_4$ (``red" lead oxide) salts employed in its manufacture (either can be used). These second generation radioclean piezos are, as a matter of fact, already dominated by the ($\alpha$,n) contribution from their $^{210}$Pb content --- 0.46 counts/year are expected from the $^{210}$Pb while only $1.1\times10^{-2}$ counts/year are expected from the rest of the $^{238}$U and $^{232}$Th chains (see Table \ref{tab:piezo_study}). These predictions are based on 4 piezos (2.5 cm$^3$ each) mounted on the neck of the bell jar, in the same location as the first generation piezos.

\begin{figure} [t!]
\centering
\includegraphics[scale=0.45]{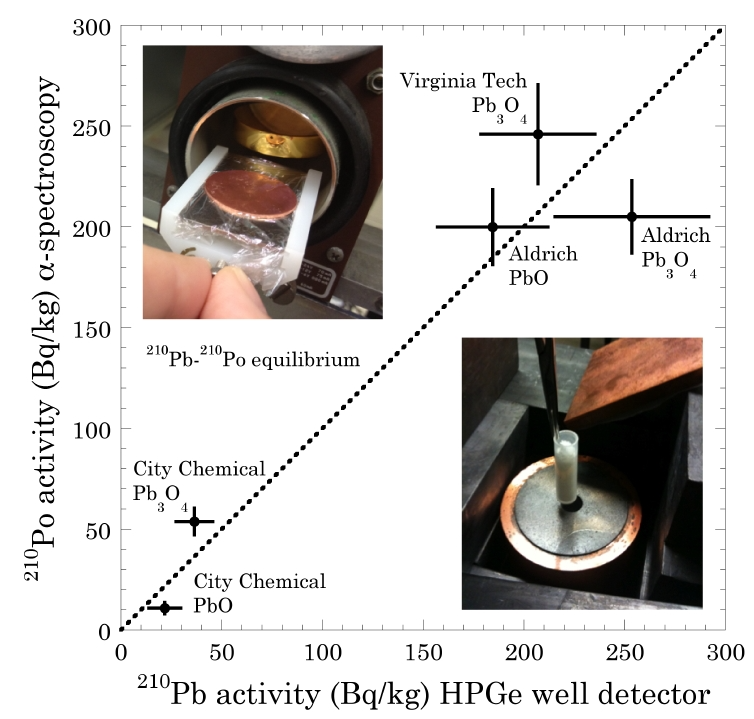}
\caption[Comparison of $^{210}$Pb in different lead oxide samples]{A comparison of the measured content of $^{210}$Pb in different lead oxide samples, using an HPGe ``well" detector (46 keV gamma emmission) and a radiochemical technique leading to the deposition of its $^{210}$Po daughter on copper planchettes, followed by low-background alpha counting using a silicon detector. The second technique was found to be much more reliable and faster than the first (hours vs. weeks), especially for the low levels of $^{210}$Pb to be used in third generation PZT (Figure \ref{fig:Pb_counts}).  Figure courtesy of J.I. Collar.}
\label{fig:Pb_alpha}
\end{figure}

\begin{figure} [t!]
\centering
\includegraphics[scale=0.072]{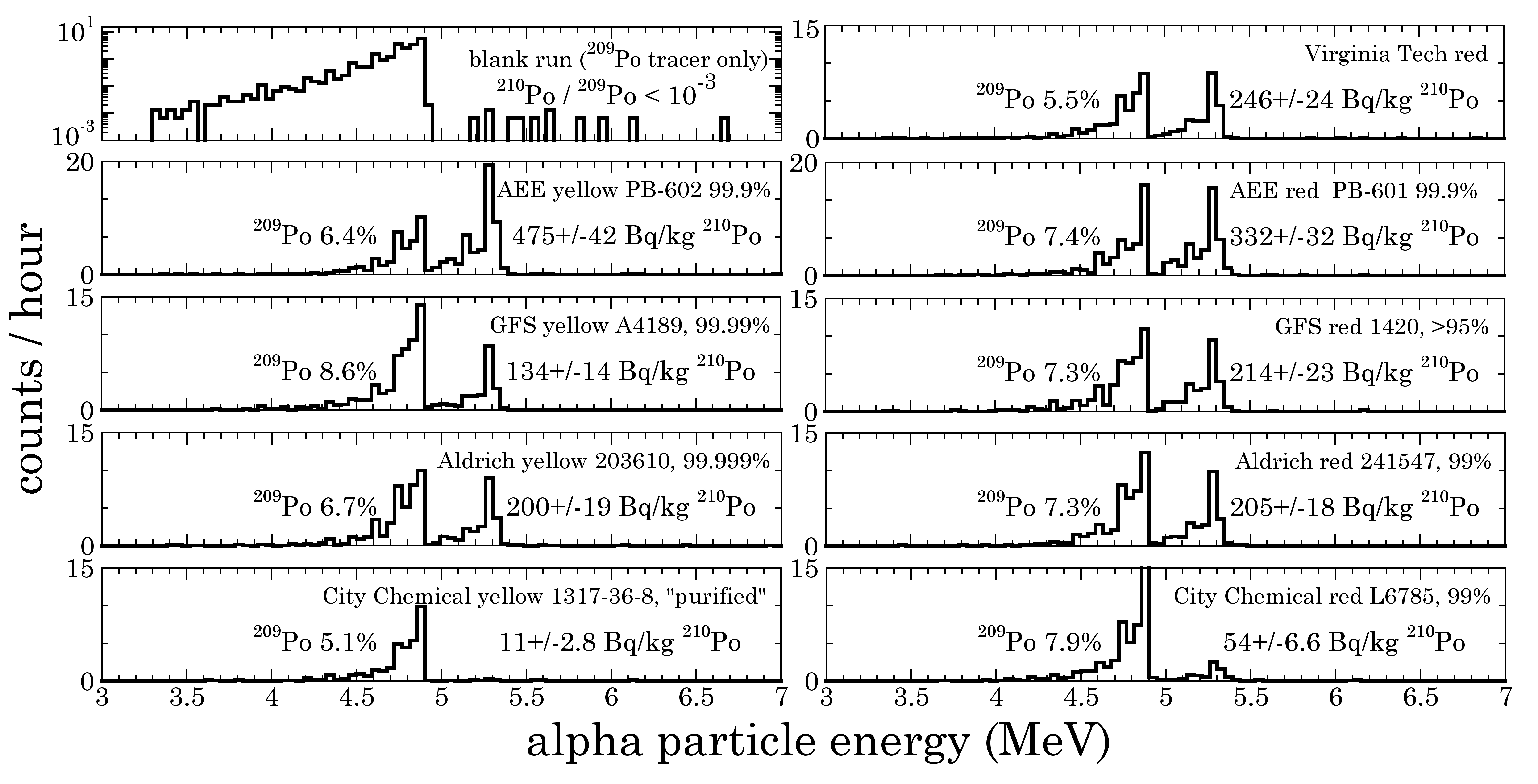}
\caption[Alpha spectroscopy of different lead oxide samples]{Alpha spectroscopy of different lead oxide samples. Chemical purity indicated by the manufacturer is shown next to the sample name: as expected, no correlation is found between this chemical purity and $^{210}$Pb content, in view of its origin (see text). The peak on the left is from a $^{209}$Po tracer used to determine efficiency of radiochemical extraction of pollonium from the salt, and alpha detection efficiency. The total efficiency is shown as a percent of the $^{209}$Po aliquot activity added. The lead oxide used in second generation radioclean PZT (top right) is a factor $\sim$25 higher in $^{210}$Po content than the newly identified yellow oxide (bottom left) to be used in 60 kg chamber third generation radioclean piezos, and beyond. The degraded energy resolution of the alpha peaks is intentional, by using a sub-optimal vacuum in the alpha counter as a safeguard against recoil-induced contamination of the low-background silicon detector (not of concern for $^{210}$Po or $^{209}$Po, but a standard precaution for this detector unit). Details of these techniques and the development of radiopure PZT will be presented in an upcoming COUPP publication. Figure courtesy of J.I. collar}
\label{fig:Pb_counts}
\end{figure}

Unfortunately, even these new piezos would introduce an inadmissible source of WIMP-like recoils in the larger COUPP 60 kg chamber to be installed at SNOLAB towards the end of 2012 --- 2.9 counts/year are expected from 4 piezos mounted adjacent to a 75 kg CF$_3$I target volume (Table \ref{tab:piezo_study}, middle piezo location). The effort to develop ``third generation" radioclean PZT is therefore ongoing. At the time of this writing, the techniques to screen lead salts in $^{210}$Pb content have been successfully developed (Figures \ref{fig:Pb_alpha} and \ref{fig:Pb_counts}),  leading to the identification of lead oxides a factor of $\sim$25 lower in $^{210}$Pb than those presently used in either the first or second generation piezos. Additional screening at SNOLAB has indicated that a further reduction in overall $^{238}$U and $^{232}$Th should be possible from these newly identified oxides, and that room for improvement  also exists with a new source of titanium oxide. We are therefore confident to be able to produce, before the summer of 2012, a new and final ``third generation" of custom PZT transducers that should allow operation of the 60 kg chamber and a planned 500 kg chamber without any concerns of an ($\alpha$,n) neutron background from their piezo elements ($\sim$1 event/year or less).

\singlespacing
\chapter{Expected Backgrounds for Larger Bubble Chambers}
\label{ch:otherchambers}
\doublespacing

The future of COUPP rests not only in improvements to the 4 kg bubble chamber, but also in scaling to larger detector volumes. The COUPP 60 kg bubble chamber recently completed a successful engineering run at the MINOS near detector hall at Fermilab. This chamber is currently being prepared for deployment to SNOLAB, where data-taking should commence near the end of 2012. A 500 kg bubble chamber is also under development. Since this thesis reports on the backgrounds found in the 4 kg chamber, a projection of these backgrounds to larger chambers is called for. Simulations for the 60 kg chamber have been carried out similar to those producing the 4 kg chamber neutron background predictions from spontaneous fission and ($\alpha$,n) sources arising from $^{238}$U and $^{232}$Th contamination of detector components (Table \ref{tab:UTh_events}). Although the MCNP-PoliMi geometry used for the 60 kg simulation is not as complete as that for the 4 kg, the predictions should still be good to within an order of magnitude or better.

The neutron background initiated by the piezos is covered in the next section. The remaining contributions to the neutron background of the 60 kg that have been calculated are dominated by the spontaneous fission of $^{238}$U in the stainless steel pressure vessel and the ($\alpha$,n) neutrons generated by radon dissolved in the water shielding. The expected increase to the neutron rate from the fact that the pressure vessel is larger for the 60 kg chamber is diminished by the fact that there is more glycol moderator between the steel and the CF$_3$I. Assuming 1 part per billion (ppb) of both $^{238}$U and $^{232}$Th in the stainless steel and the same bubble nucleation efficiency ($\eta = 0.46$, Section \ref{sec:efficiency_fits}) and data cut efficiency (79.86\%, Section \ref{sec:datasets_summary}) that were used in the 4 kg chamber, the rate of single recoil-like events generated by the stainless steel at a 15 keV recoil energy threshold is $(2.4\pm0.25)\times10^{-5}$ cts/kg/day, with a rate of multi-bubble events of $(2.3\pm0.15)\times10^{-5}$ cts/kg/day. 

For the 60 kg chamber, the entire pressure vessel is submerged in a water bath that serves to not only shield the detector from environmental neutrons, but also to regulate the temperature. The fact that the water shielding is closer to the CF$_3$I target in this chamber than it was in the 4 kg chamber, and that it has a larger volume, increases the rate from ($\alpha$,n) neutrons initiated by alpha decays of atmospheric $^{222}$Rn dissolved in the water. As calculated in Section \ref{sec:backgrounds_radon_in_water}, the number density of $^{222}$Rn atoms in the water is $15.9\pm0.8$ atoms/cm$^3$ when the water is in equilibrium with the atmosphere, which has a $^{222}$Rn density of $131.0\pm6.7$ Bq/m$^3$ \citep{snolab-handbook}. Given the amount of $^{222}$Rn in the water, the rate of single recoil-like events at a 15 keV recoil energy threshold generated by the ($\alpha$,n) neutrons from radon dissolved in the water is $(2.2\pm0.24)\times10^{-6}$ cts/kg/day, with a rate of multi-bubble events being $(2.1\pm0.14)\times10^{-6}$ cts/kg/day.

Not including the piezos, the total background in the 60 kg chamber from detector components is then $(2.6\pm0.25)\times10^{-5}$ cts/kg/day for single-bubble events and $(2.5\pm0.15)\times10^{-5}$ cts/kg/day for multi-bubble events, plus some second order corrections from sources like the natural quartz flange. Notable here is that the rate of multi-bubble events is approximately the same as the rate of single-bubble events. This is expected from the increased size of the detector volume. Also, this single-bubble rate is an order of magnitude lower than the comparable rate in the 4 kg chamber (the light blue curve in Figure \ref{fig:summary}), however with the target mass being $\sim$75 kg\footnote{The 60 kg target mass of CF$_3$I is a nominal value --- up to 80 kg can be filled in this chamber, and simulations were performed assuming 75 kg.} (versus the 4.048 kg mass for the 4 kg chamber), the total expected number of events for the 60 kg chamber is higher than for the 4 kg chamber. Still, only $\sim$0.7 single-bubble events and $\sim$0.7 multi-bubble events are expected in the 60 kg chamber is one year of operation from the combination of the steel and radon. The rate contributed by other sources (\emph{e.g.} the neutrons generated in the rock walls) will need to be considered in more detail. Figure \ref{fig:water_shield_thickness} can provide an indicative approximation of the rate from the rock walls.

\section{Piezos as Neutron Sources}
\label{sec:otherchambers_piezos}

When it comes to recording the acoustic emission from bubble formation (Section \ref{sec:bubblechambers_acoustic}), the COUPP experiment has employed PZT piezoelectric transducers affixed to the bell jar. These piezos provide information that is excellent to discriminate alpha events from those generated by nuclear recoils (Section \ref{sec:datasets_alpha}), however they also act as weak neutron sources (Section \ref{sec:backgrounds_piezos}). Efforts have been made to reduce the levels of $^{238}$U (along with $^{210}$Pb) and $^{232}$Th in materials used for manufacturing these piezos (Appendix \ref{ch:piezos}), but some residual neutron background will still remain. To determine the level of purity that these piezos must achieve to not hinder the experiment as neutron background sources, the expected event rate generated by the piezos for each ppb of $^{238}$U and $^{232}$Th and each Bq/kg of $^{210}$Pb is required.

For these predictions, the neutron yield from spontaneous fission and ($\alpha$,n) reactions in the piezo was determined by the SOURCES-4C code \citep{wilson-05} for both 1 ppb of $^{238}$U (and corresponding concentrations of each alpha emitter in the decay chain in equilibrium, which includes $^{210}$Pb) and 1 ppb $^{232}$Th (also in equilibrium). In addition, another SOURCES-4C calculation with 1 Bq/kg of $^{210}$Pb (with $^{210}$Bi and $^{210}$Po in equilibrium) was also performed. The neutron yield from $^{238}$U is ($\left.3.2\times10^{-11}\right|_{(\alpha,n)}$ + $\left.1.1\times10^{-10}\right|_\mathrm{s.f.}$ = ) $1.4\times10^{-10}$ n/s/cm$^3_\mathrm{PZT}$/ppb$_\mathrm{U}$ and from $^{232}$Th is $1.4\times10^{-11}$ n/s/cm$^3_\mathrm{PZT}$/ppb$_\mathrm{Th}$ (dominated by the ($\alpha$,n) contribution). The neutron yield from $^{210}$Pb is $2.4\times10^{-10}$ n/s/cm$^3_\mathrm{PZT}$/(Bq$_\mathrm{Pb}$/kg$_\mathrm{PZT}$), arising only from the ($\alpha$,n) contribution.

With MCNP-PoliMi simulations \citep{padovani-02}, the event rate was determined from the piezos in both the 4 kg and 60 kg bubble chambers at a recoil threshold of 15 keV. For the 60 kg chamber, separate simulations were executed for piezos attached to the very bottom of the bell jar hemisphere, on the wall of the bell jar at a vertical position at the center of the CF$_3$I, and on the wall of the bell jar adjacent to the water buffer (a height 14 cm above the CF$_3$I/water interface). The rates in a 500 kg bubble chamber should be similar to those in the 60 kg chamber, since both will have large detector volumes that fill $\sim$50\% of the solid angle of neutrons generated from the piezos. The results of these simulations are shown in Table \ref{tab:piezo_study}. \begin{table}[t!]
\centering
\small{
\begin{tabular} {| l | l | c | c | c | c |}
\hline
Bubble & Piezo & \multicolumn{4}{| c |}{Event Rate at 15 keV recoil threshold} \\
Chamber & Location & \multicolumn{4}{| c |}{(cts/kg$_\mathrm{CF_3I}$/day)/cm$^3_\mathrm{PZT}$} \\
\cline{3-6}
 & & per ppb$\left._\mathrm{U}\right|_\mathrm{s.f.}$ & per ppb$\left._\mathrm{U}\right|_\mathrm{(\alpha,n)}$ & per ppb$\left._\mathrm{Th}\right|_\mathrm{(\alpha,n)}$ & per Bq$_\mathrm{Pb}$/kg$\left._\mathrm{PZT}\right|_\mathrm{(\alpha,n)}$ \\
\hline
\hline
4 kg & nominal & $5.0\times10^{-8}$ & $1.6\times10^{-8}$ & $6.7\times10^{-9}$ & $1.2\times10^{-7}$ \\
\hline
60 kg & top & $6.3\times10^{-11}$ & $9.9\times10^{-12}$ & $5.0\times10^{-12}$ & $7.4\times10^{-11}$ \\
 & middle & $9.0\times10^{-9}$ & $5.4\times10^{-9}$ & $2.4\times10^{-9}$ & $4.1\times10^{-8}$ \\
 & bottom & $8.3\times10^{-9}$ & $5.1\times10^{-9}$ & $2.3\times10^{-9}$ & $3.9\times10^{-8}$ \\
\hline
\end{tabular} }
\caption[Event rates from piezos per ppb $^{238}$U and $^{232}$Th and per Bq/kg $^{210}$Pb]{The event rates predicted from $^{238}$U, $^{232}$Th, and $^{210}$Pb in the PZT piezoelectric transducers in the 4 kg chamber and at 3 different vertical positions in the 60 kg chamber (see text). These rates are in units of counts/kg/day per cm$^3$ of PZT per unit of concentration, so that they are scalable to different piezos.}
\label{tab:piezo_study}
\end{table} The predictions here are based on a 100\% bubble nucleation efficiency and 100\% data cut efficiency, for application to other chambers where these variables might be different. Care should be taken when making calculations to not double count the contributions from $^{210}$Po, due to its presence as part of the $^{238}$U decay chain, and as an out-of-equilibrium contamination (see Section \ref{sec:backgrounds_components}). The rate in units per ppb$_\mathrm{U}$ includes an equilibrium amount of $^{210}$Pb, $^{210}$Bi, and $^{210}$Po while the rate in units per Bq$_\mathrm{Pb}$/kg$_\mathrm{PZT}$ is calculated based on concentration of $^{210}$Pb with equilibrium amounts of $^{210}$Bi and $^{210}$Po (relative to the $^{210}$Pb), but without any concentration of isotopes above $^{210}$Pb on the $^{238}$U decay chain. An approximately accurate calculation when $^{210}$Pb is out of equilibrium with $^{238}$U is to take the rate from the $^{238}$U ($\alpha$,n) channel times 7/8 (discounting for $^{210}$Po, the only significant sub-$^{210}$Pb contributor to the alpha rate out of 8 total contributors, see Figure \ref{fig:U238_decay_chain}), and then add the rate from the $^{210}$Pb contamination.

\singlespacing
\chapter{Example Code Inputs}
\label{ch:input}
\doublespacing

\section{MCNP 4 kg Example Input}
\label{input_mcnp}

\small{
\begin{lstlisting}
Drew Fustin, 4kg COUPP chamber -- rewritten Nov 1, 2011
c ::: cell cards
c * CF3I (comment out if NaI)
 1   2  -1.99  (-1 2 -3):(-2 -4)
c * water (comment out if NaI)
 2   3  -1.00  (-1 3 -5):(5 -6 -7):(7 8 -9 -10):(10 -19 -11): &
               (19 -26 -22):(26 -27 -24):(27 -29 -22): &
               (29 -43 -32):(29 -43 -34)
c * quartz (comment out if NaI)
 3   4  -2.203 (1 -12 2 -5):(-2 4 -13):(5 -7 6 -14):(7 -10 -8 15 -9): &
               (10 -17 11 -16):(17 -19 11 -18)
c * steel inner volume (comment out if NaI)
 4   5  -8.00  (10 -17 16 -20):(17 -19 18 -20):(19 -21 -20 22): &
               (21 -28 22 -23):(28 -29 -20 22):(-20 29 -30 32 34): &
               (30 -41 32 -31):(30 -41 34 -33):(41 -42 -40 32 34): &
               (42 -43 -39 32 34)
c * replace above with this if NaI
c 4   5  -8.00  (-39 -43 42):(-40 -42 41)
c * steel bellows convolutions (comment out if NaI)
 5   6  -4.50  (26 -27 -22 24)
 6   7  -4.518 (26 -27 23 -25)
c * steel large flange through extender
 7   5  -8.00  (44 -45 -41 48):(44 -39 -48 47):(44 -40 -47 46): &
               (44 -40 -46 49):(44 -39 -49 50):(44 -45 -50 51): & 
               (-44 52 -53 55):(52 -54 -55 56)
c * steel pressure vessel body and bottom plug
 8   5  -8.00  (-58 60 -56 61):(-57 60 -61 62):(-59 60 -62 63): &
               (-60 64 -65 66 81 105): &
               (-64 67 -69 70):(-64 68 -70 71): &
               (-68 67 -70 72):(-73 68 -75 76):(-74 68 -76 77): &
               (-74 -77 78):(-73 -78 79)
c * steel windows
 9   5  -8.00  (64 -81 80 -88 95):(-85 81 87 -89):(-84 81 89 -90): &
               (-83 86 90 -91):(-85 83 92 -91):(-85 81 91 -93): &
               (-82 81 93 -94)
 10  5  -8.00  (64 -81 80 97 -95):(-85 81 -96 98):(-84 81 -98 99): &
               (-83 86 -99 100):(-85 83 -101 100):(-85 81 -100 102): &
               (-82 81 -102 103)
 11  5  -8.00  (64 -105 104 -112 119):(-109 105 111 -113): & 
               (-108 105 113 -114):(-107 110 114 -115): &
               (-109 107 116 -115):(-109 105 115 -117): &
               (-106 105 117 -118)
 12  5  -8.00  (64 -105 104 121 -119):(-109 105 -120 122): &
               (-108 105 -122 123):(-107 110 -123 124): &
               (-109 107 -125 124):(-109 105 -124 126): &
               (-106 105 -126 127)
c * Piezo (comment out if NaI)
 13  10  -8.50  (14 -15 -128 131 132 -133):(14 -15 -128 129 -132): &
                (14 -15 -128 -130 133)
 14  11  -7.80  (14 -15 -134 132 -135)
 15  12  -1.60  (14 -15 -131 134 132 -135):(14 -15 -131 135 -133)
c ***** uncomment through stars below if NaI
c * acrylic bottle
c 80  19 -1.18  (-41 303 300 -301):(-301 -303 302)
c * NaI crystal
c 81  21 -3.67   (-401 -402 403)
c * Teflon reflector
c 82  27 -2.20   (401 -406 -402 403):(-406 -403 407)
c * Quartz window
c 83   4 -2.203  (-404 402 -405)
c * Steel housing
c 84   5 -8.00   (404 -409 402 -405):(408 -409 -402 411): &
c                (404 -409 410 -411):(-404 412 -411)
c * Glass PMT envelope
c 86   9 -2.23   (-413 405 -415):(-413 414 415 -416): &
c                (419 -420 416 -418):(-422 421 418 -417 -414): &
c                (423 -424 417 -426):(-424 426 -425)
c * PMT innards
c 87  26  -1.83   (-414 415 -416):(-419 416 -418): &
c                 (422 418 -417 -414):(-423 417 -426)
c * Teflon padding
c 88  27  -2.20   (409 -427 410 -405):(-427 -410 428)
c * Steel light can
c 89  28  -7.85  (427 -429 428 -431):(-429 -428 430):(-429 431 -432)
c * Air inside NaI detector
c 85   1 -0.0012 (-300 -41 303) #80 #81 #82 #83 #84 #86 #87 #88 #89
c *****
c * glycol (switch #'s if NaI)
 16  8  -1.036 ((-44 -41 56):(-60 -56 69):(-68 -69 77): &
               (64 -81 99 -95):(64 -81 -90 95): &
               (64 -105 123 -119):(64 -105 -114 119)) & 
               #1 #2 #3 #4 #5 #6 #7 #8 #9 #10 #11 #12 #13 #14 #15
c              #7 #8 #9 #10 #11 #12 #80 #81 #82 #83 #84 #85 #86 &
c              #87 #88 #89
c * glass windows
 17  9  -2.23  (-86 90 -91)
 18  9  -2.23  (-86 -99 100)
 19  9  -2.23  (-110 114 -115)
 20  9  -2.23  (-110 -123 124)
c * polyethylene foam insulation
 21  13 -0.032 (-136 60 -63 66 81 105)
c * plumbing cavity
 22  14 -2.40  (-12 43 -137)
c * Reflectix insulation
 23  15 -0.04  (-138 57 62 141 -140):(-139 57 62 143 -142): &
               (-138 59 63 -62 141 -140):(-139 59 63 -62 143 -142): &
               (-138 -63 96 -87 81 136 109): &
               (-139 -63 120 -111 105 136 85): &
               (-85 -109 81 105 136):(-138 -63 85 87 -140): &
               (-85 82 93 -140):(-82 81 94 -140):(-81 91 -140): &
               (-85 84 89 -92):(-84 83 90 -92): &
               (-138 -63 85 -96 141):(-85 82 -102 141): &
               (-82 81 -103 141):(-81 -100 141):(-85 84 101 -98): &
               (-84 83 101 -99):(-139 -63 109 111 -142): &
               (-109 106 117 -142):(105 -106 118 -142): &
               (-105 115 -142):(-109 108 113 -116): &
               (-108 107 114 -116): (-139 -63 109 -120 143): &
               (-109 106 -126 143):(105 -106 -127 143): &
               (-105 -124 143):(-109 108 125 -122): & 
               (-108 107 125 -123): &
               (-152 138 139 136 -63 147):(-136 64 -66 71): &
               (-136 68 -71 75):(-136 73 -75 76):(-136 74 -76 78): &
               (-136 73 -78 79):(-136 -79 147): &
               (152 136 -144 -79 147): & 
               (-152 138 139 59 63 -62):(-152 138 139 57 62 -61): & 
               (-152 58 61 -56):(-152 146 56 -55):(-146 54 56 -55): &
               (-146 44 55 -51):(-146 45 51 -148):(-145 45 148 -50): &
               (-145 39 50 -49):(-145 40 49 -47):(-145 39 47 -48): &
               (-145 45 48 -41):(-145 40 41 -42):(-145 39 42 -43): &
               (-145 12 43 -149):(-9 12 149 -137):(-9 137 -150): &
               (152 -146 -56 151):(152 -144 79 -151)
c * polyethylene base above plate
 24  13 -0.96  (167 -174 181 -188 157 -147)
c * steel plate
 25  5  -8.00  (167 -174 181 -188 156 -157)
c * polyethylene base below plate
 26  13 -0.96  (167 -174 181 -188 154 -156): &
                         (164 -177 179 -191 153 -154)
c * polyethylene walls
 27  13 -0.96  (174 -175 178 -188 154 -160): &
               (175 -176 187 -188 154 -160): &
               (175 -176 178 -179 154 -160): &
               (176 -177 178 -188 154 -160): &
               (175 -176 179 -187 154 -155): &
               (175 -176 179 -187 159 -160): &
               (172 -174 181 -184 147 -158)
 28  13 -0.96  (167 -168 189 -190 154 -160): &
               (167 -177 190 -191 154 -160): &
               (167 -177 188 -189 154 -160): &
               (176 -177 189 -190 154 -160): &
               (167 -177 188 -191 154 -155): &
               (167 -177 188 -191 159 -160): &
               (171 -174 186 -188 147 -158)
 29  13 -0.96  (164 -165 181 -191 154 -160): &
               (165 -166 181 -182 154 -160): &
               (165 -166 190 -191 154 -160): &
               (166 -167 181 -191 154 -160): &
               (164 -167 181 -191 154 -155): &
               (164 -167 181 -191 159 -160): &
               (167 -169 185 -188 147 -158)
 30  13 -0.96  (164 -165 179 -180 154 -160): &
               (164 -174 178 -179 154 -160): &
               (164 -174 180 -181 154 -160): &
               (173 -174 179 -180 154 -160): &
               (164 -174 178 -181 154 -155): &
               (164 -174 178 -181 159 -160): &
               (167 -170 181 -183 147 -158)
c * polyethylene roof
 31  13 -0.96  (164 -177 179 -191 160 -161 194): &
               (164 -177 179 -191 162 -163 194): &
               (164 -177 178 -179 161 -162): &
               (164 -165 179 -190 161 -162): &
               (164 -177 190 -191 161 -162): &
               (176 -177 179 -190 161 -162): &
               (194 -195 161 -162): &
               (167 -170 181 -188 158 -160): &
               (170 -171 181 -184 158 -160): &
               (170 -171 185 -188 158 -160): &
               (171 -174 181 -188 158 -160)
c * polyethylene source pole
 32  13 -0.96  (-196 163 -193):(-194 192 -163)
c * water in walls
 33  3  -1.00  (175 -176 179 -187 155 -159)
 34  3  -1.00  (165 -173 179 -180 155 -159)
 35  3  -1.00  (165 -166 182 -190 155 -159)
 36  3  -1.00  (168 -176 189 -190 155 -159)
 37  3  -1.00  (165 -176 179 -190 161 -162 195)
c * AmBe source container
 38  17  -1.41  (-203 -204 205)
c * Cf source container
 39  16  -1.18  (-200 -201 202)
 40  17  -1.41  (-197 -198 201):(-197 200 -201 202):(-197 199 -202)
c * air
 98  1  -0.00138 ((-999 -138 140):(-999 -138 -141):(-999 -139 142): &
                 (-999 -139 -143):(-999 152 144 138 139 -151): &
                 (-999 -147 -144):(-999 152 146 138 139 151): &
                 (-999 148 -146 145):(-999 149 -145 9):(-999 150 -9)) &
                 #24 #25 #26 #27 #28 #29 #30 #31 #32 #33 #34 #35 #36 &
                 #37 #38 #39 #40
c * outside world
 99  0  999

c ::: surface cards
c * geometry center = axial center of jar, middle of windows height
c * CF3I - 4.048 kg
 1    cz  7.25
 2    pz  -1.14
 3    pz  6.3515
 4    sz  -1.14 7.25
c * water
 5    pz  6.48
 6    sz  6.48 7.25
 7    pz  10.68
 8    tz  0 0 14.93 13.05 8.30 8.30
 9    cz  12.50
 10   pz  14.93
 11   cz  4.50
c * quartz
 12   cz  7.50
 13   sz  -1.14 7.50
 14   sz  6.48 7.50
 15   tz  0 0 14.93 13.05 8.05 8.05
 16   cz  5.00
 17   pz  16.37
 18   cz  6.20
 19   pz  19.37
c * steel jar flange
 20   cz  7.49
c * steel bellows
 21   pz  20.64
 22   cz  5.09
 23   cz  5.19
 24   cz  3.89
 25   cz  6.29
 26   pz  21.91
 27   pz  30.46
 28   pz  31.73
 29   pz  33.00
c * steel lid to bellows
 30   pz  35.07
c * steel water/glycol pipes through top flange
 31   c/z 4.064 0.00 0.635
 32   c/z 4.064 0.00 0.511
 33   c/z -4.064 0.00 0.635
 34   c/z -4.064 0.00 0.511
 35   c/z 0.00 4.064 0.635
 36   c/z 0.00 4.064 0.511
 37   c/z 0.00 -4.064 0.635
 38   c/z 0.00 -4.064 0.511
c * steel top flange
 39   cz  15.875
 40   cz  10.795
 41   pz  48.567 
 42   pz  48.720
 43   pz  52.225 
c * steel extender below top flange
 44   cz  8.57
 45   cz  10.325
 46   pz  42.262 
 47   pz  42.415 
 48   pz  45.920 
c * steel flange below extender
 49   pz  41.762 
 50   pz  38.256 
 51   pz  36.682 
c * steel pipe down to big flange
 52   cz  7.703
 53   pz  41.019
c * steel big flange
 54   cz  25.88
 55   pz  32.065
 56   pz  24.765
c * steel flange below big flange
 57   cz  26.035
 58   cz  19.05
 59   cz  18.733
 60   cz  16.358
 61   pz  24.285
 62   pz  19.357
 63   pz  17.122
c * steel pressure vessel body
 64   cz  15.161
 65   pz  23.472
 66   pz  -20.955
c * steel bottom plug
 67   cz  3.899
 68   cz  4.445
 69   pz  -19.05
 70   pz  -22.065
 71   pz  -22.86
 72   pz  -28.733
 73   cz  10.478
 74   cz  6.35
 75   pz  -26.848
 76   pz  -29.566
 77   pz  -29.718
 78   pz  -29.871
 79   pz  -32.589
c * steel viewports
 80   cx  7.703
 81   cx  8.560 
 82   cx  10.264 
 83   cx  10.643
 84   cx  10.795
 85   cx  15.875
 86   cx  7.62
 87   px  18.615 
 88   px  21.563 
 89   px  22.121 
 90   px  22.273 
 91   px  24.559 
 92   px  24.521 
 93   px  28.311 
 94   px  29.819 
 95   px  0
 96   px  -18.615 
 97   px  -21.563 
 98   px  -22.121 
 99   px  -22.273 
 100  px  -24.559 
 101  px  -24.521 
 102  px  -28.311 
 103  px  -29.819 
 104  cy  7.703
 105  cy  8.560 
 106  cy  10.264 
 107  cy  10.643
 108  cy  10.795
 109  cy  15.875
 110  cy  7.62
 111  py  18.615 
 112  py  21.563 
 113  py  22.121 
 114  py  22.273 
 115  py  24.559 
 116  py  24.521 
 117  py  28.311 
 118  py  29.819 
 119  py  0
 120  py  -18.615 
 121  py  -21.563 
 122  py  -22.121 
 123  py  -22.273 
 124  py  -24.559 
 125  py  -24.521 
 126  py  -28.311 
 127  py  -29.819 
c * Piezo 
 128  1 cz  1.22
 129  1 pz  0.00
 130  1 pz  2.00
 131  1 cz  1.21
 132  1 pz  0.01
 133  1 pz  1.99
 134  1 cz  0.79
 135  1 pz  0.88
c * polyethylene foam insulation
 136  cz  17.628
c * piping in cavity above top flange
 137  pz  70.00
c * Reflectix insulation
 138  cx  20.875
 139  cy  20.875
 140  px  36.83
 141  px  -36.83
 142  py  34.29
 143  py  -43.18
 144  cz  20.875
 145  cz  22.628
 146  cz  27.8 
 147  pz  -35.565 
 148  pz  37.065
 149  pz  57.225
 150  pz  75.00
 151  pz  0
 152  so  35.92 
c * water shield
 153  pz -86.683
 154  pz -71.443
 155  pz -70.173
 156  pz -51.123
 157  pz -50.805
 158  pz 80.958
 159  pz 84.768
 160  pz 86.038
 161  pz 87.308
 162  pz 135.568
 163  pz 136.838
 164  px -96.52
 165  px -95.25
 166  px -46.99
 167  px -45.72
 168  px -44.45
 169  px -40.64
 170  px -30.48
 171  px 30.48
 172  px 40.64
 173  px 44.45
 174  px 45.72
 175  px 46.99
 176  px 95.25
 177  px 96.52
 178  py -96.52
 179  py -95.25
 180  py -46.99
 181  py -45.72
 182  py -44.45
 183  py -40.64
 184  py -30.48
 185  py 30.48
 186  py 40.64
 187  py 44.45
 188  py 45.72
 189  py 46.99
 190  py 95.25
 191  py 96.52
 192  pz 65.718
 193  pz 157.158
 194  c/z -25.4 -25.4 2.451
 195  c/z -25.4 -25.4 3.721
 196  c/z -25.4 -25.4 5.080
c * Cf source casing
 197  c/z -27.85 -25.4 1.46
 198  2 pz  4.758
 199  2 pz  -0.322
 200  c/z -27.85 -25.4 1.27
 201  2 pz  2.993
 202  2 pz  -0.187
c * AmBe source casing
 203  c/z -22.95 -25.4 1.46
 204  2 pz  51.118
 205  2 pz  45.398
c * Acrylic bottle 
 300  cz  5.24
 301  cz  5.72
 302  pz  -7.405
 303  pz  -6.135
c * NaI crystal
 401  cz  4.25
 402  3 pz  4.25
 403  3 pz  -4.25
c * Quartz window
 404  cz  3.81
 405  3 pz  4.85
c * Teflon reflector
 406  cz  4.35
 407  3 pz  -4.35
c * Steel housing
 408  cz  4.50
 409  cz  4.65
 410  3 pz  -4.65
 411  3 pz  -4.53
 412  3 pz  -4.56
c * Glass PMT envelope
 413  cz  3.90
 414  cz  3.675
 415  3 pz  5.075
 416  3 pz  7.545
 417  3 pz  10.015
 418  3 pz  8.78
 419  3 sz  6.25 3.884  $ 6.25
 420  3 sz  6.25 4.109  $ 6.25
 421  3 tz  0 0 11.2044 6.2724 3.884 3.884  
 422  3 tz  0 0 11.2044 6.2831 4.109 4.109  
 423  cz  2.35
 424  cz  2.575
 425  3 pz  16.15
 426  3 pz  15.925
c * Teflon padding
 427  cz  4.775
 428  3 pz  -4.9675
c * Steel light can 
 429  cz  5.08
 430  3 pz  -5.285
 431  3 pz  25.0325
 432  3 pz  25.6675
c * rock neutron source sphere
 777  sz  13.8584 22.50
c * gamma source sphere
 888  sz  13.8584 22.50
c * outside world
 999  so  300.00

c ::: material cards and such
c *** translation cards
 *tr1  5.72 0 11.6  67 90 157  90 0 90  23 90 67  1
 tr2  0 0 0
 tr3  0 0 -0.85
c *** modes and importances (neutrons or gammas?)
 mode  n
 imp:n  1 40r 0
c mode  p
c imp:p  1 40r 0
c imp:e  1 40r 0
c use the 42r if NaI
c imp:p  1 42r 0
c imp:e  1 42r 0
c *** material cards
 m1   7014.60c  0.78  $ air (78% N2 + 21% O2 + 1% Ar)
      8016.60c  0.21
     18000.42c  0.01
 m2   6000.60c  0.20  $ CF3I
      9019.60c  0.60
     53127.60c  0.20
 m3   1001.60c  0.67  $ water
      8016.60c  0.33
 m4   8016.60c  0.67  $ quartz (SiO2)
     14000.60c  0.33
 m5  24000.50c  0.18  $ 304L stainless steel
     25055.50c  0.02  $ (18% Cr + 2% Mn + 68% Fe + 12% Ni)
     26000.50c  0.68
     28000.50c  0.12
 m6   1001.60c  0.33  $ 50%/50% water-steel mix
      8016.60c  0.17
     24000.50c  0.09
     25055.50c  0.01
     26000.50c  0.34
     28000.50c  0.06
 m7   1001.60c  0.31  $ 50%/50% glycol-steel mix
      6000.60c  0.11
      8016.60c  0.08
     24000.50c  0.09
     25055.50c  0.01
     26000.50c  0.34
     28000.50c  0.06
 m8   1001.60c  0.62  $ propylene glycol (C3H8O2)
      6000.60c  0.23
      8016.60c  0.15
 m9   5010.60c  0.013 $ borosilicate glass
      5011.60c  0.045 $ (81% SiO2 + 13% B2O3 + 4% Na2O + 2% Al2O3)
      8016.60c  0.657
     11023.60c  0.013
     13027.60c  0.002
     14000.60c  0.270
 m10 29000.50c  0.850 $ Brass
     30000.40c  0.150 
 m11  8016.60c  0.600 $ PZT (Pb2 Zr Ti O6)
     22000.60c  0.100
     40000.60c  0.100
     82000.50c  0.200
 m12  1001.60c  0.230 $ MAS epoxy
      6000.60c  0.370
      8016.60c  0.400
 m13  1001.60c  0.667 $ polyethylene
      6000.60c  0.333
 m14  7014.60c  0.546 $ 30%/70% glycol-air mix
      8016.60c  0.147
     18000.42c  0.007
     24000.50c  0.054
     25055.50c  0.006
     26000.50c  0.204
     28000.50c  0.036
 m15  1001.60c  0.013 $ Reflectix insulation
      6000.60c  0.007 $ (97% air, 2% polyethylene, 1% Al)
      7014.60c  0.728 $ (density measured = 0.04 g/cc)
      8016.60c  0.242
     13027.60c  0.010
 m16  1001.60c  0.533 $ PMMA acrylic (C5H8O2)
      6000.60c  0.333
      8016.60c  0.134
 m17  1001.60c  0.500 $ Delrin polyoxymethylene (CH20)
      6000.60c  0.250
      8016.60c  0.250
 m19  1000   0.533 $ Acrylic
      6000   0.333
      8000   0.133
 m21  11000  0.50  $ NaI
      53000  0.50
 m26  7000   0.40  $ PMT innards (some combination of air, Si, and Cu) 
      8000   0.10
     14000   0.40
     29000   0.10
 m27  6000   0.33  $ Teflon
      9000   0.67
 m28  6000   0.10  $ low-carbon steel
     25000   0.40
     26000   0.50
c *** source cards (some examples below)
c *****
c *****
c *****
c * Cf as source (change location if AmBe) (use 1 1 1 1 or 11 0 0 0)
 sdef pos=-27.85 -25.4 1.403000 
c *****
c * Window as source (only one included) -- U238 (a,n) (use 0 0 0 0)
c sdef cel=17 axs=1 0 0 pos=23.471 0 0 rad=d1 ext=d2 erg=d3
c si1  0 7.62
c si2  -1.143 1.143
c si3  H  0.001 0.5 1.0 1.5 2.0 2.5 3.0 3.5 4.0 4.5 5.0 5.5 6.0 6.5 7.0
c sp3  D  0 4.439e-2 7.182e-2 7.207e-2 9.211e-2 1.487e-1 1.778e-1 &
c         1.613e-1 1.180e-1 6.853e-2 3.092e-2 1.124e-2 2.632e-3 &
c         3.217e-4 5.529e-5
c *****
c * Window as source (only one included) -- U238 s.f. (use 2 1 1 1)
c sdef cel=17 axs=1 0 0 pos=23.471 0 0 rad=d1 ext=d2
c si1  0 7.62
c si2  -1.143 1.143
c *****
c * Window as source (only one included) -- Th232 total (use 0 0 0 0)
c sdef cel=17 axs=1 0 0 pos=23.471 0 0 rad=d1 ext=d2 erg=d3
c si1  0 7.62
c si2  -1.143 1.143
c si3  H  0.001 0.5 1.0 1.5 2.0 2.5 3.0 3.5 4.0 4.5 5.0 5.5 6.0 6.5 &
c         7.0 7.5 8.0 8.5 9.0 9.5 10.0 10.5 11.0 11.5 12.0
c sp3  D  0 4.679e-2 7.792e-2 8.947e-2 9.880e-2 1.237e-1 1.510e-1 &
c         1.482e-1 1.154e-1 7.854e-2 4.472e-2 1.915e-2 5.360e-3 & 
c         5.924e-4 1.314e-4 4.698e-9 2.621e-9 1.451e-9 7.967e-10 & 
c         4.347e-10 2.354e-10 1.270e-10 6.801e-11 3.620e-11 1.925e-11
c *****
c * Co-60 as source -- gamma calibrations
c sdef pos=-27.85 -25.4 -3.162 erg=d1 par=2
c si1  L  .661657 1.27453 .058603 1.173237 1.332501 & 
c         .121782 .244698 .344279 .4111163 .4439650 &
c         .778904 .867378 .964079 1.085869 1.089737 &
c         1.11207 1.21295 1.29914 1.408006
c sp1     1.30977 0.00280 .000602 0.030092 0.030096 &
c         0.08000 0.02126 0.07428 0.006262 0.007907 &
c         0.03628 0.01899 0.04094 0.002861 0.004841 &
c         0.03824 0.00399 0.00455 0.058877
c *****
c * example for NaI unfolding (shows spherical surface as source)
c sdef sur=888 par=2 nrm=-1 dir=d1 erg=d2
c sb1  -21 2
c * energy bins
c si2  H  0.1 0.660 1.320 1.660 2.470 2.910 10.000
c sp2  D  0 0 0 0 0 0 1
c *****
c *****
c *****
c *** tally specifications (not really used much with PoliMi)
 f1:n  4
 e1    0 23i 12
 c1    0 1
c use f8 tally with NaI
c f8:p  81
c e8    0 0.660 1.320 1.660 2.470 2.910 10.000
c *** run specifications (specific values required for PoliMi)
 phys:n  j 20
 phys:p  0 1 1
 cut:n   j 0.001
 cut:p   2j 0
c use with NaI, can comment out others above
c phys:p  10 0 0 0 0
c *** number of histories
 nps  1e9
c *** PoliMi specifications (comment out below if not PoliMi)
c * sources defined here and/or in sdef card
c * 1 1 at the end tells PoliMi to give info for 1 cell,
c *   specifically cell number 1 <--- CF3I volume
c * 0 0 0 0 is for standard MCNP run where the source is defined
c *   in the sdef card, complete with use of erg variable.
c * 1 1 1 1 is for Cf-252, and sdef only designates position.
c *   nps gives number of fissions, not the number of neutrons.
c * 11 0 0 0 is for AmBe, and sdef only designates position.
c * 2 1 1 1 is for U-238, and sdef only designates position.
c *   nps gives number of fissions, not the number of neutrons.
 idum 1 1 1 1 2j 1 1
c * random number seed
 DBCN 9568481
c * output
 files  21 dumn1
\end{lstlisting} }

\section{SOURCES-4C Example Input}
\label{input_sources}

\small{
\begin{lstlisting}
U238 decay chain (0.513 ppm) in windows
1 2 1
5 0
 005 0.052
 008 0.643
 011 0.027
 013 0.008
 014 0.270
24 12.0 0.0
11
 0922380 2.895e15
 0922340 1.594e11
 0902300 4.883e10
 0882260 1.037e9
 0862220 6.782e3
 0842180 3.819e0
 0832140 2.451e1
 0842140 3.372e-6
 0822100 1.438e7
 0832100 8.888e3
 0842100 2.454e5
8 4000
 0050100 0.010348
 0050110 0.041652
 0080170 0.000243
 0080180 0.001287
 0110230 0.026667
 0130270 0.008000
 0140290 0.012650
 0140300 0.008348
\end{lstlisting} }

\backmatter

\newpage
\addcontentsline{toc}{chapter}{References}
\begin{singlespace}
\bibliography{biblio}
\bibliographystyle{mla}
\end{singlespace}

\end{document}